\documentclass[11pt,twoside,a4paper]{cernrep}
\usepackage{rep_common}
\usepackage{rotating}
\usepackage{multirow}
\usepackage{braket}
\newboolean{uprightparticles}
\setboolean{uprightparticles}{false} 
\usepackage{upgreek}
\usepackage{xspace}
\usepackage{wrapfig}
\usepackage{tabularx}
\usepackage{xcolor}
\usepackage{hyperref}
\usepackage{enumitem}
\usepackage{arydshln}
\newcommand{\main}{.}
\def\bibfiles{\main/bib/chapter,\main/section1/bib/section,\main/section2/bib/section,\main/section3/bib/section,\main/section4/bib/section,\main/section5/bib/section}
\providecommand{\biblio}{\nocite{article-minimal}\bibliographystyle{report}\clearpage\bibliography{\bibfiles}}  

\let\Re\undefined
\let\Im\undefined
\DeclareMathOperator{\Re}{Re}
\DeclareMathOperator{\Im}{Im}

\definecolor{Blu}{rgb}{0.,0.,1.}

\newcommand{\beq}{\begin{equation}}
\newcommand{\eeq}{\end{equation}}
\newcommand{\be}{\begin{equation}}
\newcommand{\ee}{\end{equation}}

\newcommand{\gsim}{\lower1.0ex\hbox{$\;\stackrel{\textstyle>}{\sim}\;$}}
\newcommand{\lsim}{\lower1.0ex\hbox{$\;\stackrel{\textstyle<}{\sim}\;$}}

\definecolor{oucrimsonred}{rgb}{0.6, 0.0, 0.0}
\definecolor{persianblue}{rgb}{0.11, 0.22, 0.73}
\definecolor{forestgreen}{rgb}{0.13,0.35,0.13}
\definecolor{red}{rgb}{1.,0.,0.}

\hypersetup{colorlinks, citecolor=oucrimsonred, linkcolor=persianblue, urlcolor=oucrimsonred}

\newcommand{\thetaS}{\theta_{a\eta_{s}}}
\newcommand{\thetaUD}{\theta_{a\eta_{ud}}}

\newcommand{\GeV}{{\rm GeV}}

\begin{document}
\def\biblio{}

\title{New Physics Searches at Kaon and Hyperon Factories}
\author{
Editors: Evgueni Goudzovski$^{1}$, Diego Redigolo$^{2,3}$, Kohsaku Tobioka$^{4,5}$, Jure Zupan$^{6}$
\\[3mm]
Authors: Gonzalo Alonso-\'Alvarez$^{7}$, Daniele S. M. Alves$^{8}$, Saurabh Bansal$^{6}$, Martin Bauer$^{9}$, 
Joachim Brod$^{6}$,  Veronika Chobanova$^{10}$, Giancarlo D'Ambrosio$^{11}$, Alakabha Datta$^{12}$, Avital Dery$^{13}$, Francesco Dettori$^{14}$, Bogdan A. Dobrescu$^{15}$, Babette D\"obrich$^{16}$, Daniel Egana-Ugrinovic$^{17}$, Gilly Elor$^{18}$, Miguel Escudero$^{19}$, Marco Fabbrichesi$^{20}$, Bartosz Fornal$^{21}$,  Patrick J. Fox$^{15}$, Emidio Gabrielli$^{20,22,23}$, Li-Sheng Geng$^{24}$, Vladimir V. Gligorov$^{25}$, Martin Gorbahn$^{26}$, Stefania Gori$^{27}$, Benjam{\' i}n Grinstein$^{28}$, Yuval Grossman$^{13}$, Diego Guadagnoli$^{29}$, Samuel Homiller$^{30}$, Matheus Hostert$^{17,31,32}$,  Kevin J. Kelly$^{2,15}$, Teppei Kitahara$^{33}$, Simon Knapen$^{2,34,35}$,  Gordan Krnjaic$^{36,37,38}$, Andrzej Kupsc$^{39,40}$, Sandra Kvedarait\.e$^{6}$,  Gaia Lanfranchi$^{41}$, Danny Marfatia$^{42}$,  Jorge Martin Camalich$^{43,44}$, Diego Mart\'inez Santos$^{10}$, Karim Massri$^{16}$, Patrick Meade$^{45}$, Matthew Moulson$^{41}$, Hajime Nanjo$^{46}$, Matthias Neubert$^{18}$, Maxim Pospelov$^{31,32}$,  Sophie Renner$^{2}$, 
Stefan Schacht$^{47}$,  Marvin Schnubel$^{18}$, Rui-Xiang Shi$^{24,48}$, Brian Shuve$^{49}$, Tommaso Spadaro$^{41}$, Yotam Soreq$^{50}$, Emmanuel Stamou$^{51}$, Olcyr Sumensari$^{52}$, Michele Tammaro$^{53}$, Jorge Terol-Calvo$^{43,44}$,   Andrea Thamm$^{54}$, Yu-Chen Tung$^{55}$, Dayong Wang$^{56}$, Kei Yamamoto$^{57}$, Robert Ziegler$^{58}$\\[3mm] ${}$
}

\institute{
$^{1}$School of Physics and Astronomy, University of Birmingham, Edgbaston, B15 2TT, United Kingdom\\ 
$^{2}$CERN, Theory Division, CH-1211 Geneva 23, Switzerland\\ 
$^{3}$INFN Sezione di Firenze, Via G. Sansone 1, 59100 Sesto F.no, Italy\\ 
$^{4}$Department of Physics, Florida State University, Tallahassee, FL 32306, USA\\ 
$^{5}$High Energy Accelerator Research Organization (KEK), Tsukuba 305--0801, Japan\\ 
$^{6}$Department of Physics, University of Cincinnati, Cincinnati, Ohio 45221, USA\\ 
$^{7}$McGill University Department of Physics \& McGill Space Institute, 3600 Rue University, Montr\'eal, QC, H3 2T8, Canada \\ 
$^{8}$Theoretical Division, Los Alamos National Laboratory, Los Alamos, NM 87545, USA\\ 
$^{9}$Institute for Particle Physics Phenomenology, Department of Physics Durham University, Durham, DH1 3LE, United Kingdom\\ 
$^{10}$IGFAE, Universidade de Santiago de Compostela, 15782 Santiago de Compostela, Spain\\ 
$^{11}$INFN Sezione di Napoli,  Complesso universitario di Monte S. Angelo, ed. 6 via Cintia, 80126, Napoli, Italia\\ 
$^{12}$Department of Physics and Astronomy, 108 Lewis Hall, University of Mississippi, Oxford, MS 38677-1848, USA\\ 
$^{13}$Department of Physics, LEPP, Cornell University, Ithaca, NY 14853, USA\\ 
$^{14}$Universit\`a degli studi di Cagliari and INFN, Cagliari, Italy\\ 
$^{15}$Fermilab, Particle Theory Department, PO Box 500, Batavia, IL 60510, USA\\ 
$^{16}$CERN, Esplanade des Particules 1, 1211 Geneva 23, Switzerland\\ 
$^{17}$Perimeter Institute for Theoretical Physics, Waterloo, ON N2J 2W9, Canada\\ 
$^{18}$PRISMA$+$ Cluster of Excellence \& Mainz Institute for Theoretical Physics Johannes Gutenberg University, 55099 Mainz, Germany\\ 
$^{19}$Physik-Department, Technische Universit{\"{a}}t, M{\"{u}}nchen, James-Franck-Stra{\ss}e, 85748 Garching, Germany\\ 
$^{20}$INFN Sezione di Trieste, Via Valerio 2, 34127 Trieste, Italy\\ 
$^{21}$Department of Chemistry and Physics, Barry University, Miami Shores, FL 33161, USA\\ 
$^{22}$Department of Physics, University of Trieste, Strada Costiera 11 - 34151, Trieste, Italy\\  
$^{23}$ Laboratory of High Energy and Computational Physics, NICPB, R\"avala pst 10,10143 Tallinn, Estonia \\ 
$^{24}$School of Physics, Beihang University, Beijing 102206, China\\ 
$^{25}$LPNHE, Sorbonne Universit{\' e}, Paris Diderot Sorbonne Paris Cit{\' e}, CNRS/IN2P3, Paris, France\\ 
$^{26}$Theoretical Physics Division, Department of Mathematical Sciences, University of Liverpool, Liverpool L69 3BX, United Kingdom\\ 
$^{27}$Santa Cruz Institute for Particle Physics and Department of Physics, University of California, Santa Cruz, 1156 High Street, Santa Cruz, CA 95064, USA\\ 
$^{28}$Department of Physics, University of California, San Diego, 9500 Gilman Drive, La Jolla, CA 92093, USA\\ 
$^{29}$LAPTh, CNRS et Universit\'{e} Savoie Mont-Blanc, Annecy, France\\ 
$^{30}$Department of Physics, Harvard University, Cambridge, MA 02138, USA\\ 
$^{31}$School of Physics and Astronomy, University of Minnesota, Minneapolis, MN 55455, USA\\ 
$^{32}$William I. Fine Theoretical Physics Institute, School of Physics and Astronomy, University of Minnesota, Minneapolis, MN 55455, USA\\ 
$^{33}$Institute for Advanced Research \& Kobayashi-Maskawa Institute for the Origin
of Particles and the Universe, Nagoya University, Nagoya 464–8602, Japan\\ 
$^{34}$Berkeley Center for Theoretical Physics, Department of Physics, University of California, Berkeley, CA 94720, USA\\ 
$^{35}$Theoretical Physics Group, Lawrence Berkeley National Laboratory, Berkeley, CA 94720, USA\\ 
$^{36}$Fermi National Accelerator Laboratory, Batavia, IL, USA\\ 
$^{37}$ University of Chicago, Department of Astronomy and Astrophysics, Chicago, IL, USA\\ 
$^{38}$Kavli Institute for Cosmological Physics, University of Chicago, Chicago, IL, USA\\ 
$^{39}$Department of Physics and Astronomy, Uppsala University, Uppsala, Sweden\\ 
$^{40}$National Centre for Nuclear Research, Warsaw, Poland \\ 
$^{41}$INFN Laboratori Nazionali di Frascati, 00044 Frascati RM, Italy\\ 
$^{42}$Department of Physics \& Astronomy, University of Hawaii at Manoa, 2505 Correa Rd., Honolulu, HI 96822, USA\\ 
$^{43}$Instituto de Astrof\'isica de Canarias, C/ V\'ia L\'actea, s/n E38205 - La Laguna, Tenerife,  Spain\\ 
$^{44}$Universidad de La Laguna, Departamento de Astrof\'isica - La Laguna, Tenerife, Spain\\ 
$^{45}$C. N. Yang Institute for Theoretical Physics, Stony Brook University, Stony Brook, NY 11794, USA\\ 
$^{46}$Department of Physics, Osaka University, Toyonaka, Osaka 560-0043, Japan\\ 
$^{47}$Department of Physics and Astronomy, University of Manchester, Manchester, M13 9PL, United Kingdom\\ 
$^{48}$School of Space and Environment, Beihang University, Beijing 102206, China\\ 
$^{49}$Harvey Mudd College, 301 Platt Blvd., Claremont, CA 91711, USA\\ 
$^{50}$Physics Department, Technion, Israel Institute of Technology, Haifa 3200003, Israel\\ 
$^{51}$Fakult\"at f\"ur Physik, TU Dortmund, D-44221 Dortmund, Germany\\ 
$^{52}$Universit{\' e} Paris-Saclay, CNRS/IN2P3, IJCLab, 91405 Orsay, France\\ 
$^{53}$Jo\v{z}ef Stefan Institute, Jamova 39, 1000 Ljubljana, Slovenia\\ 
$^{54}$School of Physics, The University of Melbourne, Victoria 3010, Australia\\ 
$^{55}$National Taiwan University, No. 1, Section 4, Roosevelt Rd, Da'an District, Taipei City, 10617, Taiwan\\ 
$^{56}$School of Physics and State Key Laboratory of Nuclear Physics and Technology, Peking University, Beijing 100871, China\\ 
$^{57}$Core of Research for the Energetic Universe, Hiroshima University, Higashi-Hiroshima 739-8526, Japan\\ 
$^{58}$Institut f\"ur Theoretische  Teilchenphysik, Karlsruhe Institute  of Technology, Karlsruhe, Germany \\ 
}

\maketitle

\newpage

\begin{abstract}
Rare meson decays are among the most sensitive probes of both heavy and light new physics. Among them, new physics searches using kaons benefit from their small total decay widths and the availability of very large datasets. On the other hand, useful complementary information is provided by hyperon decay measurements. We summarize the relevant phenomenological models and the status of the searches in a comprehensive list of kaon and hyperon decay channels. We identify new search strategies for under-explored signatures, and  demonstrate that the improved sensitivities from current and next-generation experiments could lead to a qualitative leap in the exploration of light dark sectors.
\end{abstract}

\setcounter{tocdepth}{2}
\tableofcontents
\newpage

\section{Introduction}

Precision measurements of rare kaon decays are among the most powerful probes of physics Beyond the Standard Model (BSM), with present and next generation kaon and hyperon factories set to significantly improve on the sensitivities achieved so far. However, there are still a number of signatures that have received little or no experimental attention, a situation that can be readily rectified as we highlight below. On the theoretical side, there is a dichotomy between the descriptions of effects due to heavy and light new physics. While the theoretical framework for interpreting the experimental results in terms of bounds on heavy new physics is well established, the situation for light new physics is more complex. For heavy new physics the BSM contributions to the various kaon branching ratios can be systematically encoded in an Effective Field Theory (EFT) Lagrangian, in a very similar way as the Standard Model (SM) contributions~\cite{Buras:2015yca}. The prime examples are the precisely predicted branching ratios for $K^+\to \pi^+\nu\bar\nu$ and $K_L\to \pi^0\nu\bar\nu$ decays~\cite{Buras:2015qea}. The present manuscript attempts to fill the gap in the literature and perform a systematic analysis  for the case of light new physics, i.e., for new physics models with new degrees of freedom lighter than the kaon mass.

The main complication in a systematic treatment of light new physics scenarios is that these are not captured by the general EFT analysis and lead to a variety of model-dependent signals. This poses the theoretical challenge of systematizing the possible signatures together with the new physics scenarios producing them, allowing a fair comparison with complementary probes from cosmology, astrophysics and accelerators other than kaon factories.  This challenge is accompanied by the more practical tension between our desire as a community to ``leave no stone unturned'' and the necessity to focus resources and time on the measurements that are likely to lead to the most insightful results, even if resulting solely in new bounds. To address this challenge, we provide in this manuscript a comprehensive overview of what can be learned about (light) new physics from kaon and hyperon decays. This also clarifies the most pressing experimental challenges for the present and the future.

Concretely, we:
{\it i)} classify the possible signals in kaon and hyperon decays, {\it ii)} explore a wide range of light new physics models that can feature these signals, reviewing the results in the literature and complementing them with new ones, and {\it iii)} compare the reach of future measurements at kaon and hyperon facilities with other experiments, including astrophysical and cosmological constraints. 
 
For completely generic flavor structures, we demonstrate that kaon experiments have typically the greatest sensitivities for new light particles in the mass range from a few MeV to a few hundred MeV, where the lower bounds on the couplings are typically set by cosmological constraints. In this mass range, new particles typically escape the detector before decaying, leading to signatures with missing energy and momentum. The same signatures are also motivated by light new particles much below the MeV scale, which circumvent the cosmological bounds by the fact that they are so long-lived and so weakly coupled to the SM to become good dark matter candidates. 
Likewise, light new physics particles promptly decaying to SM final states are also a phenomenologically viable possibility both in minimal and non-minimal new physics scenarios, generically with flavor aligned or minimally flavor violating couplings. In fact, there is a rich structure of possible models, leading to a variety of signatures characterized by resonant or non-resonant multi-lepton/photon final states, which can be completely probed experimentally in the near future. A summary of the signatures is given in Table~\ref{tab:summary}.

The theoretical promise of new physics searches in rare kaon and hyperon decays is mirrored by the active experimental program. The two currently operating kaon experiments are NA62 at CERN and KOTO at J-PARC with their main trigger lines designed for the $K^+\to\pi^+\nu\bar\nu$ and $K_L\to\pi^0\nu\bar\nu$ decays, respectively. The NA62 pre-scaled control and multi-track trigger lines, and the KOTO multiple photon triggers, are inclusive enough to make the collected data samples sensitive to a variety of new physics scenarios. One of the goals of this review is to explore to what extent the coverage of the existing triggers is sufficient to explore the variety of light new physics scenarios. 

A new proposal for a continuing ultra-rare kaon decay experiments at CERN, including experiments with high-intensity $K^+$ and $K_L$ beams, is in preparation~\cite{NA62:2020upd,Ambrosino:2019qvz} and a new experimental setup for KOTO has been proposed~\cite{KOTOstep2, KOTOstep2InWhitePaperOfHEFex}. We comment throughout the paper on how these proposals could ameliorate the reach on light new physics. 

Interestingly, the LHCb experiment using $pp$ collisions at LHC offers a complementary  program with leading sensitivity for decays of short-lived strange particles ($K_S$ and hyperons) into final states with charged particles. We also consider the impact of hyperon searches at the BESIII experiment at BEPCII~\cite{BESIII:2021cxx} as well as at future super charm-tau factories~\cite{Epifanov:2020elk,Vorobiev:2021sct,Luo:2018njj}.  

The paper is structured as follows. In Section~\ref{sec:models} we provide a list of models that can lead to interesting signatures in rare kaon and hyperon decays, and perform the phenomenological study of present and future constraints in the relevant parameter regions. Section~\ref{sec:facilities} contains a brief description of present and future experimental facilities, followed by a comprehensive discussion of experimental signatures and expected reach in kaon decays in Section~\ref{sec:exp:kaons}. Section~\ref{sec:hyppheno} discusses the signatures that are unique to hyperon decays and can give complementary probes of new physics. Section \ref{sec:flagship} distills the detailed analyses given in the bulk of paper into a set of flagship measurements that would most probably have a high impact on the BSM phenomenology. Section~\ref{sec:conclusions} contains conclusions.

\newpage

\section{Representative models}
\label{sec:models}

We start by reviewing the different BSM models featuring light new states that can be produced in rare hyperon and kaon decays. For each of the model we recast the current experimental constraints and comment briefly on the future reach, with a more detailed discussion of experimental issue relegated to Section \ref{sec:exp:kaons}. Whenever possible, we set well-motivated target branching ratios, explore their phenomenological implications, and comment on the implication of kaon and hyperon measurements on new physics scenarios addressing the shortcomings of the SM.

\subsection{Higgs portal scalar} 
\label{sec:higgs}
{\em Authors: Gori, Egana-Ugrinovic, \underline{Homiller}, Knapen, Meade}%
\footnote{For each section we list the contributing authors, with the corresponding author underlined.}

The Higgs portal is the minimal extension of the SM by one additional singlet scalar particle, $S$~\cite{Veltman:1989vw, McDonald:1993ex, OConnell:2006rsp, Clarke:2013aya, Winkler:2018qyg}.
In addition to its simplicity, the Higgs portal has received considerable attention both because it can serve as a simple, renormalizable dark sector portal~\cite{Krnjaic:2015mbs, Evans:2017kti}
and because it is similar to UV-motivated models with light dilatons or radions~\cite{Gildener:1976ih, Goldberger:1999uk, Goldberger:2007zk, Damour:2010rp, Bellazzini:2012vz, Abe:2012eu, Chacko:2012sy, Coradeschi:2013gda, Csaki:2020zqz, Gershtein:2020mwi}. The Lagrangian for the minimal Higgs portal model is given by
\begin{equation}
\mathcal{L} = \mathcal{L}_{\textrm{SM}} + \frac{1}{2}\partial_{\mu} S \partial^{\mu} S - V(S, H) .
\end{equation}
The full potential $V(S,H)$ can be found e.g. in Ref.~\cite{Egana-Ugrinovic:2015vgy}. For rare kaon and hyperon decays the only relevant feature is that the potential gives rise to mixing in the mass matrix between the singlet $S$ and the CP-even, neutral scalar component of the Higgs doublet, $H$. The resulting mass eigenstates are identified as the 125~GeV Higgs boson $h$, and a new scalar boson $\varphi$ with mass $m_{\varphi}$. The couplings of $\varphi$ to the SM fields are equal to the SM Higgs couplings up to a multiplication by a universal mixing angle $\theta$ (or rather by $\sin\theta \simeq \theta$). Therefore, both the production and decay rates of the scalar are each proportional to $\theta^2$. Despite its simplicity, the Higgs portal scalar has interesting phenomenology --- in particular at beam dumps, meson factories and other low energy experiments --- since its decay is governed by very small inherited couplings proportional to the SM electron and muon Yukawa couplings, while its production is governed by inherited couplings to heavier quarks.\footnote{The potential $V(H,S)$ also leads to couplings between the new singlet and the SM-like Higgs, leading for instance to exotic Higgs decays to a pair of scalars. Searches for these decays may be relevant for heavier singlets, but on general ground it can be shown that a large amount of fine-tuning would be required to have sizeable ${\cal B}( h \to \varphi\varphi)$ for a singlet in the mass range of interest for rare kaon decays~\cite{Gershtein:2020mwi}.}

\begin{figure}[t]
\centering
\includegraphics[width=12cm]{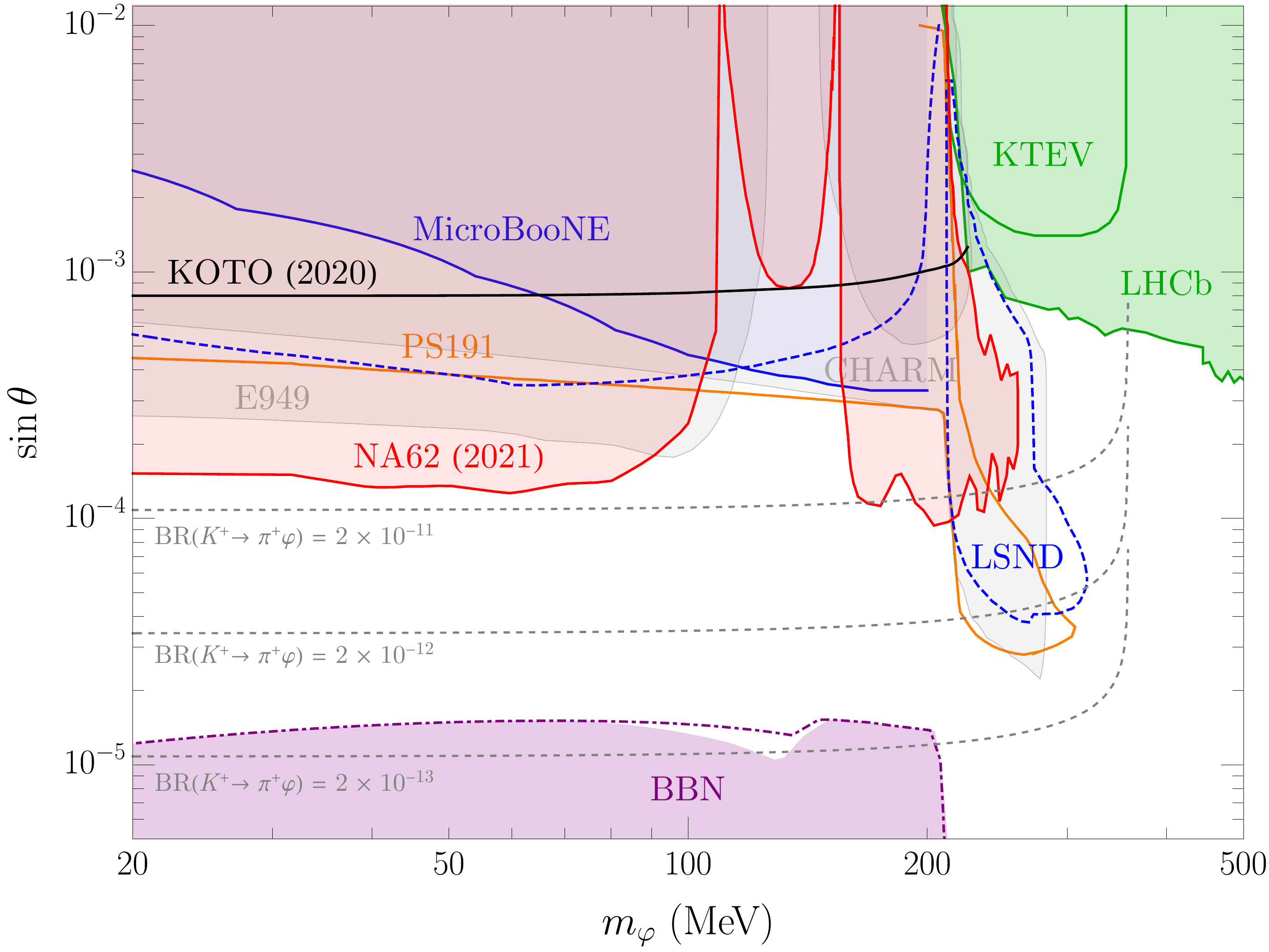}
\caption{\small Constraints on the Higgs portal scalar in the mixing angle vs mass plane. The NA62 and E949 limits extend down to $m_\varphi=0$ with almost no deterioration.}
\label{fig:higgs_portal_summary}
\end{figure}

At kaon factories, the new scalar would be produced in the two-body $K^+\to\pi^+\varphi$ and $K_L\to \pi^0 \varphi$ decays via its coupling to up-type quarks and gauge bosons in SM-like penguin diagrams~\cite{Leutwyler:1989xj, Gunion:1989we, Bezrukov:2009yw}.
The corresponding branching ratios for $m_X<300$~MeV are approximately equal to
\begin{equation}
{\cal B}(K^+ \to\pi^+\varphi) = 1.7\times 10^{-3}\sin^2\theta, \quad
{\cal B}(K_L \to\pi^0\varphi) = 5.7\times 10^{-3}\sin^2\theta,
\label{eq:BrKpiX}
\end{equation}
and decrease for larger $m_X$ values (for the explicit formulae, see Eqs. 4.4 and 4.5 of Ref.~\cite{Bezrukov:2009yw}). Contours of constant ${\cal B}(K^+\to\pi^+\varphi)$ are shown in Fig.~\ref{fig:higgs_portal_summary}.

In the $m_\varphi<m_K$ regime relevant for kaon experiments,  decays of the new scalar particle to charged lepton and pion pairs ($e^+e^-$, $\mu^+\mu^-$, $\pi^+\pi^-$, $\pi^0\pi^0$) should be considered, above the corresponding thresholds. Their partial decay widths are given by~\cite{Batell:2019nwo}
\begin{equation}
\label{eq:higgs_portal_widths}
\Gamma(\varphi \to \ell^+ \ell^-) = \sin^2\theta \frac{m_{\ell}^2 m_{\varphi}}{8\pi v^2} \bigg( 1 - \frac{4 m_{\ell}^2}{m_{\varphi}^2}\bigg)^{3/2},
\quad
\Gamma(\varphi \to \pi \pi) = \sin^2\theta \frac{3 | G_{\pi}(m_{\varphi}^2)|^2 }{32 \pi v^2 m_{\varphi}^2} \bigg( 1 - \frac{4 m_{\pi}^2}{m_{\varphi}^2}\bigg)^{1/2},
\end{equation}
where $v = 246$~GeV is the electroweak scale, the $\pi\pi$ decay width is computed using chiral perturbation theory, and the relevant form factor is $G_{\pi}(x) = \frac{2}{9}x + \frac{11}{9}m_{\pi}^2$. The $\varphi\to\gamma\gamma$ decay is always subdominant and can be neglected in practice.

For $2m_e < m_\varphi < 2m_\mu$, the scalar decays primarily to electron-positron pairs and is long-lived with a lifetime $c\tau = \mathcal{O}(10^{-2}~\textrm{m} / \sin^2\theta)$, therefore being largely invisible and experimentally indistinguishable from the $K\to\pi\nu\bar\nu$ decays. Above the di-muon threshold, $\varphi\to\mu^+\mu^-$ becomes the dominant decay mode and lifetime drops to $c\tau= \mathcal{O}(10^{-6}~\textrm{m} / \sin^2\theta)$. Above the di-pion threshold, $\varphi\to\pi\pi$ take over as the dominant decay modes, and the lifetime drops by another order of magnitude. As a result, the rest-frame decay length reaches $10.4~\textrm{m}$ for $m_{\varphi} = 300~\textrm{MeV}$ and $\sin\theta = 10^{-4}$. Further discussion of the lifetime and decay branching fractions can be found in Ref.~\cite{Batell:2019nwo}.

Experimental constraints on the Higgs portal are aggregated in the $\sin\theta$ vs $m_{\varphi}$ plane in Fig.~\ref{fig:higgs_portal_summary}. Constraints from $K\to\pi X_{\rm inv}$ searches  obtained by the NA62, KOTO and E949 experiments are discussed in Section~\ref{sec:sig:Kpi} where $X_{\rm inv}$ represents a particle invisible in each detector, while the prospects for searches for visible scalar decays ($\varphi\to\mu^+\mu^-$) are discussed in Section~\ref{sec:sig:Kpill}. Constraints from the CHARM beam dump experiment at the SPS~\cite{Bergsma:1985qz} are recast following Ref.~\cite{Egana-Ugrinovic:2019wzj}, including the effects of absorption in the target~\cite{Winkler:2018qyg},  geometric efficiency and $\eta$ decay contributions.
A similar but slightly weaker constraint from the NuCal/U70 experiment~\cite{Blumlein:1990ay, Blumlein:2011mv, Blumlein:2013cua} is omitted for clarity.
In a similar fashion to beam dumps, complementary limits can be placed using neutrino experiments by searching for the visible decays of $\varphi$. Taking advantage of the large number of kaon decays at rest at the NuMI absorber, the MicroBooNE collaboration set strong limits on $\theta$ with a dedicated search for $\varphi\to e^+e^-$~\cite{MicroBooNE:2021ewq}. Similar limits were obtained with a rescast of LSND~\cite{Foroughi-Abari:2020gju} as well as of PS191 data~\cite{Gorbunov:2021ccu}. Although not shown explicitly in the figure, T2K can marginally improve on the latter by using the existing search for HNLs~\cite{Arguelles:2021dqn}.
Constraints arising from searches for displaced vertices in $B \to \varphi(\mu^+\mu^-)$ decays at LHCb~\cite{Aaij:2015tna, Aaij:2016qsm}, adapted from Ref.~\cite{Winkler:2018qyg}, are the strongest in the $m_{\varphi} > 2m_{\mu}$ range. Searches in $B$ decays from Belle~\cite{Grygier:2017tzo} and BaBar~\cite{Lees:2015rxq}, as well as constraints from the $K_L \to\pi^0\mu^+\mu^-$ decay at KTeV~\cite{AlaviHarati:2000hs} are sub-leading in the parameter space considered.
Finally, the proposed SHiP experiment~\cite{Anelli:2015pba} is expected to be able to cover much of the parameter space for $m_\varphi > 2m_\mu$ down to mixing angles $\sin\theta\sim 10^{-5}$, and reach a sensitivity of $\sin\theta\sim 10^{-4}$ for smaller masses~\cite{Winkler:2018qyg}.

There are also cosmological and astrophysical constraints on light scalars. If the singlet lifetime is too large, the decay of scalars after nucleosynthesis spoils the light element abundances. The resulting Big Bang nucleosynthesis (BBN) constraint adapted from Ref.~\cite{Fradette:2017sdd} is shown in Fig.~\ref{fig:higgs_portal_summary}. The BBN constraint is sensitive to the metastable abundance of singlets, set via thermal interactions with the SM bath at temperatures above an MeV. This introduces a very mild dependence with the singlet-higgs quartic $S^2 H^\dagger H$, which is in one-to-one correspondence with the $h\to \varphi\varphi$ branching ratio. The shaded region indicates the weakest constraint from Ref.~\cite{Fradette:2017sdd} at a given mass, while the dot-dashed line indicates the constraint for ${\cal B}(h\to\varphi\varphi)=10^{-3}$, which allows the singlet mass to not be fine tuned. 
For even smaller values of the mixing angle, there are also constraints from the observation of SN1987a neutrinos, which sets a bound on the amount of energy that could have been carried away by scalars produced in the core. An older calculation of this limit in Refs.~\cite{Krnjaic:2015mbs, Evans:2017kti} was recently revisited in Ref.~\cite{Dev:2020eam}, finding substantial disagreement.
The SN1987a constraints are sub-leading to the BBN bounds in the entire parameter space shown, and are not included in Fig.~\ref{fig:higgs_portal_summary}.

Thus far we have focused on the {\em minimal} Higgs portal scenario, when the scalar inherits all its couplings via mixing with the Higgs. 
In many well-motivated UV scenarios, however, dimension-5 couplings between the scalar and gauge bosons (in particular, the photon) can arise. 
This is the situation in models of light radions or dilatons, or in theories where there are additional vector-like fermions charged under the SM gauge group.
These dimension-5 couplings can compete with the small, inherited couplings of the scalar to light fermions, and significantly change the phenomenology, especially the decay patterns, of the scalar. 
While the minimal Higgs portal for $m_{\varphi} < 2 m_{\mu}$ is long lived, with only a percent-level branching ratio to $\gamma\gamma$ compared to $e^+e^-$, more general scalars could decay predominantly to $\gamma\gamma$ and have much shorter lifetimes that significantly change the constraints from beam dump and fixed-target experiments.
While the underlying models in the scalar case are very different from the general pseudoscalar (or axion-like particle, ALP) case, the phenomenological signatures at kaon and hyperon experiments are virtually identical. 
We thus leave a detailed discussion of light particles with varying couplings to photons to the next section.  

In case $\varphi$ is a pure scalar, the $K\to\pi\pi\varphi$ is generically suppressed compared to $K\to\pi\varphi$ and the search in this channel can be ignored. If $\varphi$ is not purely CP even, as is the case in some well motivated theories~\cite{Flacke:2016szy}, then $K\to\pi\pi\varphi$ probes a different chiral structure of the light scalar couplings to the SM. This will be discussed in detail in Sec.~\ref{sec:fv_alp}. 


\subsection{QCD axion and axion-like particles (ALPs)}\label{sec:ALP}
\label{subsec:QCDaxion:ALP}
{\em Authors: Alonso-Alvarez, Alves, Bauer,  Datta,  Gori, Kvedarait\.e, Martin Camalich, Neubert, \underline{Redigolo}, Renner, Schnubel, Soreq, Thamm, \underline{Tobioka}, Ziegler}

Axion-like particles (ALPs) are ubiquitous in BSM models. Whenever an approximate global symmetry is spontaneously broken, there is an associated pseudo Nambu-Goldstone boson (pNGB) which is often called an ALP in the recent literature. Being a pNGB, the ALP mass can naturally be much lighter than the UV scale of symmetry breaking. As such they are the phenomenologically interesting remnants of the high scale new physics dynamics, potentially accessible at experiments. Arguably the most motivated example of an ALP is the QCD axion, which can simultaneously solve the strong CP problem~\cite{Peccei:1977hh,Wilczek:1977pj,Weinberg:1977ma} and account for the observed dark matter abundance~\cite{Preskill:1982cy,Abbott:1980zj,Dine:1982ah}. Already early on, in the early 1980s, it was realized that the QCD axion paradigm strongly motivates searches for massless ALPs at kaon factories~\cite{Wilczek:1982rv}, and even more so if it is connected to a model of flavor~\cite{Calibbi:2016hwq,Ema:2016ops} (for earlier discussions of flavor-violating axions see Refs.~\cite{Bardeen:1977bd,Davidson:1981zd,Reiss:1982sq,Davidson:1983fy, Davidson:1984ik,Berezhiani:1990wn,Berezhiani:1990jj,Babu:1992cu,Feng:1997tn,Albrecht:2010xh,Ahn:2014gva,Celis:2014iua}).  When experimental searches are performed, however, one should keep an open mind and not needlessly limit the potential of such searches. For instance, non-minimal implementations allow for the strong CP problem to be solved by heavier axions than one might naively expect~\cite{Dimopoulos:1979pp,Holdom:1982ex,Holdom:1985vx,Dine:1986bg,Flynn:1987rs,Choi:1988sy,Rubakov:1997vp,Choi:1998ep,Berezhiani:2000gh,Choi:2003wr,Hook:2014cda,Fukuda:2015ana,Dimopoulos:2016lvn,Agrawal:2017ksf,Agrawal:2017eqm,Gaillard:2018xgk,Hook:2019qoh,Gherghetta:2020keg,Kitano:2021fdl,Gupta:2020vxb}. Moreover, other theoretically motivated scenarios such as low scale supersymmetry, composite Higgs models, models of dark matter freeze-out, and models of electroweak baryogenesis may contain light ALPs (see for instance~\cite{Kilic:2009mi,Bellazzini:2012vz,Ferretti:2013kya,CidVidal:2018blh,Jeong:2018ucz}). The plethora of diverse theory motivations calls for a wide experimental exploration of the accessible ALPs masses and their couplings to the SM particles.

Let us assume that a global $U(1)$ symmetry is spontaneously broken by a condensate of size $f_a$ (such as the vacuum expectation value of a scalar field) and results in an ALP at low energies. The effective Lagrangian describing ALP interactions with the SM at lower energies
 is given by
\begin{equation}
\mathcal{L}_{\text{ALP}}=\frac{(\partial a)^2}{2}-\frac{m_a^2a^2}{2}+\mathcal{L}_{\text{ALP-gauge}}+\mathcal{L}_{\text{ALP-F}} \label{L_ALP}\, .
\end{equation}
The interactions with the SM fields, $\mathcal{L}_{\text{ALP-gauge}}$, $\mathcal{L}_{\text{ALP-F}}$, are of dimension 5, and are obtained after the heavy fields with masses of the order $\Lambda_{\text{UV}}\sim 4\pi f_a$ are integrated out. 
The ALP interactions include  in general couplings to all the SM gauge bosons: the gluons, $G^a$, the $SU(2)_L$ gauge bosons, $W^i$, and the hypercharge boson, $B_\mu$, 
\begin{equation}\label{eq:alpgauge}
   \mathcal{L}_{\text{ALP-gauge}}=\frac{N_3\alpha_s}{8\pi f_a} a G_{\mu\nu}^a\tilde{G}^{a\mu\nu}+\frac{N_2\alpha_2}{8\pi f_a} a W_{\mu\nu}^i\tilde{W}^{i\mu\nu}+\frac{N_1\alpha_1}{8\pi f_a} a B_{\mu\nu}\tilde{B}^{\mu\nu}\ .
\end{equation}
Here $\tilde G^{a\mu\nu}=1/2\epsilon^{\mu\nu\rho\sigma} G_{\rho\sigma}^a$, with $\epsilon^{0123} = 1$, and similarly for $\tilde W^{i\mu\nu}$, $\tilde B^{\mu\nu}$. The $N_i$ coefficients are  scale independent since they are related to the mixed 't Hooft anomalies involving the spontaneously broken global $U(1)$ symmetry and the SM gauge group.\footnote{In the QCD axion literature the $N_3$ coefficient is often absorbed in the definition of the decay constant $f_a$, $f_a/N_3\to f_a$, so that $\mathcal{L}_{\text{QCD axion}}=\frac{\alpha_s}{8\pi f_a} a G_{\mu\nu}^a\tilde{G}^{a\mu\nu}$.  We keep the $N_3$ factor explicit.} 
In general, these receive contributions from both the SM fermions and from the heavy fermions that were integrated out at $\Lambda_{\text{UV}}$. After electroweak symmetry breaking the couplings to $W^i$ and $B_\mu$ gauge bosons result in ALP couplings to photons
\beq
   \mathcal{L}_{\text{ALP-photon}}=\frac{ \alpha}{8\pi f_a} C_{\gamma\gamma} a F_{\mu\nu}\tilde{F}^{\mu\nu},
\eeq 
with $C_{\gamma\gamma}=N_1+N_2$, see also Eq. \eqref{eq:cgamgameff} below.

The ALP couples derivatively to the SM fermions,
\begin{equation}\label{eq:ALP-f}
   \mathcal{L}_{\text{ALP-f}} = 
\frac{\partial_\mu a}{2 f_a} \, \bar f_i \gamma^\mu \big( C^V_{f_i f_j} + C^A_{f_i f_j} \gamma_5 \big) f_j  \, , 
\end{equation}
where $f=u,d,e,$ and the sum over repeated generational indices, $i,j=1,2,3$, is implied. $C^{V,A}_{f_i f_j}$ are hermitian matrices in flavor space which we take to be real for simplicity. Because of gauge invariance $C^V_{d_i d_j} - C^A_{d_i d_j} = V^*_{u_k d_i } (C^V_{u_k u_l} - C^A_{u_k u_l}) V_{u_l d_j}$, so that in total the ALP interactions with the charged SM fermions and gauge bosons are specified by 33  parameters. Below we also use the notation $F_{f_i f_j}^{V/A}\equiv 2 f_a/C_{f_i f_j}^{V/A}$.

For kaon decays the $s\to d a$ transition is the dominant ALP production mode even if the flavor violating couplings, $C_{sd}^{V/A}$, are highly suppressed compared to the flavor diagonal ones. The dominant kaon decay channels resulting in ALP production are thus 
\begin{equation}
K\to \pi a, \quad    
K\to \pi \pi a.    
\end{equation}
The two-body (three-body) decays probe the vectorial (axial) ALP coupling $C_{sd}^V$ ($C_{sd}^A$), as can be easily shown using  the parity conservation of QCD interactions.

If we take the couplings in Eq.~\eqref{eq:ALP-f} to be defined at the EW scale, $\mu_W = {\mathcal O}(M_Z)$, the renormalization group (RG) evolution from  $\Lambda_{\text{UV}}$ to $\mu_W$ should be taken into account and can be extracted from Refs.  \cite{Choi:2017gpf,MartinCamalich:2020dfe,Chala:2020wvs,Bauer:2020jbp}. The one loop corrections generate the CKM-suppressed flavor violating ALP couplings at the EW scale, even if the ALP couplings in the UV are flavor diagonal. The log-enhanced contribution, which dominates the one loop correction for $\Lambda_{\text{UV}}\gg M_Z$, follows from the anomalous dimension controlling the RG running and is given by 
\begin{equation}
C_{ds}^{A,V}(M_Z)= \frac{y_t^2}{16\pi^2}V_{td}^*V_{ts}C_{tt}^{A} \log\frac{\Lambda_{\text{UV}}}{M_Z} \simeq 2\times 10^{-6} C_{tt}^{A} \log\frac{\Lambda_{\text{UV}}}{M_Z}. \label{eq:FVRGE1}
\end{equation}
The above leading-log expression for $C_{ds}^{A,V}(M_Z)$ is of only limited phenomenological importance. Due to the numerically large CKM suppression the current experiments probe the loop induced $C_{ds}^{A,V}(M_Z)$ only for values of $f_a$ relatively close to the weak scale. The finite loop corrections are therefore of similar size as the  leading log term in Eq.~\eqref{eq:FVRGE1}; note that $\Lambda_{\text{UV}}$ can be at most $4\pi f_a$. The complete expression for $C_{ds}^{V}(M_Z)$, including finite corrections, is given in Eq.~\eqref{eq:CdsV}.


\subsubsection*{ALP decays}

To be produced in rare kaon decays, the ALP needs to be lighter than $m_K-m_\pi \approx 350$~MeV. This has phenomenological implication for the ALP total width, possible decay channels and parameter space in general. The most important decay channels are $a\to \gamma\gamma$, and, if kinematically allowed, the decay modes $a\to \ell^+\ell^-$, $\ell=e,\mu$. The corresponding decay widths are given by
\beq\label{eq:ALP-widths}
\Gamma(a\to \gamma\gamma)
= \frac{\alpha^2\, m_a^3}{256\pi^3 f_a^2} |C_{\gamma\gamma}^\text{eff}|^2 \,,
\qquad
\Gamma(a\to\ell^+\ell^-) 
=   \frac{m_a\, m_\ell^2}{8\pi f_a^2}\,\big(C_{\ell\ell}^{A} \big)^2\,\Big(1-\tfrac{4m_\ell^2}{m_a^2}\Big)^{1/2} \,, 
\eeq
where the effective coupling of the ALP to photons is \cite{Bauer:2017ris, Bauer:2020jbp}
\beq
\begin{split}
\label{eq:cgamgameff}
   C_{\gamma\gamma}^\text{eff}(m_a)
   &= N_1+N_2 +\sum_{q}\, 6\, Q_q^2\,C_{qq}^A(\mu_0)\,B_1(\tau_q)
    + 2\sum_{\ell} C_{\ell\ell}^A\,B_1(\tau_\ell)
    \\
   &\quad\mbox{}- (1.92\pm 0.04)\,N_{3} - \frac{m_a^2}{(m_\pi^2-m_a^2)} 
    \left[ N_{3}\,\frac{m_d-m_u}{m_d+m_u} + C_{uu}^A-C_{dd}^A \right] \,.
    \end{split}
\eeq
The loop function $B_1(\tau_f)$ depends on $\tau_f\equiv 4m_f^2/m_a^2$, and is $B_1\simeq 1$ for light fermions ($m_f\ll m_a$) while it decouples as $B_1\simeq -m_a^2/(12 m_f^2)$ for heavy fermions ($m_f\gg m_a$). The explicit form of $B_1(\tau)$ can be found in Ref.~\cite{Bauer:2020jbp}.
We neglected the threshold corrections coming from $W$-loops since these are negligible for $m_a\ll m_W$. The second line in Eq.~(\ref{eq:cgamgameff}) encodes the leading order chiral perturbation theory (ChPT) contribution of $N_3$ to the ALP di-photon coupling. Similarly, the ALP coupling to electrons is generated at one-loop from the ALP couplings to gauge bosons~\cite{Bauer:2017ris}.  A common notation in axion literature is also $g_{\gamma\gamma}=C_{\gamma\gamma}^\text{eff} \alpha /(2\pi f_a)$.

For $m_a<2 m_e$, ALP decays are dominated by $a\to \gamma\gamma$, and the decay length is given by
\begin{align}
    c\tau_a 
    \approx 2.9\times10^7\,{\rm meters} \left( \frac{f_a}{\rm TeV} \right)^2 
    \left(\frac{1\,{\rm MeV}}{m_a} \right)^3
    \left(\frac{1}{|C_{\gamma\gamma}^\text{eff}|}\right)^2  \, .
    \label{eq:ALPlifetime:photon}
\end{align}
For heavier ALPs, the decay channel to lepton pairs opens up and typically dominates over the diphoton decay mode, as long as $C^A_{\ell\ell}\gtrsim  C_{\gamma\gamma}^\text{eff} m_a/{\rm GeV}$. We can then estimate the ALP decay length as 
\begin{equation}
\label{eq:ctau_a:leptons}
 c\tau_a 
    \approx
    \left\{
    \begin{aligned}
        & 1.9\text{ meters}\left( \frac{f_a}{\rm TeV} \right)^2 
    \left(\frac{10\,{\rm MeV}}{m_a} \right)
    \left(\frac{1}{C_{ee}^A}\right)^2\ , & 2 m_e<m_a< 2 m_\mu,
        \\
        & 1.5\times 10^{-6}\text{ meters}\left( \frac{f_a}{\rm TeV} \right)^2 
    \left(\frac{300\,{\rm MeV}}{m_a} \right)
    \left(\frac{1}{C_{\mu\mu}^A}\right)^2\ ,& 2 m_\mu<m_a.
    \end{aligned}
    \right.
\end{equation}
The above scalings show that an ALP lighter than the di-electron threshold is unlikely to decay promptly on detector scales, while a heavier ALP can be  decaying promptly for $f_a\lesssim \text{TeV}$.   

The hadronic ALP decay modes are typically negligible. The $a\to 3\pi$ decay is kinematically forbidden, if rare kaon decays with an ALP in the final state are kinematically allowed. The $a\to \pi\pi\gamma$ decay could be important in scenarios where the ALP mixes with $\eta'$, due to ${\cal B}(\eta'\to\pi\pi\gamma)\approx 30\%$. In practice, however, the $a-\eta'$ mixing is suppressed by $(m_a/m_{\eta'})^2$, and for $m_a<m_K-m_\pi$ the $a\to \pi\pi\gamma$ decay channel is always subdominant compared to $a\to\gamma\gamma$~\cite{Aloni:2018vki,Cheng:2021kjg}. The remaining kinematically allowed decay modes are $a\to\mu^\pm e^\mp$, which is lepton flavor violating (Section~\ref{sec:FV:ALP:mue}), and $a\to\pi^0 e^+e^-$, which is CP violating and hence expected to be highly suppressed.   


\subsubsection*{ALP production in kaon decays}
\label{sec:kaondecay}

The decay amplitudes for charged and neutral kaon decays into a pion and an ALP receive an IR contribution at energies $\mu<2$~GeV from the matching of the ALP to the ChPT  Lagrangian~\cite{Bauer:2021wjo,Bauer:2021mvw}, and UV contributions which are encoded in $C_{ds}^V$. The latter can be induced at tree level~\cite{Ema:2016ops,Calibbi:2016hwq} or at one-loop via ALP couplings to the up-quarks~\cite{Hall:1981bc,Freytsis:2009ct,Bauer:2020jbp} or to the $W$-boson~\cite{Izaguirre:2016dfi}. The value of $C_{ds}^V$ at the charm threshold, $\mu_c=2$ GeV, is therefore given by 
\begin{equation}
\label{eq:CdsV}
C_{ds}^V(\mu_c)=C_{ds}^V(\Lambda_{\text{UV}}) + \sum_{F=t,c}\frac{y_F^2V_{Fd}^*V_{Fs}C_{FF}^{A}}{16\pi^2}\left(\log\frac{\Lambda_{\text{UV}}}{\mu_F}-f_F(x_F)\right)-\frac{g_2^4N_2}{256\pi^4}V_{td}^*V_{ts}f_W(x_t),
\end{equation} 
where $\mu_t=M_Z$, while $x_F=m_F^2/m_W^2$ for $F=t,c$. The loop functions are
\beq
f_F(x) = \frac{3\left(1-x+\log x\right)}{2(1-x)^2}+\frac{1}{4},
\qquad
f_W(x) = \frac{3 x\left[1-x+x\log x\right]}{2(1-x)^2}.
\eeq
All the Wilson coefficients on the r.h.s. of Eq.~(\ref{eq:CdsV}) are defined at the scale $\Lambda_{\rm UV}$. The log enhanced term in Eq.~(\ref{eq:CdsV}) is the RGE contribution already discussed in Eq.~\eqref{eq:FVRGE1}, where we used the fact that no resummation of electroweak logs is needed. The other two contributions are the finite threshold corrections from the top, charm and the $W$ boson running in the loop, with the ALP either coupling to the quark or the $W$, respectively. In concrete UV completions, the UV threshold corrections might also be important~\cite{Hall:1981bc,Freytsis:2009ct}. For simplicity we assume these to be subdominant. Note that only the diagonal axial couplings to quarks contribute to the one loop correction of $C_{ds}^V$. 

Numerically, the sizes of the different contributions in Eq.~(\ref{eq:CdsV}) are given by (dropping the trivial UV contribution) 
\begin{equation}
C_{ds}^V\supset\!\!	
-10^{-6}\! \left[(2+i)C_{tt}^A\!\left(\log\frac{\Lambda_{\text{UV}}}{\mu_t} +0.02
\right)\!+\! 0.08 \, C_{cc}^A\!\left(\log\frac{\Lambda_{\text{UV}}}{\mu_c}+11\right)\!-(4+2i)10^{-3} N_2\right]. \label{eq:EWestimate}
\end{equation}

If the UV ALP couplings are flavor diagonal, the FV coupling $C_{ds}^V$, relevant for the kaon decays, is highly suppressed by the product of loop and CKM factors. This means that in the case of UV flavor universality, the $K\to\pi a$ decays probe significantly lower values of $f_a$ than for the case of flavor violating UV ALP couplings. Interestingly, the Yukawa suppression of the charm contribution relative to the top gets partially compensated by the less severe CKM suppression, so that the two terms can be of similar numerical size. 

\begin{figure}[t!]
\begin{center}
\includegraphics[width=2.9in]{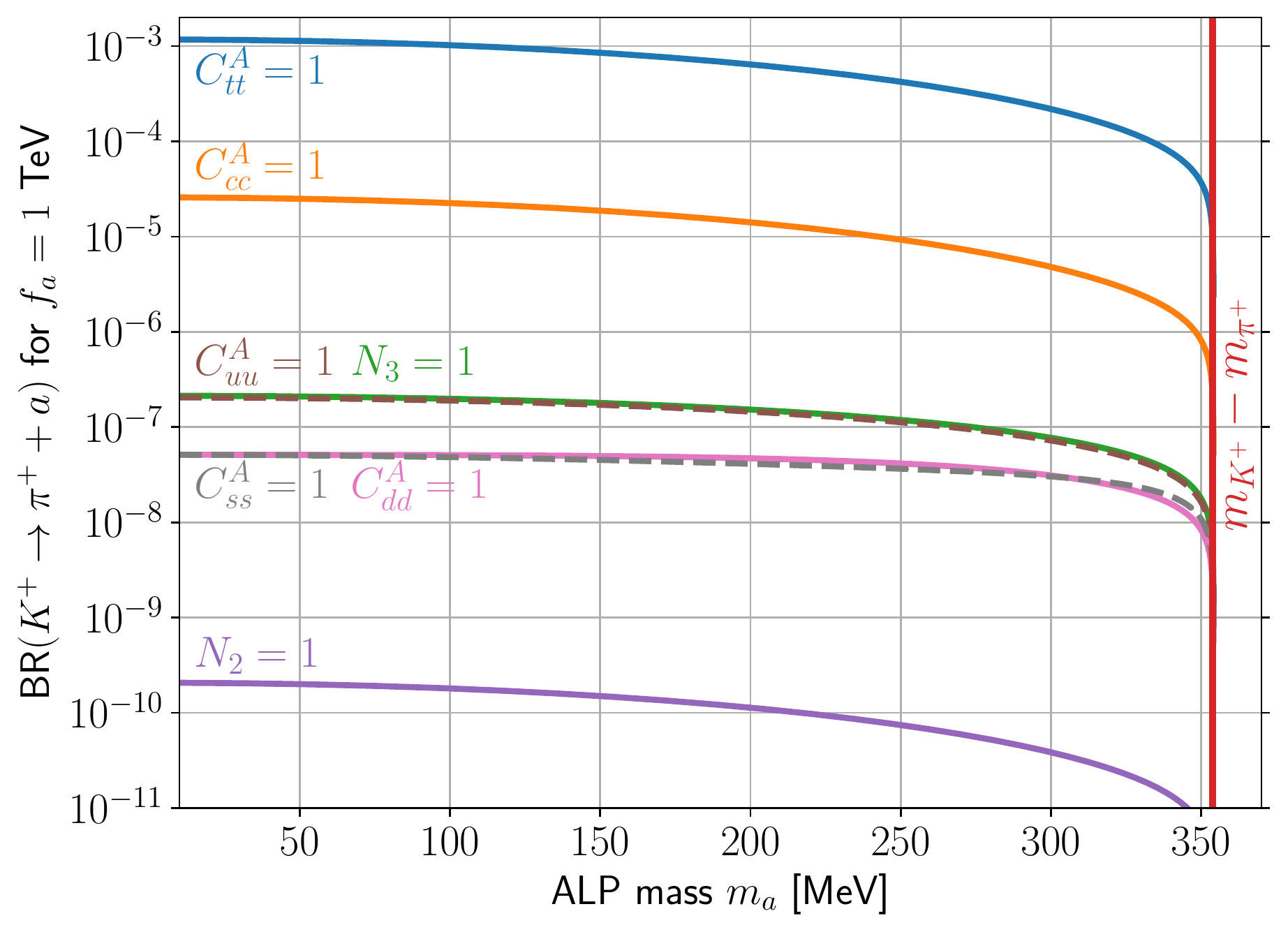} ~~~~ \includegraphics[width=2.9in]{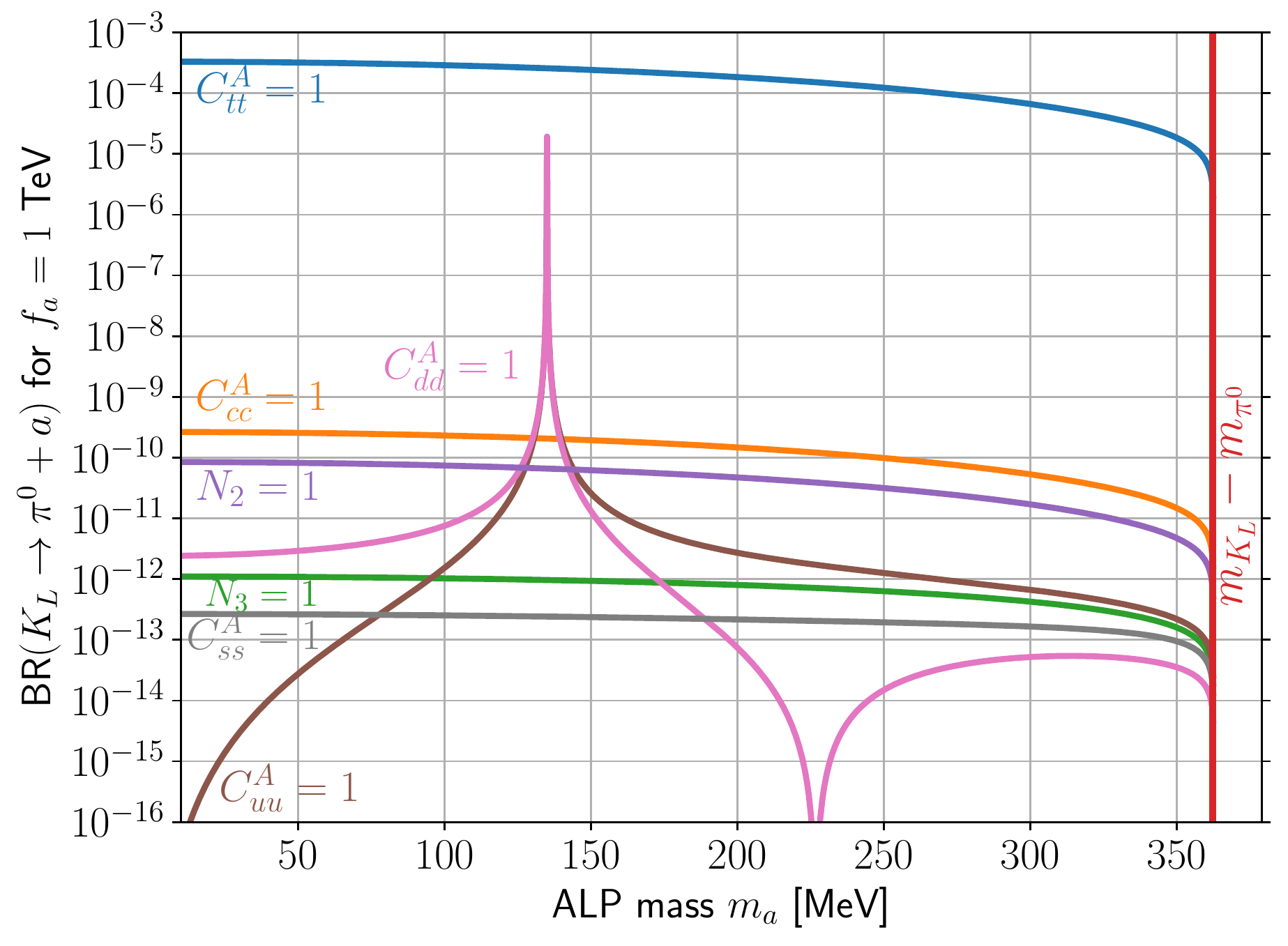}
\caption{\small Branching ratios of the $K^+\to\pi^+a$ (left) and $K_L\to\pi^0a$ (right) decays as  functions of the ALP mass, taking only a single coupling in the ALP Lagrangian~\eqref{eq:alpgauge}, \eqref{eq:ALP-f} to be nonzero as shown by the labels, and setting $f_a=1$~TeV and $\Lambda_{\rm UV}= 4\pi f_a$. For a different value of the ALP decay constant, the branching ratios should be rescaled by $1/f_a^2$.}
\label{fig:Prompt_ALP} 
\end{center}
\end{figure}

Following Refs.~\cite{Bauer:2021wjo,Bauer:2021mvw}, we can write down the $K\to\pi a$ decay amplitudes. The decay widths are given by
$\Gamma(K\to\pi a) = \left|\mathcal{A}(K\to\pi a) \right|^2/ 16\pi m_K $, ignoring the phase space factor due to nonzero pion and ALP masses. Furthermore, for $K_L$ decays the $K^0$ and $\bar K^0$ decay amplitudes interfere, since $K_L = \big[(1+\epsilon) K^0 + (1-\epsilon)\bar K^0\big]/\big[2(1+|\epsilon|^2)\big]^{1/2}$.
At leading order in the ChPT expansion, the decay amplitudes are given by
\begin{align}
\label{eqn:chargedKtopia}
\begin{split}
\!\! i{\cal A}(K^-\to\pi^- a)
&\ \ = \frac{N_8}{4f_a}\,\Big[
    8\, N_{3}\,m_{K-\pi}^2 \xi_a  + \big(4\, C_{ss}^A+6 \xi_a C_{uu+dd-2ss}^A\big) \,m_a^2 
    \\
   & \quad  
   + C_{2uu+dd+ss}^A\,m_{K-\pi-a}^2  +C_{dd-ss}^V\,m_{K+\pi-a}^2 \Big]  - \frac{m_{K-\pi}^2}{2 f_a} C_{ds}^V \,, 
  \end{split}
   \\ 
   \begin{split}
   \label{eqn:neutralKtopia}
 \!\! - i\sqrt2{\cal A}(\bar K^0\to\pi^0 a)    
  & = \frac{N_8}{4f_a}\,\Big[
   \big( 8\, N_{3} \xi_a + C_{3dd+ss}^A\big)\,m_{K-\pi}^2  
    + \big( C_{2uu-dd-ss}^A - 2 \xi_a C_{uu+dd-2ss}^A\big)m_a^2 
    \\
  &   - 2 C_{uu-dd}^Am_a^2 \,\frac{\,m_{K-a}^2}{m_{\pi-a}^2} 
    + C_{dd-ss}^V m_{K+\pi-a}^2 \Big]  - \frac{m_{K-\pi}^2}{2 f_a} C_{ds}^V ,
   \end{split}
\end{align}
while the $K^0\to\pi^0a$ decay amplitude is obtained via the replacement  ${\cal A}(K^0\to\pi^0 a) = - {\cal A}(\bar K^0\to\pi^0 a) \big|_{C_{ds}^V \to C_{ds}^{V*} = C_{sd}^V}$. Above, we use the shorthand notation, $C_{c_1 uu+c_2 dd+c_3 ss}^{V}=c_1 C_{uu}^V +c_2 C_{dd}^{V}+c_3 C_{ss}^{V} $,  and similarly for $C_i^A$, for instance, $C_{2uu-dd-ss}^{A}=2 C_{uu}^A- C_{dd}^{A}-C_{ss}^{A} $, with all the Wilson coefficients defined at $\mu_c=2$~GeV. We also defined $m_{M_1-M_2}^2=m_{M_1}^2-m_{M_2}^2$, $m_{K\pm \pi-a}^2=m_K^2\pm m_\pi^2-m_a^2$, and $\xi_a= {(m_K^2-m_a^2)}/{(4m_K^2-m_\pi^2-3m_a^2)}$. 
The overall  prefactor multiplying the first terms in Eqs. \eqref{eqn:chargedKtopia}, \eqref{eqn:neutralKtopia}, is given by  
\begin{equation}
N_8 = -{G_F} V_{ud}^* V_{us} g_8 f_\pi^2/\sqrt{2}\simeq - 1.5 \times 10^{-7}\ .\end{equation}
If the $K\to \pi a$ transitions are induced by the flavor diagonal couplings of ALP to quarks, with no large hierarchies between the couplings, then the ALP couplings to light quarks and gluons (the terms proportional to $N_8$) in Eqs. \eqref{eqn:chargedKtopia}, \eqref{eqn:neutralKtopia} and the contributions generated via EW loops from ALP coupling to top, charm and $W$, Eq.~\eqref{eq:EWestimate}, are of roughly similar sizes. A more quantitative comparison is shown in Fig.~\ref{fig:Prompt_ALP}, where the branching ratios for $K\to \pi a$ decays are plotted as functions of the ALP mass for a number of benchmark scenarios, in each of which only one ALP coupling at a time is taken to be nonzero, as indicated. 

For a massless ALP we can write the $K^+\to \pi^+ a$ branching ratio as 
\begin{equation}
\label{eq:K+:pi+:ALP:numerical}
{\cal B}(K^+\to \pi^+ a)=1.1\times 10^7 \left(\frac{\text{TeV}}{f_a}\right)^2\big[(C_{ds}^V)^2+2 N_8C_{ds}^V \tilde N_3+ N_8^2 \tilde N_3^2\big],   
\end{equation}
where we neglected $m_\pi$ and defined $\tilde N_3 =N_3+\big(C_{2uu+dd+ss}^{A}+C_{dd-ss}^V\big)/2$. The form of Eq. \eqref{eq:K+:pi+:ALP:numerical} motivates us to distinguish two limiting cases.
\begin{itemize}
\item If $C_{ds}^V(\Lambda_{\rm UV})$ is nonzero this implies strong limits on the symmetry breaking scale $f_a$. For $C_{ds}^V\sim {\mathcal O}(1)$ the value of $f_a\sim 10^7~\text{GeV}$ would imply very large modifications of kaon decay branching ratios, ${\cal B} (K\to \pi a)\sim \mathcal{O}(1)$. As discussed in Section~\ref{eq:longlived}, much larger decay constants $f_a$ are already probed by the experiments, thanks to the fact that the irreducible SM background of $K^+\to\pi^+ a$ is $K^+\to\pi^+\nu\bar\nu$ which is highly suppressed in the SM. In this scenario, the ALP is naturally long-lived and, if light enough, can be the Dark Matter.
\item If $C_{ds}^V(\Lambda_{\rm UV})=0$ then ${\cal B}(K^+\to \pi^+ a)\sim 10^{-6}-10^{-7}$ for $f_a\sim \text{TeV}$. In this scenario the ALP can decay promptly into the SM particles, a possibility discussed further in Section~\ref{sec:prompt:ALP}. Intriguingly, for $m_a<m_K-m_\pi$ the region where the ALP decays promptly can be fully probed in the near future. 
\end{itemize}

\subsubsection{Long-lived ALPs from kaon decays}
\label{eq:longlived}

Flavor-violating ALP decays obey the generalized Grossman-Nir bound~\cite{Grossman:1997sk,Kitahara:2019lws}
\begin{equation}
\label{eq:2body:GN}
{\cal B}(K_L\to\pi^0 a)\leq4.3\cdot{\cal B}(K^+\to\pi^+a).
\end{equation}
The main effect in the numerical factor is due to the kaon lifetime ratio $\tau_L/\tau_+=4.1$. This makes the $K^+\to\pi^+ a$ decay more sensitive to invisible ALPs than the $K_L\to\pi^0 a$ decay (which is also more challenging experimentally). In the remainder of this section we therefore focus on ALP production in the $K^+\to\pi^+a$ decays. Note however that the above rule does not apply generally. Certain new physics models feature an enhanced signal in rare $K_L$ and not $K^+$ decays, as discussed in Section~\ref{sec:GNVgamma}.

\paragraph{Experimental status and projections}

\begin{figure}[t]
\centering
\includegraphics[width=11cm]{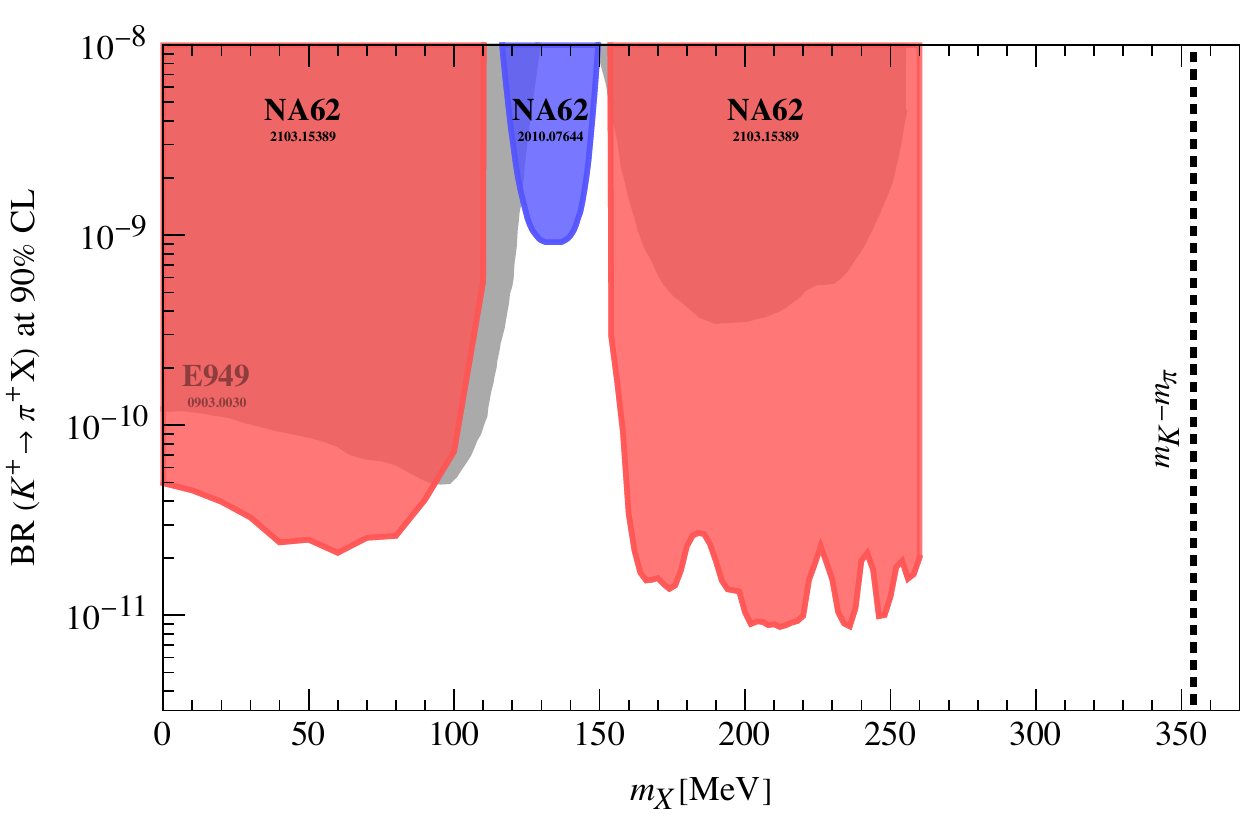}
\caption{\small Constraints on ${\cal B}(K^+\to\pi^+X)$ with the $X$ particle decaying invisibly or escaping the detector as a function of $X$ mass up to the kinematic endpoint (vertical dotted line). Red regions are excluded by the NA62 search for $K^+\to\pi^+X$~\cite{NA62:2021zjw}, the blue region by $\pi^0\to{\rm inv}$~\cite{NA62:2020pwi}, and grey regions by the BNL E949 experiment~\cite{Artamonov:2009sz}.  \label{fig:BRKpiinv}}
\end{figure}

\begin{figure}[t]
\centering
\includegraphics[width=8cm]{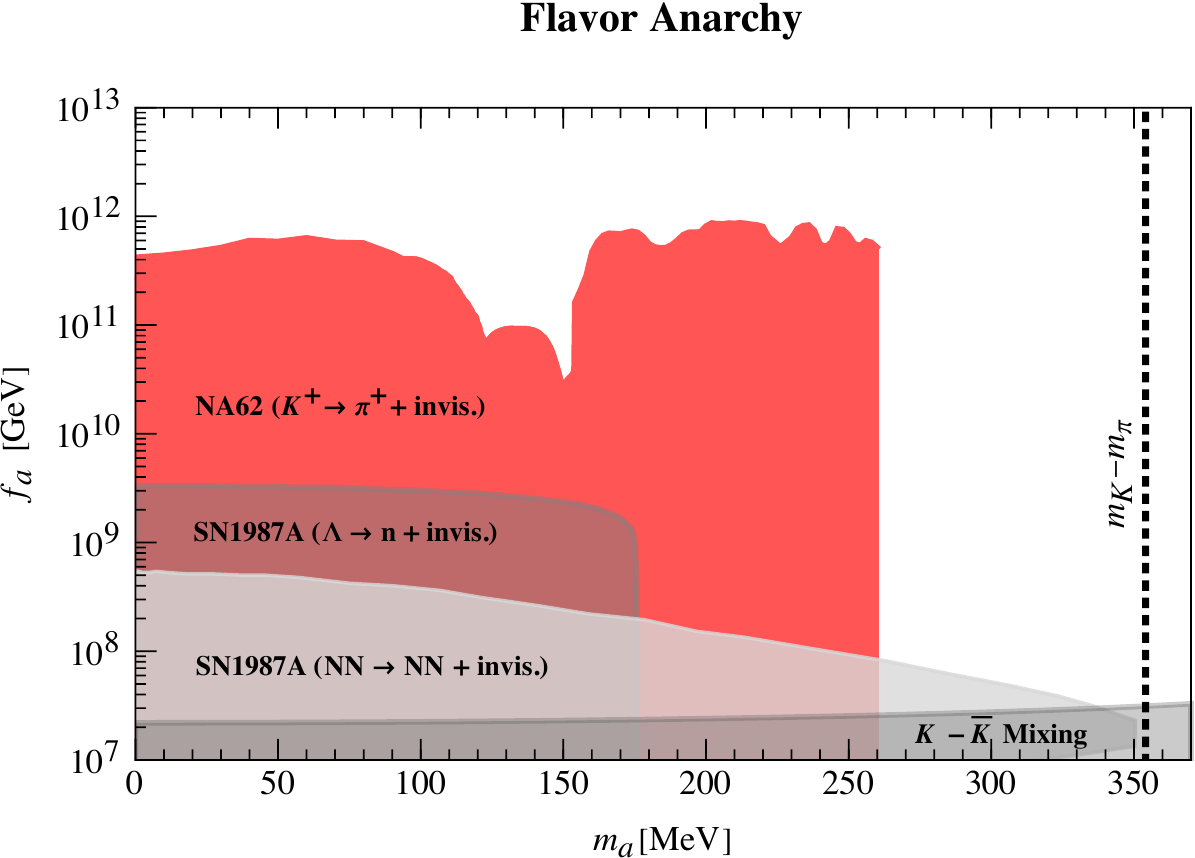}~~~
\includegraphics[width=7.9cm]{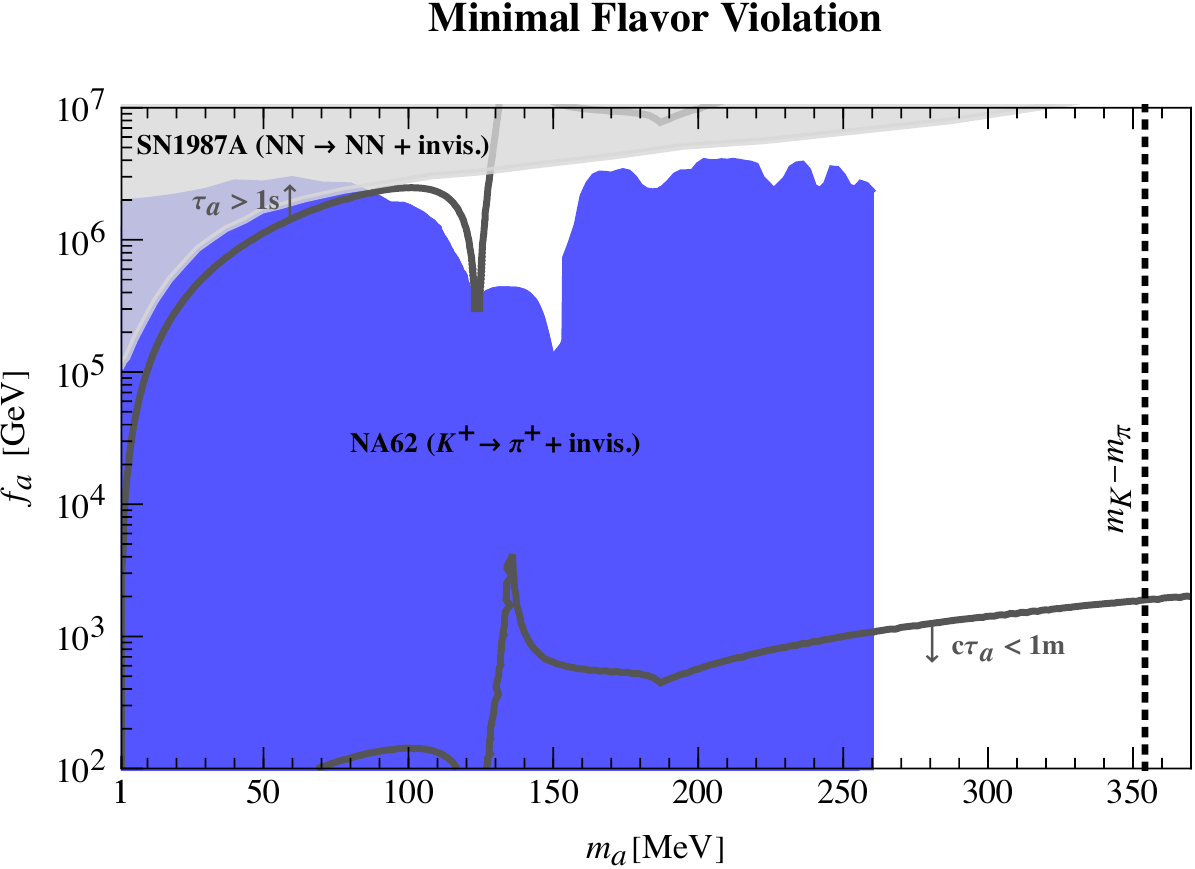}
\caption{\small Constraints on the axion decay constant $f_a$ for the flavor anarchic (left) and minimal flavor violation (right) benchmarks. The constraints from NA62 are showed as colored regions, cf. Fig.~\ref{fig:BRKpiinv}. The SN1987A constraints (grey region) are taken from Ref.~\cite{Lee:2018lcj} for SN cooling through nucleon bremsstrahlung $NN \to NN +{\rm inv}$, and from Ref.~\cite{MartinCamalich:2020dfe} for cooling via hyperon decays $\Lambda\to n +{\rm inv}$. The $K-\bar K$ mixing constraints assume that the dimension-6 UV contributions are small. Also shown in the right panel are the contour lines below which the ALP is no longer long-lived ($c \tau_a < 1$~m) and above which its lifetime is stringently constrained by cosmology ($\tau_a > 1$~s).  
}
\label{fig:plotma}
\end{figure}

The experimental status and prospects of $K\to\pi X$ searches, with $X$ escaping the detector or decaying invisibly, are discussed in Section~\ref{sec:sig:Kpi}. For the $K^+$ case, the upper limits of ${\cal B}(K^+\to\pi^+X)$ at 90\% CL as a function of $m_X$ up to the kinematic endpoint $m_{K}-m_{\pi} = 354$~MeV are shown in Fig.~\ref{fig:BRKpiinv}. Three distinct kinematic regions are determined by the experimental conditions.
\begin{itemize}
\item The regions 1 and 2 of the $K^+\to\pi^+\nu\bar\nu$ search (results shown as red shaded areas). The largest background comes from the ultra-rare $K^+\to\pi^+\nu\bar\nu$ decay, leading to very stringent bounds on ${\cal B}(K^+\to\pi^+a)$. The deterioration of the bound at very low $m_a$ masses owes to the definition of the $K^+\to\pi^+\nu\bar\nu$ signal region.
\item The $K^+\to\pi^+\pi^0$ region, where the results are obtained from a dedicated search for the $\pi^0$ decays into a fully invisible final state (results shown as a blue shaded area).
\item The region $m_{a}>260$~MeV, where the background is dominated by the $K^+\to\pi^+\pi^+\pi^-$ decays. This region serves as a control region for the $K^+\to\pi^+\nu\bar\nu$ measurement (and thus is not masked during analysis), and has not yet been  explored by the experiments. 
\end{itemize}
%
The constraints in Fig.~\ref{fig:BRKpiinv} can be re-interpreted as bounds on the ALP couplings entering the $K^+\to \pi^+ a$ decay amplitude, see Eq.~\eqref{eqn:chargedKtopia}. This is usually done either in specific UV completions or by assuming that only a single coupling is nonzero (a conventional approach for obtaining bounds on couplings in the effective field theory). 
As illustrative benchmarks we consider two limiting cases:
\begin{itemize}
\item \emph{Flavor Anarchy:} The FV couplings are of the same order as the diagonal ones. In numerical estimates we set at the UV scale $C_{sd}(\Lambda_{\text{UV}})=1$, and similarly for the diagonal couplings, $C^V_{qq} = C^A_{qq} = N_3 = 1$.  
\item \emph{Minimal Flavor Violation:} The FV couplings vanish at the UV scale, $C_{sd}(\Lambda_{\text{UV}})=0$. For flavor diagonal couplings we take $C^V_{qq} = C^A_{qq} = N_3 = 1$ at $\mu=\Lambda_{\rm UV}$.
\end{itemize}
The resulting constraints on the axion decay constant $f_a$ from the NA62 searches are shown for the flavor anarchic and minimal flavor violating benchmarks in Fig.~\ref{fig:plotma}, overlaid with the SN1987A constraints. In the case of nonzero flavor-violating couplings the most stringent constraints arise mainly from Supernova (SN) cooling through hyperon decays~\cite{MartinCamalich:2020dfe, Camalich:2020wac}, while for the flavor-conserving benchmark the main cooling mechanism is via nucleon bremsstrahlung~\cite{Lee:2018lcj,Carenza:2019pxu}. For the flavor anarchic benchmark the SN cooling constraints are roughly comparable with the constraints from $K-\bar K$ mixing due to tree level ALP exchanges~\cite{MartinCamalich:2020dfe}. These constraints are UV sensitive, since UV dependent dimension-6 operators suppressed by the UV scale $\Lambda_{
\text{UV}}\sim 4\pi f_a$ are expected to give contributions of the same parametric size as the IR contribution from ALP exchanges. The constraints from $K-\bar K$ mixing in Fig.~\ref{fig:plotma} should thus be viewed with this caveat in mind. 

Fig.~\ref{fig:plotma} (left) illustrates the fact that for generic ALP couplings, with no peculiar flavor structure, the strongest constraints on the axion decay constant are provided by NA62 and are $f_a\gtrsim 10^{12}~\GeV$ for ALP couplings of ${\mathcal O}(1)$. Remarkably, the NA62 constraints are the leading ones in the phenomenologically very interesting region where $m_a\lesssim {\mathcal O}(\text{keV})$, and the ALP can be cold DM produced through the misalignment mechanism. Most importantly, the ALP lifetime in this region, $\tau_a\gtrsim~10^{26}~\textrm{s}$, is sufficiently long to satisfy the stringent bounds on decaying DM. In this context, see also the more detailed discussion of the QCD axion in Section~\ref{sec:QCD}.

For the minimal flavor violation benchmark (Fig.~\ref{fig:plotma}, right), the bounds on $f_a$ are a factor of ${\cal O}(10^{-5})$ less stringent. This is a direct consequence of the CKM suppression required for the flavor violating transition, cf. Eq.~\eqref{eq:EWestimate}. In this case the constraints from $K^+\to\pi^+a$ decays are relevant exactly in the SN trapping regime $(f_a \lesssim 10^6~\GeV)$ where SN cooling through ALP emissions is no longer possible, since ALPs get trapped inside the proto-neutron star. At low masses ($m_a\lesssim 1$~MeV) this region is excluded by cosmological constraints coming from CMB and BBN~\cite{Cadamuro:2012rm,Cadamuro:2011fd,Depta:2020wmr}. At higher masses, a detailed study of the BBN constraints for the ALP couplings considered here is still lacking. However, as an indication of roughly where the BBN constraint are expected to be applicable we delineate in Fig.~\ref{fig:plotma} with the solid line the region below which the ALP lifetime is $\tau<1$~s. This as a plausible guess of the parameter space that would survive the BBN constraints without invoking extremely small reheating temperatures or other highly non-standard cosmologies. See Ref.~\cite{Depta:2020wmr} for a careful assessment of the dependence of BBN constraints on the cosmological history.  

\paragraph{Consequences for the QCD axion}
\label{sec:QCD} 

The prototypical example of a light ALP is the QCD axion addressing the strong CP problem in the SM. As a consequence, the mass of the QCD axion predominantly arises from its couplings to SM gluons even though exceptions are possible in more elaborate models. In minimal scenarios the QCD axion mass can be directly related to the axion decay constant~\cite{GrillidiCortona:2015jxo} (redefining $f_a\to N_3 f_a$ one recovers the standard QCD axion notation):
\begin{align}
m_a = 5.7~\mu{\rm eV} \left( N_3\frac{10^{12}~\GeV}{f_a} \right).  
\end{align}
Phenomenological constraints in explicit models usually imply $f_a \gg 10^6~\GeV$, so that the QCD axion is essentially massless and stable for collider purposes. By definition the QCD axion couples to gluons, $N_3 \ne 0$, while couplings to fermions are model dependent. Flavor-violating couplings generically arise when Peccei-Quinn (PQ) charges are flavor non-universal. This possibility is especially well motivated if the PQ symmetry is (partially) responsible for explaining the  hierarchies in the SM Yukawa couplings~\cite{Ema:2016ops,Calibbi:2016hwq}. 

\begin{figure}[t]
\centering
\includegraphics[width=9.5cm]{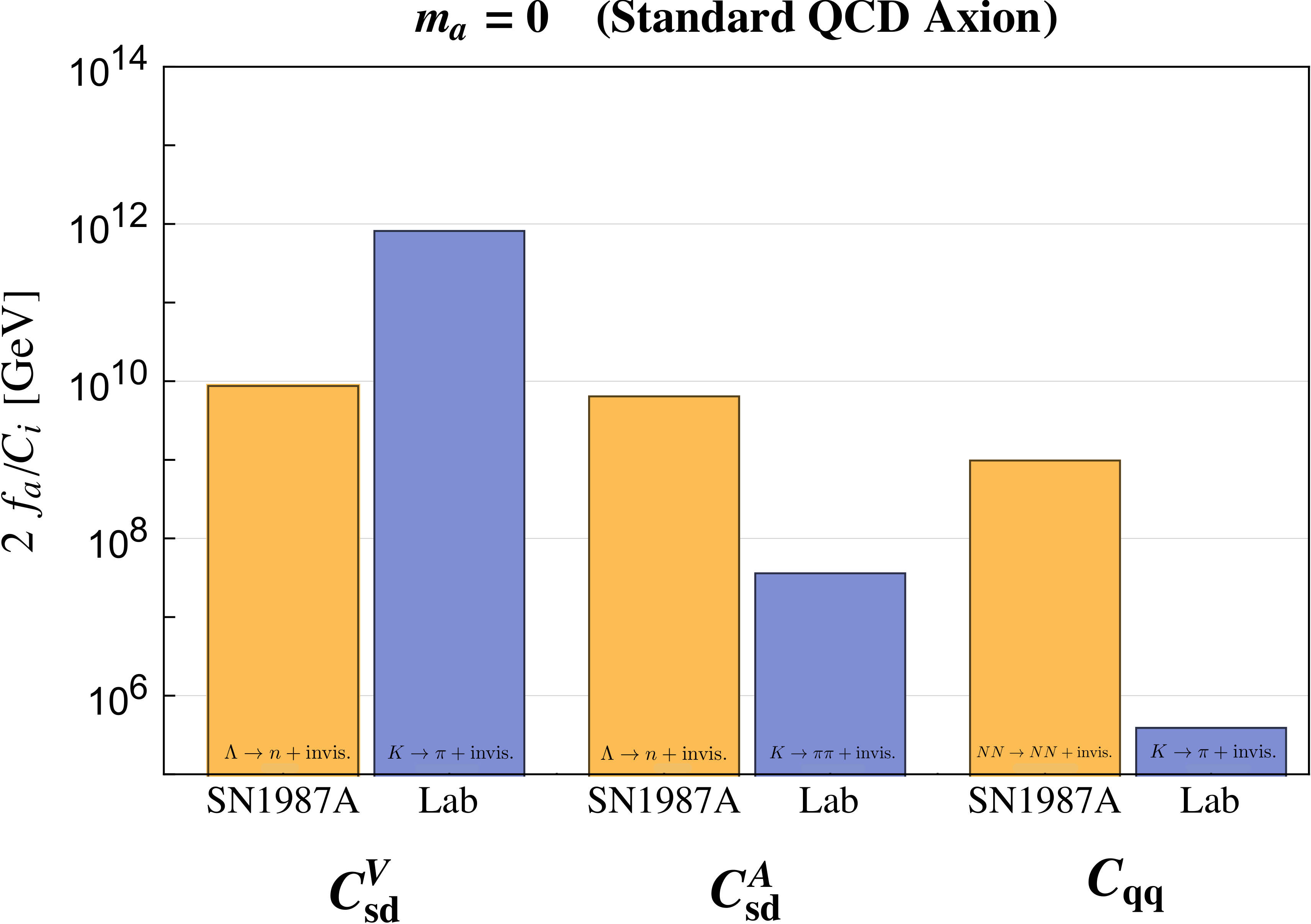}
\caption{\small Laboratory and astrophysiccal constraints on the inverse axion couplings for the case of a massless ALP, $m_a = 0$. These  apply unchanged also to the standard QCD axion. The SN1987A bounds are taken from Ref.~\cite{MartinCamalich:2020dfe} for the case of $C^V_{sd}$ and $C^A_{sd}$ (cooling via hyperon decays) and from Ref.~\cite{Carenza:2019pxu} for the case of $C_{qq}$ (cooling via nucleon bremsstrahlung). 
In contrast to the bounds from hyperon decays the nucleon bremsstrahlung has a trapping regime for smaller values of $f_a$. These values are excluded by Kamiokande constraints~\cite{Engel:1990zd}, 
leaving an open parameter space only for very small values of $f_a$, outside the plotted range. The laboratory bounds for $C^V_{sd}$ and $C_{qq}$ ($K \to \pi$ decays) are discussed in the main text, 
while the bounds on $C^A_{sd}$ ($K \to\pi\pi$ decays)~\cite{MartinCamalich:2020dfe} are from the ISTRA+ experiment~\cite{Tchikilev:2003ai}.}
\label{fig:plotma0}
\end{figure}

In Fig.~\ref{fig:plotma0} we compare the most stringent constraints on the QCD axion couplings at $\mu = \mu_c$. As illustrative examples we consider three distinct cases: i) only the FV vector coupling $C^V_{sd}$ is nonzero, ii) only the FV axial-vector coupling  $C^A_{sd}$ is nonzero, and iii) the combination of flavor-diagonal couplings $C_{qq} \equiv N_3 = C^V_{qq} = C^A_{qq}$ are all nonzero. For all three cases we show  the constraints both from SN cooling (``SN1987A'') as well as the constraints from the laboratory experiments (``Lab''). The SN cooling constraints are due to hyperon decays (for flavor-violating cases) and from nucleon bremsstrahlung (for the flavor-diagonal scenario). For nucleon bremsstrahlung the constraints come from observations of the SN neutrino flux, implying that proto-neutron star (PNS) did not have effective non-standard cooling mechanisms. The bounds from SN cooling apply for values above $f_a \gtrsim 15~{\rm TeV}$, since otherwise axions become trapped inside the PNS~\cite{Carenza:2019pxu}. 
Note that the constraints from axion production through hyperon decays do not have a trapping regime~\cite{MartinCamalich:2020dfe} and therefore are valid also in the large coupling (small $f_a$) regime. The laboratory constraints are due to the bounds on $K^+\to\pi^+X_{\rm inv}$ for the $C^V_{sd}$ and $C_{qq}$ cases, and from $K^+\to\pi^+\pi^0 X_{\rm inv}$ for the $C^A_{sd}$ case~\cite{MartinCamalich:2020dfe}, using the results of the ISTRA+ experiment~\cite{Tchikilev:2003ai}. The stronger constraints on $K_L\to\pi^0\pi^0 X_{\rm inv}$ decays from E391a~\cite{E391a:2011aa} do not apply to the $m_a=0$ case.
 
Fig.~\ref{fig:plotma0} shows that NA62 searches are sensitive to the QCD axion (or more generally, a very light invisible ALP) if it has sizable flavor-violating vector couplings. For flavor-violating axial couplings or the flavor-diagonal case the constraints from SN1987A are instead stronger than the NA62 reach by several orders of magnitude. This model-independent observation allows to test the way Peccei-Quinn $U(1)$ symmetry is embedded in the SM flavor group. In particular, if this abelian symmetry is identified with the symmetry explaining the Yukawa textures \`a la Froggatt-Nielsen then the resulting QCD axion could be probed simultaneously as DM by the cavity experiments such as ADMX and  via the $K^+\to \pi^+ a$ decays at NA62~\cite{Ema:2016ops,Calibbi:2016hwq}. If  $U(1)_{\rm PQ}$ is part of a non-abelian flavor group such as $U(2)$~\cite{Barbieri:1995uv, Barbieri:1997tu, Dudas:2013pja, Linster:2018avp, Linster:2020fww} the flavor-violating axion couplings can instead be strongly suppressed~\cite{Linster:2018avp}. 
 
\begin{figure}[t]
\centering
\includegraphics[width=11cm]{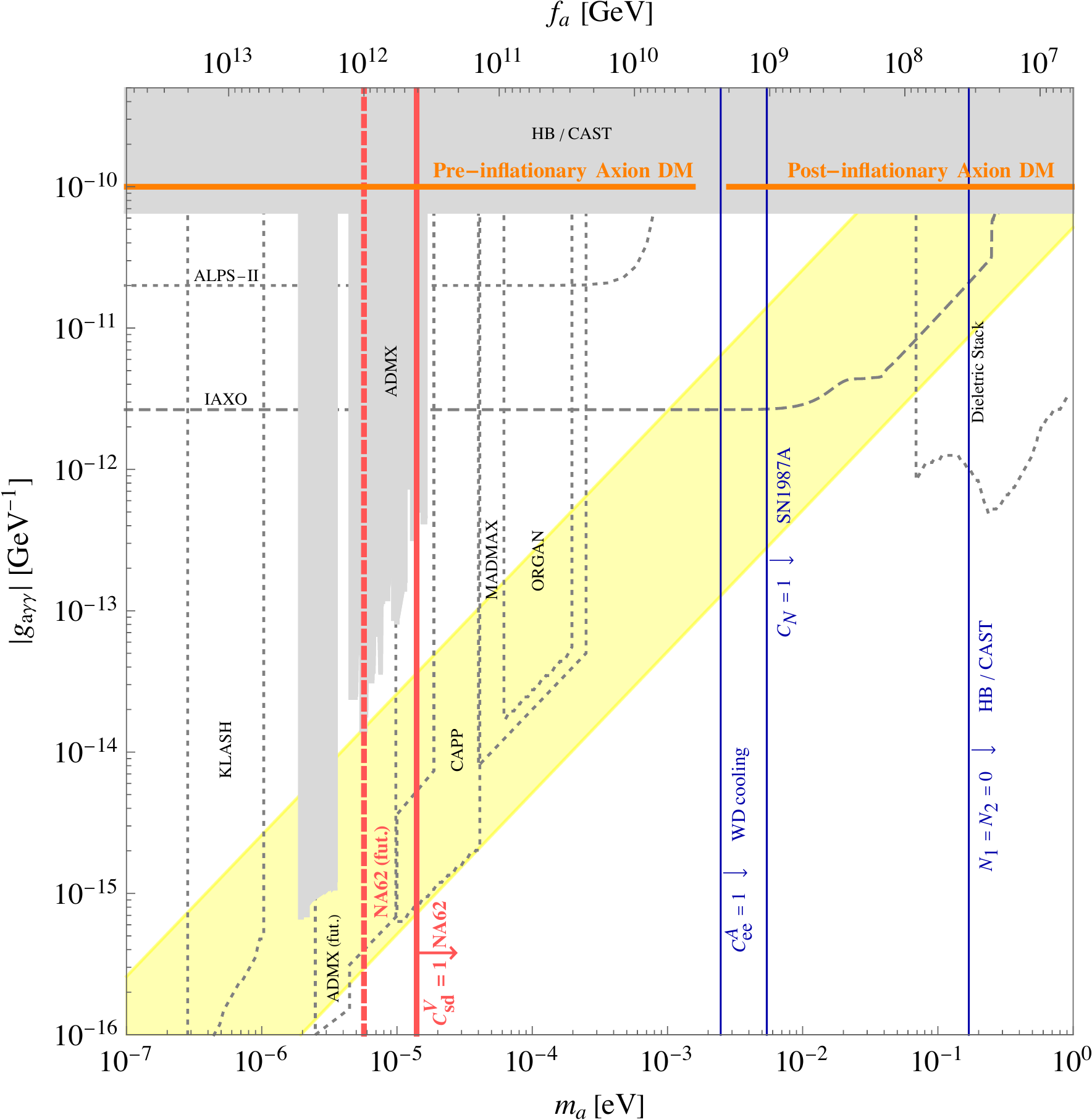}
\caption{\small Constraints on the QCD axion in the $m_a(f_a)$--$g_{a \gamma \gamma}$ plane. Shown in gray are the usual constraints from existing and planned halo- and helioscopes and from HB star cooling, overlaid is the standard QCD axion band in yellow. In addition, we show the constraints on $m_a/f_a$ for various couplings of order unity; displayed in blue are the HB/CAST bounds~\cite{CAST:2017uph, Ayala:2014pea} for $N_1 = N_2 = 0$ (i.e. the ``hadronic axion'' which couples to photons through its mixing with $\pi^0$), the bounds from star cooling via nucleon bremsstrahlung in SN1987A for $C_N =1$ \cite{Carenza:2019pxu}, and via electron bremsstrahlung in white dwarfs (WD) for $C^A_{ee} =1$ \cite{MillerBertolami:2014rka}. Red solid (dashed) lines show the present (future) constraints from NA62 for $C^V_{sd} =1$. In orange we indicate the values of $m_a$ (and $f_a$) for which the QCD axion can fully account for the observed dark matter abundance, depending on whether the PQ  symmetry is broken before or after inflation (taken from Ref.~\cite{Gorghetto:2020qws}). \label{fig:axionplot}}
\end{figure}

The NA62 sensitivity in the ($m_a$, $g_{a\gamma\gamma}$) plane obtained setting $C^V_{sd} = 1$ is shown in Fig.~\ref{fig:axionplot}. In this scenario NA62 can provide constraints on the axion decay constant $f_a$ of the order of $10^{12}$~GeV, which is about two orders of magnitude more stringent than the astrophysical bounds from white dwarf (WD) cooling (for $C^A_{ee} = 1$) and SN1987 (for $C_N = 1$), and about four orders of magnitude stronger than the constraints from CAST/HB stars (for $N_1 = N_2 = 0$). Therefore NA62 can potentially constrain the region in the parameter space for which the QCD axion yields the observed DM abundance via misalignment in the pre-inflationary scenarios, complementing the standard axion searches with haloscopes.

\paragraph{Probing axial flavor violating ALP couplings} 
\label{sec:fv_alp}

Three-body decays $K^+\to\pi^+\pi^0 a$ and $K_L\to\pi^0\pi^0 a$ are crucial laboratory probes of the axial FV coupling of the ALP. Assuming for simplicity $m_a\ll m_\pi$, the decay widths are given by~\cite{MartinCamalich:2020dfe} 
\begin{align}
\label{eq:Gamma:3body:K+}
& \frac{\text{d}\Gamma(K^+\to \pi^+\pi^0 a)}{\text{d}s}=\frac{(C_{sd}^A)^2}{f_a^2}\frac{(m_{K^+}^2-s)^3}{6144\pi^3m_{K^+}^5}\beta(F_p^2+\beta^2G_p^2+2\beta F_p G_p),\\
\label{eq:Gamma:3body:KL}
& \frac{\text{d}\Gamma(K_L\to \pi^0\pi^0 a)}{\text{d}s}=\frac{(C_{sd}^A)^2}{f_a^2}\frac{(m_{K_L}^2-s)^3}{4096\pi^3m_{K_L}^5}\beta F_s^2,
\end{align}
where $s=(p_{\pi_1}+p_{\pi_2})^2$, $\beta=1-4m_\pi^2/s$, and the functions $F_s$, $F_p$ and $G_p$ can be extracted from the $K^+\to\pi^+\pi^-e^+\nu$ decay form factors. The decay widths satisfy a generalized Grossman-Nir bound
\begin{equation}
{\cal B}(K_L\to\pi^0\pi^0 a)< 31\cdot {\cal B}(K^+\to\pi^+\pi^0 a),
\end{equation}
where the numerical factor on the r.h.s. is enhanced compared to the two-body generalized Grossman-Nir bound, Eq.~\eqref{eq:2body:GN}, by phase space effects and form factors. 

The experimental status is summarized in Section~\ref{sec:sig:K2pi}. The strongest limit on the massless ALP comes from the ISTRA+ experiment~\cite{Tchikilev:2003ai}:
\begin{equation}
\label{eq:K+:3body}
{\cal B}(K^-\to\pi^-\pi^0 a)<0.9\times 10^{-5}\quad \Rightarrow\quad 2f_a/C^A_{sd}>3.6\times 10^7\text{ GeV}.
\end{equation}
This is expected to be improved with the NA62 Run~1 dataset. The projected NA62 reach would still be below the nominal bounds from the SN cooling due to hyperon decays~\cite{MartinCamalich:2020dfe} (Fig.~\ref{fig:plotma0}).
However the SN cooling bounds carry an important caveat that they 
are likely affected by large systematic uncertainties related to the SN modeling (it is even possible to argue that the bound may be completely absent~\cite{Bar:2019ifz}).  
In addition to the above bound on the $K^-$ decay, there is a phenomenologically interesting experimental bound on ${\cal B}(K_L\to \pi^0\pi^0 a)$ from the E391a experiment~\cite{E391a:2011aa}, which is set to be improved by KOTO. This bound does not cover the massless ALP (see Section~\ref{sec:sig:K2pi}), while it leads to $2 f_a/C_{sd}^A\gtrsim 1.7\times 10^7$~GeV for a light, yet still massive, ALP~\cite{MartinCamalich:2020dfe}.

For massive ALP, the expressions for the decay widths in Eqs. \eqref{eq:Gamma:3body:K+}, \eqref{eq:Gamma:3body:KL} need to be modified, and include a new form factor that cannot be extracted from data (see Appendix C in Ref.~\cite{MartinCamalich:2020dfe}). While a prediction for this form factor can be made using chiral perturbation theory~\cite{Weinberg:1966zz,Bijnens:1989mr}, understanding of theoretical errors for the massive ALP case could benefit from further studies. 
In summary, experimental exploration of the axial flavor violating coupling of the ALP calls for updates on the experimental and on the theoretical fronts, both of which can be achieved  in the near future.

\subsubsection{Prompt ALPs from kaon decays}
\label{sec:prompt:ALP}

Experimentally distinct phenomenology is obtained if ALPs decay promptly within the  instrumented volume of a kaon experiment. This requires relatively low values of the ALP decay constant $f_a$.
For $m_a$ in the MeV to GeV range the ALP decay constant  should be within a few orders of magnitude of the EW scale,  $f_a\sim{\mathcal O}(100~\text{GeV})$. In this range of couplings the ALP is in a thermal equilibrium in the early Universe, and its mass is bounded to be heavier than roughly 1~MeV by BBN constraints on the number of relativistic degrees of freedom~\cite{Depta:2020wmr}. 

Depending on the specific model, different ALP decays are allowed. For light ALPs, the $a\to\gamma\gamma$ or $a\to e^+e^-$ decays dominate, with at most one loop factor hierarchy between the electron and the photon couplings. At one loop, the ALP-electron and ALP-photon couplings mix, cf. Eq.~\eqref{eq:cgamgameff}. If kinematically allowed, the $a\to\mu^+\mu^-$ decay should also be considered. 

Promptly decaying ALPs can be produced in kaon decays via their couplings to gluons, $W$-bosons, or light quarks. ALP flavor violating couplings can be safely ignored in this context because FCNC constraints from $K-\bar K$ mixing require $f_a\gtrsim$ $10^{7}$~GeV and hence imply an invisible ALP on detector scales (see Section~\ref{sec:FV:ALP:mue} for further discussion). Flavor diagonal couplings with a small enough decay constant would change appreciably the total kaon decay width. 
Summing the measured $K^+$ exclusive branching ratios gives a model-independent bound at 95\% CL\footnote{The PDG performs a constrained fit to $K^+$ branching ratios that includes the total decay width and also the requirement that the SM channels sum up to one, giving ${\cal B}(K^+\to\text{SM})=(100.00\pm0.15)\%$. Ignoring that a constrained fit was used this would imply  ${\cal B}(K^+\to\text{SM}+\text{new physics})<2.9\times 10^{-3}$ at 95\% CL. We expect that repeating the PDG fit, but without imposing the constraint, would give a constraint of a similar size. Until such a fit is performed the constraint in \eqref{eq:modelind} should be viewed as only indicative.}
\begin{equation}
{\cal B}(K^+\to\text{SM}+\text{new physics})\lesssim 3 \times 10^{-3}.
\label{eq:modelind}
\end{equation}
This provides a lower bound on the ALP decay constant that is independent of the ALP decay mode, and only depends on the production mechanism. For instance, for the ALP that couples only to gluons this is denoted with a dark green shaded exclusion region in Fig.~\ref{fig:cggcwwplots}~(left). 

As we will see below, dedicated experimental searches on exclusive channels already exclude significant parts of the prompt ALP parameter space. However, certain regions of parameter space still remain unexplored and deserve further attention. Typically, the parameter space corresponding to $c\tau_a\gtrsim 10$~cm is constrained by the beam-dump experiments (even though a complete study of the implications of these measurements on the ALP parameter space is still lacking), while for shorter lifetimes the searches for rare $K^+$ and $K_L$ decays become relevant. It is in the latter parameter region that the searches for {\it visible} ALP signatures at kaon experiments can make significant improvements, potentially completely covering  the parameter space left unexplored at small decay constant.

\begin{figure}[t]
\centering
\includegraphics[width=3in]{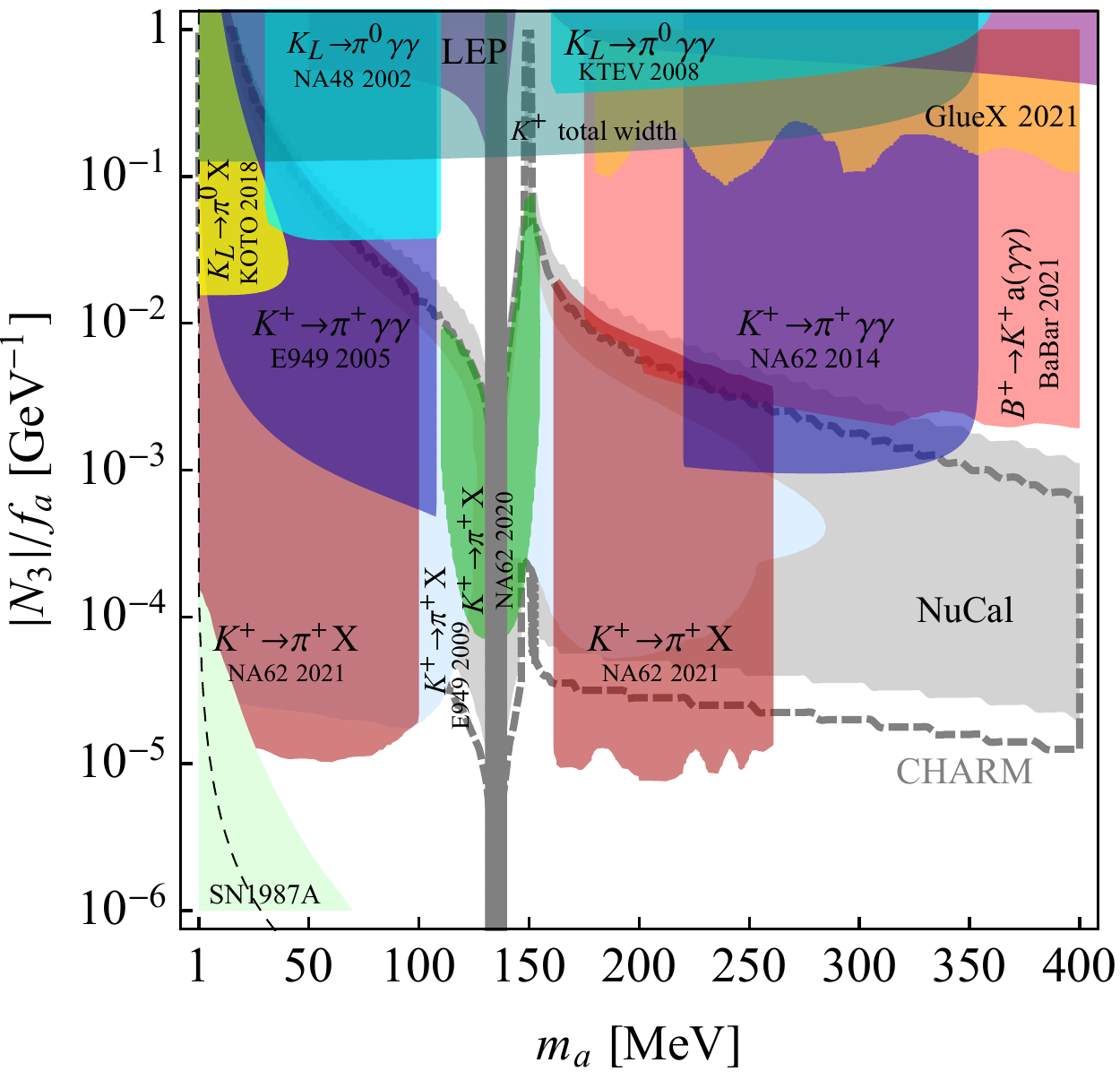}~~~~
\includegraphics[width=3in]{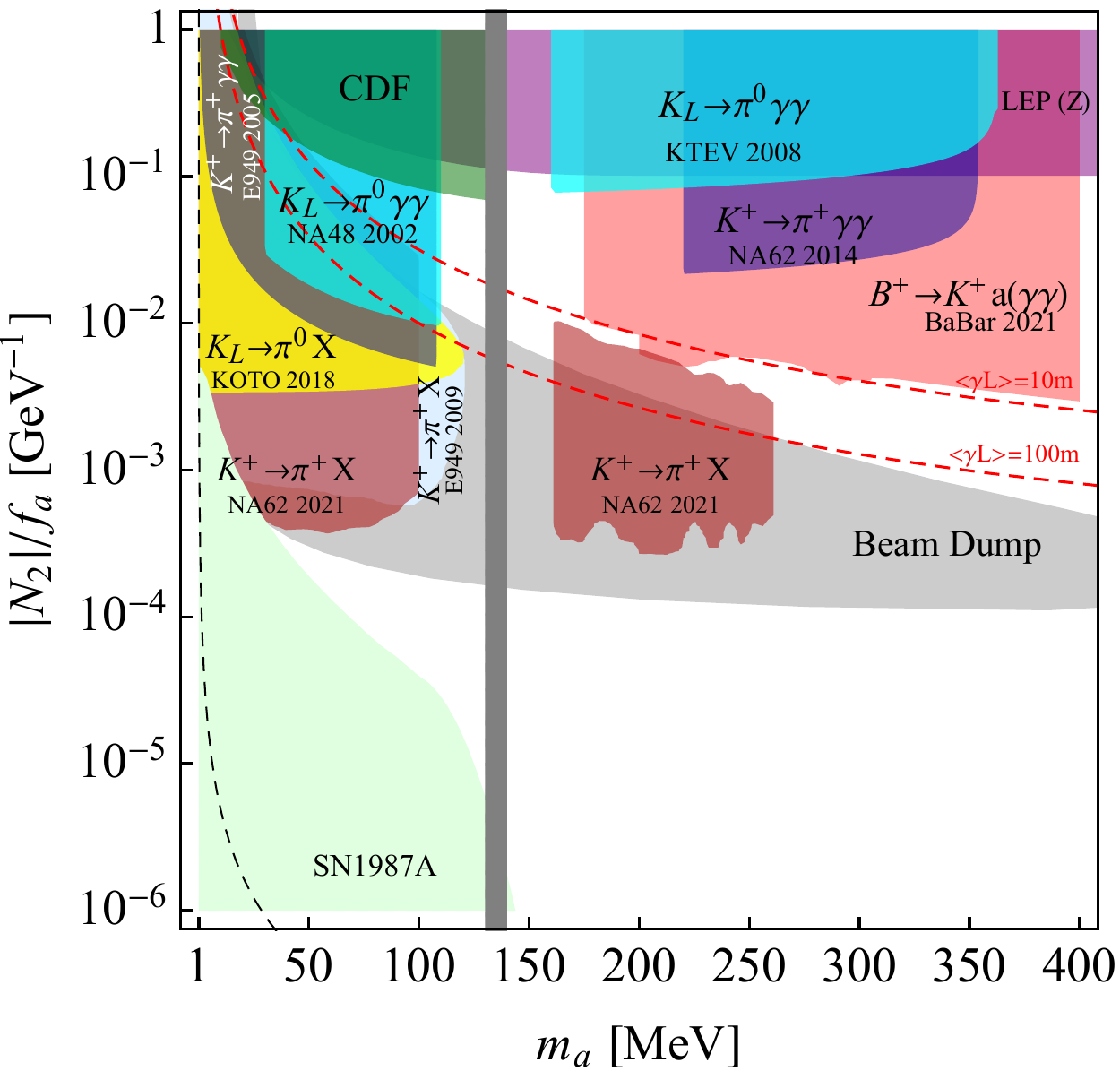}
\caption{\label{fig:cggcwwplots}
\small Left: Bounds on the ALP coupling only to gluons ($N_3(\Lambda_{\rm UV})\ne 0$ only) scenario. The coupling of ALP  to photons is induced at one loop, cf. Eq.~\eqref{eq:cgamgameff}, allowing for the $a\to \gamma\gamma$ decays. 
{The Babar bound was derived assuming $\Lambda_{\rm UV}=1$~TeV.}
Right: Bounds on the ALP coupling to only  $SU(2)_L$ gauge bosons ($N_2(\Lambda_{\rm UV})\ne 0$ only)  scenario~\cite{Gori:2020xvq}. In this case $a\to \gamma\gamma$ is generated already at tree level. In the parameter space region below the dashed line in both plots, the ALP has proper lifetime longer than a second, and is therefore disfavoured by Big Bang Nucleosynthesis considerations. The red dashed lines on the right-hand plot trace the points at which the ALP has an average lab frame decay length of 100~m and 10~m at NA62. These illustrate why the $K^+\to\pi^+X$ bound where $X$ is invisible in the laboratory does not extend to larger couplings, since the ALP becomes too prompt for the search to be sensitive. }
\end{figure}

\begin{figure}[t]
\centering
\includegraphics[width=8cm]{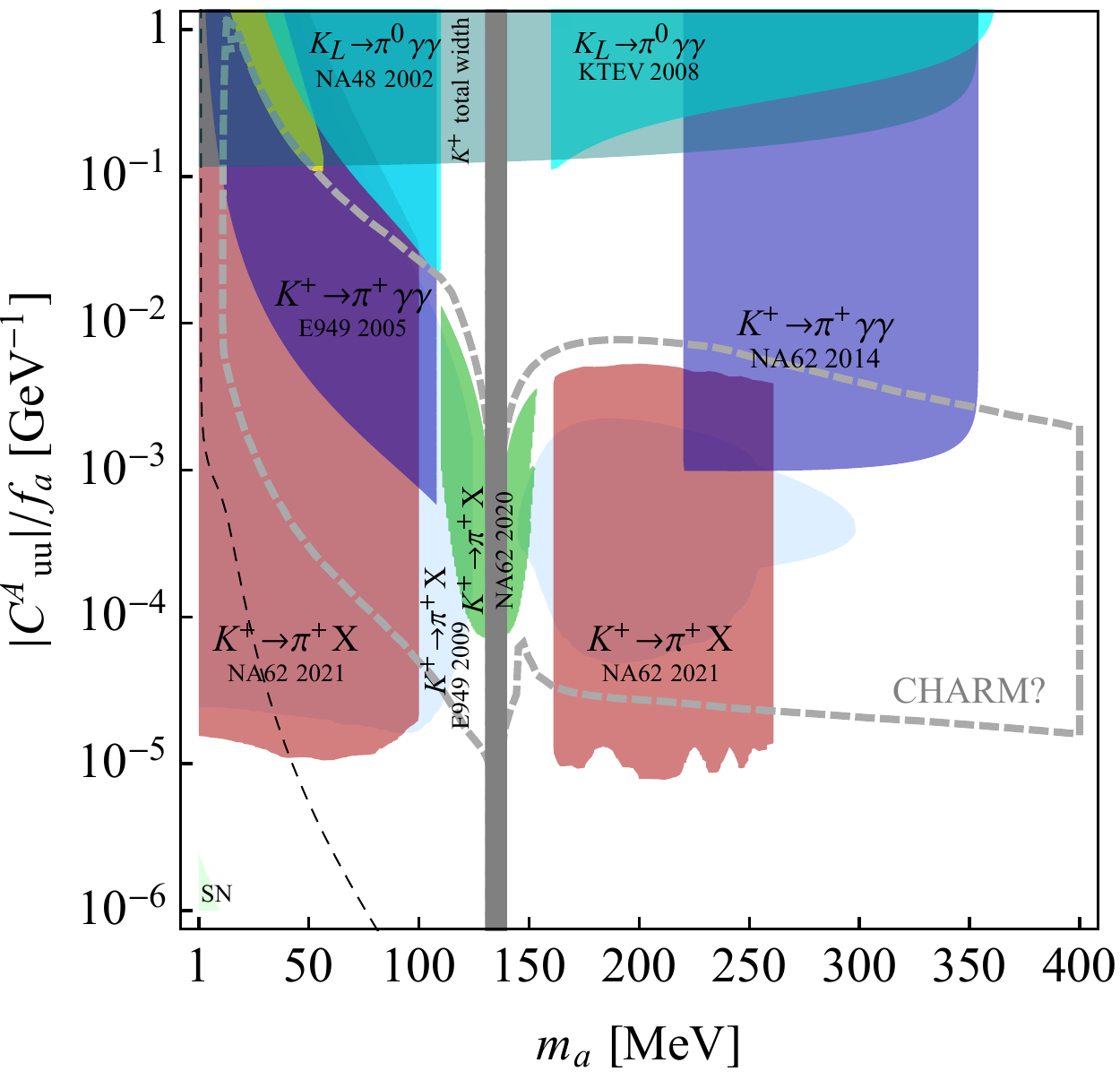}~~~~
\includegraphics[width=8cm]{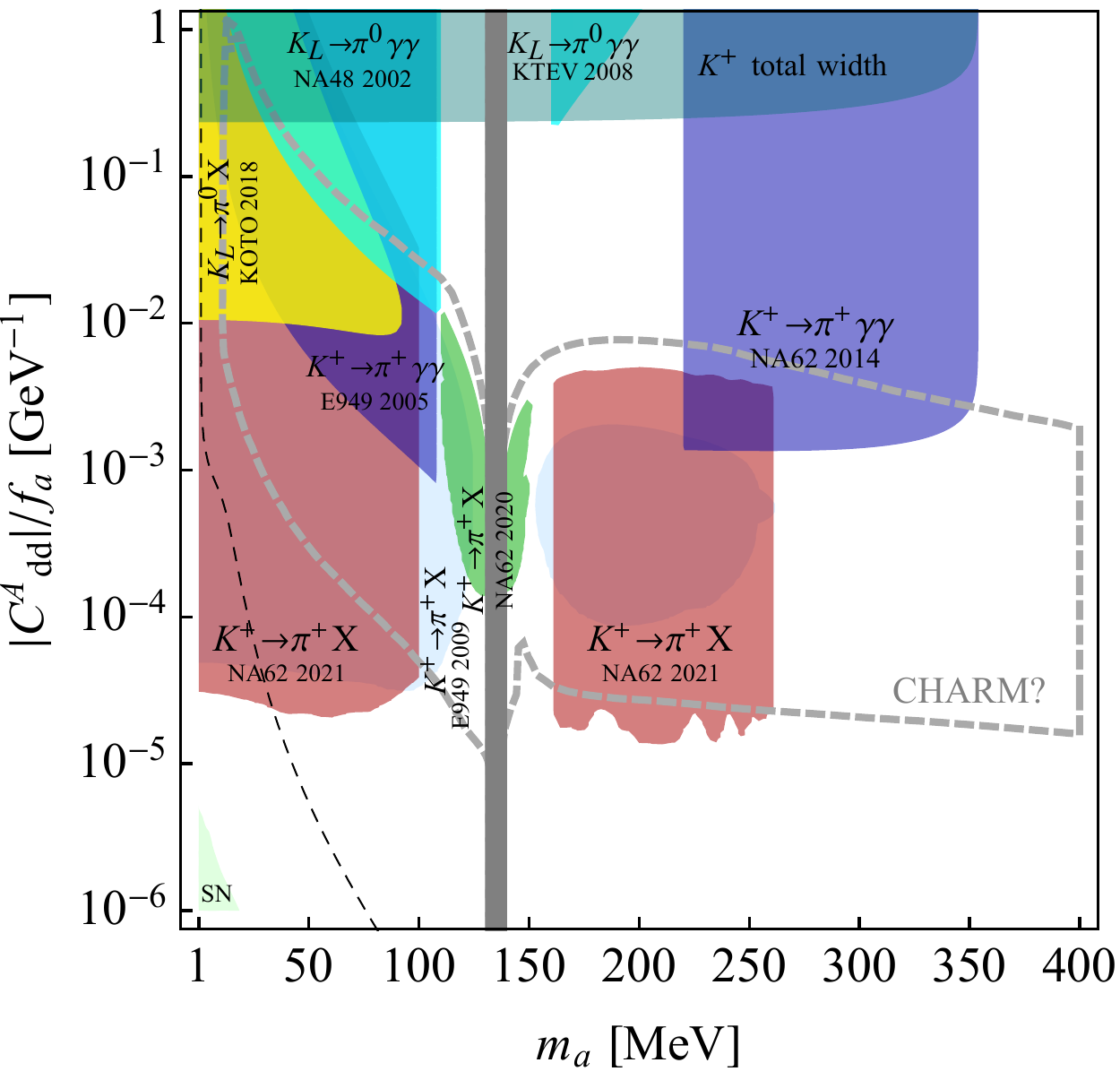}
\caption{\label{fig:quarkplots}
\small Bounds on flavor-diagonal pseudoscalar quark couplings of ALP to the first generation quarks: coupling to up quarks (left) and down quarks (right). In each case, it is assumed that only the relevant coupling is nonzero in the EFT Lagrangian. The coupling to photons is induced at loop level through Eq.~\eqref{eq:cgamgameff}, allowing the ALP to decay to two photons. In both plots,  the ALP has proper lifetime longer than a second in the parameter space region below the dashed line,  and is therefore disfavoured by Big Bang Nucleosynthesis.}
\end{figure}

\subsubsubsection{ALP promptly decaying to $\gamma\gamma$}
\label{sec:ALPdiphotondecay}
Light ALPs with couplings to gluons and quarks can be produced in kaon decays,  with the decay amplitudes given in Eqs.~\eqref{eqn:chargedKtopia} and \eqref{eqn:neutralKtopia}. Such an ALP then subsequently decays into photons, electrons or muons. If the ALP couples only to gluons or first-generation quarks at tree level, sizeable couplings to photons are generated at loop level, cf. Eq.~\eqref{eq:cgamgameff}, such that the ALP decays predominantly to photons. For large enough couplings, for which the ALP decay is prompt, the ALP parameter space can be constrained by the $K\to\pi\gamma \gamma$ measurements (Section~\ref{sec:sig:Kpigg}). In contrast, if the couplings are small enough, the ALP can be sufficiently long-lived to escape the detector, in which case $K\to\pi X$ searches with a missing energy signature (Section~\ref{sec:sig:Kpi}) become relevant.

In the following we consider 
four distinct limiting cases where a single coupling of ALP to the SM dominates at the scale $\mu=\Lambda_{\rm UV}=4\pi f_a$:
\begin{itemize}
\item[(a)] coupling to gluons, $N_3\ne 0$, only~\cite{Aloni:2018vki, Gori:2020xvq,Bauer:2021wjo,Bauer:2021mvw}, shown in Fig.~\ref{fig:cggcwwplots}~(left);
\item[(b)] coupling to $W^i$-bosons, $N_2\ne 0$ only~\cite{Izaguirre:2016dfi,Gori:2020xvq,Bauer:2021mvw}, shown in Fig.~\ref{fig:cggcwwplots}~(right); 
\item[(c)] $C_{uu}^A$ only: tree level coupling to only the first generation up quarks, shown in Fig.~\ref{fig:quarkplots}~(left);
\item[(d)] $C_{dd}^A$ only: tree level coupling to only the first generation down quarks, shown in Fig.~\ref{fig:quarkplots}~(right).
\end{itemize}
In each of the above four scenarios the ALP couplings to leptons are generated only through two loop radiative corrections and, if prompt, the ALP decays almost exclusively to photons so that we can take ${\mathcal B}(a\to\gamma\gamma)\simeq 1$. The experimental constraints for the four scenarios in the plane of ALP mass and relevant coupling are shown in Figs.~\ref{fig:cggcwwplots}, \ref{fig:quarkplots}. For large $f_a$ (small ALP couplings), the ALP becomes long-lived, and the bounds on ${\mathcal B}(K\to\pi X_{\rm inv})$ apply (Section~\ref{sec:sig:Kpi}). For smaller $f_a$ (larger ALP couplings), the ALP decays promptly to photons, and thus the bounds on  ${\mathcal B}(K\to\pi\gamma\gamma)$ apply (Section~\ref{sec:sig:Kpigg}). If ALP couples to charm or top quarks, the flavor violating $K\to \pi a$ decay is induced at one loop via the one loop generated flavour changing coupling, see Eq.~\eqref{eq:CdsV} and Fig.~\ref{fig:Prompt_ALP}. In this case, a large one-loop coupling to electrons is also induced~\cite{Bauer:2020jbp,Chala:2020wvs,MartinCamalich:2020dfe}, so that the ALP  decays predominantly to electrons.  Consequently, this scenario is already strongly constrained, as discussed below, in Section~\ref{sec:lldecaysAlves}. More exotic scenarios, such as the  lepton flavor violating ALP decays, are discussed in Section~\ref{sec:LPV}.

In many of the $m_a< m_K$ scenarios discussed here, the dominant constraint at small decay constant $f_a$ is the model independent bound on ${\mathcal B}(K^+\to\text{SM}+\text{ anything})$,  cf. Eq.~\eqref{eq:modelind}. Complementary bounds from other collider probes such as  LEP~\cite{L3:1995nbq,OPAL:2002vhf}, CDF~\cite{CDF:2013lma}, GlueX~\cite{GlueX:2017zoo,GlueX:2021myx} and $B$-factories \cite{Izaguirre:2016dfi, Aloni:2018vki, Chakraborty:2021wda, BaBar:2021ich, Bertholet:2021hjl}\footnote{The Babar search for $B\to K a(\to \gamma\gamma)$ excludes the large parameter space for the cases (a) and (b). The bound for (a) $N_3\neq 0$ is derived based on the calculation of \cite{Chakraborty:2021wda} which assumes $\Lambda_{\rm UV}=1$~TeV, not $4\pi f_a$. The bound will be stronger if $\Lambda_{\rm UV}=4\pi f_a$ because the RGE effect is enhanced. The projection of this channel at Belle~II is discussed in Ref.~\cite{Bertholet:2021hjl}.
}
 can also be relevant. For the large decay constant, the ALPs are long-lived, so different bounds associated with the long lifetime should be considered:  
{\it i)} from beam dump experiments, {\it ii)} $K\to \pi +$inv bounds, {\it iii)} supernova cooling bounds, {\it iv)} BBN bounds.
For an ALP coupled to gluons we present the bound from CHARM~\cite{CHARM:1985anb} and NuCal~\cite{Blumlein:1990ay} derived in Ref.~\cite{Jerhot:2022chi} by considering only the production mode through meson-ALP mixings.
We expect a similar bound to hold for the first generation up and down quark couplings. 
For an ALP coupled to EW gauge bosons we take the results from Ref.~\cite{Dolan:2017osp}.

\begin{figure}[t]
\begin{center}
\includegraphics[width=3in]{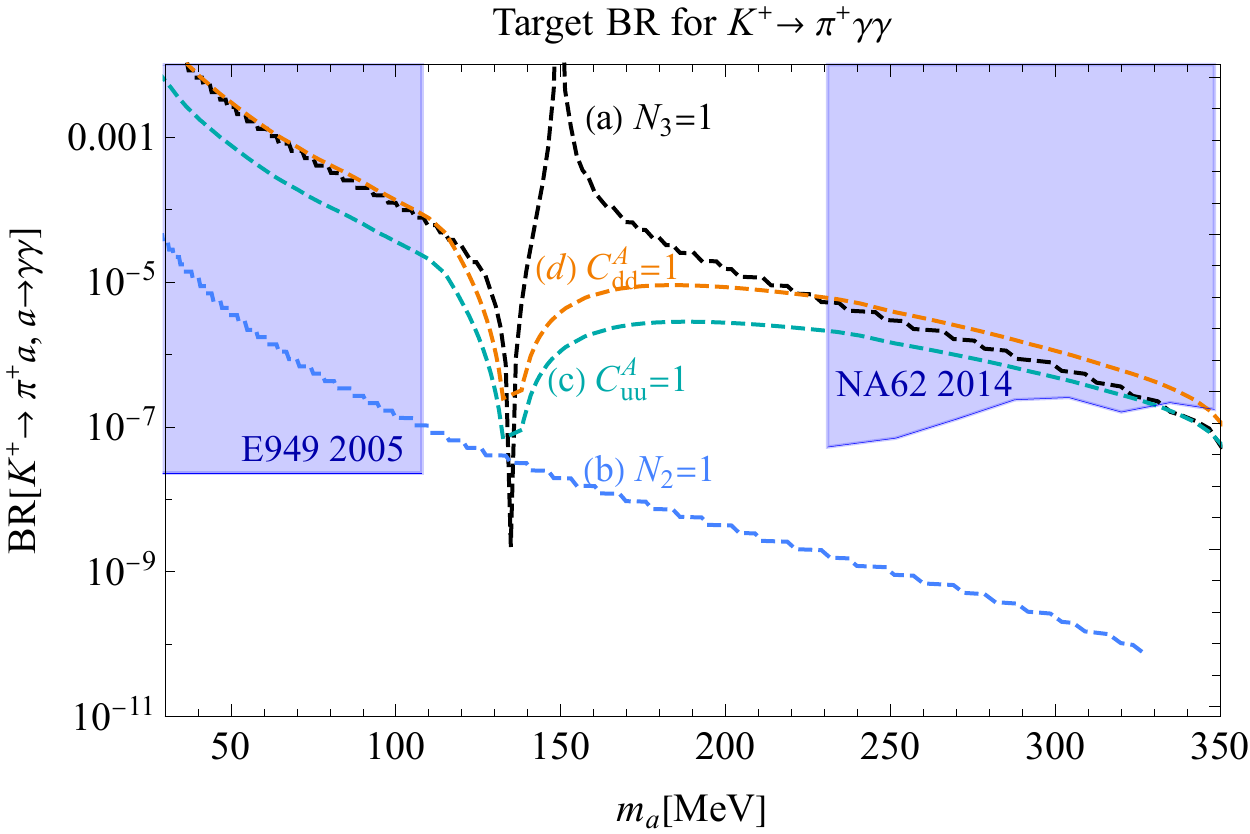}   \ \ 
\includegraphics[width=3in]{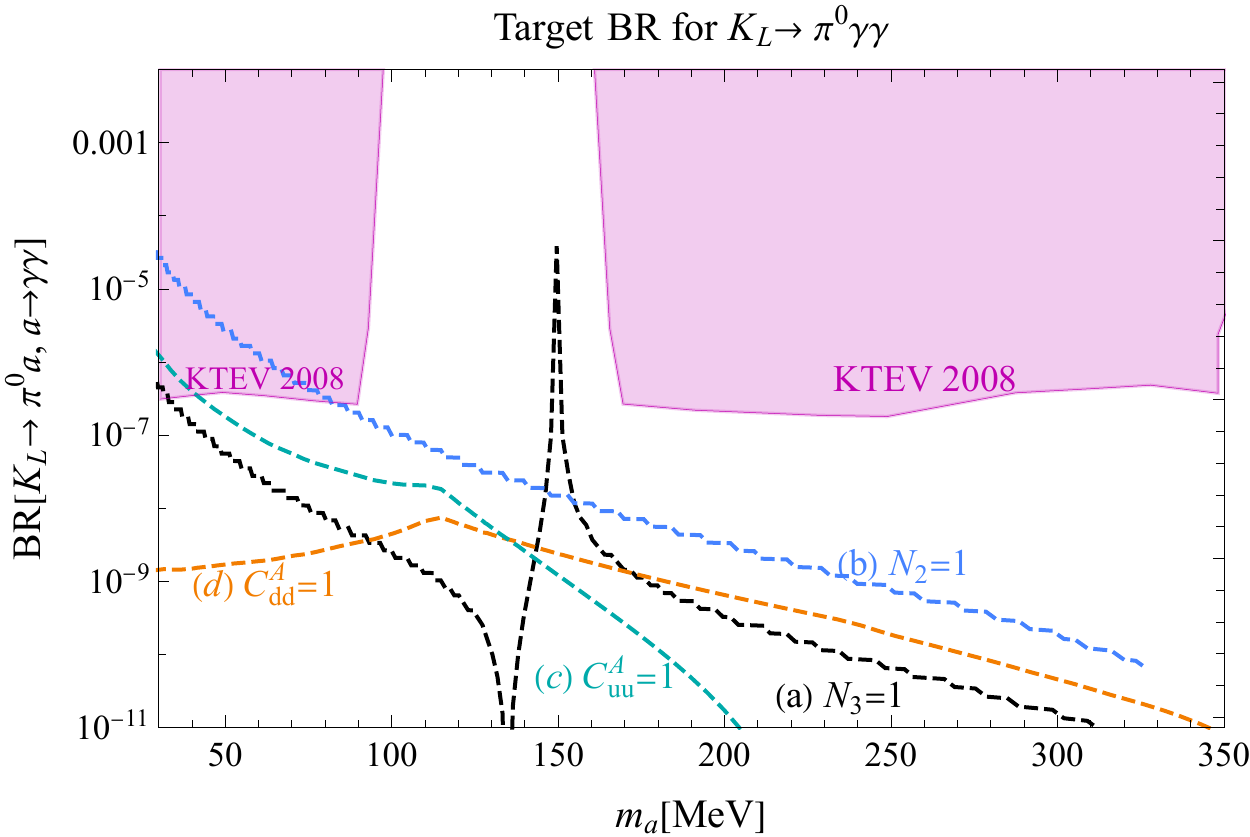}   
\caption{\small Target branching ratios for $K^+\to\pi^+a(\to \gamma\gamma)$ (left) and $K_L\to \pi^0 a(\to \gamma\gamma)$ (right) decay searches for four different benchmark scenarios: the ALP coupling only to gluons (scenario (a), black dashed line), only to $SU(2)_L$ gauge bosons ((b), blue dashed line),  to up quarks ((c), light blue dashed line) and to down quarks ((d), orange dashed line). In each case the regions below the dashed lines are excluded by the beam-dump constraints. The shaded regions are excluded by prompt searches, as indicated. 
   }
\label{fig:ALPvis} 
\end{center}
\end{figure}

In the window of intermediate $f_a$ the ALP decays promptly to photons, and thus the limits on ${\cal B}(K\to\pi\gamma\gamma)$ apply. The experimental status of these searches is discussed in Section~\ref{sec:sig:Kpigg}. As shown in Figs.~\ref{fig:cggcwwplots} and \ref{fig:quarkplots}, the visible signature of the ALP in $K\to \pi a (\to \gamma\gamma)$ is essential to close the gap between the model independent bound in Eq.~\eqref{eq:modelind} and the invisible searches. 
{We assume the ALP resonance is narrow for simplicity.}\footnote{However, the signal is broadened when the ALP decays with a significant displacement. For example, in $K^+\to \pi^+\gamma\gamma$ search, the diphoton invariant mass shifts to a lower value than the ALP mass as $m_{\gamma\gamma}\simeq m_a(1-d/L)$ where $L$ is the distance from the $K^+$ decay point to the ECAL, and $d$ is the displacement of ALP decay. Similar issue was discussed in the context of $B^\pm\to K^\pm a(\to \gamma\gamma)$ \cite{Bertholet:2021hjl, BaBar:2021ich}.}
To set target sensitivities for the future experiments, we evaluate in each scenario the value of the $K\to \pi a(\to\gamma\gamma)$ branching ratio corresponding to the boundary of the region already excluded by the beam dump experiments. These target branching ratios for $K^+$ and $K_L$ decays are shown in Fig.~\ref{fig:ALPvis}.
For $m_a\lesssim 100~\text{MeV}$, the promptly decaying ALP that couples predominantly to either gluons or to electroweak gauge bosons (scenarios (a) and (b)) is already excluded by the search at BNL-E949~\cite{E949:2005qiy}. For $m_a\gtrsim 230$~MeV, an ALP coupling to gluons is excluded by the $K^+\to\pi^+\gamma\gamma$ measurement at NA48/NA62~\cite{NA482:2013faq, NA62:2014ybm}, while a $10^{-10}$ sensitivity in the branching ratio is necessary for a definitive test of the electroweak scenario. Note that the recent Babar analysis excludes the large paramater space in a mass range $175 \lesssim m_a \lesssim 260$~MeV for the gluon  and  electroweak scenarios~\cite{BaBar:2021ich,Bertholet:2021hjl}. Similarly, reducing the bounds on ${\cal B}(K_L\to \pi^0\gamma\gamma)$ to $\sim 10^{-10}$ at KOTO or KLEVER would test the scenarios (a) and (b) for $m_a\gtrsim 170$~MeV.

A particularly difficult mass window to cover is the mass region $100~\text{MeV}<m_a<200~\text{MeV}$, where the $K\to\pi a(\to \gamma\gamma)$ searches are affected by the $K\to 2\pi$ background. This mass interval is characterized by large mixing between the ALP and the pion which could be in tension with measurements of the neutral pion lifetime. For this reason a tiny mass window around the neutral pion mass is shaded in the plots. A more detailed theoretical study of this window would be required to properly assess the experimental chances of closing it. Interestingly, $B$ meson decays might close this mass window for an ALP coupled to gluons or electroweak gauge bosons by using the lifetime information to separate the ALP from the pion. Conversely, for an ALP coupled solely to up or down generation quarks the $b\to s$ transition will be highly suppressed and the $100~\text{MeV}<m_a<200~\text{MeV}$ window will have to be closed by kaon experiments.


\subsubsubsection{ALP promptly decaying to di-leptons}
\label{sec:ALPdilepton}
\begin{figure}[t]
\centering
\includegraphics[width=3in]{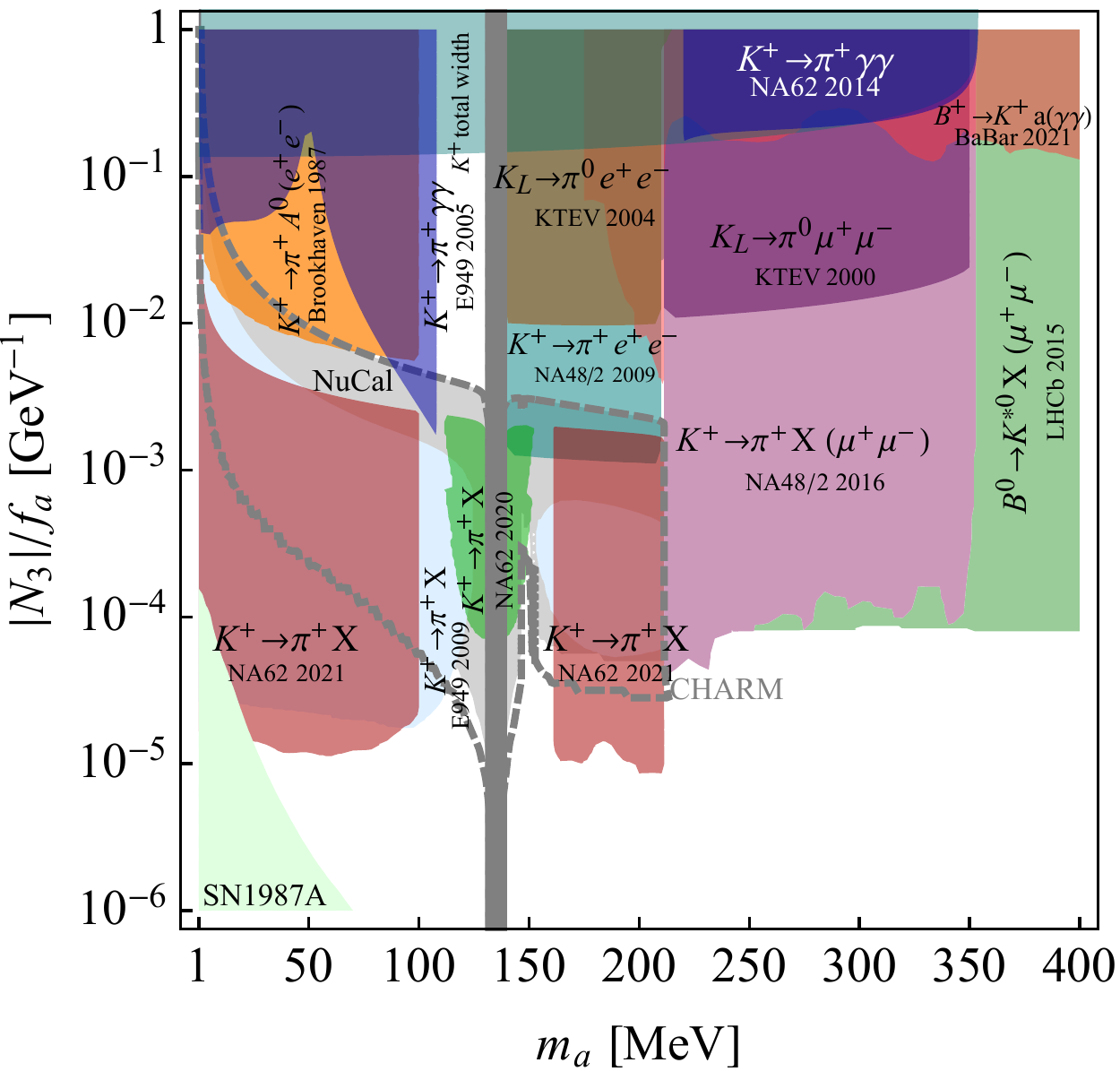}~~~~
\includegraphics[width=3in]{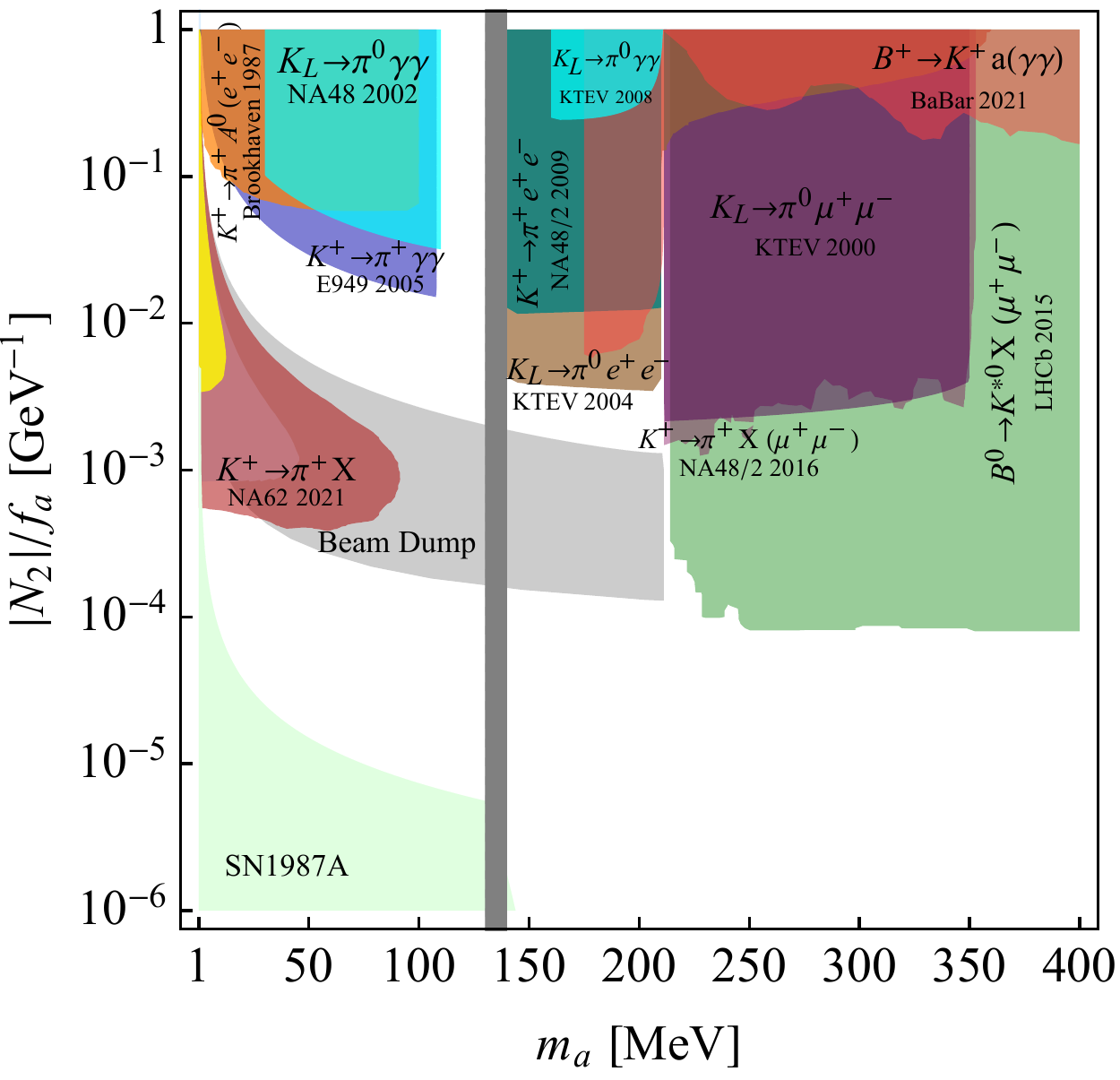}
\caption{
\label{fig:cggcwweeplots}
\small Left: Bounds on the coupling of an ALP to gluons for a scenario in which only the couplings to gluons and to leptons are nonzero in the Lagrangian. The coupling to leptons is set by $C^A_{ll}=N_3$ and is lepton universal. The coupling to photons is induced at loop level through equation~\eqref{eq:cgamgameff}, allowing the ALP to decay to two photons. Right: Bounds on the coupling of an ALP to $SU(2)_L$ gauge bosons for a scenario in which only this coupling and the coupling to leptons are nonzero in the Lagrangian~\cite{Gori:2020xvq}. The coupling to leptons is set by $C^A_{ll}=N_2$ and is lepton universal. The ALP decay to photons occurs at tree level through the $N_2$ coupling. }
\end{figure}
\begin{figure}[ht]
\begin{center}
\includegraphics[width=2.9in]{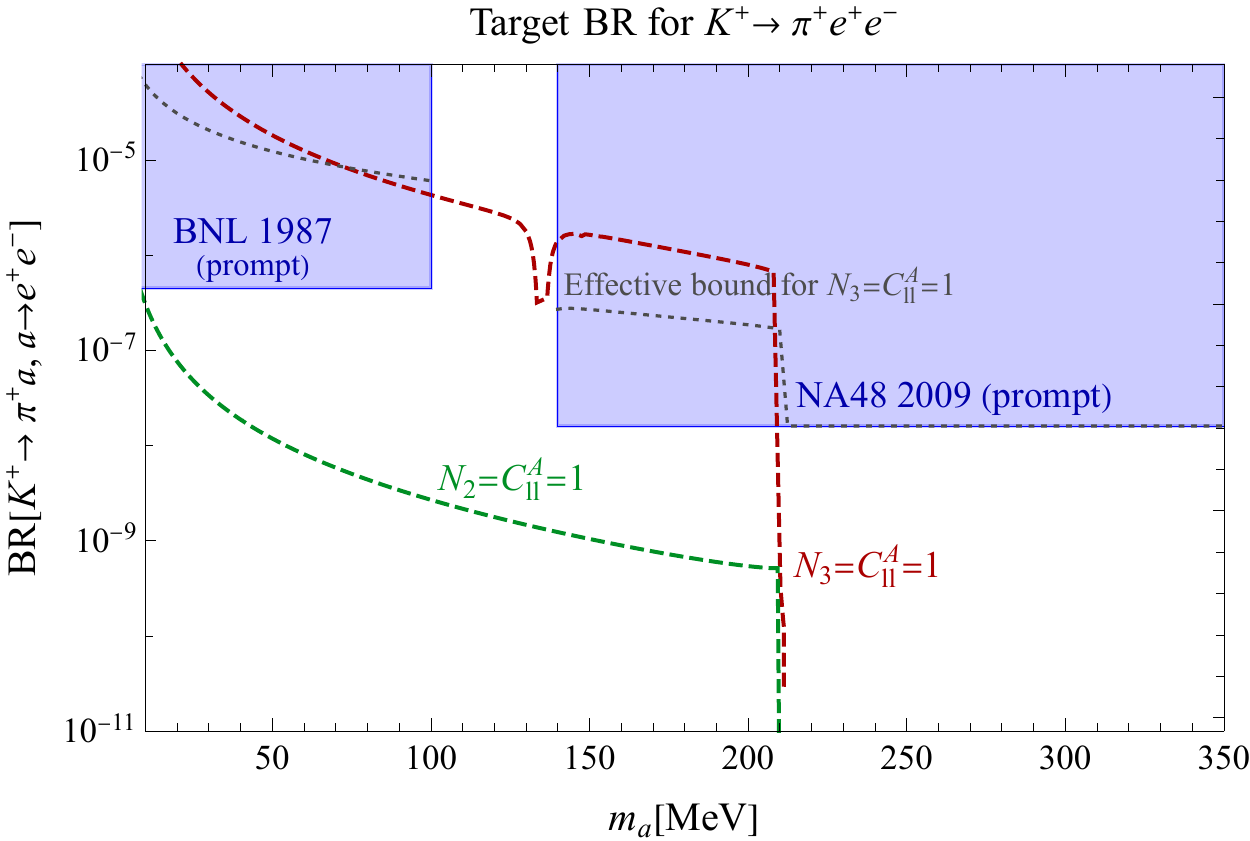}   \ \ 
\includegraphics[width=2.9in]{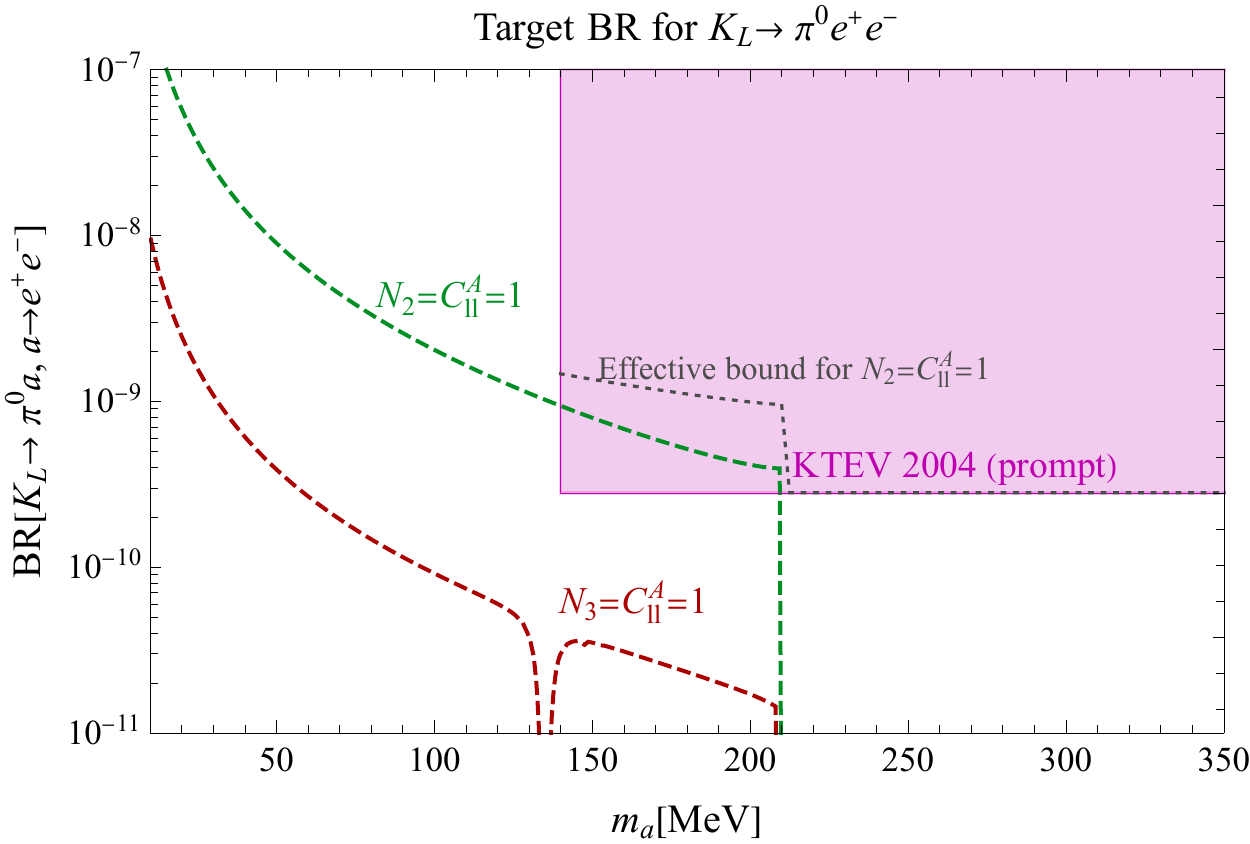}   
\caption{
\small Target branching ratios for $K^+\to\pi^+a(\to e^+e^-)$ (left) and $K_L\to \pi^0 a(\to e^+e^-)$ (right) decay searches for two different benchmark scenarios: the ALP couplings to gauge bosons, either $N_2$ or $N_3$, where the ALP also couples (diagonally and flavour-universally) to leptons. In each case, the regions below the dashed lines are excluded by the beam-dump searches. The shaded regions are excluded by prompt decay searches, as indicated. Due to the displacement of the ALP decay, there is sometimes a discrepancy between the bound assuming prompt decay and the effective bound on each scenario. The dotted lines show the effective bounds for the $N_3=C_{ll}^A$ scenario of $K^+\to\pi^+ e^+e^-$, and the $N_2=C_{ll}^A$ scenario of $K_L\to\pi^0 e^+e^-$.}
\label{fig:ALPtoeetarget} 
\end{center}
\end{figure}

Measurements of $K^+\to\pi^+e^+e^-$~\cite{E865:1999ker, NA482:2009pfe} and $K^+\to\pi^+\mu^+\mu^-$~\cite{NA482:2010zrc,Bician:2020ukv} decays, dedicated peak searches in the $m_{ee}$~\cite{PhysRevLett.59.2832} and $m_{\mu\mu}$~\cite{NA482:2016sfh} spectra of these decays, and searches for the $K_L\to\pi^0 e^+ e^-$~\cite{KTeV:2003sls} and $K_L\to\pi^0\mu^+\mu^-$~\cite{KTEV:2000ngj} decays lead to constraints on the ALP parameter space.
  The details and prospects for these measurements are discussed in Section~\ref{sec:sig:Kpill}. To obtain the corresponding constraints on ALP production and decay, we use  the two sigma upper range on the total number of measured events in the measurements and searches of Refs.~\cite{E865:1999ker, NA482:2009pfe,KTeV:2003sls,KTEV:2000ngj}, and use the peak-search results from Refs.~\cite{PhysRevLett.59.2832,NA482:2016sfh}. The ALP lifetime tends to exceed 10~ps, especially for $m_a<2m_\mu$. To estimate the fraction of ALP decays that pass the event selections we assume that the ALP energy is half of the mean kaon energy, and keep the ALP signal that has a displacement less than $1\,\text{m}$ at KTeV and NA48/2 ($1.4\,\text{cm}$ at BNL E865).

In Fig.~\ref{fig:cggcwweeplots} we show two scenarios in which at $\Lambda_{\rm UV}$ in addition to the ALP couplings to gauge bosons, either $N_2$ or $N_3$, the ALP also has a (diagonal and flavour-universal) coupling to leptons, $C_{\ell \ell}^A=N_{2,3}$, large enough to dominate the ALP decay width. In the two scenarios, the ALP branching ratio to photons ($a\to\gamma\gamma$) is reduced compared to Fig.~\ref{fig:cggcwwplots} due to the availability of the leptonic decay channels.

In Fig.~\ref{fig:ALPtoeetarget} we show the target $K\to\pi a (\to e^+e^-)$ branching ratios for future experiments, required to reach the lower boundary of the parameter space already excluded by the beam dump experiments (while the constraints from kaon invisible searches are weaker).
The sensitivity of the $K\to\pi a(\to e^+e^-)$ searches offered by the currently available samples is potentially sufficient to cover the gap to the region of parameter space in which ALPs have very displaced decays and where ALPs thus appear mainly as missing momentum in the experiments. In previous experimental analyses,  the decay products were assumed to originate from a single prompt decay vertex, which reduces the acceptance to the actual signal events that comprise of displaced $a\to e^+e^-$ decay vertices. The effective bounds for the $N_3=C_{ll}^A$ scenario from $K^+\to\pi^+ e^+e^-$ decays and for the $N_2=C_{ll}^A$ scenario from $K_L\to \pi^0 e^+e^-$ searches (shown as dotted lines in Fig.~\ref{fig:ALPtoeetarget}, with regions above the lines excluded) take the reduced acceptance into account, and  differ significantly from the bounds that assume prompt decays (shaded regions in Fig.~\ref{fig:ALPtoeetarget}). The unexcluded parameter space above the beam-dump constraints (dashed lines in Fig.~\ref{fig:ALPtoeetarget}) corresponds to ALP decays with large vertex displacements. Future searches with rare kaon decays would therefore optimally relax the decay vertex requirement, and perform scans in both ALP mass and lifetime.

Note that beam-dump constraints weaken above the di-muon threshold, thus the $K\to\pi a (\to\mu^+\mu^-)$ searches represent a unique opportunity to cover the unexplored parameter space for $m_a>210$~MeV. In Fig.~\ref{fig:cggcwweeplots} we show the present bounds from $B\to K a(\gamma\gamma)$ derived in Ref.~\cite{Chakraborty:2021wda} and we expect searches for $B\to K a(e^+e^-)$ and $B\to K a(\mu^+\mu^-)$ to also have an impact on this parameter space. Currently, we include the $B^0\to K^{0*} a(\to \mu^+\mu^-)$ search at LHCb~\cite{LHCb:2015nkv}, which is one of the leading constraints for $m_a>2m_\mu$.


\subsubsection{A simplified model of prompt ALP decays to $\ell^+\ell^-$ final states} \label{sec:lldecaysAlves}

Prompt leptonic ALP decays (with $c\tau_a\lesssim\mathcal{O}(\text{m})$) are only possible for sufficiently low decay constants $f_a/C_{ee}^A\lesssim\mathcal{O}$(TeV), see Eq.\,\eqref{eq:ctau_a:leptons}. Such low $f_a$ values face stringent constraints. Most notably, the flavor off-diagonal couplings $C_{q_i q_j}^A$ ($i\neq j$) must be strongly suppressed to be compatible with bounds on flavor violation, see Fig.\,\ref{fig:plotma}.  In this case, it is natural to express the flavor-diagonal ALP couplings in the Yukawa basis with the ALP appearing as a phase of the fermion masses in the form $e^{i\,Q^{_\text{PQ}}_{f}\,a/f_a}m_f\,ff^c$, where $Q^{_\text{PQ}}_f$ 
is the PQ charge of the fermion species $f$. Such flavor alignment without flavor universality could be achieved, for instance, in UV completions where only right-handed SM fermions are charged under PQ symmetry. Performing a chiral rotation of the right-handed fermions $f^c$, these couplings can be mapped into the basis in Eqs.~(\ref{eq:alpgauge}--\ref{eq:ALP-f}) to yield $C^V_{f_i f_j}=C^A_{f_i f_j}=\delta_{ij}\,Q^{_\text{PQ}}_f$, $N_3 =N_3\big|_{\rm UV}+\sum_{i=1}^6 Q^{_\text{PQ}}_{q_i}$,~~~$N_2 = N_2\big|_{\rm UV}$,~~~$N_1 = N_{1}\big|_{\rm UV}-2\,\sum_{i\,=1}^3\big(Q^{_\text{PQ}}_{\ell_i}+\frac{4}{3}\,Q^{_\text{PQ}}_{u_i}+\frac{1}{3}\,Q^{_\text{PQ}}_{d_i}\big)$.

Strong constraints apply to this ALP parameter space for the mass range  $2m_e\lsim m_a \lsim 2m_\mu$.
\begin{itemize}
\item A PQ charge for the top quark larger than $\big|Q^{_\text{PQ}}_t\big| \approx f_a/(100\,\text{TeV})$ would lead to experimentally excluded rates for $b\to s a$ and $s\to d a$ transitions via one loop penguin diagrams. 
\item The PQ charges of charm and bottom quarks, $Q^{_\text{PQ}}_{c}$ and $Q^{_\text{PQ}}_{b}$, are significantly constrained by bounds on axionic decays of $B$, $D$, $\Upsilon$ and $J/\Psi$ mesons, as well as their contributions to ALP-pion mixing.
\item
ALP-pion mixing is strongly constrained by bounds on ${\cal B}(\pi^{+}\to e^{+}\nu\,(a\to e^{+}e^{-}))$ \cite{Eichler:1986nj} and by the observed value of ${\cal B}(\pi^{0}\to e^{+}e^{-})$ \cite{Abouzaid:2006kk}, which require the $a-\pi^0$ mixing angle $\big|\theta_{a\pi}\big|\lesssim 10^{-4}$. At LO in ChPT, $\theta_{a\pi}$ is given by:
 \begin{equation}
\theta_{a\pi}\,\approx\,
-\frac{(m_u\, Q^{_\text{PQ}}_{u}-m_d\,Q^{_\text{PQ}}_{d})}{(m_u+m_d)}\frac{f_\pi}{f_a}-
\frac{1}{2}\,\bigg(\!Q^{_\text{PQ}}_{s}+N_3\big|_{\rm UV}\!\bigg)\,\frac{(m_u-m_d)}{(m_u+m_d)}\frac{f_\pi}{f_a}\,+\,
\mathcal{O}\bigg(\!\frac{m_{u,d}}{m_s}\!\bigg)\frac{f_\pi}{f_a}\,,
\end{equation} 
where $f_\pi=92$ MeV and we assume $m_a\ll m_{\pi^{_0}}$. ALP-pion mixing could then be made compatible with experimental bounds if $Q^{_\text{PQ}}_{u}=2\,Q^{_\text{PQ}}_{d}=\mathcal{O}(1)$ and $N_3\big|_{\rm UV}=Q^{_\text{PQ}}_{s}=0$ ~\cite{Alves:2017avw}.
\end{itemize}
 
\begin{figure}[t]
\begin{center}
\includegraphics[width=1.075 \textwidth]{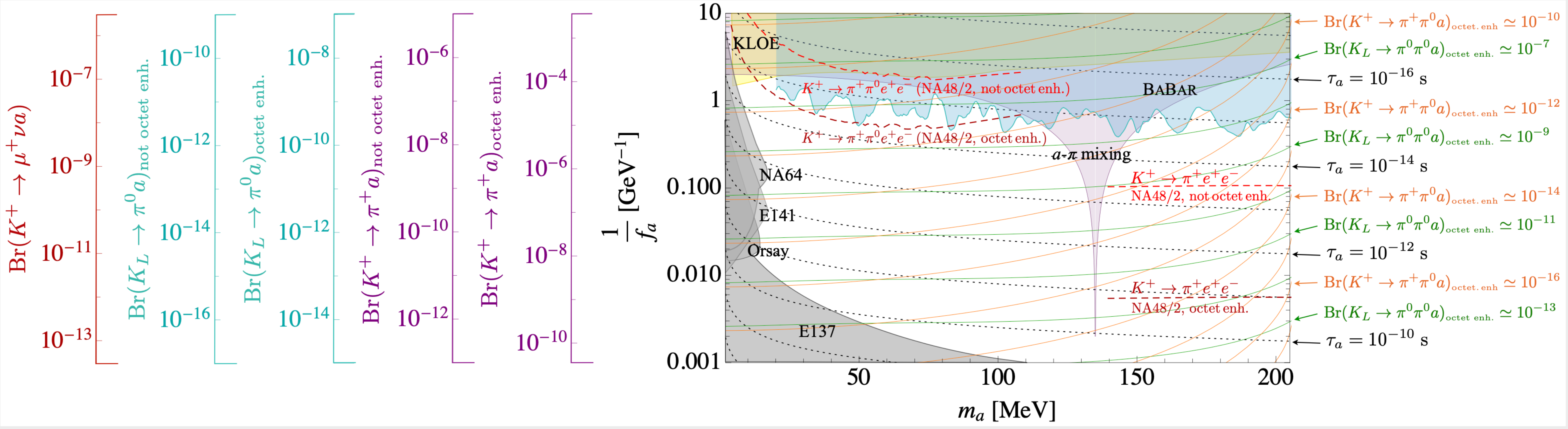}
\end{center}
\caption{\small
Order-of-magnitude range of ${\cal B}(K^+\to\pi^+a)$ (left purple axes), ${\cal B}(K_L\to\pi^0 a)$ (left cyan axes), ${\cal B}(K^+\to\mu^{+}\nu a)$ (left red axis), and contours of ${\cal B}(K^+\to\pi^+\pi^0 a)$ (orange curves), and ${\cal B}(K_L\to\pi^0\pi^0a)$ (green curves) in the parameter space of ALP mass ($m_a$) vs inverse decay constant ($1/f_a$) for the prompt $a\to e^+e^-$ benchmark scenario \eqref{paramBench}. Contours of ALP lifetime are shown as dotted black curves. Shaded regions are excluded by the indicated searches. Expected upper limits on ${\cal B}(K^+\to\pi^+(\pi^0)a)$ from the full NA48/2 dataset are shown as dashed dark (light) red curves. 
In the left axes we ignore the $\sim 30\%$ (factor of $\sim\!\!3$) variation of ${\cal B}(K\to\pi a)$ (${\cal B}(K^+\to\mu^+\nu a)$) with $m_a$. 
For $K\to\pi a$ decays we show the estimated branching ratios with and without the assumption of $\Delta S=1$ octet-enhancement of these amplitudes (see main text and Ref. \cite{Alves:2020xhf}).}
\label{fig:K_to_ee_prompt}
\end{figure}

The suppression of $\theta_{a\pi}$ combined with poorly determined $\mathcal{O}(1)$ corrections to $a-\eta$ and $a-\eta^{_\prime}$ mixing at second order in the chiral expansion imply that the naive LO ChPT estimation of hadronic ALP couplings is not reliable \cite{Alves:2017avw}. Therefore, in this subsubsection we consider a simplified model in which we parametrize the low energy hadronic ALP couplings via a generalized Bardeen-Tye current \cite{Bardeen:1977bd,Alves:2017avw,Alves:2020xhf}, which is defined by the ALP-meson mixing angles $\theta_{a\pi}$, $\theta_{a\eta_{ud}}$, and $\theta_{a\eta_{s}}$. 

Axionic kaon decay amplitudes can then be calculated via $\mathcal{A}(K\to aX)=\sum_\phi\theta_{a\phi}\;\mathcal{A}(K\to \phi^* X)\big|_{p^2_{\phi}=m_a^2}$, with $\phi=\pi^0, \eta_{ud}, \eta_s$ (the SM amplitudes can be found in~\cite{Alves:2020xhf}). 
As illustrative target benchmarks for experimental searches, we take
\begin{equation}\label{paramBench}
 \big|Q^{_\text{PQ}}_e\big|=1~,\qquad \big|\theta_{a\pi}\big|\sim\mathcal{O}\left(\frac{50\text{ keV}}{f_a}\right)~,\qquad \big|\thetaUD\big|\sim\big|\thetaS\big|\sim\mathcal{O}\left(\frac{\text{MeV}}{f_a}\right),
\end{equation}
in anticipation that improved ChPT predictions will map these low energy couplings into a nonvanishing region of the UV parameter space.

The resulting branching ratio predictions can still show significant variation due to the possibility of interference between the amplitudes $\mathcal{A}(K\to\phi^*X)$. Therefore in plotting order-of-magnitude estimates for the various axionic kaon decay channels in Fig.~\ref{fig:K_to_ee_prompt} we assume the following branching ratios, consistent with (\ref{paramBench}):
\begin{subequations}\label{BRee}
\begin{alignat}{3}
{\cal B}(K^+\to\pi^+a)~&\simeq&&~\left(\frac{\text{GeV}}{f_a}\right)^2\,\frac{\,|\vec{p}_a|\,}{|\vec{p}_{\pi^{_+\!}}|}~ \times~
	\begin{cases}
  	    ~3.6\times10^{-5}\,&\text{(octet-enhanced)},\\
  	    ~~~~~~~~~~~10^{-7}\,&\text{(non octet-enhanced)},
   	 \end{cases}\label{BR2bodyKplus}\\
{\cal B}(K_L\to\pi^0a)~&\simeq&&~\left(\frac{\text{GeV}}{f_a}\right)^2\,\frac{\,|\vec{p}_a|\,}{|\vec{p}_{\pi^{_0\!}}|}~ \times~
	\begin{cases}
  	    ~10^{-9}\,&\text{(octet-enhanced)},\\
  	    ~10^{-11}\,&\text{(non octet-enhanced)},
   	 \end{cases}\\
{\cal B}(K^+\to\mu^+\nu a)~&\simeq&&~~3 \times 10^{-8}\;\left(\frac{\text{GeV}}{f_a}\right)^2\;\times\;\Phi(m_a)/\Phi(m_a=10\;\text{MeV}), \label{BrKaxioleptonic}
\end{alignat}
\end{subequations}
where $\Phi(m_a)$ is the Dalitz phase-space integral. The predictions for ${\cal B}(K^+\to\pi^+a)$ and ${\cal B}(K_L\to\pi^0a)$ differ in whether $\Delta S=1$ octet-enhancement (a.k.a. $\Delta I=1/2$ enhancement) is assumed to arise at $\mathcal{O}(p^{_{^2}}\!)$ (octet-enhanced) or at $\mathcal{O}(p^{_{^4}}\!)$ (non-octet-enhanced) in ChPT power counting (see \cite{Alves:2017avw,Alves:2020xhf} for details). Fig.~\ref{fig:K_to_ee_prompt} shows branching ratio contours for the three-body decays $K\to\pi\pi a$, where we assume the same relative sign for $\theta_{a\pi}$, $\thetaUD$ and $\thetaS$, and we neglect the contributions of non-octet-enhanced amplitudes.

Sensitivity projections at 90\% CL for ${\cal B}(K^+\to\pi^+a)$ and ${\cal B}(K^+\to\pi^+\pi^0a)$ from the NA48/2 analyses of $K^+\to\pi^+e^+e^-$~\cite{Batley:2009aa} and $K^+\to\pi^+\pi^0e^+e^-$~\cite{NA482:2018gxf} are also shown in Fig.~\ref{fig:K_to_ee_prompt} (see Sections~\ref{sec:sig:Kpill}, \ref{sec:sig:Kpipiee}). These projections assume that: (i) the signal acceptance for $K^+\to\pi^+(a\to e^+e^-)$ is the same as in Fig.~4 of~Ref.\cite{Batley:2009aa}; (ii) the signal acceptance for $K^+\to\pi^+\pi^0(a\to e^+e^-)$ is the same as the acceptance for the IB decay shown in Fig.~3 of~Ref.\cite{NA482:2018gxf}; and (iii) the expected backgrounds are accounted for using the full background simulations in Refs.\cite{Batley:2009aa, NA482:2018gxf}. 
KTeV also placed a bound on the SM decay ${\mathcal B}(K_L\to\pi^0\pi^0e^+e^-)<6.6 \cdot 10^{-9}$ at 90\% CL~\cite{KTeV:2002tpo} (see Section \ref{sec:sig:Kpipiee}). However, due to the low signal efficiency resulting from background rejection cuts, recasting this analysis into bounds on ${\mathcal B}(K_L\to \pi^0\pi^0 a)$ would require a dedicated simulation of the ALP signal.

\begin{figure}[t]
\begin{center}
\includegraphics[width=1.075 \textwidth]{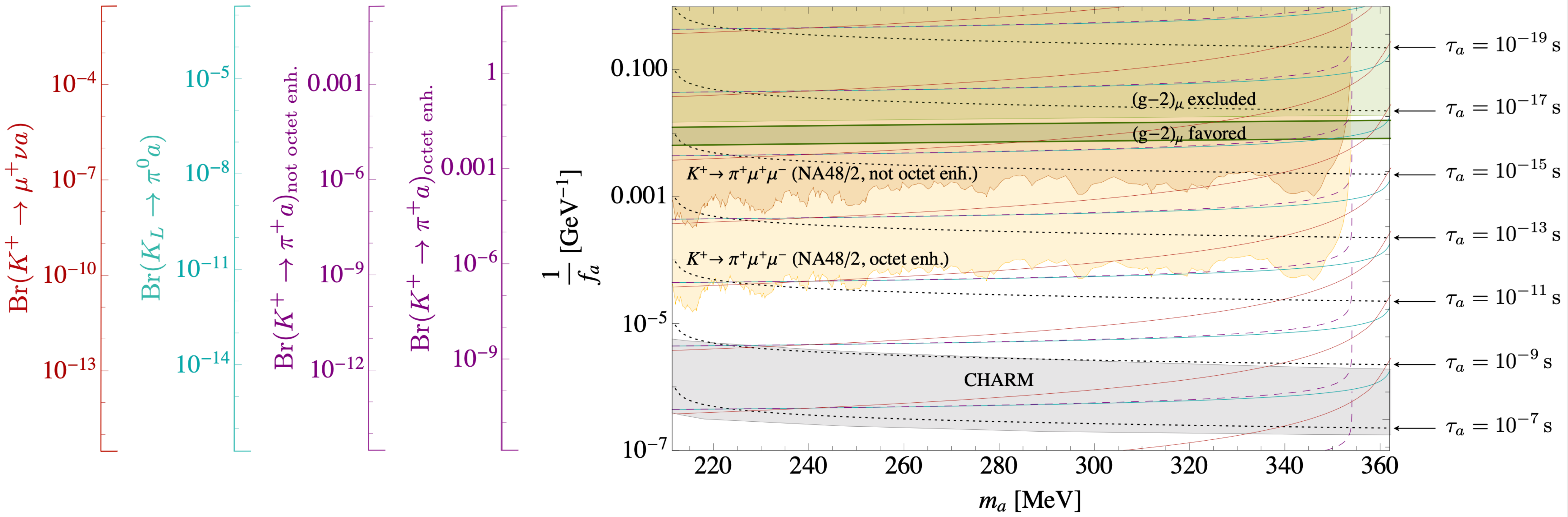}
\end{center}
\caption{\small
Order-of-magnitude ranges of axionic kaon decays in the parameter space of ALP mass ($m_a$) vs inverse decay constant ($1/f_a$) assuming dominance of the prompt $a\to\mu^+\mu^-$ channel. Shaded regions are excluded. Contours of ALP lifetime are shown as dotted black curves. Contours for ${\cal B}(K^+\to\pi^+a)$, ${\cal B}(K_L\to\pi^0a)$ and ${\cal B}(K^+\to\mu^+\nu a)$ are shown as dashed purple, solid cyan, and solid red curves, respectively (see text for assumptions). For $K^+\to\pi^+a$, we show the estimated branching ratios with and without the assumption of $\Delta S=1$ octet-enhancement of the decay amplitude (see~\cite{Alves:2020xhf} for details).}
\label{fig:K_to_mumu_prompt}
\end{figure}

We turn next to the ALP mass range $2\,m_\mu<m_a<m_K-m_\pi$ and consider the case of prompt $a\to\mu^+\mu^-$ decays. If the ALP couples to heavy (top and/or bottom) quarks, constraints from $B\to K(a\to\mu^+\mu^-)$ decays are severe enough to preclude any viable parameter space in the prompt region. Therefore we restrict our discussion to a special region of parameter space for which $Q^{_\text{PQ}}_t=Q^{_\text{PQ}}_b=0$. On the other hand, $\mathcal{O}(1)$ ALP couplings to $G\tilde{G}$ and second generation quarks are still viable as long as constraints on $\theta_{a\pi}$ can be evaded, which can be obtained by setting $Q^{_\text{PQ}}_e=0$. With these assumptions, we have:
\begin{equation}\label{paramBenchmumu}
\big|Q^{_\text{PQ}}_\mu\big|=1~,\qquad \big|\theta_{a\pi}\big| \sim \big|\thetaUD\big|\sim\big|\thetaS\big|\sim\mathcal{O}\left(\frac{f_\pi}{f_a}\right).
\end{equation}

In Fig.~\ref{fig:K_to_mumu_prompt} we show existing constraints on the prompt $a\to\mu^+\mu^-$ parameter space derived from NA48/2 measurements (Refs.~\cite{NA482:2010zrc,NA482:2016sfh}; Section~\ref{sec:sig:Kpill}), as well as $K\to\pi a$ and $K^+\to\mu^+\nu a$ branching ratio contours. The same considerations and caveats for the discussion of $a\to e^+e^-$ apply in this case: the estimated range of branching ratios can vary significantly due to possible constructive or destructive interference between amplitudes. For the contours in Fig.\,\ref{fig:K_to_mumu_prompt}, we assume the following branching ratios consistent with Eq.~(\ref{paramBenchmumu}):
\begin{subequations}\label{BRmumu}
\begin{alignat}{3}
{\cal B}(K^+\to\pi^+a)~&\simeq&&~~\left(\frac{f_\pi}{f_a}\right)^2\,\frac{\,|\vec{p}_a|\,}{|\vec{p}_{\pi^{_+\!}}|}~ \times~
	\begin{cases}
  	    ~~~32.2\,&\text{(octet-enhanced),}\\
  	    ~0.072\,&\text{(non octet-enhanced),}
   	 \end{cases}\\
{\cal B}(K_L\to\pi^0a)~&\simeq&&~~~0.72\times 10^{-3}\,\left(\frac{f_\pi}{f_a}\right)^2\,\frac{\,|\vec{p}_a|\,}{|\vec{p}_{\pi^{_0\!}}|}~~~\text{(octet-enhanced),}\\
{\cal B}(K^+\to\mu^+\nu a)~&\simeq&&~~~3.36\;\left(\frac{f_\pi}{f_a}\right)^2\;\times\;\Phi(m_a)/\Phi(m_a=10~\text{MeV}), \label{BrKaxioleptonicmumu}
\end{alignat}
\end{subequations}
where $\Phi(m_a)$ in (\ref{BrKaxioleptonicmumu}) is the $m_a$-dependent integral over the Dalitz phase-space.


\subsection{Heavy neutral leptons}    \label{sec:HNL}
{\em Authors: Fox, Hostert,  \underline{Kelly}}

The existence of heavy neutral leptons (HNLs) has been proposed as one possible solution to the smallness of the observed neutrino masses. In these scenarios, the observed neutrino masses are so small due to one of several ``seesaw'' mechanisms, where a heavier neutrino drives the mass well below that of the charged fermions of the Standard Model. While these heavy neutral leptons can span orders of magnitude in masses, we are interested here in those below the kaon mass to produce observable signatures in kaon factories.

If one or more heavy neutral leptons $N$ exist, they can interact with the other fields of the SM via the Lagrangian
\begin{eqnarray}
- \mathcal{L} \, \supset \,  -y_N \left( L H\right) N\ + \ {\rm h.c.},
\end{eqnarray}
along with a Majorana mass term generating a mass for the heavy neutral lepton. The existence of the Yukawa coupling $y_N$, along with this mass, will generate mixing between the heavy, mostly-sterile state and the light, mostly-active neutrino states. The mixing, in the case where one $N$ is assumed, is parameterized in terms of the elements of an extended, $4\times 4$ leptonic mixing matrix: $\left\lvert U_{eN}\right\rvert^2$, $\left\lvert U_{\mu N}\right\rvert^2$, and $\left\lvert U_{\tau N}\right\rvert^2$ (see, e.g., Refs.~\cite{deGouvea:2015euy,Bolton:2019pcu,Bryman:2019bjg} for constraints on these angles as a function of $N$ mass). In most phenomenological studies, one of these mixing angles is taken to be nonzero at a time, however, combinations of two or three can yield less trivial signatures.

In this minimal ``mass plus mixing'' scenario, the $N$ can be produced in any process that yields SM neutrinos, as long as its mass is small enough to be kinematically accessible in the process. At kaon factories, this leads to the two most relevant channels for HNL searches~\cite{Shrock:1980vy,Shrock:1980ct}:
\begin{equation}
K^+\to\ell^+N, \quad \ell=e,\mu.
\end{equation}
The rates of the above decays are proportional to $|U_{\ell N}|^2$, with explicit expressions available e.g. in Ref.~\cite{NA62:2020mcv}.
In most of the parameter space of interest, $N$ is long-lived on detector timescales and so the signature is a single charged lepton and missing mass. The HNL lifetime (for $m_N<m_K$) exceeds $10^{-4}/|U_4|^2~\mu$s, where $|U_4|^2$ is the largest of the three mixing parameters $|U_{\ell N}|^2$, $\ell=e,\mu,\tau$~\cite{Bondarenko:2018ptm}.

The HNLs can modify the early Universe evolution, either by contributing excessively to the energy density of the Universe or through decay products. In particular, if HNLs are sufficiently long lived to survive until the time of BBN, the decays to SM particles can modify the precisely-measured abundance of light elements by altering the proton to neutron ratio~\cite{Sarkar:1995dd,Dolgov:2000jw,Ruchayskiy:2012si,Hufnagel:2017dgo,Sabti:2020yrt}. This leads to an upper bound on the HNL lifetime  $\tau^0 < 0.1$~s (or $\tau^0 < 0.023$~s when the HNL can decay to pions~\cite{Boyarsky:2020dzc}), which then sets lower limits on the mixings of HNLs with the SM neutrinos. The BBN limits are complementary to laboratory searches. The combination of the two covers extensively  the parameter space below the kaon mass, with some parameter space  still open in particular due to weak constraints on $|U_{\tau 4}|^2$~\cite{Bondarenko:2021cpc}. It should be noted, however, that the BBN limits are model-dependent, and can be avoided in models where HNLs decay through forces other than the weak force~\cite{Arguelles:2021dqn}. In such models both the decay-in-flight as well as the BBN constraints are modified, depending on the new branching ratios and lifetimes of the HNLs, while the kaon constraints remain largely unchanged (see
Sec.~\ref{sec:2X} for some examples).

\begin{figure}[t]
\begin{center}
\includegraphics[width=3.1in]{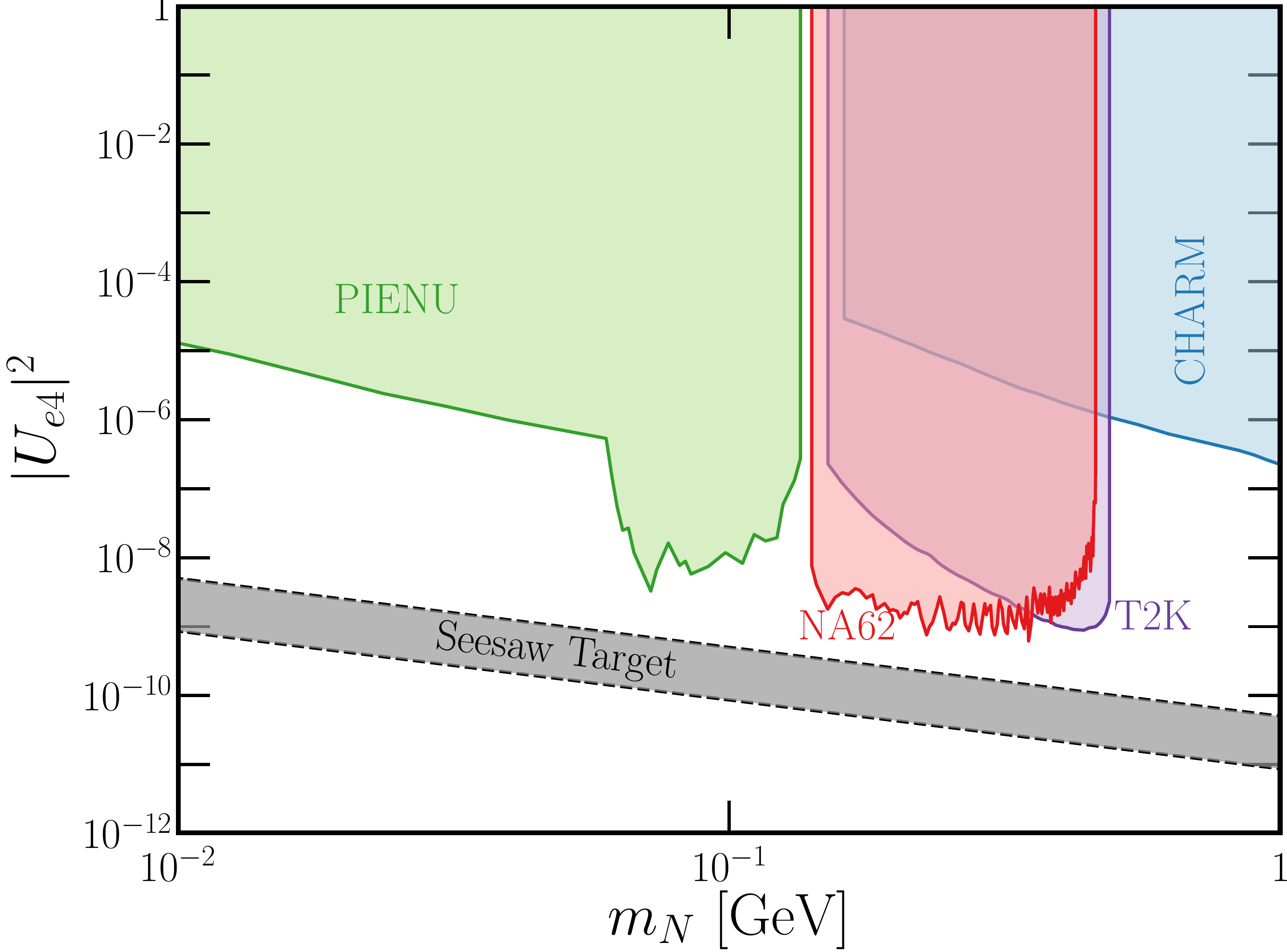}   
\includegraphics[width=3.1in]{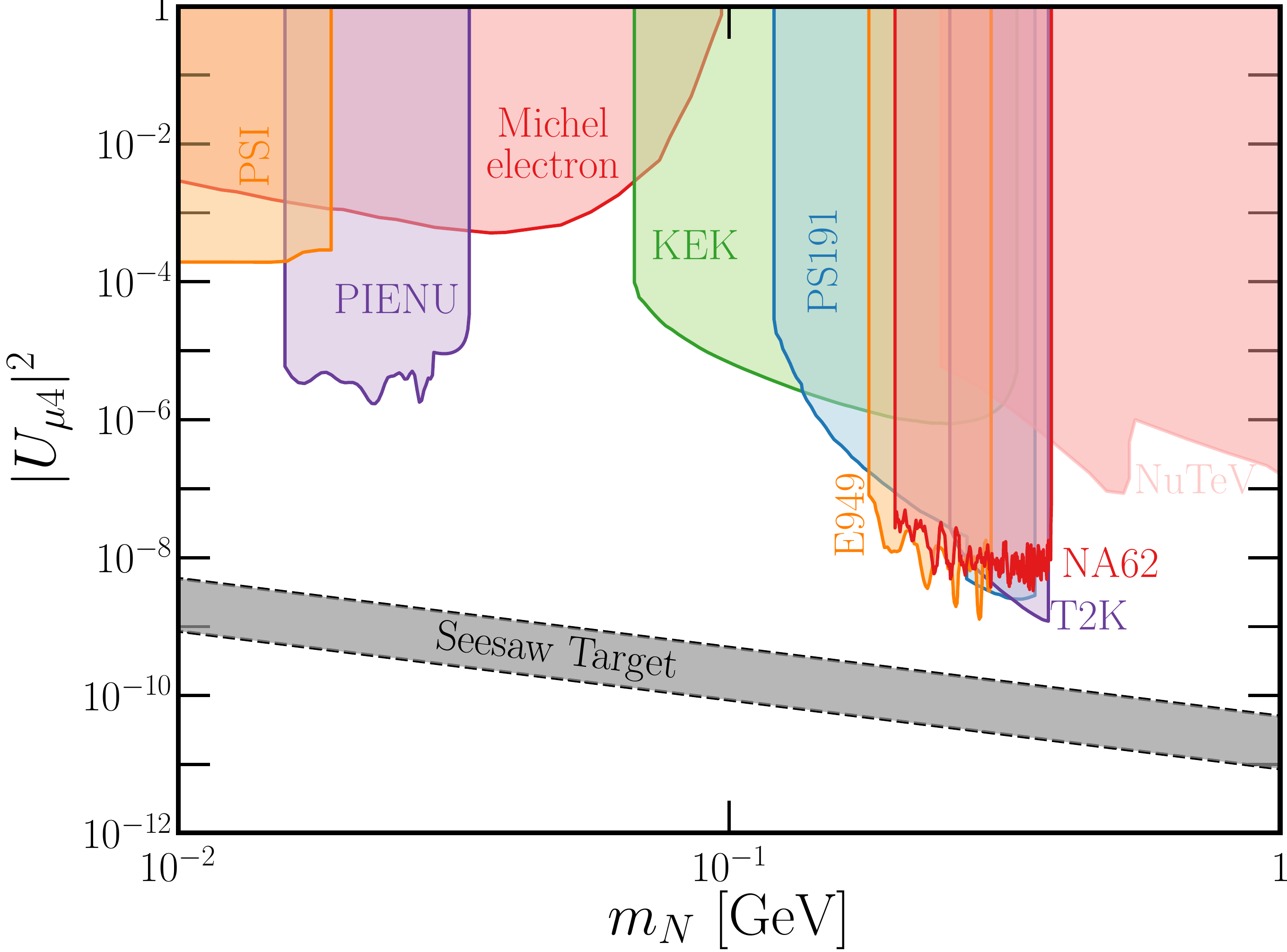}
\caption{\small Experimental constraints on electron-coupled (left) and muon-coupled (right) heavy neutral leptons as functions of their mass and the mixing $|U_{\ell N}|^2$, along with the na\"{i}ve seesaw target (grey bands).}
\label{HNLLims} 
\end{center}
\end{figure}

The NA62 collaboration has performed world-leading searches for HNL production in kaon decays, leading to some of the most stringent limits on $\left\lvert U_{eN}\right\rvert^2$ and $\left\lvert U_{\mu N}\right\rvert^2$ in the $m_N < m_K$ regime~\cite{NA62:2020mcv,NA62:2021bji}. These constraints are shown as red excluded regions in Fig.~\ref{HNLLims}, compared to other experimental searches for the minimal ``mass plus mixing'' scenario. The grey bands in each panel correspond to the na\"{i}ve target for a seesaw solution to the light neutrino masses. These are calculated assuming a simple one-light-one-heavy-neutrino scenario, where the light neutrino acquires a mass equal to $m_\nu = |U_{\alpha 4}|^2 m_N$, where we take $m_\nu = \sqrt{\Delta m_{21}^2}\approx 9\ \mathrm{meV}$ and $\sqrt{|\Delta m_{31}^2|}\approx 50\ \mathrm{meV}$ from neutrino oscillation measurements~\cite{Esteban:2020cvm} as the minimum mass of two of the three light-neutrino eigenstates in nature. When considering a three-light-two-heavy-neutrino scenario (the minimum required to explain oscillation results in a seesaw mechanism), additional mixing parameters break the simple dependence between light-neutrino masses, heavy-neutrino masses, and mixing, making this target broader. Experimental details and future prospects of the $K^+\to\ell^+N$ measurements are discussed in Section~\ref{sec:sig:Kl}. 

Depending on the HNL mass, the branching ratio of the $K^+\to \pi^0\ell^+ N$ decay can be comparable to that of the $K^+\to\ell^+N$ decay, both of which are proportional to $\left \lvert U_{\ell N}\right\rvert^2$. This is particularly true for the electron-coupled final state with $m_N \lesssim 20$ MeV~\cite{Berryman:2019dme,Tastet:2020tzh,Abada:2016plb}. If experiments can constrain ${\cal B}(K^+\to\pi^0 e^+N)$ at comparable levels as ${\cal B}(K^+\to e^+N$), then smaller $\left \lvert U_{eN}\right\rvert^2$ can be probed. However the currently available datasets only offer a limited sensitivity (Section~\ref{sec:Klgg+inv}).

In the minimal Type-I seesaw model $N$ is a Majorana fermion with a mass term that violates lepton number by two units. After production in kaon decays, the charged-current decays of $N$ would lead to lepton number violating final states~\cite{Zuber:2000vy,Atre:2005eb,Atre:2009rg,Abada:2017jjx}, such as $K^+\to\ell^+ N$, $N \to \ell^+\pi^-$. Therefore searches for signatures such as $K^+ \to\pi^-(\pi^0)\ell_1^+\ell_2^+$ represent a unique probe of LNV in kaon facilities (Section~\ref{sec:exp:lnv}).

Beyond the minimal scenarios, there is also the possibility that $N$ has additional interactions, for instance, if it is gauged under a new $U(1)^\prime$ symmetry~\cite{Bertuzzo:2018itn,Ballett:2018ynz,Ballett:2019cqp,Ballett:2019pyw}. This can lead to short-lived $N$ which decay within the detector into, for instance, a neutrino and a charged lepton pair:
\begin{equation}
K^+\to\mu^+ N, \quad N\to\nu e^+ e^-. 
\end{equation}
With the additional interactions considered, HNL lifetime becomes effectively an additional free parameter. Peak searches in this three-track signature (Section~\ref{sec:3l:inv}) could constrain neutrino mixing directly under different HNL lifetime assumptions. More exotic HNL production channels are discussed in Section~\ref{sec:miss:eng:manybody}.

\subsection{Dark photon}    \label{sec:darkphoton}

{\em Authors: \underline{Dobrescu}, Fabbrichesi, Gabrielli, Kvedarait\.e,  Martin Camalich, Terol-Calvo}

The dark photon is a gauge boson of a secluded $U(1)_{\rm d}$ gauge group under which the dark sector states are charged. If the  $U(1)_{\rm d}$ symmetry is exact, the dark photon is massless.  If the symmetry on the other hand is  either spontaneously broken by the Higgs mechanism or explicitly broken, as in the Stueckelberg Lagrangian, the dark photon is massive.  

The dark photon interacts with SM particles via kinetic mixing with the photon \cite{Holdom:1985ag}, or via dipole-like operators \cite{Dobrescu:2004wz}. The most general kinetic part of the $U(1)_{\rm em}\times U(1)_{\rm d}$ Lagrangian is given by
\be
{\cal L}_0=- \frac{1}{4} F_{{\rm em}, \mu\nu} F^{\mu\nu}_{\rm em}- \frac{1}{4} F_{{\rm d} \mu\nu} F_{\rm d}^{\mu\nu} -\frac{\varepsilon} {2} F_{{\rm d} \mu\nu} F^{\mu\nu}_{\rm em} \, , \label{kinetic}
\ee
where the term proportional to $\varepsilon$ introduces the mixing. The gauge boson $A^{\mu}_{\rm em}$ couples  to  the ordinary electromagnetic matter current $J_{\mu}$, while  $A_{\rm d}^{\mu}$ couples to the dark-sector matter current $J_{\mu}^\prime $,
\be
{\cal L}= e \, J_{\mu} A^{\mu}_{\rm em} + e^\prime J_{ \mu}^\prime A^{\mu}_{\rm d} \, , \label{int}
\ee
with $e$ and $e^\prime$ the respective gauge couplings.

For a {\it massless} dark photon, the kinetic terms in Eq.(~\ref{kinetic}) can be conveniently diagonalized by a rotation and a field redefinition of the gauge fields such that the  interaction Lagrangian  takes the following form~\cite{Holdom:1985ag}
\be
{\cal L_{\rm int}} = e^\prime  J^{\prime}_{\mu} A^{\prime\mu} 
+ \frac{1}{\sqrt{1 -\varepsilon^2}} \left( - \varepsilon e^\prime   J_{\mu} ^\prime + e J_{\mu} \right) A^{\mu}\, .
\label{Lmassless} 
\ee
The redefined dark photon $A^{\prime \mu}$ then does not couple to the SM matter, while $A^{\mu}$ couples to the SM and represents the ordinary photon. As a result, the dark sector particles carry electric charges, which are small (``milli-charges'') for $\varepsilon \ll 1$, being proportional to $\varepsilon e'/e$.

For a {\it massive} dark photon, the physical states are obtained through a field redefinition that both diagonalizes the kinetic term and leaves the mass matrix diagonal. This implies that the ordinary photon does not couple to the dark sector, while the dark photon couples to the electrically-charged SM fermions, proportionally to the mixing parameter $\varepsilon$:
\be
\mathcal{L_{\rm int}^\prime} =  -\frac{\varepsilon \, e}{ \sqrt{1 - \varepsilon^2}}   J_{\mu}  A^{\prime \mu}  \simeq - \varepsilon\, e\, J_{\mu}A^{\prime\mu}. \label{coupling}
\ee

\begin{figure}[t]
\begin{center}
\includegraphics[width=4in]{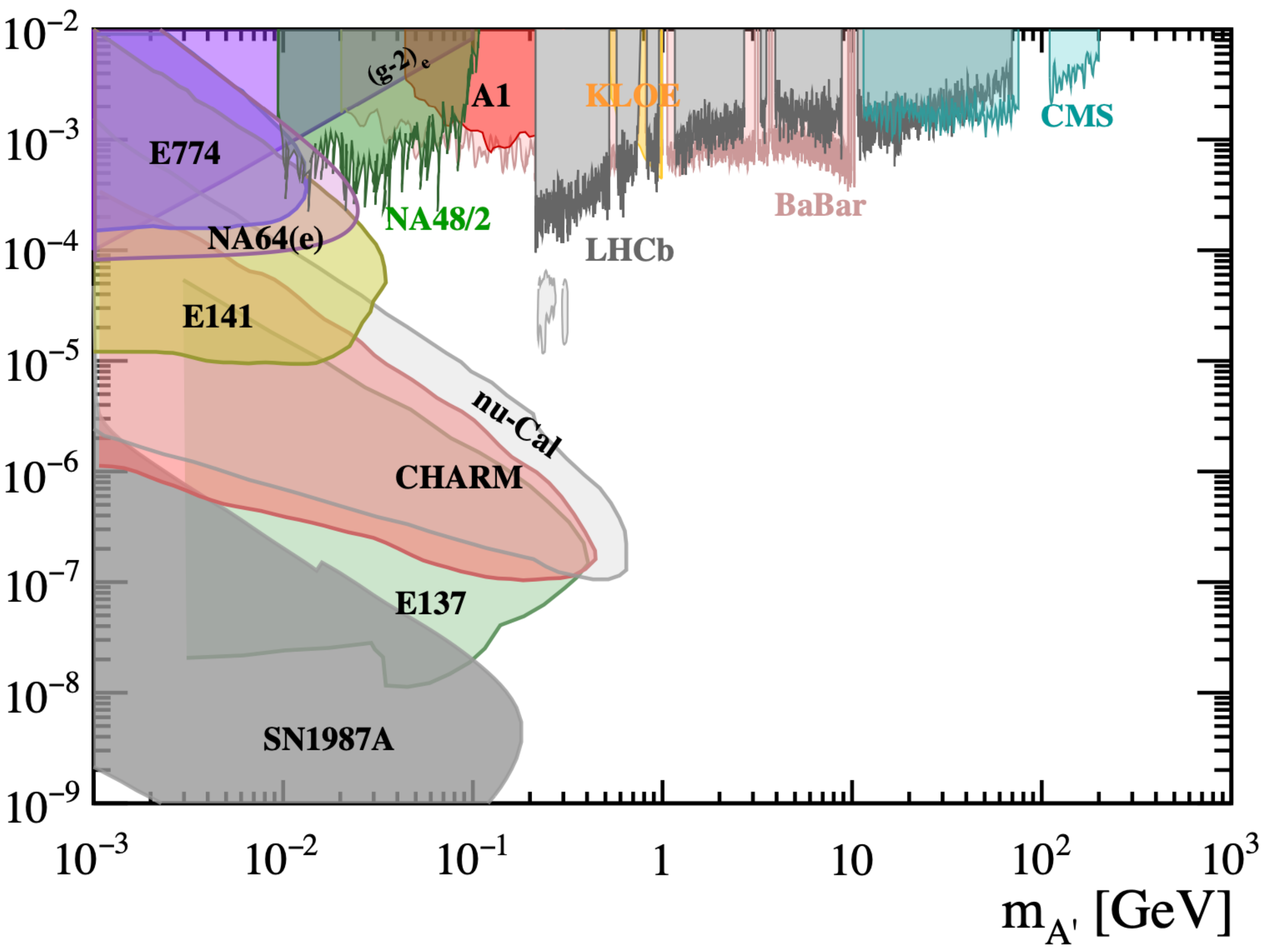}
\caption{\small Experimental limits on the kinetic mixing  $\varepsilon$ as a function of the dark photon mass for $m_{A'}>1$~MeV  \cite{Fabbrichesi:2020wbt}.}
\label{fig:darkphoton_bounds}
\end{center}
\end{figure}

Although the massless dark photon (also referred to as the ``photon-prime" \cite{Dobrescu:2004wz}) does not interact at tree level with the SM, such interactions can be generated radiatively if the full UV theory contains messengers coupled to the SM and also to states charged under $U(1)_d$. At low energies 
the interactions of the dark photon with the SM fermions are described by higher-dimensional operators, starting with the dimension-5 dipole operators \cite{Gabrielli:2016cut,Fabbrichesi:2017vma,Dobrescu:2004wz}:
\begin{equation}
\label{dipole}   
   \mathcal L_{\rm dim-5}= \dfrac{1}{\Lambda}\bar{\psi}_i\sigma^{\mu\nu}\left(\mathbb{D}_M^{ij}+i \,  \mathbb{D}_E^{ij}\gamma_5\right)\psi_j F^{\prime}_{\mu\nu} ~~,
\end{equation}
where $\sigma_{\mu\nu}=i/2 \, [\gamma_{\mu},\gamma_{\nu}]$, $\psi^i$ are the SM fermions with $i=1,2,3$ labeling the generations, and $F^{\prime}_{\mu\nu}$ is the  dark photon field strength.
The effective couplings $\mathbb{D}_M$ ($\mathbb{D}_E$) multiplying the magnetic (electric) dipole moment  depend on the parameters of the underlying UV model and are proportional to $e'$. They are suppressed by the scale $\Lambda$, representing the typical mass of the dark sector states. 
Note that for $i\ne j$ the neutral current transitions induced by \eqref{dipole}  are flavor violating. Such flavor violating dipole transitions are possible also for massive dark photon. This is in contrast to the tree-level couplings of the massive dark photon to the SM which are proportional to the electromagnetic SM current, and thus flavor diagonal.
For $s\to d $ transitions the measured kaon mass mixing imposes a strong constraint on the parameters of the UV model and therefore on the scale $\Lambda$ suppressing the dipole operator  \eqref{dipole}.

\begin{figure}[t]
\begin{center}
  \includegraphics[width=4in]{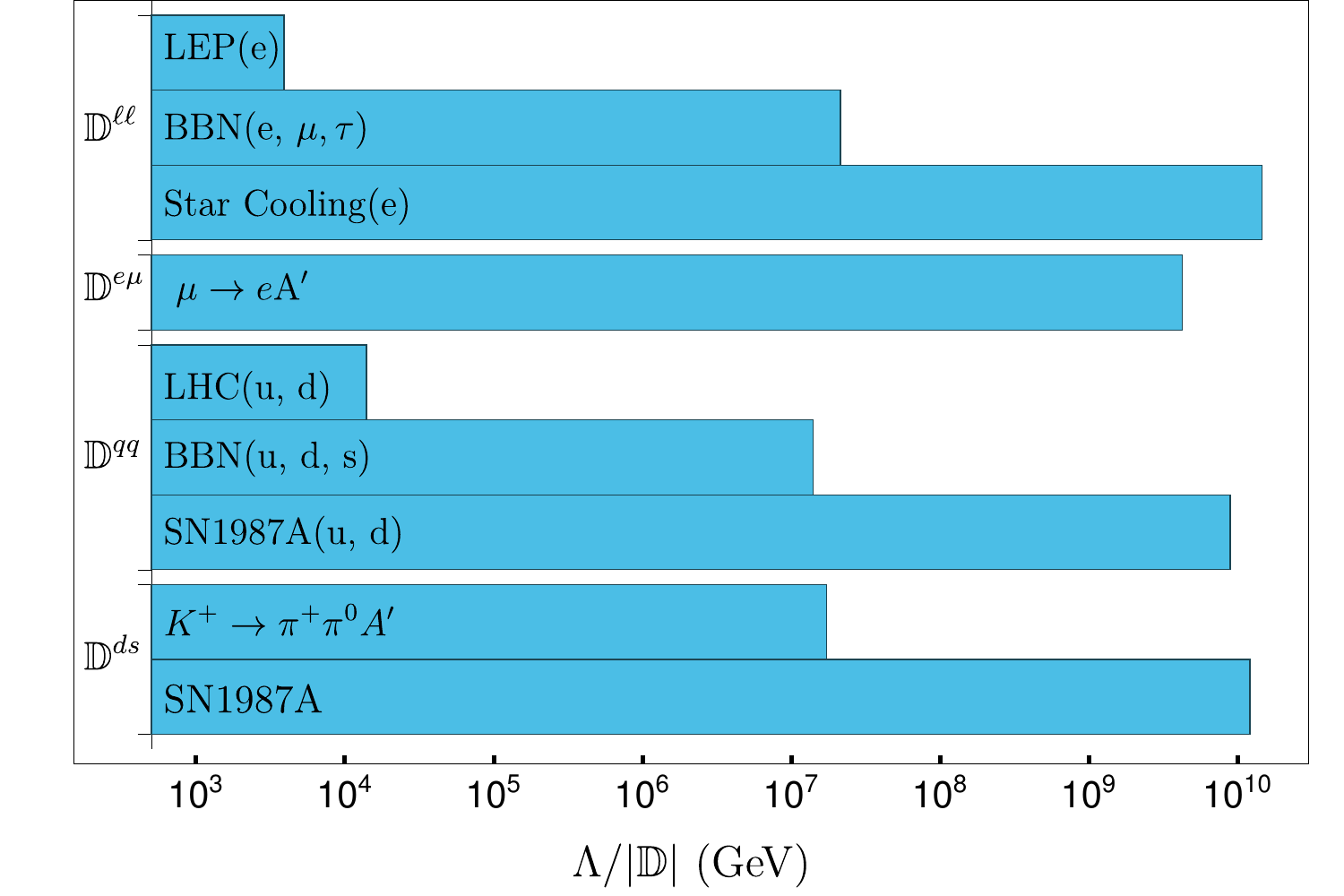}
\caption{\small Experimental limits on the  parameter space of the  massless  dark photon, with $\Lambda$ the new physics scale and $\mathbb{D}$ the effective dipole coupling (adapted from Refs.~\cite{Camalich:2020wac}, \cite{Fabbrichesi:2020wbt}). 
   }
\label{fig:dp-limits1} 
\end{center}
\end{figure}

Dark photons can be searched for in various experiments and observational data. The current limits on $\Lambda/\mathbb{D}^{ij}$ for the massless dark photon are shown in Fig.~\ref{fig:dp-limits1},
where $\mathbb{D}^{ij}\equiv\big((\mathbb{D}^{ij}_M)^2+(\mathbb{D}^{ij}_E)^2\big)^{1/2}\!$ is the effective dipole coupling to leptons or light quarks. 
Even though some of the limits are as stringent as $\Lambda/\mathbb{D}^{ij} \gtrsim 10^{10}$ GeV, the effective dipole coupling may be small enough to allow the new particles charged under $U(1)_{\rm d}$ to have masses at the TeV scale \cite{Dobrescu:2004wz}. In the next subsection we discuss in details the different probes of dipoles interactions of the massless dark photon.

For the massive  scenario,  the current limits on the photon-dark photon mixing are shown in Fig.~\ref{fig:darkphoton_bounds}, with $m_{A'}\simeq 1$~MeV delineating the regions where the dark photon decays into either ``visible'' SM particles (for larger masses). As we see from the figure, rare kaon decays play an important role in constraining the massive dark photon in the region between 10 and 100 MeV. The current best limit are set by the NA48 experiment~\cite{NA482:2015wmo}, looking for $\pi^0\to A'\gamma$ with the dark photon $A'$ further decaying into $e^+e^-$ pairs. The pion dataset is collected from identified kaon decays (e.g. $K^{\pm}\to \pi^{\pm}\pi^0$, $K^{\pm}\to \pi^0\mu^+\nu$ and the dark photon yield is controlled by the mixing squared $\sim\varepsilon^2$. Similarly, one could hunt for the dark photon at NA62 in all the channels containing SM photons, performing a bump hunt on the dielectron spectrum of the collected data. For instance one could estimate 
\begin{equation}
{\cal B}(K^+\to \mu^+ \nu A' (e^+ e^-))\sim 10^{-3}\times \varepsilon^2,
\end{equation}
which should yield a competitive bound on the dark photon mixing angle. 

\subsubsection{Searches for dark photons dipoles in kaon and hyperon decays}
\label{subsec:darkphoton}

Flavor-changing neutral current (FCNC) processes involving the dark photon are induced by off-diagonal magnetic- and electric-like dipole operators in Eq.~(\ref{dipole}). In particular, the $\Delta S=1$ transitions provide the leading contribution to the gauge-invariant FCNC interactions with SM quarks, for both massless and massive dark photon scenarios.


\begin{itemize}
\item {\bf Dark photon production in charged kaon decays:} The $K^+\to \pi^+\gamma A'$ and $K^+\to\pi^+ \pi^0A'$ decays induced by the dipole operators~\eqref{dipole} are analyzed for the massless dark photon in Refs.~\cite{Fabbrichesi:2017vma,Su:2020xwt}. 
The bounds on allowed NP contribution to $\Delta M_{K}$ in kaon mixing, assuming no cancellations with other UV contributions, limit the branching ratios to be at most ${\cal B}(K^+ \to\pi^0\pi^+ A')\lsim 1.6\times 10^{-7}$ and ${\cal B}(K^+ \to \pi^+\gamma A')\lsim 1.1\times 10^{-8}$ for the UV model and hadronic inputs of~\cite{Fabbrichesi:2017vma} with $\alpha_D\equiv e^{\prime 2}/4 \pi=0.1$. Using the same UV model and $\alpha_D$ but the hadronic inputs of~\cite{Su:2020xwt} one obtains, instead, the upper limits  ${\cal B}(K^+ \to\pi^0\pi^+ A')\lsim 1.3\times 10^{-7}$ and ${\cal B}(K^+ \to \pi^+\gamma A')\lsim 3.2\times 10^{-8}$.
These branching ratios are within the reach of the NA62 experiment (Sections~\ref{sec:sig:K2pi}, \ref{sec:Kpiginv}). Similar results are expected to apply for a massive dark photon in the limit of small mass. The $K^+\to\pi^+A'$ decay is not allowed by angular momentum conservation for a massless dark photon, while being allowed for a massive dark photon (Section~\ref{sec:sig:Kpill}), with a branching ratio expected to be of the same order as for $K^+\to \pi^+\pi^0A'$. Production of multiple dark photons in kaon decays is discussed in Section~\ref{sec:2X}. 


\item
\noindent
{\bf Dark photon production in neutral kaon decays:}
Dark photon emission in FCNC decays of $K_L$ and $K_S$ have been explored for the massless case~\cite{Su:2020xwt,Camalich:2020wac}, in particular in the  $K_L\to\gamma A'$, $K_L\to\pi^0 \gamma A'$ and $K_L\to\pi^+\pi^- A'$ processes and their $K_S$ counterparts. The $K_L$ decays have larger branching ratios compared to $K_S$ because  of the longer lifetime of the decaying state. Using upper bounds on $\Delta M_{K}$ from kaon mixing, and assuming a vanishing electric-dipole contribution in the $s\to d A'$ transitions, the upper limits on branching ratios are expected to be~\cite{Su:2020xwt}: ${\cal B}(K_{L(S)}\to \gamma A') < 6.8 (1.2) \times 10^{-5(-7)}$, ${\cal B}(K_{L(S)} \to\pi^0\gamma A') < 5.9 (1.0)\times 10^{-8(-10)}$,  ${\cal B}(K_{L(S)}\to\pi^+\pi^- A') < 5.6(9.7) \times 10^{-7 (-10)}$. These $K_L\to(\pi^0)\gamma A'$ decay rates are within the sensitivity reach of the KOTO experiment (Sections~\ref{sec:Kpiginv}, \ref{sec:KL:gamma:inv}).

\item
{\bf Dark photon production in hyperon decays:}
FCNC dipole operators may mediate $s\to dA'$ transitions, which induce hyperon decays into dark photons such as  $\Lambda\to n A^\prime$, $\Sigma^+\to p A^\prime$, $\Xi^{(0,-)}\to \Sigma^{(0,-)} A^\prime $, and $\Omega^{-}\to \Sigma^{-}A^\prime$. These can be used efficiently to search for dark photons~\cite{Su:2019ipw,MartinCamalich:2020dfe}.

\item
{\bf Dark photon production in $\pi^0$ decays:}
Kaon factories serve as the principal tools for studies of rare $\pi^0$ decays. The main source of tagged $\pi^0$ mesons in $K^+$ experiments is the $K^+\to\pi^+\pi^0$ decay with a branching fraction ${\cal B}\approx 21\%$. Searches for the $\pi^0\to\gamma A'$ decay followed by either invisible (Section~\ref{sec:Kpiginv}) or prompt visible (Section~\ref{sec:sig:Kpill}; see Ref.~\cite{Alexander:2016aln} for a review) dark photon decays have been performed in kaon experiments.

\item
{\bf Decays into  invisible states:}
The rare FCNC decays $K^+\to\pi^+ \nu\bar\nu$ and $K_L\to\pi^0\nu\bar\nu$ are processes that may be mimicked by dark photon production followed by an invisible decay~\cite{Fabbrichesi:2019bmo}. For instance, a tree-level exchange of a virtual dark photon can  mediate $K^+\to \pi^+ Q\bar{Q}$ and $K_L\to \pi^0 Q\bar{Q}$ decays, where $Q$  are 
dark fermions which escape the detector, resulting in the missing energy signature. Due to the long-distance nature of the dark-photon mediated FCNC interactions, the GN bound on ${\cal B}(K_L\to\pi^0 Q\bar{Q})$ set by the constraint on ${\cal B}(K^+\to \pi^+ Q\bar{Q})$, cf. Section \ref{sec:GNVgamma}, can be seemingly violated by the asymmetric effect of the different kinematic cuts enforced by the NA62 and KOTO experiments, allowing for a branching ratio for $K_L \to \pi^0 Q \bar{Q}$ within the reach of the KOTO sensitivity~\cite{Fabbrichesi:2019bmo}. The $K_L\to$ invisible decay can be searched for by the NA64 experiment up to the $10^{-6}$ level~\cite{Gninenko:2015mea,Gninenko:2014sxa}. The $K_L\to Q\bar Q$ decay would be an example of such process, with $K-\bar K$ mixing bounds still allowing branching ratios well above the projected reach of NA64~\cite{Barducci:2018rlx}.

\item
{\bf Dark photon production via hyperon decays in supernovae:}
 The proto-neutron star created right after the supernova explosion contains a thermal population of $\Lambda$ hyperons. The $\Lambda\to n A^\prime$ decays can cool the proto-neutron star via a flux of escaping dark photons, modifying the observed SN1987A neutrino flux. 
This sets the most stringent bound on the strength of the $s\to d A'$ dipole operator,
$\Lambda/\big|\mathbb{D}^{sd}\big|\gtrsim1.2\times10^{10}$ GeV~\cite{MartinCamalich:2020dfe,Camalich:2020wac}. The above limit can be taken as a useful benchmark for the sensitivity that one would like to achieve by the next generation of experiments. Note however that the bound does rely on the validity of the standard core-collapse interpretation of the SN1987~A and thus may potentially even not apply at all (see refs.~\cite{Bar:2019ifz,Page:2020gsx} for recent discussions). It does not apply also in the case of a massive dark photon with $m_{A^\prime} > 1$~MeV because of its decay into electron pairs and the re-absorption of its energy.
\end{itemize}

\begin{table}[t]
\centering
\small
\setlength{\tabcolsep}{1.1em}
  \begin{tabular}{cccc}
  \hline\hline
\multicolumn{2}{c}{\textbf{Kaons}}&\multicolumn{2}{c}{\textbf{Hyperons}}\\
Decay mode & Max branching ratio  & Decay mode  & Max branching ratio \\
    \hline  
     $K^+ \to\pi^+\pi^0 A^\prime$ & $1.3\times 10^{-7}$   $\; \big(7.8\times 10^{-11}\big)$ & $\Lambda\to n A^\prime$  &    $1.4\times 10^{-5}$  $\; \big(8.0\times 10^{-9}\big)$\\ 
     $K^+ \to \pi^+ \gamma A^\prime$ & $3.2\times 10^{-8}$ $\; \big(1.9\times 10^{-11}\big)$ &   $\Sigma^+\to p A^\prime$  &  $8.3\times 10^{-7}$ $\; \big(4.9\times 10^{-10}\big)$\\   
      $K_L \to \pi^0 \gamma A^\prime$ & $5.9\times 10^{-8}$ $\; \big(3.4\times 10^{-11}\big)$ &  $\Xi^{0}\to \Sigma^{0} A^\prime $  &    $5.2\times 10^{-6}$ $\; \big(3.0\times 10^{-9}\big)$\\
     $K_L \to \pi^+\pi^- A^\prime$ & $5.6\times 10^{-7}$ $\; \big(3.2\times 10^{-10}\big)$ &   $ \Xi^{0}\to \Lambda A^\prime$  &    $2.4\times 10^{-6}$ $\; \big(1.4\times 10^{-9}\big)$\\ 
     $K_L\to \gamma A^\prime$ & $6.8\times 10^{-5}$ $\; \big(4.0\times 10^{-8}\big)$ & $\Xi^{-}\to \Sigma^{-} A^\prime $  &    $6.1\times 10^{-6}$ $\; \big(3.6\times 10^{-9}\big)$\\
      $K_L\to \mbox{inv.}$ & $1\times 10^{-4}$ $\;\;\; \big( 6.5 \times 10^{-10}\big) $     &  $ \Omega^{-}\to \Sigma^{-}A^\prime$  & $6.1 \times 10^{-5} $  $\; \big(3.6 \times 10^{-8} \big)$\\
   \hline\hline
  \end{tabular}
\caption{Upper bounds on the branching ratios for different radiative kaon and hyperon decay channels using the maximal $s\to d A^\prime$ dipole transition allowed by $\Delta M_K$ from kaon mixing. If the stronger SN constraints apply, the maximal branching ratios are those listed in the parentheses. For the hadronic inputs we have used~\cite{Su:2020xwt}.\label{tab:Kaon decay}}
\end{table}

Kaon and hyperon decays induced by the $d\to s A'$ dipole transition, and upper limits on their branching ratios set by $\Delta M_K$ and the SN1987A cooling constraints, are listed in Table~\ref{tab:Kaon decay}.

\subsection{Leptonic force mediators}
\label{sec:muonicf}
{\em Authors: Datta, Krnjaic, Marfatia, Redigolo, \underline{Shuve}}\\
A leptonic force involves a scalar or vector mediator that couples predominantly to leptons. Due to stringent constraints from experiments with electron initial states, the most interesting viable leptonic mediator scenarios couple predominantly to muons, taus, and/or their neutrinos. Such models lead to signatures that are distinct from canonical hidden sector scenarios for dark photons and dark scalars, and are motivated by:
   \begin{itemize}
\item {\bf Muon $g-2$:}~One of the leading motivations for leptonic forces is the longstanding discrepancy between the predicted and measured $(g-2)_\mu$ values~\cite{Aoyama:2020ynm,Abi:2021gix}. If this discrepancy is indeed the result of new physics, then we expect new states coupled to muons. The new physics models are of two types \cite{Calibbi:2018rzv,Arcadi:2021cwg}: if the chirality flip in $(g-2)_\mu$ is due to the external muon leg, the NP needs to be light, while NP can be heavy if the chirality flip is due to the states running in the loop. Rare kaon decays can shed light on the light NP models. Direct searches exclude the simplest hidden sector scenarios, such as dark Higgs bosons and dark photons, from accounting for the observed muon $g-2$ assuming the mediator decays either invisibly or to pairs of SM fermions (for a recent summary, see Ref.~\cite{Beacham:2019nyx})\footnote{We remark, however, that scenarios where a dark photon undergoes cascade or semi-visible decays could still be viable candidates for the muon $g-2$ anomaly \cite{Mohlabeng:2019vrz}.}. If, however, the new particles have an enhanced coupling to muons relative to electrons and quarks, many of the direct searches can be evaded. For a point of reference, if there exists a 100 MeV leptonic mediator, the coupling to muons needed to account for $(g-2)_\mu$ is approximately $6\times10^{-4}$ for a scalar and  $1\times 10^{-3}$ for a vector.

\boldmath  
\item {\bf $B$ anomalies:}
\unboldmath Several measurements in charged current and neutral current semileptonic $B$ decays  differ from the standard model predictions. An  observable which has received much attention is $R_K \equiv {\cal B}(B^+ \to K^+ \mu^+ \mu^-)/{\cal B}(B^+ \to K^+ e^+ e^-)$ as this is a clean test of lepton universality with tiny hadronic corrections. There are measurements of similar ratios with a $K^*$ in the final state. 
The latest LHCb measurement of $R_K$
is~\cite{Aaij:2021vac}
\beq
R_K(1.1 < q^2 < 6.0 ~ {\rm{GeV^2}})  = 0.846^{+0.044}_{-0.041},
\label{RKmeas}
\eeq
differing from the SM prediction of by $\sim 3.1\sigma$. This measurement could indicate new interactions coupling preferably to  leptons of the second and third generation.
In many models proposed to address the muon $g-2$ measurement, FCNC (Flavor Changing Neutral Current) decays of the $b$ quark are generated  with the involvement of either SM particles \cite{Batell:2009jf,Datta:2019bzu} or new particles\cite{Fuyuto:2015gmk,Datta:2017ezo,Datta:2018xty,Darme:2020hpo,Dutta:2021afo}. These new FCNC contributions may resolve the $B$ anomalies in the  measurement of the semileptonic $ b \to s \ell \ell$ decays.
   
\item {\bf Dark matter:}~Viable GeV-scale thermal dark matter models predict the existence of new, low-mass mediators that allow for efficient DM annihilations at temperatures  $T\lesssim M_{\rm DM}$. In the simplest models, the mediators have minimal ``portal'' couplings, such as mixing with the Higgs boson in the case of a scalar or mixing with the photon in the case of a vector. In recent years, there have been extensive tests of the minimal portals at accelerators, colliders, and beam dump experiments, and these tests are expected to constrain dark matter models coupled through such portals \cite{Beacham:2019nyx}.
   
Leptonic force mediators serve equally well as conduits between the DM and SM, but are much less studied. As a result, for a conclusive test of thermal DM models at or below the GeV scale, it is important to test also leptonic mediators~\cite{Krnjaic:2019rsv}.  
   
\end{itemize}
Kaon factories are ideal experiments for probing low-mass leptonic mediators as the dominant charged kaon decay mode is $K^+\to\mu^+\nu$. Thus the dominant leptonic mediator production mode is through final-state radiation off the muon or muon neutrino, $K^+\to\mu^+\nu X$. In this section we argue that kaon factories can be sensitive to promising regions of parameter space that are favored by the above motivations.

\subsubsection{Models}

\paragraph*{Flavor non-universal gauge boson:} Low-mass vectors can couple directly to SM leptons if lepton flavor symmetries are embedded in gauge symmetries. The model is UV complete if the gauge symmetry is anomaly free. A simple example of this is gauged differences of lepton flavor number, such as $L_\mu-L_\tau$ or $L_e-L_\mu$ \cite{He:1990pn,He:1991qd}. Since vectors coupled to electron number face stringent constraints from $B$-factories and other experiments with electron initial states, we focus on $L_\mu-L_\tau$ as a representative model with $Z'$ mediator  (Ref. \cite{Greljo:2021npi} showed that only small deviations from the $L_\mu-L_\tau$ model are phenomenologically allowed). Its interaction Lagrangian is
\begin{equation}
    \mathcal{L} = -g'Z'_\alpha \left(\bar\mu \gamma^\alpha \mu - \bar\tau\gamma^\alpha\tau+
    \bar\nu_{\mu\mathrm{L}} \gamma^\alpha \nu_{\mu\mathrm{L}} - \bar\nu_{\tau\mathrm{L}}\gamma^\alpha\nu_{\tau\mathrm{L}}\right). 
\end{equation}
For all $Z'$ masses, there is an appreciable invisible branching fraction due to its coupling to neutrinos. Between the dimuon and ditau thresholds, the $Z'$ has an approximately equal branching fractions to muons and invisible products, while below the dimuon threshold the invisible branching fraction is approximately 100\%. For masses $m_{Z'}\gtrsim210$ MeV, the model is constrained by direct searches at BaBar \cite{TheBABAR:2016rlg}, and indirect constraints from neutrino trident processes \cite{Altmannshofer:2014pba}. However the model is relatively unconstrained below the dimuon threshold, which is precisely the region of greatest sensitivity for kaon experiments \cite{Krnjaic:2019rsv}.

\paragraph*{Leptophilic scalar:}
A leptophilic scalar, $\phi$, couples predominantly to leptons with suppressed or absent quark couplings~\cite{Chen:2015vqy,Batell:2016ove}:
\begin{equation}
\mathcal{L} = -y_{ij}\,\bar\ell_i\phi \ell_j+\mathrm{h.c.}.
\end{equation}
To evade stringent constraints from charged lepton flavor violation experiments, we only consider flavor-diagonal couplings, $y_{ij}=y_i\delta_{ij}$. There are several possible textures for the Yukawa couplings:
\begin{itemize}
    \item Mass proportional coupling, $y_i\propto m_i/v$. This interaction occurs in models with leptophilic Higgs sectors:~for example, in two-Higgs doublet models where one of the doublets gives mass exclusively to the leptons \cite{Batell:2016ove}. Mixing between the leptophilic doublet and a low-mass singlet scalar, $\phi$, gives rise to this structure of interactions. In this case, the scalar $\phi$ couples predominantly to $\tau$ leptons and gives suppressed coupling to electrons, allowing the model to evade stringent constraints from accelerators and $e^+e^-$ colliders.
    \item Muonphilic scalar, $y_e=y_\tau=0$. This can arise in a model with a new weak-scale vectorlike lepton, $\Psi$, with the same gauge charges as the SM RH muons \cite{Chen:2015vqy}.  The couplings $\phi\bar\Psi\mu_{\rm R}$ and $H^*\bar\Psi \mu_{\rm L}$ generate the $\phi\bar\mu\mu$ coupling when the vectorlike leptons are integrated out. This scenario can account for $(g-2)_\mu$ with a minimum of couplings to other SM leptons.
    \item Generic texture. This can arise in models with multiple vectorlike leptons. The principal requirement is that $y_e\ll y_\mu ,y_\tau$ to avoid constraints from experiments with electron initial states.
\end{itemize}


\subsubsection{Branching fraction benchmarks}

Because the production of leptonic mediators $X$ predominantly arises from muon FSR in $K^+\to\mu^+\nu X$ decays, the branching fraction is determined by the coupling to muons. The same coupling determines the $X$ contribution to $(g-2)_\mu$, therefore it is possible to map the models that account for $(g-2)_\mu$ to specific kaon branching fractions. The only model dependence arises in the spin of $X$, since the scalar and vector FSR rates are different.

There is more ambiguity in mapping the DM parameter space to the kaon branching fraction. Even assuming that DM thermalizes with the SM and its abundance is determined through thermal freeze-out, the specific process that sets the relic abundance depends on the mediator/DM mass hierarchy. If the mediator is lighter than DM (the ``secluded scenario'' \cite{Pospelov:2007mp}), then DM annihilates predominantly into the mediator and the muon coupling plays no role in establishing the DM abundance. Conversely, if the mediator is heavier than DM, then with specific assumptions about the magnitude of the DM-mediator coupling  we can determine the coupling to muons needed for viable DM\footnote{For the mass range probed at kaon factories, the coupling to taus is unimportant because taus are largely absent from the early-universe plasma at temperatures of $\mathcal{O}(100~\mathrm{MeV})$.}. In determining the range of parameters consistent with DM, we consider a vector mediator coupled to three possible types of DM (scalar, Majorana, and pseudo-Dirac).

\begin{figure}[t]
\begin{center}
\includegraphics[width=2.9in]{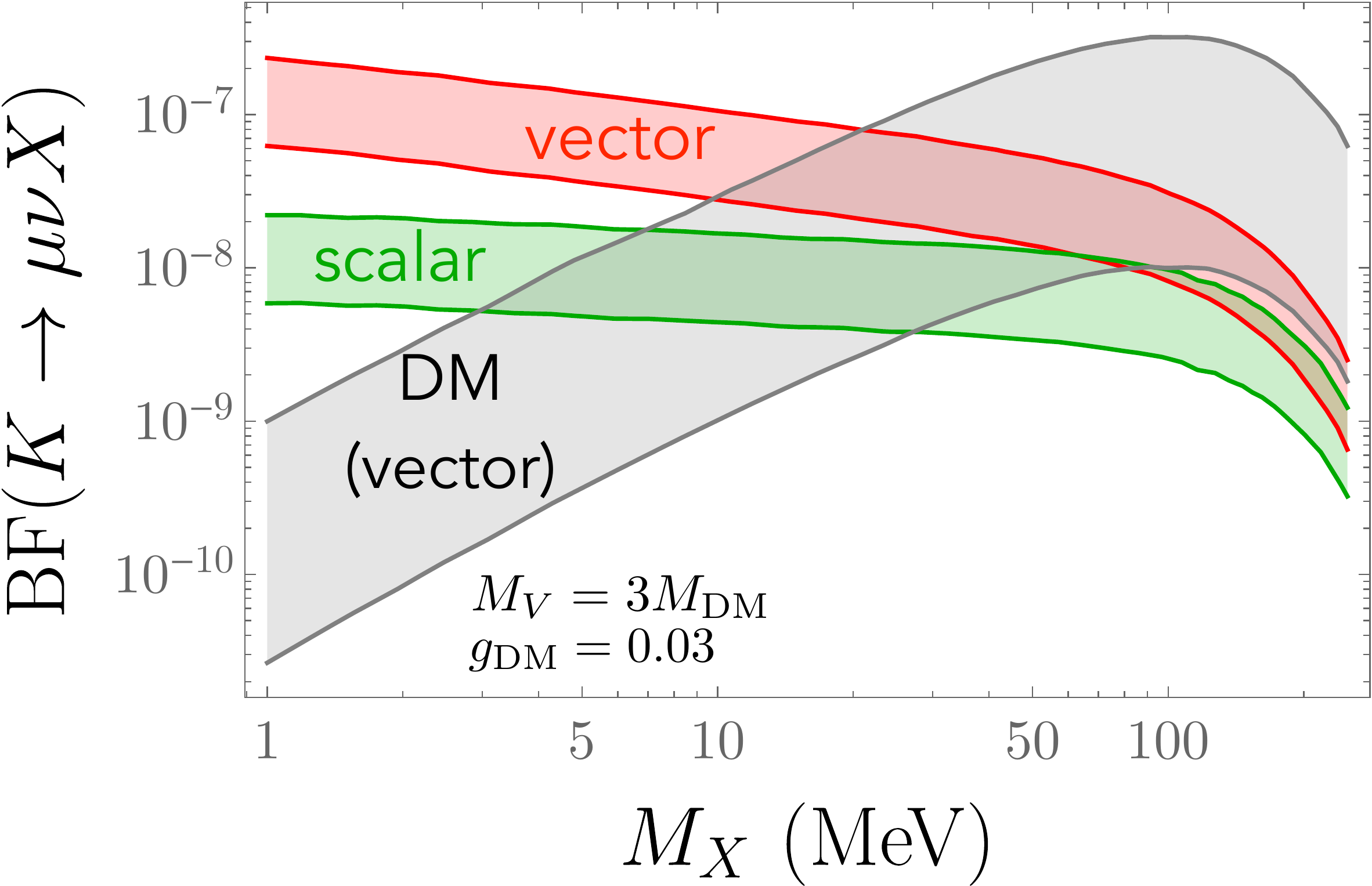}\hspace{0.5cm}
\includegraphics[width=2.9in]{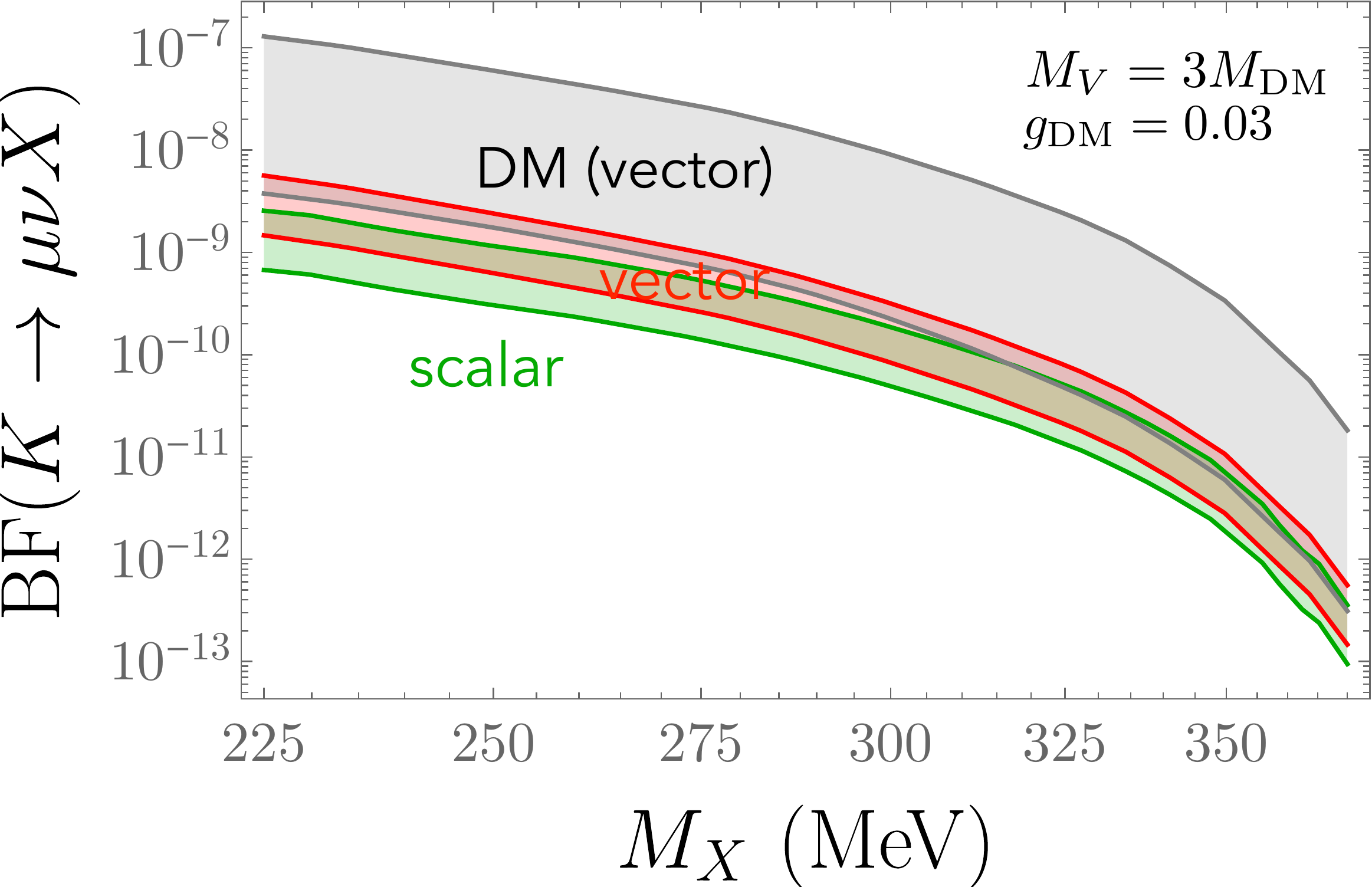}
\caption{\small Branching fraction benchmarks for $K^+\to \mu^+ \nu_\mu X$, where $X$ is a leptonic force mediator for (left) $m_X < 225$ MeV; (right) $m_X\ge 225$ MeV. The red (green) band shows the branching fraction consistent with the observed discrepancy in $(g-2)_\mu$ for a vector (scalar) $X$. The grey shaded range indicates the preferred parameter space for the DM relic abundance for a vector mediator, with a mediator that is three times heavier than DM and with a coupling of 0.03. The band indicates the parameter space spanned by models where DM is a scalar, Majorana fermion, or Dirac fermion \cite{Krnjaic:2019rsv}.
}
\label{gm2_targets_DM_paper} 
\end{center}
\end{figure}

Using the results of Ref.~\cite{Krnjaic:2019rsv}, we derive the target branching fractions motivated by $(g-2)_\mu$ and DM, showing our results in Fig.~\ref{gm2_targets_DM_paper}. Obtaining a sensitivity to branching fractions $\ge10^{-9}$ would cover the most interesting parameter space for both DM and $(g-2)_\mu$. For masses smaller than 10 MeV, thermal DM models can run into constraints from $\Delta N_{\rm eff}$, the number of additional light degrees of freedom in the early universe,  while for larger mediator masses near the kinematic endpoint, Belle II and other higher-energy experiments can provide complementary sensitivity to kaon factories. Even a null result is interesting:~since leptonic mediators are among the last viable models explaining $(g-2)_\mu$ at low masses, branching fraction exclusions of order $10^{-9}$ would decisively exclude these light mediator models.

Model-dependent constraints from a BaBar search for a leptonic scalar produced in association with $\tau$ leptons and decaying to $e^+e^-$ already rules out nearly the entire $(g-2)_\mu$ parameter space in the case that the scalar couples mass proportionally to leptons \cite{BABAR:2020oaf}, but these constraints do not apply if the mediator has a reduced coupling to taus.

\subsubsection{Signatures}

The predominant production mode for the leptonic force mediator is $K^+\to\mu^+\nu X$. The experimental signatures are driven by the decay mode of $X$, and the best-motivated decay modes are the following.
\begin{itemize}
\item $X\to\mu^+\mu^-$ (Section~\ref{sec:3l:inv}): because the $X$ coupling to muons is always present for leptonic mediators, this decay mode is always present at some level. It is the dominant decay mode in models where $X$ is lighter than all other hidden sector particles, and $2m_\mu \le m_X\le m_K-m_\mu$. The dimuon decay is typically prompt for the couplings accessible at kaon factories. The projected sensitivity of NA62 in this channel has been derived in Ref.~\cite{Krnjaic:2019rsv}.
\item $X\to\rm inv$ (Section~\ref{sec:sig:Kl}): this decay mode is present if $X$ couples to neutrinos, and dominates if $X$ can decay into invisible dark matter states. This decay mode is also relevant if $X$ does not couple to electrons:~while $X\rightarrow\gamma\gamma$ proceeds at one loop, the lifetime can be long enough that $X$ typically decays outside of the detector. Another option is $X$ decaying to neutrinos, discussed in more detail in Section \ref{sec:neutrinos}.
\item $X\to \gamma\gamma$ (Section~\ref{sec:Klgg+inv}): this loop-induced decay mode dominates if $X$ does not couple to electrons and is lighter than the dimuon threshold. The partial width depends on the UV completion of the model and can lead to prompt or displaced diphoton decays. If $X$ is a scalar that couples only to muons, the proper decay length to diphotons in the unexcluded parameter space below the dimuon threshold is in the 1 cm--1 m range \cite{Krnjaic:2019rsv} if the photon coupling is generated purely from muon loops. However further UV contributions to the photon coupling could sensibly change this prediction so that a comprehensive range of lifetimes should be searched for experimentally. 
\item $X\to e^+e^-$ (Section~\ref{sec:3l:inv}): while not present in every leptonic mediator model, this decay mode dominates below the dimuon threshold in models with mass-proportional couplings of $X$. This is therefore an important channel for discovering or constraining leptonic mediators. Searches for both prompt and displaced decays are of interest, with proper decay lengths in the 0.1--10 cm range predicted in models with mass-proportional couplings of $X$ \cite{Batell:2016ove,BABAR:2020oaf}.
\end{itemize}

   
\subsection{Strongly interacting neutrinos}
\label{sec:neutrinos}
{\em Authors: Datta, Marfatia, \underline{Zupan}}

Sterile neutrinos interacting with light particles have been proposed as an explanation for the anomalies in neutrino experiments such as the low energy MiniBoone excess, see e.g.~\cite{Bertuzzo:2018itn,Ballett:2018ynz,Ballett:2019pyw,Datta:2020auq,Abdullahi:2020nyr}.\footnote{The goal of the MiniBooNE experiment was to address the 3.3$\sigma$ LSND anomaly in electron-like events seen in the $\bar\nu_e$ channel~\cite{Aguilar:2001ty}. After 15 years of data taking the MiniBoone dataset itself now shows a 4.8$\sigma$ excess in the low energy part of electron spectra in both the neutrino and antineutrino channels~\cite{Aguilar-Arevalo:2018gpe}, possibly not inconsistent with the LSND anomaly.} 
Similar interactions/models have been used to explain the XENON1T electron excess \cite{Bally:2020yid}. The light mediator would also induce neutrino self-interactions that can potentially explain the Hubble tension \cite{Cyr-Racine:2013jua,Kreisch:2019yzn,Blinov:2019gcj,Esteban:2021ozz,Brinckmann:2020bcn,Das:2020xke,Archidiacono:2020yey} or allow for sterile neutrino dark matter \cite{Johns:2019cwc,deGouvea:2019phk}. The mediator that is light enough would be produced in rare kaon decays, $K^+\to \mu^+\nu X$, via bremsstrahlung of $X$ off the final state neutrino, so that rare kaon decays offer a unique opportunity to test for such neutrino interactions (similar signatures where $X$ is radiated off the charged lepton are discussed in Section~\ref{sec:muonicf}).

\begin{figure}[t]
\begin{center}
  \includegraphics[width=2.9in]{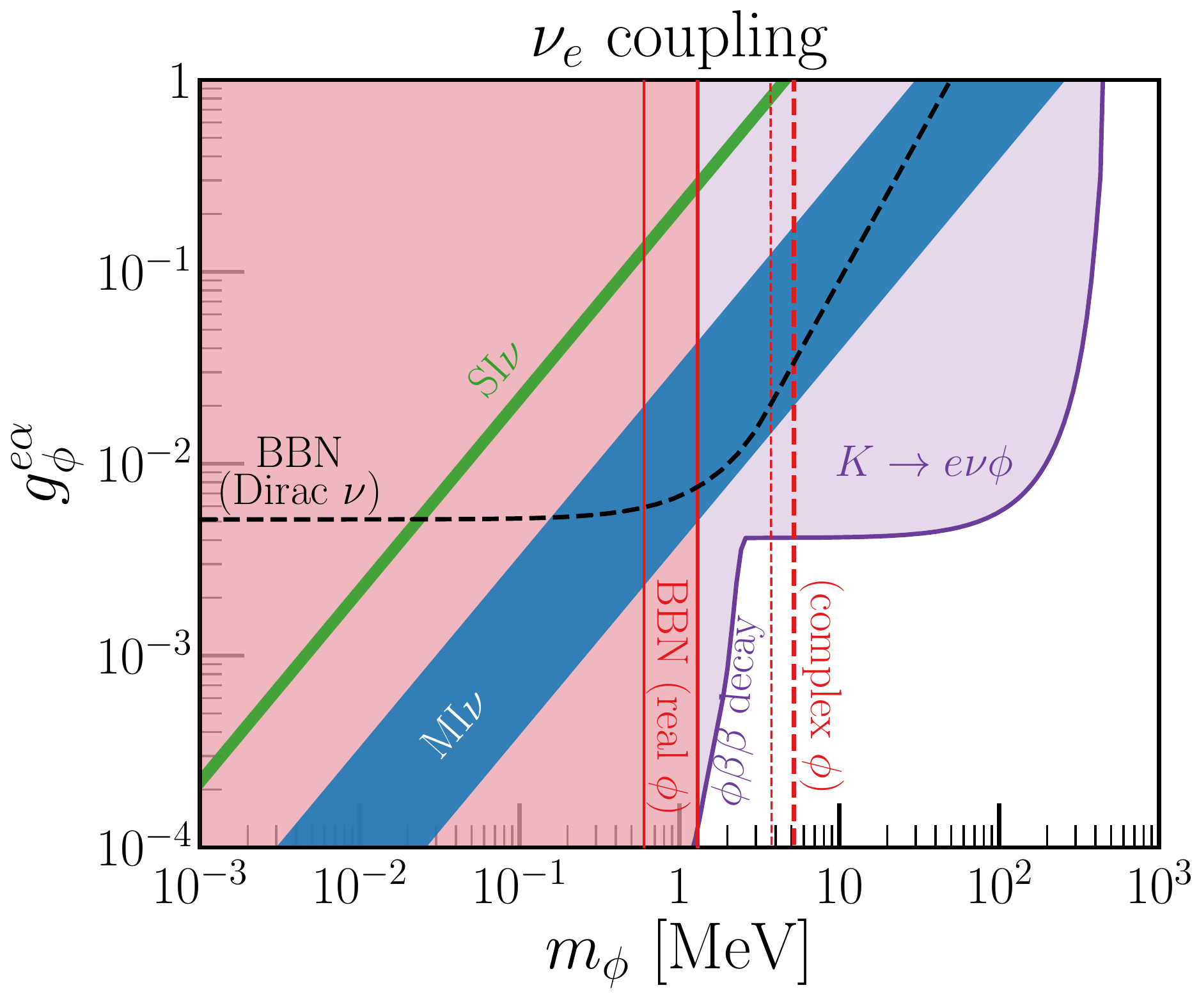}\hspace{0.5cm}
 \includegraphics[width=2.9in]{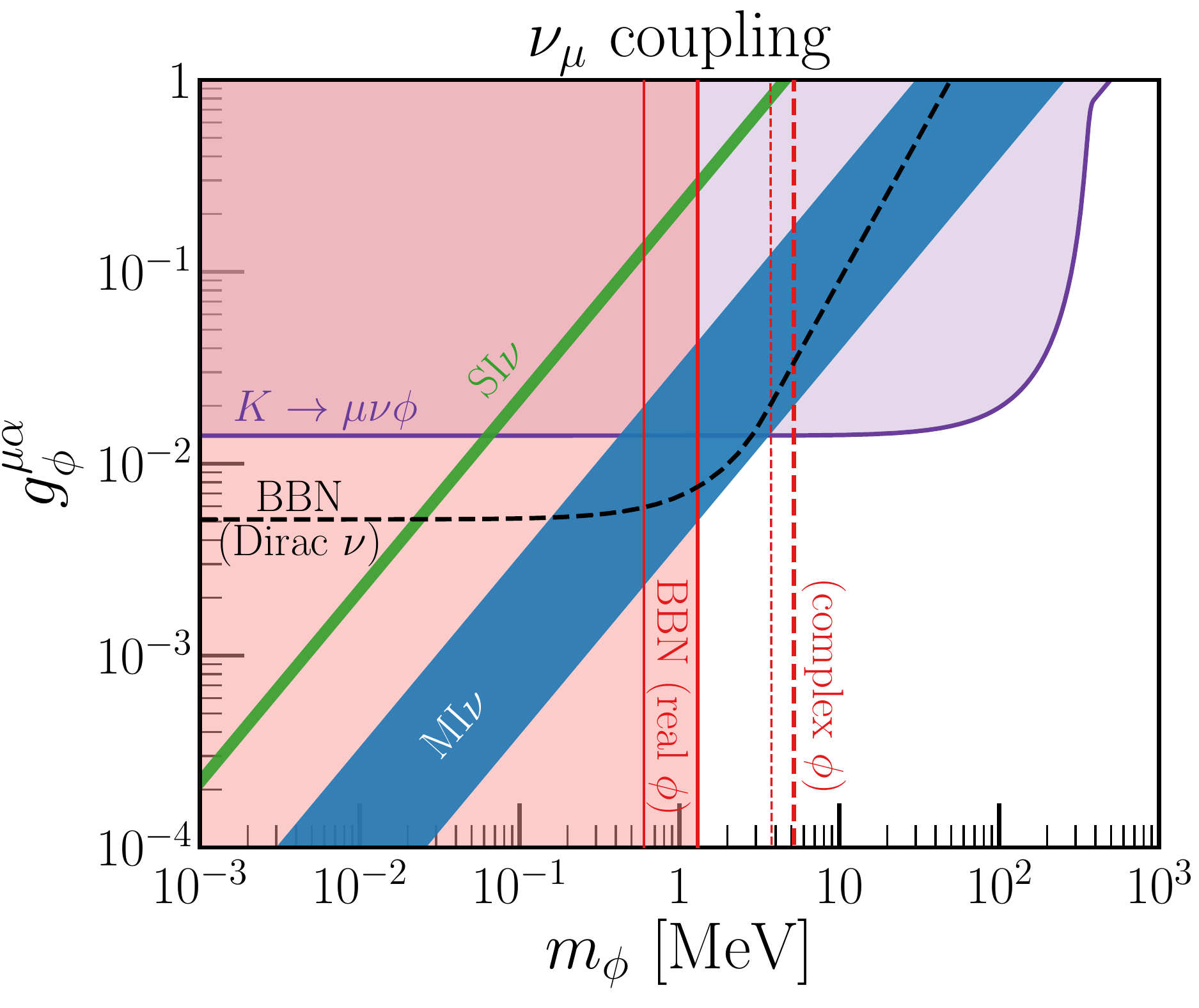}
\caption{\small Bounds (purple shaded regions for constrains from $\tau$ and rare meson decays, red for BBN constraints) on light neutrino-coupled mediators with flavor- specific couplings to either $\nu_e$ (left) or $\nu_\mu$ (right). The bands labeled MI$\nu$ and SI$\nu$ are the  regions in the $g_\phi-m_\phi$ plane preferred by the Hubble tension.  If neutrinos are Dirac, their right-handed components equilibrate before BBN above the dashed black line. Plots taken from \cite{Blinov:2019gcj}.
}
\label{fig:Hubble:tension:neutrinos} 
\end{center}
\end{figure}

The phenomenology depends on whether the SM neutrinos are Majorana or Dirac, and on the properties of the mediator. The two popular choices for a light mediator are a light vector  boson, $V_\mu$, or a (pseudo-)scalar, $\phi$. If the SM neutrinos are Majorana, the interactions with the scalar mediator take the form \cite{Berryman:2018ogk,Blinov:2019gcj}
\beq
\label{eq:strong:int:nu}
{\cal L}_S=\frac{1}{2} g_{\phi}^{\alpha\beta}\bar \nu_{L\alpha}^c \nu_{L\beta} \phi+{\rm h.c.},
\eeq
with the summation over generational indices $\alpha,\beta=1,2,3$ understood. The couplings $g_\phi^{\alpha\beta}$ are not necessarily related to neutrino masses and can be sizable. This is the case, for instance, if $\phi$ is the Majoron from the inverse see-saw model where the lepton number breaking vev is small. The MiniBoonE low energy excess could be explained for $g_{\phi}^{e\alpha}={\mathcal O}(1)$ or $g_{\phi}^{\mu\alpha}= {\mathcal O}(1)$ and $m_\phi= {\mathcal O}(50~{\rm MeV})$. However such large couplings are already excluded by the upper bounds of the ${\cal B}(K^+\to\ell^+\nu X_{\rm inv})$ decay rates (Section~\ref{sec:sig:Kl})~\cite{Pasquini:2015fjv,Berryman:2018ogk}. Improvements on ${\cal B}(K^+\to\ell^+\nu X_{\rm inv})$, on the other hand, could probe the self-interacting neutrino explanation of the Hubble tension \cite{Blinov:2019gcj}. The possibility of $\phi$ coupling to only electron neutrino is excluded (Fig.~\ref{fig:Hubble:tension:neutrinos}, left), while an improvement of a factor of a few on ${\cal B}(K^+\to \mu^+\nu X_{\rm inv})$ would probe fully the preferred region in the case of $\phi$ coupling exclusively to the muon neutrinos (Fig.~\ref{fig:Hubble:tension:neutrinos}, right). The possibility of coupling to tau neutrinos remains wide open, and is not probed by rare kaon decays. The SM neutrinos in \eqref{eq:strong:int:nu} can also be replaced by light, MeV or keV, sterile neutrinos, $N_i$, if these couple to light scalars (see also discussion on HNLs in Section \ref{sec:HNL}). The couplings of $N_i$ to $\phi$ are in general only very weakly constrained.

Similar considerations apply to the Dirac neutrinos. The coupling between $\phi$ and the SM neutrinos can be generated from the $\phi$ interactions with sterile neutrinos, ${\cal L}=- g_D \phi \bar {\nu}_D  \nu_D + \ldots,$ if the sterile neutrinos mixes with the lighter SM neutrinos. The phenomenology can be more involved, if $\phi$ mixes with the Higgs, and resembles the phenomenology of the Higgs portal scalar in Section~\ref{sec:higgs}, but with the large couplings of $\phi$ to neutrinos allowed. In particular, the ${\cal B}(K\to\ell \nu_\ell \phi)$ can be followed by decays to visible SM particles as well as by decays to neutrinos, with the relative branching ratios controlled by the size of $g_D$ and the mixing angles. A UV complete realization of such a model was discussed in Refs.~\cite{Datta:2019bzu, Liu:2020qgx, Batell:2016ove}, where a light scalar singlet was added to a two-Higgs-doublet model. In addition decays through either on-shell or off-shell sterile neutrino, such as $K^+\to\ell \bar\nu_D \to\ell \bar\nu\phi$ with $\phi$ decaying to visible/invisible final states are possible. 

The phenomenology of the light vector mediator also follows the above considerations in broad strokes. The light $Z'$ can couple to either left-handed or right-handed neutrino components. For instance if the sterile neutrino carries a hidden $U(1)$ gauge charge, with $Z'_\mu$ the corresponding gauge boson, the interaction Lagrangian is given by $\mathcal{L}_{V} \supset  g' Z'_\mu \bar {\nu}_D \gamma^\mu  \nu_D= g' Z'_\mu \left(\bar {\nu }_{DL} \gamma^\mu  \nu_{DL}+\bar{\nu} _{DR} \gamma^\mu  \nu_{DR} \right)$. The sterile-active neutrino mixing then leads to couplings of $Z'$ to the SM neutrinos, with similar phenomenology to the one for the scalar mediator discussed above (see also Section~\ref{sec:miss:eng:manybody} for more complex decays involving HNLs). An explicit model of a light $Z'$ coupling predominantly to neutrinos can be found in \cite{Berbig:2020wve}.

\subsection{Grossman-Nir violating models}
\label{sec:GNVgamma}
{\em Authors: Grossman, Hostert, Kitahara, Pospelov,  \underline{Tobioka}}

There is a robust theoretical relation between
${\cal B}(K_L \to \pi^0 \nu \bar\nu)$ and ${\cal B}(K^+\to \pi^+\nu\bar\nu)$, valid in any model that satisfies some mild assumptions: isospin symmetry between $K_L$ and $K^+$ decays, the dominance of the $\Delta I=1/2$ amplitude, and the smallness of CP violation in the decay~\cite{Grossman:1997sk}.
From this one obtains
\beq
\begin{split}
\frac{\Gamma(K_L\to\pi^0\nu\bar\nu)}
{\Gamma(K^+\to\pi^+\nu \bar\nu)}
 \simeq &  \frac{\left| p A - q \bar{A}\right|^2}{\left|\sqrt{2} A\right|^2}
= \frac{\left|p\right|^2}{2} \left| 1 - \lambda\right|^2
= \frac{\left|p\right|^2}{2}
\left( 1+\left|\lambda\right|^2 - 2 \textrm{Re}\lambda \right)
 \\
\simeq & \sin^2 \left[\frac{\textrm{Arg}(\lambda)}{2} \right]
\leq   1\,,
\label{eq:GNderivation}
\end{split}
\eeq
where $K_L = p K^0 - q \bar{K}{}^0$,
$A = \langle \pi^0 \nu \bar \nu |\mathcal{H}|K^0 \rangle$, $\bar A = \langle \pi^0 \nu \bar \nu |\mathcal{H}|\bar{K}{}^0 \rangle$, and $\lambda = (q/p)(\bar{A}/A)$.
 In the approximation in the first line 
the isospin breaking effects are discarded and  $\Delta I=1/2$ dominance is assumed.
The approximation in the second line assumes that  
CP violation is negligible and thus 
$|p| =|q|= 1/\sqrt 2$ and $|\lambda|=1$.
The last inequality in \eqref{eq:GNderivation} translates into a  
 Grossman-Nir (GN) bound,
\begin{equation}
{\cal B}(K_L\to\pi^0\nu\bar\nu) \leq 
4.3\cdot{\cal B}(K^+\to\pi^+\nu \bar\nu),
\label{eq:GNbound}
\end{equation}
where summation over different neutrino species 
is understood \cite{Grossman:2003rw,Li:2019fhz}.
The numerical factor accounts 
for the difference in the $K_L$ and $K^+$
total decay widths,
the isospin breaking effects, and QED radiative corrections~\cite{Mescia:2007kn,Buras:2015qea}.
Since the SM satisfies all of the above assumptions, 
the SM branching ratios fulfill the GN bound
\cite{Buras:2015qea,Brod:2021hsj}.

The latest NA62 analysis of Run 1 data set results in the definitive measurement of this branching ratio: ${\cal B}(K^+ \to\pi^+\nu \bar\nu) = (10.6^{+4.0}_{
-3.4}|_{\rm stat} \pm 0.9_{\rm syst}) \times 10^{-11}$ \cite{NA62:2021zjw}. The most stringent limit from the KOTO experiment is ${\cal B}(K_L \to \pi^0\nu\bar\nu)
<3.0\times 10^{-9}$ at 90\% CL~\cite{KOTO:2018dsc}, while the standard GN bound~\eqref{eq:GNbound} indicates ${\cal B}(K_L\to\pi^0 \nu \bar\nu)<7.7\times 10^{-10}$.

Similarly to the $K_L \to \pi^0 \nu \bar\nu$ case,
a generalized version of the GN bound holds for the production of light NP state(s) $X$,
\begin{equation}
{\cal B}(K_L\to\pi^0 X ) \leq
4.3\cdot{\cal B}(K^+ \to \pi^+ X),
\label{eq:generalGNbound}
\end{equation}
if the assumptions are satisfied for the NP decay (for $K\to \pi\pi X$ see \cite{Geng:2020seh}). The inequality can be saturated
in case $X$ is a CP-even scalar~\cite{Leutwyler:1989xj}. 
If $X$ is a single neutral particle, the current NA62 limit is ${\cal B}(K^+ \to \pi^+ X_{\rm inv})<\mathcal{O}(10^{-11})$,
where the exact numerical value has a strong dependence on the particle mass $m_X$, as well as on its lifetime $\tau_X$ when it decays into visible SM particles~\cite{NA62:2021zjw}.
The generalized GN bound~\eqref{eq:GNbound} then leads to ${\cal B}(K_L \to \pi^0 X_{\rm inv})<\mathcal{O}(10^{-10})$.

Because of the inequality~\eqref{eq:generalGNbound} rare $K^+$ decays set stronger constraints on the parameter space of most NP models, compared to the rare $K_L$ decays. Nevertheless,  there exist also NP models that violate the assumptions of the (generalized) GN bound, for instance, $K_L$ can mix directly with dark sector states, while the charged $K^+$ cannot. In such cases the $K_L$ decays place stronger constraints. The violation of the generalized GN bound may be a fundamental feature of the NP sector, or it can just be due to the experimental loopholes.  

The interest in NP models violating the generalized GN bound increased after the KOTO
experiment observed three events in their signal region~\cite{Ahn:2020opg}, where 
the initial preliminary background estimate was $0.05$ events, later revised to $1.22\pm 0.26$ after the unblinding. Since KOTO is not able to measure the $\gamma\gamma$ invariant mass and is, therefore, unable to determine if they come from a $\pi^0$ or not, the GN violation can also be just apparent, from enhanced $K_L$ decays to $\gamma\gamma+{\rm inv}$ final states, without intermediate $\pi^0$. Besides the interest in the KOTO excess, which might end up being explained by an underestimated background, we will try to emphasize the different physics mechanisms that were put forward. The latter typically points to new signals, or different analysis to be performed at kaon factories. Also, as we will show below, reasonable scenarios require the existence of new light particles, which is in accord with the concept of this review.  

There are five known ways in which the generalized GN violation is achieved in the NP scenarios:
\begin{enumerate}
\item[$(i)$] CP violation in the decay;
\item[$(ii)$] $\Delta I=3/2$ dominance;
\item[$(iii)$] exploiting charge conservation in the decay;
\item[$(iv)$] exploiting mass difference;
\item[$(v)$] effective violation due to experimental loopholes. 
\end{enumerate}

The first four items in the above list 
violate the GN bound fundamentally, in the sense that the corresponding $K^+$ decay is either absent or does not obey the relation in Eq.~\eqref{eq:generalGNbound}. The effective violation can avoid Eq.~\eqref{eq:generalGNbound} by exploiting loopholes in the experimental searches. 
We consider both the cases of heavy ($\Lambda_{\rm NP} \gg m_W$) and light
($\Lambda_{\rm NP} < m_K$) NP.
If the GN violation is fundamental, the heavy new physics is implausible, and instead {\it  light new states} are favored. For the effective violation, we only need one light dark state, which would also be testable with improved $K^+$ analyses.


\subsubsection{Fundamental Violation} 
\label{sec:fundamendal:viol}

\subsubsection*{$(i)$ CP violation in the decay} 

CP violation in the decay, in particular, that implies $|\lambda |\gg 1$ in Eq.~\eqref{eq:GNderivation}, can significantly violate the GN bound. In that case, we also have large CP violation in the charged mode, that is, the $K^+$ and $K^-$ decay rates are different.  Since NA62 measures only one type of charge ($K^+$), this idea is not experimentally ruled out. It can be tested in principle by measuring the charged modes also with $K^-$.
 
Theoretically, this idea requires both strong and weak phases to be present, and the strong phases from QCD are generically expected to be
suppressed  as the final states involve neutrinos (or other SM
singlets). Thus, this idea cannot work with heavy NP. In principle,
light NP can generate a strong phase via intermediate particles that are on-shell. We are unaware of any realistic model that can do it, but that might be due to the fact that no investigation in this
direction was made, and we cannot conclude that it is impossible.

\subsubsection*{$(ii)$ $\Delta I=3/2$ dominance} 

The GN bound assumes $\Delta I=1/2$ dominance. Without new light states, making the $\Delta I=3/2$ transition dominant is the only known way to violate the GN bound~\cite{He:2018uey, He:2020jzn, He:2020jly}. The SMEFT operators that can violate the GN bound are required to contain at least all the relevant fermions (four quarks and two neutrinos), and are thus at least of dimension nine. While the numerical coefficient of the GN bound in Eq.~\eqref{eq:GNbound} can be brought from $4.3\to 20$, the suppression scale of the SMEFT operators needs to be as low as few GeV~\cite{He:2018uey} and hence naively incompatible with the SMEFT framework. Therefore, it seems implausible the GN bound can be appreciably violated only using heavy new physics.

The $\Delta I=3/2$ transition may become dominant more easily
if the SM neutrinos in the final state are replaced with a new light singlet state, $S$~\cite{He:2020jly}. The dimension of the $\Delta I=3/2$ operator is then reduced by two, and even for the suppression scale as high as a TeV, the GN bound is still significantly relaxed, 
\begin{align}
{\cal B}(K_L\to \pi^0 S ) \leq
51 \cdot {\cal B}(K^+ \to \pi^+ S).
\label{GNboundDim7}
\end{align}
The singlet scalar is expected to be invisible in kaon experiments. 
The current NA62 limit on ${\cal B}(K^+\to\pi^+X_{\rm inv})$ is at the ${\mathcal O}(10^{-11})$ level~\cite{NA62:2021zjw} (Fig.~\ref{fig:BRKpiinv} and Section~\ref{sec:sig:Kpi}), which implies that ${\cal B}(K_L\to \pi^0 X_{\rm inv}) \lesssim 5 \times 10^{-10}$ for these types of NP models, well within reach of future KOTO and KLEVER sensitivities.

\begin{table}[t]
\renewcommand{\arraystretch}{1.3}
\centering
  \resizebox{\textwidth}{!}{
\small
\setlength{\tabcolsep}{.15 em}
  \begin{tabular}{ccccc}
\hline\hline
    $K_L$ decay & Model  &   
\begin{tabular}{@{}c@{}}
$K_L \to \gamma\gamma +\textrm{inv}$\\
search strategy 
\end{tabular} 
&
\begin{tabular}{@{}c@{}}
$K^+\to \pi^+ +\textrm{inv}$\\implication
\end{tabular}    
& Other predictions  \\
  \hline  
%
\multirow{3}{*}{$ \to \pi^0 X$} &(d)  \cite{Ziegler:2020ize}   
&
\multirow{3}{*}{current  $\pi^0 X_{\rm inv}$} & SM-like &     \begin{tabular}{c}
    $K^0$--$\bar K^0$ mixing 
    \\[-0.7ex]
    $X_{1,2}$  at beam-dump
    \end{tabular}
\\
    \cdashline{2-2}\cdashline{4-5}
    & $\Delta I=\frac{3}{2}$ \cite{He:2020jly} 
    &  & 

\begin{tabular}{@{}c@{}}
\\[-3ex]
${\cal B}(K_L \to \pi^0 X ) 
$\\[-0.75ex]
$\leq51 {\cal B}(K^+ \to \pi^+ X)$
\\[-3ex]\phantom{a}
\end{tabular} 
& None 
\\
\hline
\multirow{5}{*}{$ \to \pi^0 X_1  X_1 $} & \multirow{3}{*}{(c)  \cite{Hostert:2020gou}}  & 
\multirow{5}{*}{current $\pi^0 \nu\bar\nu$} &
    \begin{tabular}[t]{@{}c@{}}
    \\[-3ex]
${\cal B} (K_L \to \pi^0 X_1 X_1)$\\[-0.7ex]
$\sim 4 (4 \pi)^2 {\cal B} (K^+ \to \pi^+ X_i X_j)$
    \\[-0.7ex]
    if $m_{X_i}+m_{X_j} < m_{K} - m_{\pi}$
    \\[-3ex] \phantom{a}
    \end{tabular} 
& 
    \multirow{3}{*}{\begin{tabular}[t]{@{}c@{}}
    \\[-3ex]
    $B\to K (\pi^0) XX$ (in MFV)
    \\[-0.7ex]
    $X_2$ at beam dumps 
    \\[-3ex]\phantom{a}
    \end{tabular}}
\\  
\cdashline{2-2}\cdashline{4-5} & (e)  \cite{Ziegler:2020ize} &    & SM-like   & 
    \begin{tabular}{@{}c@{}}
    \\[-3ex]
    $K^0$--$\bar K^0$ mixing 
    \\[-0.7ex]
    $X_{2}$  at beam-dump
    \\[-3ex]\phantom{a}
    \end{tabular}
\\   
\hline  
\multirow{3}{*}{$\to \gamma\gamma X_1$} & (a) \cite{Gori:2020xvq}  
      & \multirow{3}{*}{\begin{tabular}{@{}c@{}} optimize $\gamma\gamma$\\[-1.5ex] selection \end{tabular}}
    &  
    \multirow{3}{*}{\begin{tabular}{@{}c@{}}
    \\[-4ex]
  ${\cal B}(K_L \to   \gamma\gamma X_1)$\\[-0.7ex]
  $\sim 4 (4 \pi)^2 {\cal B}(K^+ \to \pi^+ \gamma\gamma X_1 ) $\\[-0.7ex]
   $\sim 4 (4 \pi)^2 {\cal B}(K^+ \to \pi^+ X_1 X_1 ) $\\[-0.7ex]
if $m_{X_i}+m_{X_j} < m_{K} - m_{\pi}$
    \\[-3ex]\phantom{a}
    \end{tabular}}
       & 
\begin{tabular}{@{}c@{}}
\\[-3ex]
${\cal B}(K_L\to \pi^0 XX) 
$\\[-0.7ex]
$\sim 10^{-2}{\cal B}( K_L \to \gamma\gamma X )$ 
\\[-3ex]\phantom{a}
\end{tabular} 

     \\  \cdashline{2-2}\cdashline{5-5}
      &  (a) \cite{Hostert:2020gou}  &  &  & 
\begin{tabular}{@{}c@{}}
\\[-3ex]
$B\to K (X_2 \to \gamma\gamma) X_1$ (in MFV)
\\[-0.7ex]
$X_2$ at beam dumps
\\[-3ex] \phantom{a}
\end{tabular} 
\\   
   \hline
     $ \to \gamma\gamma X_1 X_1  $ & (b) \cite{Hostert:2020gou}   & \begin{tabular}{@{}c@{}} optimize $\gamma\gamma$\\[-1ex] selection \end{tabular} &  
    \begin{tabular}{@{}c@{}}
    \\[-3ex]
 ${\cal B} (K_L \to \gamma\gamma X_1 X_1)$\\[-0.7ex]
 $\sim 4 (4 \pi)^2 {\cal B} (K^+ \to \pi^+ \gamma\gamma X_1 X_1)$\\[-0.7ex]
    if $m_{X_1}+m_{X_2} < m_{K} - m_{\pi}$
    \\[-3ex]\phantom{a}
    \end{tabular}  
& 
    \begin{tabular}{@{}c@{}}
    $B\to K(\gamma \gamma) XX$ (in MFV)
    \\[-0.7ex]
    $X_2$ at beam dumps 
    \end{tabular} 
\\ 
\hline\hline
\end{tabular}
}
\caption{The NP scenarios that lead to violations of the generalized GN bound at the fundamental level. In all cases $X_1(X)$  is an invisible particle. 
Some of the $K^+$ decays are kinematically allowed only for $2m_X<m_K-m_\pi$.
\label{tab:GNV}}
\end{table}
\renewcommand{\arraystretch}{1.0}

\subsubsection*{ $(iii)$ Exploiting Charge Conservation in the Decay} 
Without relying on $\Delta I=3/2$ dominance, one can violate the generalized GN bound if there are new light dark sector particles in the sub-GeV mass range. The basic idea is to exploit the charge difference between $K_L$ and $K^+$. The dark sector particles $X_i$ can only couple to electromagnetic neutral operators of the SM, such as $\bar s d$, meaning that operators like  $X_i X_j \bar s d$ would lead to neutral kaon decays to the dark sector without necessarily predicting analogous decays in charged kaons. These types of scenarios are discussed in Refs.~\cite{Gori:2020xvq, Hostert:2020gou}.

 Typically, these models require at least two states in order to generate the GN-violating $K_L\to \pi^0/\gamma\gamma+\textrm{inv}$ decays. After the direct decay of $K_L$ to the dark sector via $K_L\to X_2 X_{1,2}$, the dark states ought to transition back to pion-like final states. This can easily be achieved through subsequent decays such as $X_2\to \gamma\gamma$, $X_2\to X_1 \gamma$, or $X_2\to  X_1 \pi^0$, where $X_1$ is stable and invisible. These scenarios can be categorized as follows,
\begin{align}\label{eq:pair-production-categories}
& \text{(a) $\pi^0$ impostor \cite{Gori:2020xvq, Hostert:2020gou}}
&&K_L\to X_2 X_1\ \&\ X_2\to \gamma\gamma, 
\\
&\text{(b) dipole portal \cite{Hostert:2020gou}} 
&&K_L\to X_2 X_2 \ \&\ X_2\to X_1\gamma,
\\
&\text{(c) $\pi^0$ production \cite{Hostert:2020gou}}
&&K_L\to X_2 X_1 \ \&\ X_2\to X_1\pi^0.
\end{align}
In Ref.~\cite{Hostert:2020gou}, UV completions for all of the above scenarios are provided using either a Higgs or a $Z'$ portal, all the while assuming Minimal-Flavor-Violation (MFV). If this assumption holds, $b\to s,d$ transitions will also be sensitive to potential signals in $K_L$ decay. Of particular relevance is the process $B\to K X_2 X_{1,2}$, which can be searched for at Belle-II, with $B\to K + \textrm{inv}$ constituting the most promising channel when $X_2$ particles decay outside the detector. Another model example is provided by Ref.~\cite{Gori:2020xvq}, which introduces a complex scalar with {\it a half strangeness} whose real and complex parts correspond to $X_1$ and $X_2$, respectively. This generates the diphoton signal in scenario (a) of Eq.~\eqref{eq:pair-production-categories} via the $X_2F\tilde F$ coupling, which is responsible for $X_2\to \gamma\gamma$ decays. 

Another way to exploit the charge difference is through the mixing of $X_2$ with light neutral mesons. If there is only a linear coupling $X_2  \bar s d$,  both $K^+ \to \pi^+ X_2$ and $K_L \to \pi^0 X_2$ are allowed and the generalized GN bound holds. However, if $X_2$ is heaver than $m_K-m_\pi$  and couples to another light dark sector particle $X_1$, such as $X_1 X_2^2$ ($X_1^2 X_2^2$), the authors of \cite{Ziegler:2020ize} pointed out $K_L \to \pi^0 X_1(\pi^0 X_1X_1)$ has a contribution from mixing:
\begin{align}
&\text{(d) Model 1  \cite{Ziegler:2020ize}}
&&K_L\to X_2^* \to  X_1 \pi^0,
\\
&\text{(e) Models 2\&3 (scalar or fermion $X_1$)  \cite{Ziegler:2020ize}}
&&K_L\to X_2^* \to  X_1X_1\pi^0,   
\end{align}
while $K^+\to\pi^+ X_1(\pi^+ X_1X_1)$ does not have a similar mixing contribution. The prediction of this scenario is  $X_2\bar d d$ operator for $X_2$--$\pi^0$ mixing, and $K^+ \to \pi^+ X_1(\pi^+ X_1X_1)$ is not kinematically forbidden.
The violation of the GN bound is achieved if a diagram  with mixing between the neutral mesons ($K^0, \pi^0$) and $X_2$ dominates in the neutral kaon decay. The $X_2$ cannot be too heavy because the mixing contribution decouples fast as $m_{X_2}^{-8}$. As long as the intermediate scalar is light enough, the correction to $K_L\to \pi^0 +\textrm{inv}$ can be orders of magnitude above the SM while the correction to $K^+\to \pi^+ +\textrm{inv}$ is order one of the SM branching ratio. Since the mixing between $X_2$ and $K^0$ is essential, $K^0$--$\bar K^0$ oscillation sets some stringent constraint, but viable parameter space still exists. Another essential coupling, $X_2 \bar d d$, predicts signals from $X_2$ production or double $X_1$ production at beam-dump experiments. 

So far we only discussed models where the NP couples to flavor-non-diagonal operators. This need not be the case, as pointed out in~\cite{Hostert:2020gou}, as $X$ particles could be produced from flavor-diagonal couplings to the first generation of quarks. This can happen via long-distance ($\Delta S = 1$) induced $K_L$ transitions to virtual unflavored mesons ($\pi^0, \eta$, and $\eta^\prime$), which in turn couple to the dark sector in a flavor diagonal fashion ($\pi^0 X_1 X_2$, for instance). The decays $X_2 \to X_1 \pi^0$ are then unavoidable, if kinematically allowed, and will naturally lead to scenario (c) in Eq.~\eqref{eq:pair-production-categories}.

The scenarios discussed above are categorized in the experimental perspectives in Table~\ref{tab:GNV}. At KOTO/KLEVER, the standard $\gamma\gamma+\text{inv}$ analysis assumes that the two photons are from a $\pi^0$, so the origin of two photons is one criterion. Another criterion would be the number of the invisible particles. The scenarios (c,d,e) and $\Delta =3/2$ dominance scenario will be well tested by the standard $K_L\to \gamma\gamma+\text{inv}$ search, however the standard searches are not optimized for scenarios (a,b). While in scenario (a) it would be possible to adapt the searches to different values of $m_{\gamma \gamma}$, in scenario (b) the two gammas are uncorrelated and will be selected by the standard analyses much less often. Nevertheless, the efficiencies for all cases will be nonzero, depending on the masses, and are shown in the left panel of Fig.~12 of Ref.~\cite{Gori:2020xvq} and in Fig.~8 of Ref.~\cite{Hostert:2020gou}.

We emphasize that in all of the scenarios discussed above, $K^+\to \pi^+XX$ would be allowed only if $2 m_{X_{1}}<m_K-m_\pi$, and the branching fraction would be smaller when compared to $K_L \to X_1\gamma\gamma$ due to the reduced phase space and the three-body nature of the decay.

The models presented here, with simple modifications, can predict  other neutral kaon decay patterns, while still having new physics effects in $K^+$ decays either suppressed or forbidden. For example, if   $X_2$ in scenario (a) can decay through a dilepton channel, the model then predicts the $K^0 \to \ell^+\ell^-+{\rm inv}$ signature.  Instead, if $X_1$ is unstable and decays to leptons, then the $K^0 \to \ell^+\ell^-\gamma\gamma$ signature is predicted. The experimental searches of these channels are discussed in Section~\ref{sec:KL2l:inv}.  
If both $X_1$ and $X_2$ decay to leptons this gives the  $K^0\to \ell_\alpha^+\ell_\alpha^-\ell^+_\beta\ell_\beta^-$ signature, with the  experimental status discussed in Section~\ref{sec:K0:4l}. See also Section~\ref{sec:2X} for more concrete realizations of such models.

\subsection*{$(iv)$ Exploiting isospin dependence of light meson masses}
Within the SM, radiative corrections produce a mass difference between charged and neutral mesons. As a result, the difference between $K$ and $\pi$ masses is different between the charged and neutral modes. If a new particle $Q$ satisfies the following mass relation:
\begin{equation}
    m_{K^+} - m_{\pi^+} =354\,{\rm MeV} < 2 m_{Q}< m_{K_L} - m_{\pi^0} = 363\,{\rm MeV},
    \label{eq:kinematicsrange}
\end{equation}
the three-body kinematics maximally violates the isospin symmetry and the generalized GN bound \cite{Fabbrichesi:2019bmo}.
While $K_L \to \pi^0 Q \bar{Q}$ remains open, 
$K^+ \to \pi^+  Q \bar{Q}$ is kinematically forbidden.
However, the predicted $\pi^0$ momentum is  $p_{T}^{\pi^0} < 60\,$MeV, and in most the events the pions are expected to decay down the beam hole and remain unobservable in the current design of the KOTO setup. In addition, for the decays that do get reconstructed, the measured $p_{T}^{\pi^0}$ will lie much below current signal selection cuts, where backgrounds, such as $K_L \to \gamma \gamma$, are expected to be large.

\subsubsection{Effective violation due to experimental loopholes}
    
We move on to discuss effective violation of the GN bound due to the specific experimental setups of the NA62 and KOTO experiments. We have identified several possibilities, discussed below.

\subsubsection*{$(v$-$i)$ $\pi^0$ blind spot} 
An additional neutral state $X$ is fundamentally constrained by the generalized GN bound in Eq.~\eqref{eq:generalGNbound}. 
However this bound can be effectively violated by resorting to differences in experimental efficiency and background levels.
When one considers the case of $m_{X}\approx m_{\pi^0}$, with $X$ decaying invisibly, constraints
from $K^+\to\pi^+X_{\rm inv}$ searches at the E949 and NA62 experiments are significantly loosened due to the $K^+\to\pi^+\pi^0$ background~\cite{Fuyuto:2014cya}, leading to the \textit{$\pi^0$ blind spot}. Even under these circumstances, an upper bound on ${\cal B}(\pi^0 \to $ \text{inv}) in the $K^+\to\pi^+\pi^0$ channel,
\begin{equation}
{\cal B}
(\pi^0 \to {\rm inv}) < 4.4 \times 10^{-9} \quad (90\%\,{\rm CL}),
\end{equation}
has been adapted to constrain
${\cal B}(K^+\to\pi^+X)$ for the case with $m_{X} = m_{\pi^0}$
as done recently by the NA62 collaboration~\cite{NA62:2020pwi}:
\begin{equation}
{\cal B}(K^+ \to \pi^+ X) \lesssim 10^{-9} \quad (\tau_X = \infty).
\end{equation}
Compared to the current NA62 limit of ${\cal B}(K^+ \to \pi^+ X_{\rm inv} ) < \mathcal{O}( 10^{-11})$ for
$\tau_X = \infty$ at 90\% CL~\cite{NA62:2021zjw}, 
the GN bound is weakened by two orders of magnitude at the $\pi^0$ blind spot. This simple idea can be applied to a wide range of new physics models. The experimental aspects are discussed in Section~\ref{sec:sig:Kpi}.


\subsubsection*{$(v$-$ii)$ Lifetime gap} 
The generalized GN bound \eqref{eq:generalGNbound} is changed when the invisible particle $X$ is unstable and can decay into the visible particles such as photons.
Then, the photons from $X$ decay are vetoed or go to different search categories
where the bound on branching ratio is significantly weaker due to large SM contributions of 
$K_L(K^+)\to \pi^0(\pi^+)\pi^0$ or $\pi^0(\pi^+)\gamma\gamma$  \cite{Artamonov:2005ru,Abouzaid:2008xm,Ceccucci:2014oza}.
In this NP scenario, the signal efficiency is determined by $X$ lifetime $\tau_X$.
A crucial point is that the efficiency also depends on the boost factor $p_X/m_X$ and the 
effective detector size $L$ of the experimental setup. The efficiency from the finite $X$ lifetime  gives \cite{Kitahara:2019lws}
\begin{equation}
{\cal B}(K \to \pi X_{\rm inv} ;\,\textrm{detector})
= \exp\left(- \frac{L}{c \tau_X}\frac{m_X}{p_X}\right) {\cal B}(K\to \pi X),
\end{equation}
where $L$ and $p_X$ depend on the experimental setup. By substituting this into the GN bound \eqref{eq:generalGNbound}, one obtains the following inequality 
\begin{align}
    {\cal B}(K_L \to \pi^0 X_{\rm inv};\,\textrm{KOTO})
        \leq 4.3 \exp\left[ \frac{m_X}{ c\tau_X} \left( \frac{L \big|_{\rm NA62}}{p_X\big|_{\rm NA62}} -\frac{L \big|_{\rm KOTO}}{p_X\big|_{\rm KOTO}}\right)\right]
        {\cal B}(K^+ \to \pi^+ X_{\rm inv};\,\textrm{NA62}).
\end{align}
Since the NA62 has a large decay volume [$L({\rm NA62}) \gg L({\rm KOTO})$], the above parenthesis is always positive and the GN bound is amplified by an exponential factor.
For KOTO, the exponential factor is evaluated  
through a dedicated MC simulation using the selected event samples in
the signal region \cite{Kitahara:2019lws}.
To a good approximation, one can use 
$L({\rm  NA62}) =150\,$m, $p_X({\rm NA62}) = 37\,$GeV and $L({\rm KOTO}) \simeq 3\,$m, $p_X({\rm KOTO}) \simeq 1.5\,$GeV.

As a reference input value, 
when  $m_X = 50\,$MeV and $\tau_X = 0.1\,$ns are taken, 
the relation between the effective branching ratios becomes
\begin{align}
       {\cal B}(K_L \to \pi^0 X_{\rm inv};\,\textrm{KOTO})  \lesssim 130\cdot {\cal B}(K^+ \to \pi^+ X_{\rm inv};\,\textrm{NA62}).
\end{align}
In this way, the GN bound can be violated effectively by orders of magnitude. 
Detailed studies based on this mechanism have been done in Refs.~\cite{Kitahara:2019lws,Egana-Ugrinovic:2019wzj,Liu:2020qgx, Liao:2020boe}.
Furthermore, such a light long-lived particle 
[$m_X=\mathcal{O}(100)$\,MeV and $\tau_X = \mathcal{O}(0.1)$\,ns] is a  good target for  
the FASER experiment \cite{Kling:2020mch}.


\subsubsection*{$(v$-$iii)$ Kinematics}
Finally, the GN bound can  effectively violated by change of the decay particle distributions.
Such a situation occurs when the NP sector contains very light invisible mediator. In Ref.~\cite{Fabbrichesi:2019bmo}, 
a virtual dark photon production followed by the invisible decay is considered: $K^+\to \pi^+ Q\bar{Q}$ and  $K_L \to \pi^0 Q\bar{Q}$, where $Q$ are dark fermions (see  Section~\ref{sec:darkphoton} for details) and the dark photon has a negligible mass.
This light dark photon provides a long-distance contribution to $K \to \pi Q\bar{Q}$. 
One cannot distinguish the $\pi Q\bar{Q}$ and $\pi\nu\bar\nu$ final states, however the decay particle distributions deviate from the standard Dalitz distributions.
If the signal efficiency of KOTO analysis is enhanced by the change of the decay distributions relative to that of NA62 analysis, an effective violation of the GN bound could be realized. 
This possibility is  confirmed numerically: a slight violation of the GN bound is possible when the dark-fermion mass is only around 128~MeV. Quantitatively, the effective violation of the GN bound turns out to be limited compared to what has been estimated in Ref.~\cite{Fabbrichesi:2019bmo}.\footnote{We had a private discussion with the authors of Ref.~\cite{Fabbrichesi:2019bmo} and the source of discrepancy in our findings can be tracked back to their overestimate of the signal efficiency for the dark-photon contribution in the KOTO detector.}


\subsection{Models leading to production of two dark sector particles}
\label{sec:2X}
{\em Authors: Pospelov, \underline{Hostert}, Shuve}
\newcommand{\mh}[1]{\textcolor{purple}{#1}}

Dark sectors may contain a rich spectrum of particles, which brings the question of whether $s\to d$ transitions in the SM may also lead to the production of not one, but multiple dark sector states. This is the case when, for example, a mediator particle couples to the SM FCNC and to pairs of lighter particles in the dark sector. If the mediator is heavier than the kaon, it may be integrated out to generate higher-dimensional operators that are responsible for the direct production of the lighter dark degrees of freedom in kaon decays. This is discussed in Section~\ref{sec:GNVgamma} in the context of GN violating models, where $K \to X_1 X_2\to\gamma\gamma +$inv decays are generated by scalar or vector boson mediators~\cite{Hostert:2020gou}. On the other hand, if the mediator is sufficiently light to be produced on-shell, it may decay to lighter dark states via a cascade of fast decays in the dark sector. This cascade could eventually result in decays to visible SM particles, resulting in a high multiplicity of visible final states in kaon decays, typically much larger than discussed in the previous sections. Decays with multi-lepton final states, for instance, could then be a more interesting strategy to search for these dark sectors than existing searches for missing mass or for a single visible resonance~\cite{Hostert:2020xku}.

Although several SM extensions can lead to the production of more than one new particle, we focus on specific examples which are both minimal and theoretically appealing. As a general rule, these can be categorized according to how the pair production of new particles takes place: \emph{i)} via a mediator particle, or \emph{ii)} via non-linear couplings of the new states to the SM. The latter scenario includes examples such as $K\to \pi X X$ from contact interactions, see Section~\ref{sec:GNVgamma}, as well as the more exotic possibility of new pseudo-scalars that mix with the light SM mesons, such as an MeV axion~\cite{Alves:2017avw}. For the production via mediator particles, we restrict the discussion to mediators coupled to the $\overline{s} d$ transitions (as opposed to the emission of mediators from the final state leptons, discussed in Sections~\ref{sec:muonicf} and \ref{sec:neutrinos}).


\subsubsection{Higgsed U$(1)^\prime$}
\label{sec:HiggsedU1-0}

\begin{figure}[t]
    \centering
    \includegraphics[width=\textwidth]{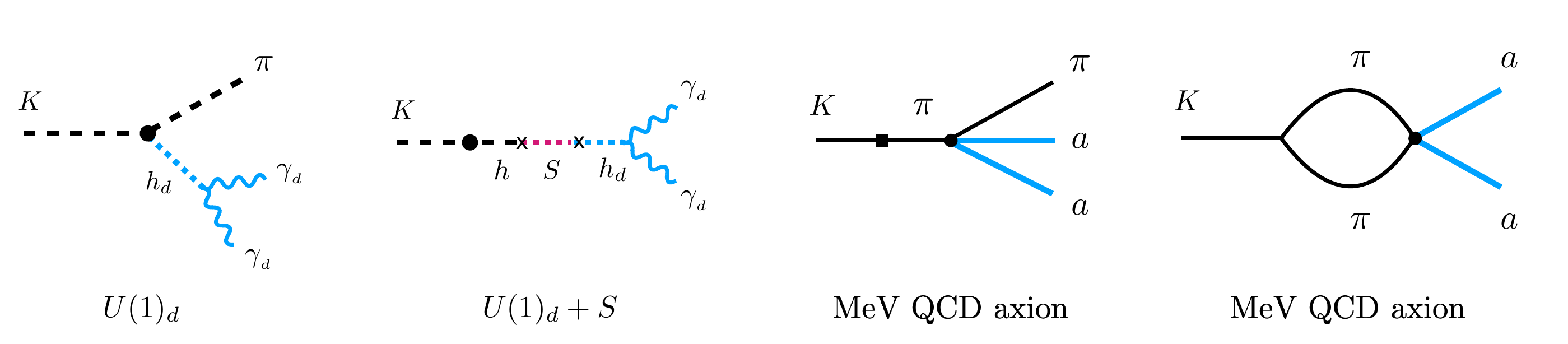}
    \caption{\small Kaon decay diagrams in minimal and extended $U(1)_d$ dark sectors, as well as in the MeV QCD axion model. The subsequent prompt decays of $A'=\gamma_d$ and $\varphi=h_d$ lead to observable four-lepton signatures.\label{fig:DS_diagrams}}
\end{figure}

One of the simplest extensions of the SM that predicts multiple dark particle production is a higgsed dark $U(1)_d$ which couples to the SM both via kinetic mixing and scalar mixing. The Lagrangian is given by~\cite{Ahlers:2008qc,Pospelov:2008zw}
\begin{align}\label{eq:higgsedU1}
    \mathcal{L}_{\rm DS} &= |D^\mu \Phi|^2 - \frac{1}{4} F_{d}^{\mu \nu}F_{d\,\mu \nu}  - \frac{\varepsilon}{2} F_{d}^{\mu \nu} F_{\mu\nu}  -\mu^2(\Phi^\dagger \Phi) - \lambda(\Phi^\dagger \Phi)^2 - \lambda_d (\Phi^\dagger \Phi)(H^\dagger H), 
\end{align}
where $F_d$ is the field strength tensor of the dark photon, $\varepsilon$ is the kinetic mixing parameter, and $\lambda_d$ the mixed quartic coupling of the scalars. After spontaneous symmetry breaking $\Phi$ and the SM Higgs boson mix with a mixing angle $\theta\simeq \lambda_d v_d/\lambda v_{\rm EW}$.
Similarly to Section~\ref{sec:higgs}, the dark scalar $\varphi$ can be produced in $K\to\pi\varphi$ decays by mimicking the Higgs-mediated FCNC in the SM. If the dark photon is lighter than $m_\varphi/2$, then $\varphi \to A^\prime A^\prime$ decays proceed promptly, followed by two individual $A^\prime \to e^+e^-$ decays. The latter decays are prompt even for small kinetic mixing,
\begin{equation}
c\tau^0_{A^\prime} \simeq 70~\mu\text{m}\times\left( \frac{2\times10^{-4}}{\varepsilon} \right)^2 \times\left( \frac{30~\text{MeV}}{m_{A'}} \right).
\end{equation}
The final signature is that of $K\to \pi \varphi \to \pi 2(e^+e^-)$ as shown in the leftmost panel of Fig.~\ref{fig:DS_diagrams}. The total invariant mass of the four leptons reconstructs the $\varphi$ mass and the two individual invariant masses of the pairs reconstruct the dark photon mass for a given combination.

Note that the constraints shown in Fig.~\ref{fig:higgs_portal_summary} for the case of an invisible dark scalar mixed with the Higgs do not apply as shown. For the parameter space relevant for this section, both the scalar and the dark photon decays are much faster than the decays of the dark Higgs in the minimal Higgs-portal model. These decays, therefore, do not run into problems with BBN, even for very small mixing angles $\theta$. For the same reason, limits from decay-in-flight searches at beam dump and neutrino experiments are no longer relevant since the scalars decay well before reaching the detector. The LHCb constraints on $\varphi\to \mu^+\mu^-$ are also modified as $\varphi$ decays predominantly to dark photons, which may or may not decay back to muons, depending on their mass. Regarding the remaining constraints from $K\to\pi \varphi$, we note that the scalar and the subsequent dark photon decays lead to  additional energy deposition in the kaon detectors. In the current searches, these multi-lepton final states are mostly vetoed, depending on the detector acceptance, and thus the bounds are relaxed compared to the minimal Higgs-portal model. Even though no complete analysis of the relevant parameter space of the higgsed dark  $U(1)_d$  model, Eq. \eqref{eq:higgsedU1}, has been performed, mixing angles $\theta$ as large as $\mathcal{O}(10^{-3})$ are not excluded. Naturally, this section is concerned with the searches for such additional visible signatures. 

With the largest allowed values of $\theta$, we find kaon decay branching ratios of~\cite{Hostert:2020xku}
\begin{align}\label{eq:BRKtopiX}
\mathcal{B}\big({K_L\to\pi^0 2(e^+e^-)}\big) &\simeq 2 \times 10^{-7} \times \left(\frac{\sin\theta}{5\times 10^{-3}}\right)^2,
\\\label{eq:BRKtopiX2}
\mathcal{B}\big({K^+\to\pi^+ 2(e^+e^-)}\big) &\simeq 5 \times 10^{-8}\times \left(\frac{\sin\theta}{5\times 10^{-3}}\right)^2.
\end{align}
The branching ratio for the corresponding $K_S$ decay is suppressed by the ratio of neutral kaon lifetimes, $\tau_{K_S}/\tau_{K_L}\simeq 1.8 \times 10^{-3}$, $ \mathcal{B}\big(K_S\to \pi^0\varphi\big)=\big(\tau_{K_S}/\tau_{K_L}\big) \mathcal{B}\big(K_L\to \pi^0\varphi\big)$.
As long as the dark photon decays promptly, the branching ratio is only sensitive to the scalar potential parameters. The prompt decay assumption is clearly valid in the region of $m_{A'}=\mathcal{O}(100)$~MeV and $\varepsilon= \mathcal{O}(10^{-4})$, which is very challenging to probe experimentally (Fig.~\ref{fig:darkphoton_bounds}). These decay modes are also discussed in the multi-axion processes of Section~\ref{sec:nonlinear_axion}. The experimental prospects are discussed in Section~\ref{sec:sig:5track}.

\subsubsection{Super-renormalizable Higgs portal in the higgsed U$(1)^\prime$ sector}
\label{sec:HiggsedU1}

Generically, the mixed quartic coupling in the Higgs portal model is strongly constrained by $h\to \varphi\varphi$ decays. It can be shown that Higgs to invisible constraints preclude any signatures in $K_{S,L}\to A^\prime A^\prime$ in a purely dimension-4 Higgs portal model. Nevertheless, super-renormalizable portals can avoid such strong constraints provided no direct $h\varphi^\dagger \varphi$ coupling is present. A simple extension of the higgsed $U(1)^\prime$ model above by a dark charge-neutral scalar $S$ can avoid that. Consider, for instance, and extension of \eqref{eq:higgsedU1} by the Lagrangian 
\begin{equation}
\label{eq:LS:extendU1}
\mathcal{L}_{S} = \partial_\mu S\partial^\mu S - m_S^2 S^2 -A_H H^\dagger H S - A_{\phi} \Phi^\dagger \Phi S.
\end{equation}
In the limit $\lambda_d\to 0$, the only portals between the SM and the dark sectors are kinetic mixing $\epsilon$ in \eqref{eq:higgsedU1}, which guarantees fast $A^\prime\to e^+e^-$ decays, and the super-renormalizable portals of $S$, namely the couplings $A_H$ and $A_\phi$ in \eqref{eq:LS:extendU1}. The $A_{H,\phi}$ couplings mediate the production of dark photons in kaon decays without violating $h\to \varphi\varphi$ or $h\to SS$ constraints. The kaon decay mode into two dark photons is showed in the second panel of Fig.~\ref{fig:DS_diagrams}. More specifically, neglecting the mass of $A^\prime$ in Higgs decays,
\begin{align}\label{eq:KLto2Adark}
\mathcal{B}({K_L\to A^\prime A^\prime}) &\simeq 5\times10^{-8} \times  \left(\frac{\mathcal{B}({h\to A^\prime A^\prime })}{10\%}\right)\times\left( \frac{m_K^4}{(m_K^2-m_S^2)^2 + m_S^2\Gamma_S^2}\right)\times f\left(\frac{m_{A^\prime}}{m_{K}}\right),
\end{align}
where $f(x) = (1 - 4x^2 + 12x^4)\sqrt{1-4x^2}$.
For masses of $m_S=\mathcal{O}(1)$~GeV, we obtain experimentally interesting branching ratios of $\mathcal{O}(10^{-9})$. As long as $m_{A^\prime}<2 m_\mu$, the dark photon decays to $e^+e^-$ pairs $100\%$ of the time. For $K_S$ decays, the rate is enhanced but the branching ratio is smaller, 
\begin{equation}
\frac{\mathcal{B}(K_S\to A' A')}{\mathcal{B}(K_L\to A'A')} \simeq \left(\frac{\Re{V_{ts}^*V_{td}}}{\Im{V_{ts}^*V_{td}}}\right)^2\times \left(\frac{\tau_{K_S}}{\tau_{K_L}}\right)\simeq 9\times 10^{-3}.
\end{equation}

Compared to $K\to\pi X$ decays, the complete kaon annihilation into dark sector particles allows for the production of heavier $X$ particles. For example, decays into four muons $K_{S,L}\to 2 (A^\prime \to \mu^+\mu^-)$ are kinematically allowed, as opposed to $K\to \pi 2(A^\prime \to \mu^+\mu^-)$, which are not. As a benchmark point for the dark photon in the super-renormalizable portal model, we may take $m_{A^\prime} = 230$~MeV and $m_S = 800$~MeV, enhancing the kaon decay rates by the fact that $m_S$ is close to $m_K$. For the above parameter values, considering that $\mathcal{B}(A^\prime\to e^+e^-) = 0.78$ and $\mathcal{B}(A^\prime\to \mu^+\mu^-) = 0.22$, we expect
\begin{align}
&\mathcal{B}(K_L\to e^+e^-e^+e^-) \simeq 3.2 \times 10^{-9},
\quad
&\mathcal{B}(K_S\to e^+e^-e^+e^-) \simeq 2.9 \times 10^{-11},\\
&\mathcal{B}(K_L\to e^+e^-\mu^+\mu^-) \simeq 8.9 \times 10^{-10},
\quad
&\mathcal{B}(K_S\to e^+e^-\mu^+\mu^-) \simeq 8.0 \times 10^{-12},\\
&\mathcal{B}(K_L\to \mu^+\mu^-\mu^+\mu^-) \simeq 2.5 \times 10^{-10}, 
\quad 
&\mathcal{B}(K_S\to \mu^+\mu^-\mu^+\mu^-) \simeq 2.2 \times 10^{-12}.
\end{align}
The $\mu^+\mu^-$ decays are expected to be smaller than or comparable to the $e^+e^-$ ones. These decays are constrained experimentally by the measurements of the analogous process $K_{L}\to e^+e^-e^+e^-$ at KTeV and NA48 leading to a PDG averaged branching ratio of $(3.56\pm 0.21)\times 10^{-8}$~\cite{ParticleDataGroup:2020ssz}, and the KTeV measurement of $\mathcal{B}(K_L \to e^+e^- \mu^+\mu^-)= (2.69\pm 0.27)\times 10^{-9}$~\cite{KTeV:2002kut}. The experimental status is further discussed in Section~\ref{sec:K0:4l}: in general, the experimental sensitivity to $K_L$ decays is at the ${\cal O}(10^{-10})$ level. These measurements can be considered as an upper limit for annihilation to dark sector particles that decay to a total of four leptons. Finally, we note that a measurement of the SM rates for these channels is also of theoretical interest in order to constrain the sign of amplitude $\mathcal{M}(K_{S,L}\to\gamma\gamma)$ and improve the predictions for $K_{S,L}\to\mu^+ \mu^-$~\cite{Isidori:2003ts,DAmbrosio:2013qmd}.

\subsubsection{Non-linear interactions of MeV axions}
\label{sec:nonlinear_axion}
New particles may also be pair-produced if they couple non-linearly to the SM mesons, for instance, in MeV-scale QCD axion models. As argued in Refs.~\cite{Alves:2017avw,Alves:2020xhf}, a QCD axion is possibly still allowed to have MeV masses, if it is electron-philic and pion-phobic, see also Section \ref{sec:lldecaysAlves}. In this case the axion decays promptly, $c\tau=\mathcal{O}(10^{-6})$ meters, and predominantly into $e^+e^-$ pairs. For $m_a \simeq 17$ MeV it may also explain the anomalous results by the ATOMKI experiment~\cite{Krasznahorkay:2015iga}. 

While in these models the axion-pion mixing $\theta_{a \pi}$ is small, other terms involving the SM light mesons and higher-order terms of the axion field will not be. For instance, even for $\theta_{a \pi} \to 0$, the terms 
\begin{align}\label{eq:coupling_pi2a2}
    \mathcal{L} \supset &\frac{m_{a}^2}{4F_\pi^2} aa \pi^0\pi^0 + \frac{m_{a}^2}{2F_\pi^2} aa \pi^+\pi^-
    +\frac16 \frac{m_a^2}{F_\pi f_a} a^3\pi^0,
\end{align}
are still present and have their coefficient fixed for any given QCD axion mass. For a $17$~MeV axion with $f_a=1$~GeV decay constant the non-linear interactions are rather strong. Analogous terms with the $\eta$ mesons are not shown but are also present. 

By virtue of the non-linear interactions above, the axion can be pair produced in kaon decays~\cite{Hostert:2020xku} in the decay modes showed in Fig.~\ref{fig:DS_diagrams}.  
For a $17$~MeV axion with $f_a=1$~GeV the axion pair production in association with a final state pion has large branching ratios, 
\begin{align}
\mathcal{B}(K_L\to\pi^0 a a \to \pi^0 2(e^+e^-))&\simeq7 \times 10^{-5},\\
\mathcal{B}(K^+\to\pi^+ a a\to \pi^+ 2(e^+e^-))&\simeq1.7 \times 10^{-5}.
\end{align}
Closing the pion lines in $K\to \pi\pi$ gives a one-loop generated $K_{S,L}\to a a$ decays. Taking into account only the absorptive part of the amplitude gives \begin{equation}
\mathcal{B}(K_S\to aa \to 2(e^+e^-)) \simeq 2.6 \times 10^{-7},    
\end{equation}
for the $17$~MeV axion. These modes are all within reach of current experiments and may conclusively rule out or confirm the QCD MeV axion models, including its explanation of the ATOMKI anomaly. It should be noted that even if $K-\pi$ mixing generated by the $\Delta S =1$ chiral Lagrangian were to be anomalously small, the measured $K\to 3\pi$ amplitude would still generate the decays above at loop-level at an appreciable rate.

\subsubsection{Dark fermions}
\label{sec:miss:eng:manybody}

We now return to the U$(1)_d$ models, this time considering also the production of dark fermions.
In particular, if some of these fermions eventually decay back to SM particles, then their production in kaon decays can lead to exotic final states.
Two examples of U$(1)_d$ extensions of the SM with unstable dark fermions include dark seesaw models~\cite{Ballett:2019pyw,Abdullahi:2020nyr} as well as co-annihilating dark matter particles~\cite{Tucker-Smith:2001myb}. The main difference between these lies in whether or not these fermions mix with SM neutrinos. If they do, they can be produced in charged-current (CC) kaon decay processes via neutrino mixing and the lightest stable fermion produced in the decay chains will correspond to an SM neutrino. In addition, they may also decay via CC (e.g. $N\to\pi^{\pm}\ell^\mp$). For the purposes of this section, however, these can be safely ignored as their width is much smaller than those generated by the U$(1)_d$ interactions. If CC decays were to dominate, then, given current experimental constraints, the HNL would be too long-lived to decay inside the detector and the HNL would constitute an invisible particle.

Consider, for instance, an extension of the higgsed dark $U(1)^\prime$ model by a pair of neutral fermions, one of which is charged under the new symmetry~\cite{Ballett:2019pyw},
\begin{align}\label{eq:nudL}
    \mathcal{L}_{\rm DS-\nu} &= \overline{N_R}i\slashed{\partial}N_R + \overline{N_L}i\slashed{D}N_L  + y_d \overline{N_L}\varphi N_R + + y_N \overline{L}\widetilde{H}N_R).
\end{align}
If multiple generations of $N$ particles exist, the heaviest states can decay down to the lightest ones by emission of on- or off-shell mediators, $N^\prime\to N A^\prime$ or $N^\prime\to N\varphi$. As the notation implies, $N$ can also mix with SM neutrinos and generate light neutrino masses through the inverse seesaw mechanism. For all purposes, we assume that at low energies there are one or multiple generations of $N$ particles that interact with SM neutrinos as described in Section~\ref{sec:HNL}.

\begin{figure}[t]
    \centering
    \includegraphics[width=0.78\textwidth]{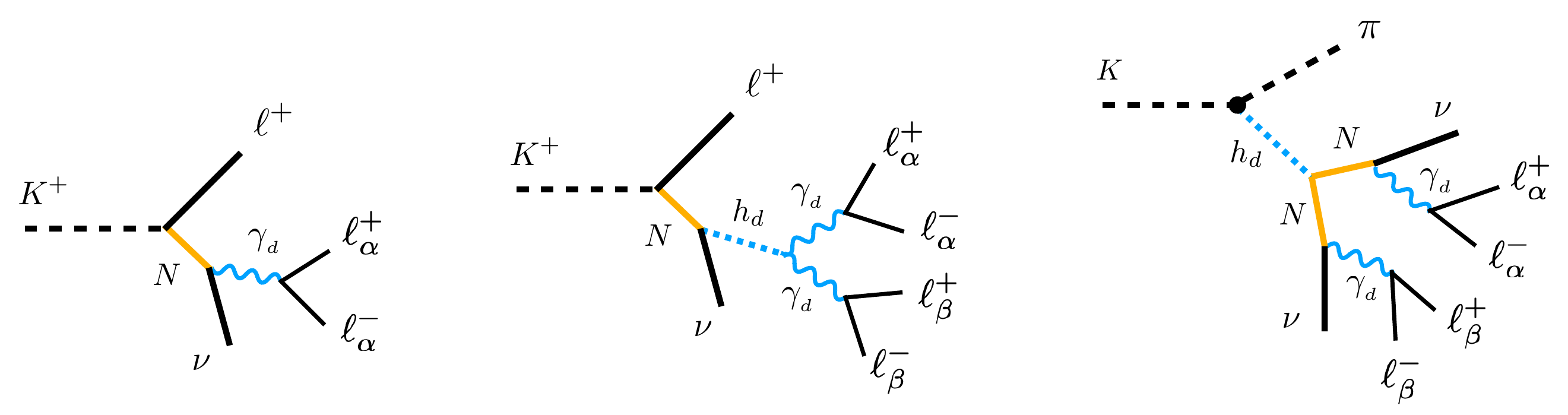}
    \caption{Kaon decays to heavy neutral leptons with multi-lepton final states. All dark sector particles, $N$, $A'=\gamma_d$, and $\varphi=h_d$, are produced on shell. Experimental reconstruction of the dark sector masses is possible for the first two diagrams. The flavor indices $\alpha$ and $\beta$ can be $e$ or $\mu$, depending on the process. \label{fig:hnl_multilepton}}
\end{figure}

We start by discussing the production of the mediators. When they are heavier than the dark fermions $N$, they will decay to pairs of $N$. For example, if $m_{A^\prime} < m_N < m_{\varphi}/2$, we can expect the following decays
\begin{align}
    K_{S,L}&\to\pi^0 \big(\varphi \to N N \to 2(\nu A^\prime) \to 2(\nu e^+e^-)\big),
    \\
    K^+&\to\pi^+ \big(\varphi \to N N  \to 2(\nu A^\prime) \to 2(\nu e^+e^-)\big),
\end{align}
with branching ratios as in Eqs.~\eqref{eq:BRKtopiX}, \eqref{eq:BRKtopiX2} provided that the dark sector decay chain has a branching ratio of $1$. 
The multiple invisible particles in the final states above make for a challenging experimental signature. For multiple generations of $N$ particles, decays of the type $\varphi \to N^\prime N^\prime \to 2 (N A^\prime) \to 4(\nu e^+e^-)$ can also take place, illustrating how fast the multiplicity of leptons can grow in more elaborate dark sectors. If $N$ has sizeable mixing with SM neutrinos, it can be produced in CC kaon decays, $K^+\to \ell^+ N$, with $\ell = e$ or $\mu$, depending on the hierarchy of the electron and muon mixing. If $N$ is long-lived, the signature is identical to that discussed in Section~\ref{sec:HNL}: a single invisible peak in $m_{\rm miss}^2 = (p_K - p_\ell)^2$. However, if the HNL is heavier than the dark mediators, it subsequently rapidly decays as $N\to\nu A^\prime$ or $N\to \nu\varphi$. These two channels have comparable branching ratios as a consequence of symmetry breaking. 
\begin{figure}
    \centering
    \includegraphics[clip, trim=0.1cm 0.3cm 0.1cm 0.1cm, width=0.49\textwidth]{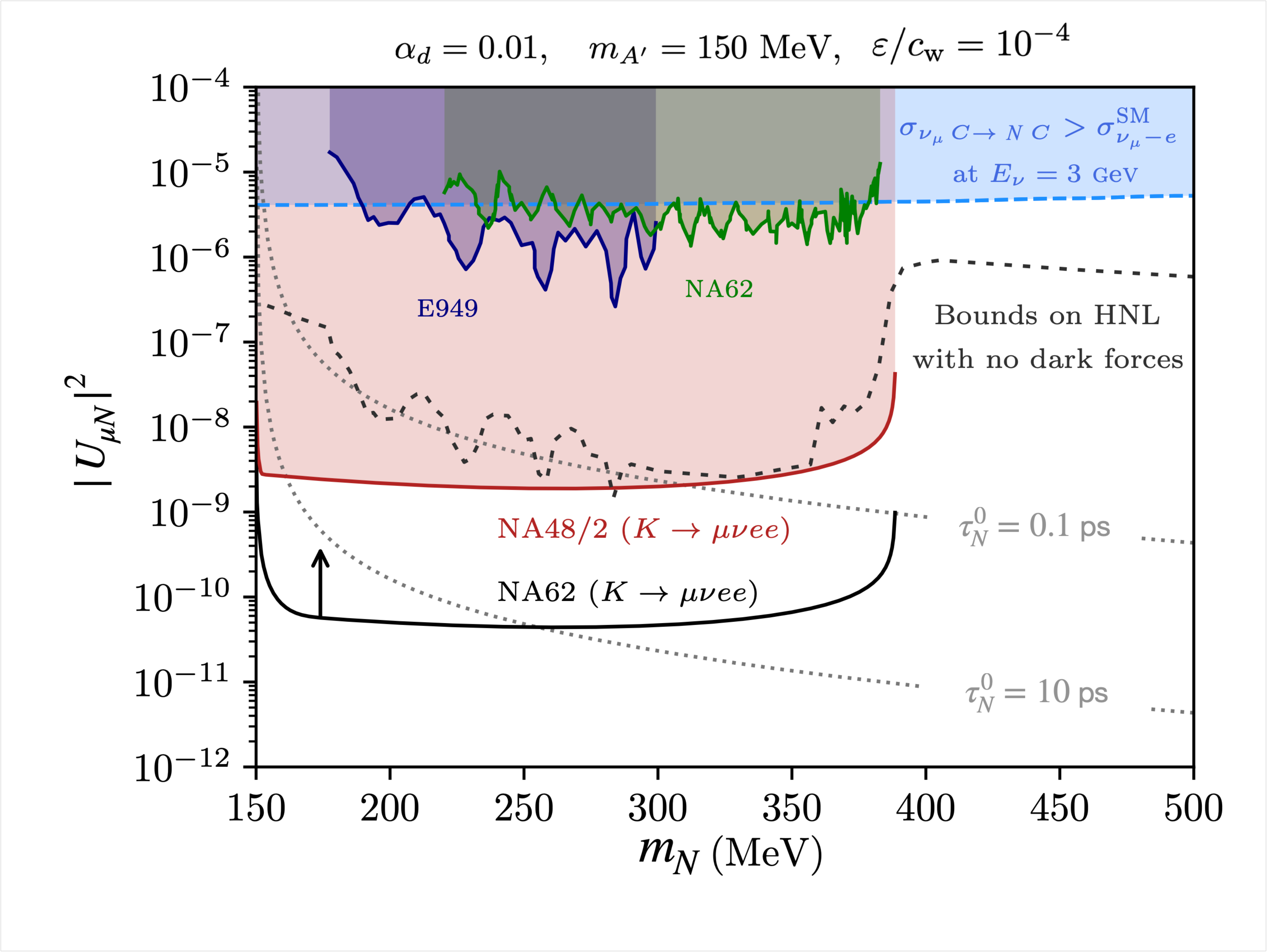}
    \includegraphics[clip, trim=0.1cm 0.1cm 0.1cm 0.2cm, width=0.49\textwidth]{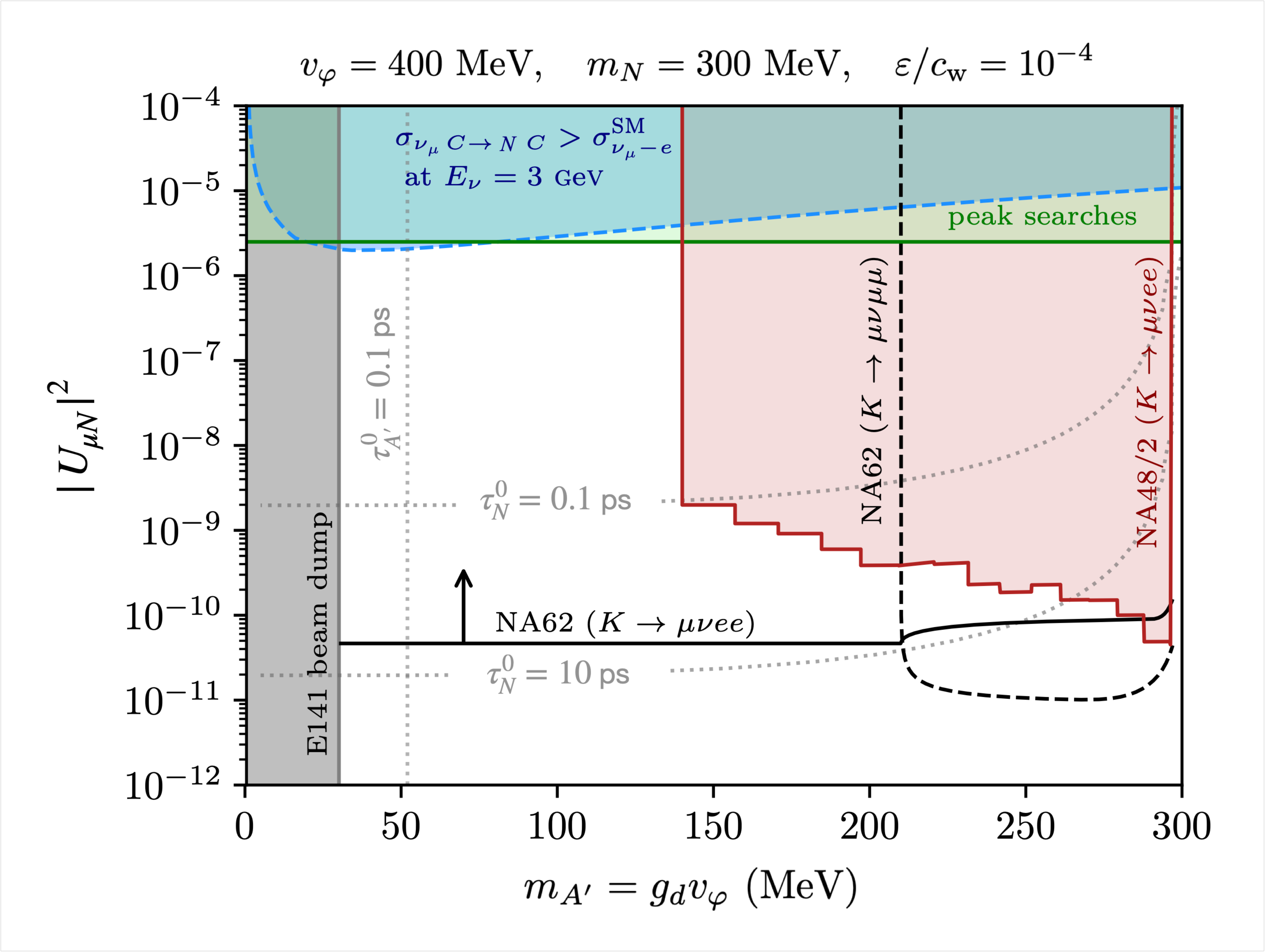}
    \caption{The mixing of the HNLs with muon neutrinos as a function of the HNL mass (left) and the dark photon mass (right) for a given choice of dark sector parameters~\cite{Ballett:2019pyw}. We show the NA62 sensitivity to $K^+\to \mu^+ \nu e^+e^-$ as well as $K^+\to \mu^+ \nu \mu^+\mu^-$ decays arising from CC kaon decays followed by prompt HNL decays $N\to \nu (A^\prime\to\ell^+\ell^-)$. \label{fig:dark_HNL_ps}}
\end{figure}

The resulting cascade process may be searched for due to the leptonic decays of the mediators as follows:
\begin{align}
&K^+\to\ell^+N, \quad N\to \nu A^\prime \to \nu \ell^+\ell^-,
\\
&K^+\to\ell^+N, \quad N\to \nu \varphi \to \nu A^\prime A^\prime \to \nu 2(e^+e^-).
\end{align}
The former decay chain results in three charged tracks (similarly to the muonic force mediator model discussed in Section~\ref{sec:muonicf}), while the latter decay chain leads to a total of five charged tracks (Fig.~\ref{fig:hnl_multilepton}).
Since only a single particle in the process is invisible, it is possible to experimentally constrain the masses of all dark particles involved. The HNL emission produces a peak in the missing mass, while the dark scalar and dark photon masses appear as resonances in combinations of the lepton four-momenta. No searches for these processes have been performed to date; future experimental prospects are discussed in Section~\ref{sec:sig:5track}.

The three-track signature is particularly interesting and is in general always present whenever the five-track process is possible. It also provides a direct test of dark-neutrino explanations to the MiniBooNE excess~\cite{Bertuzzo:2018itn,Ballett:2018ynz,Ballett:2019pyw,Datta:2020auq,Dutta:2020scq,Abdallah:2020biq,Liu:2020ser,Abdullahi:2020nyr}, where HNLs are produced in muon neutrino interactions inside the detector due to the exchange of a new force. A slice of the dark sector parameter space is shown in Fig.~\ref{fig:dark_HNL_ps} together with the NA62 sensitivity to $K^+\to \mu^+ (N\to \nu \ell^+\ell^-)$ decays~\cite{Ballett:2019pyw}. Analogous decays with positrons are also possible, $K^+\to e^+ N$, providing sensitivity to $|U_{e N}|^2$ instead of $|U_{\mu N}|^2$. Depending on the dark sector couplings, the HNL lifetime can be larger than a few picoseconds, and therefore may lead to displaced decays. Further discussion of the experimental sensitivity can be found in Section~\ref{sec:3l:inv}.

\label{sec:decays}
\renewcommand{\arraystretch}{1.4}
\begin{table*}[t]
    \centering
    \begin{tabular}{lccc}
        \hline\hline
        Kaon decay & Model & Branching ratio & Additional resonances \\
        \hline
        $K_{S}\to 2(e^+e^-)$ & $U(1)_{d}+S$ & $2\times 10^{-10}$ & $-$ 
\\
        $K_{S}\to (\mu^+\mu^-)(e^+e^-)$ & $U(1)_{d}+S$ & $ 8\times 10^{-12} $ & $-$ 
\\
        $K_{S}\to 2(\mu^+\mu^-)$ & $U(1)_{d}+S$ & $ 2\times 10^{-12} $ & $-$ 
\\
\hline
     $K_{L}\to 2(e^+e^-) $ & $U(1)_{d}+S$ & $2\times 10^{-8}$ & $-$
\\
  $K_{L}\to (\mu^+\mu^-)(e^+e^-)$ & $U(1)_{d}+S$ & $ 9\times 10^{-10} $ & $-$ 
\\
     $K_{L}\to 2(\mu^+\mu^-) $ & $U(1)_{d}+S$ & $ 2\times 10^{-10} $ & $-$
\\
        $K_{L}\to \pi^0 \, 2(e^+e^-)$ & $U(1)_{d}$ & $2\times 10^{-7}$ & 
        $m_{4e}=m_{\varphi}$
\\
        $K_L\to \pi^0 \, 2(\nu e^+e^-)$ & $U(1)_d +$HNL & $2\times 10^{-8}$ & $-$
\\\hline
        $K^+\to \pi^+ \, 2(e^+e^-)$ & $U(1)_{d}$ & $5\times 10^{-8}$ & $m_{4e}=m_{\varphi}$
\\
       $K^+ \to e^+ \nu \, e^+e^-$ & $U(1)_d +$HNL & 
       $6 \times 10^{-8}$ & $m_{ee} = m_{A^\prime}$, $m_{\nu ee } = m_N$ 
\\
        $K^+ \to \mu^+ \nu \, e^+e^-$ & $U(1)_d +$HNL & $9 \times 10^{-7}$ & $m_{ee} = m_{A^\prime}$, $m_{\nu ee } = m_N$ 
\\
       $K^+ \to e^+ \nu \, 2(e^+e^-)$ & $U(1)_d +$HNL & 
       $6 \times 10^{-8}$ & $m_{4e} = m_\varphi$, $m_{\nu 4e } = m_N$ 
\\
        $K^+ \to \mu^+ \nu \, 2(e^+e^-)$ & $U(1)_d +$HNL & $9 \times 10^{-7}$ & $m_{4e} = m_\varphi$, $m_{\nu 4e } = m_N$ 
\\
%
        $K^+\to \pi^+ \, 2(\nu e^+e^-)$ & $U(1)_d +$HNL & $5\times10^{-9}$ & $m_{2\nu 4e} = m_\varphi$ \\\hline
\hline
        \end{tabular}
    \caption{Multi-lepton kaon decays channels with branching ratios in several representative new physics models. For decays into electrons we take for the dark sector model parameters,  $s_\theta \sim 5\times10^{-3}$, $m_{\varphi} = 100$~MeV, $m_{A^\prime} = 30$~MeV, and $m_S=0.8$~GeV. For channels involving muons, we assume instead $m_{A^\prime} = 230$~MeV. For the dark sector models with heavy neutral leptons, we take $|U_{\mu 4}|^2 \sim 10^{-6}$, $|U_{\mu 4}|^2\sim10^{-7}$, and $m_N=150$~MeV. In all new physics scenarios, $m_{ee}=m_{ee}^\prime = m_{A^\prime}$. In the last column we show additional kinematical conditions that may be explored in experimental searches.
     }
    \label{tab:BRstable}
\end{table*}
\renewcommand{\arraystretch}{1.0}

\subsection{Dark-baryon models} \label{subsec:dark-baryon}
\label{sub:dark:baryon:models}
{\em Authors: Alonso-Alvarez, Elor, Escudero, Fornal, Grinstein, \underline{Martin Camalich}}

The energy density of cold dark matter and of baryons in the Universe is remarkably similar, $\Omega_{\rm cdm}/\Omega_{\rm b} \simeq 5.38$~\cite{Aghanim:2018eyx}.
This similarity suggests that there could be a direct connection between the dark sector and baryon number, thus motivating the existence of dark sector particles charged under U(1)$_B$ with masses at the GeV scale, see Refs.~\cite{Petraki:2013wwa,Zurek:2013wia} for reviews. 
Recently, the idea of (anti)baryonic dark sectors has been further motivated by an anomaly in the measurements of the neutron lifetime which could be resolved if the neutron decays into dark states carrying baryon number with a branching fraction $\sim 1\%$~\cite{Fornal:2018eol}.
In addition, dark baryons play a key role in the newly proposed \emph{B-Mesogenesis} mechanism, in which baryogenesis and dark matter production arise from the CP-violating oscillations and subsequent decays of $B$ mesons in the early Universe~\cite{Elor:2018twp,Nelson:2019fln,Alonso-Alvarez:2021qfd}.

The studies~\cite{Elor:2018twp,Nelson:2019fln,Alonso-Alvarez:2021qfd} (see also~\cite{Aitken:2017wie,Heeck:2020nbq,Fajfer:2020tqf,Fajfer:2021woc}) have highlighted that hadrons could have rather large branching fractions into final states containing light dark baryons, which would appear as missing energy in a detector. It is quite remarkable that, in the light of current data, branching fractions as large as $1\%$ for the neutron and $0.5\%$ for the $B$ meson are allowed if the decays occur into states in the dark sector \emph{which carry baryon number}. Given this possibility, it is timely to explore the possible reach of current and upcoming experiments to similar decays in the hyperon sector. In fact, in the models of~\cite{Fornal:2018eol} and~\cite{Elor:2018twp,Nelson:2019fln,Alonso-Alvarez:2021qfd}, one generically expects all hadrons to decay into these types of modes at a non-negligible rate. Therefore, hyperons offer the opportunity to test, in a precise manner, the existence of such particles, which could very well be linked to dark matter or the origin of the baryon asymmetry of the Universe. In what follows, we address the minimal particle content and interactions needed to trigger such decays. We identify the decay channels that appear more amenable for experimental searches, find theoretically motivated benchmarks, and put the parameter space into perspective by also considering LHC and flavor mixing constraints on the mediators needed to trigger the decay. We refer to~\cite{Alonso-Alvarez:2021oaj} for further details and precise calculations of the decay rates.

\subsubsection{Lagrangian, interactions and particle content}

Let us consider a new heavy-colored boson which couples to two quarks, as well as to a quark and a dark baryonic fermion $\chi$. There are three realizations of this scenario, given by
\begin{align}
\mathcal{L}_{1}&\supset-y_{d_ad_b} \epsilon_{ijk} \Psi^i d_{Ra}^j d_{Rb}^{k}  - y_{\chi u_c}\Psi^{*}_i {\chi}u_{Rc}^i  + {\rm h.c.}, \label{eq221}
\\
\mathcal{L}_{2}&\supset-y_{u_ad_b} \epsilon_{ijk} \Phi^i u_{Ra}^j d_{Rb}^k  - y_{\chi d_c} \Phi^{*}_i {\chi}d_{Rc}^i  -\,y_{Q_aQ_b} \epsilon_{ijk}\epsilon_{\alpha\beta} \Phi^i Q_{La}^{j\alpha} Q_{Lb}^{k\beta}  +{\rm h.c.},\label{eq22}
\\
\mathcal{L}_{3}&\supset-y_{Q_ad_b} \epsilon_{ijk}\epsilon_{\alpha\beta} X_{\mu}^{i\alpha} Q_{La}^{j\beta}\gamma^\mu d_{Rb}^k- y_{\chi Q_c} X_{\mu}^{\dagger i\alpha} Q_{Lc}^{i\alpha}\gamma^\mu {\chi}   + {\rm h.c.}, \label{eq222}
\end{align}
where $u_R$, $d_R$ and $Q_L$ denote the SM quark fields, $i$, $j$ and $k$ the color indices, $a$, $b$ and $c$ the generation indices and $\alpha$ and $\beta$ the $SU(2)_L$ indices. Note that in $\mathcal{L}_{1}$ the two $d_{R}$ quarks belong to different families due to the antisymmetry of the color indices. The field $\chi$ is a SM singlet Dirac fermion with baryon number $B_\chi=1$. The possible new heavy colored mediators are color triplet scalars $\Psi = (3,1)_{\frac23}$ and $\Phi = (3,1)_{-\frac13}$, and a color triplet vector $X_\mu = (3,2)_{\frac16}$, where the first (second) number in parenthesis indicates the $SU(3)_c$ ($SU(2)_L$) representation and the subscript the hypercharge. 
The latter two  particles carry the quantum numbers of leptoquarks and as such one must impose baryon number conservation in order to forbid proton decay. In the first model, couplings inducing proton decay are not allowed by gauge invariance. 

Direct LHC searches for new colored states place a lower bound $M \gtrsim 1.2~\text{TeV}$ on the $\Psi$, $\Phi$ and $X_\mu$ masses~\cite{Aad:2020aze,Sirunyan:2019ctn}. Therefore, at low energies,  the heavy colored states can be integrated out leading to the following  effective Lagrangian:
\beq
\begin{split}
\label{eq:LEEFTchi}
\mathcal L_{\rm eff}\,\, \supset \,\, &C^{R}_{ud,d}\,\mathcal O^{R}_{ud,d}+ C^{L}_{ud,d}\,\mathcal O^{L}_{ud,d}\\
+&C^{R}_{ud,s}\,\mathcal O^{R}_{ud,s}+C^{R}_{us,d}\,\mathcal O^{R}_{us,d}+C^{L}_{ud,s}\,\mathcal O^{L}_{ud,s}+C^{L}_{us,d}\,\mathcal O^{L}_{us,d}\\
+&C^{R}_{us,s} \,\mathcal O^{R}_{us,s}+C^{L}_{us,s}\,\mathcal O^{L}_{us,s}.
\end{split}
\eeq
The operators in the first, second and third line correspond to  $\Delta S=0$, 1, and 2 strangeness changing transitions, respectively, with
\begin{align}
\mathcal O_{ud_a,d_b}^{R}=\epsilon_{ijk}(u_R^i d_{aR}^j)(\chi_R d_{bR}^k),\hspace{0.8cm}
\mathcal O_{ud_a,d_b}^{L}=\epsilon_{ijk}(u_L^i d_{aL}^j)(\chi_R d_{bR}^k).
\end{align}
the corresponding Wilson coefficients are parametrically suppressed by $C_{ud_a,d_b}^{R(L)}\sim1/M^2$. We have used Fierz relations to express the operators 
produced by the exchange of $\Psi$ (model 1) and $X^\mu$ (model 3) in terms of the scalar operators in this Lagrangian~\cite{Weinberg:1979sa,Abbott:1980zj,Alonso-Alvarez:2021oaj}. The contributions of the three models to this effective Lagrangian can be found in~\cite{Alonso-Alvarez:2021oaj}. Note that $\mathcal L_{\rm eff}$ exhausts the possible dimension-six operators that couple the SM fields to the singlet $\chi$ in an $SU(2)_L\times U(1)$-invariant way. 

On top of the effective couplings in Eq.\,(\ref{eq:LEEFTchi}), the dark particle $\chi$ may have interactions with additional dark sector particles. A minimal model for such a scenario is provided by including the Lagrangian term
\begin{align} \label{eq:Lag_dark}
    \mathcal{L} \ &\supset \  y_{\xi \phi}\, \bar{\chi} \xi \phi \ + \ \text{h.c.}\,,
\end{align}
where $\xi$ is a Dirac fermion and $\phi$ is a scalar, both singlets under the SM gauge group. We shall assume baryon number conservation which implies $B_\phi + B_\xi = 1$. In addition, assuming a $Z_2$ symmetry under which $\xi$ and $\phi$ are odd and all other particles are even assures that either $\xi$, $\phi$, or both are stable, making them good dark matter candidates. 

\subsubsection{Hyperon decays}
\label{sec:darkbaryons}
The hyperon decays induced by the above effective Lagrangians are summarized in Table~\ref{tab:hyperon:dark:sector:decays}. The three simplest types of decay  with different final state particles are:
\begin{enumerate}
    \item \textit{Missing energy + pion}: $\qquad$ $B_1 \to \chi + \pi$, $\,$ for $m_\chi < m_{B_1}-m_{\pi}$.
    \item \textit{Missing energy + photon}: $\quad$ $B_1 \to \chi + \gamma$,  $\,$ for $m_\chi < m_{B_1}$.
    \item \textit{Missing energy only}:$\quad$ $\quad \,\,\,\,\,\,\,$ $B_1 \to \xi + \phi$, $\,$ for $m_\xi+m_\phi < m_{B_1}$. 
\end{enumerate}
The states $\chi$, $\xi$, and $\phi$ are inert and do not interact with the detector. Above, $B_1$ denotes any initial hyperon. Importantly, lighter baryons including the proton could undergo similar decays as well. Given the stringent limits on the proton lifetime, $\tau_p \gtrsim 10^{29}\,\text{years} $~\cite{pdg}, we restrict our study to sufficiently heavy dark baryons, $m_\chi > m_p - m_e$, and $m_\xi + m_\phi > m_p - m_e \simeq 937.76\,\text{MeV}$, for which proton decay is kinematically forbidden.

\begin{table}[t]
\renewcommand{\arraystretch}{1.5}
  \setlength{\arrayrulewidth}{.25mm}
\centering
\small
\setlength{\tabcolsep}{0.8 em}
\begin{tabular}{  c  c  c c  c }
    \hline
    \hline
    Initial state & Final state & ${\rm Max}[m_\chi]$ (MeV)   & ${\cal B} \times \left[ {C_{ud_a,d_b}^{R(L)}}/{(0.5\,\text{TeV}^{-2})}\right]^2$  & BESIII limit \\ 
    \hline 
    $\Lambda^0\, (uds)$ &  $\chi + \gamma$ & $1116$ &  $3.5\times 10^{-6}$ & -- \\
     $\Lambda^0\, (uds)$ &  $\xi + \phi$ & $1116$ &  $5.0 \times 10^{-4} \, y_{\xi \phi}^2$ & $<7\times 10^{-5}$~\cite{BESIII:2021slv} \\
     $\Lambda^0\, (uds)$ &  $\chi + \pi^0$ & $981$ &  $1.1 \times 10^{-3} $ & -- \\
    $\Sigma^+\, (uus)$ &  $\chi + \pi^+$ & $1050$ &   $1.6\times 10^{-3}$ & -- \\
    $\Sigma^-\, (dds)$ &  $\chi + \pi^-$ & $1058$ &   $1.2\times 10^{-3}$ & -- \\
    $\Xi^0\, (uss)$ & $\chi + \gamma$ & $1315$  &   $ 1.5\times 10^{-5}$ & -- \\
     $\Xi^0\, (uss)$ & $\xi + \phi$ & $1315$  &   $3.7\times 10^{-4}\,y_{\xi \phi}^2$  & -- \\
     $\Xi^0\, (uss)$ & $\chi + \pi^0$ & $1180$  &   $6.2\times 10^{-3}$  & -- \\
    $\Xi^-\, (dss)$ &  $\chi + \pi^-$ & $1182$ &  $7.1\times 10^{-3}$  & -- \\
    \hline
    \hline
\end{tabular}
\caption{The most relevant hyperon decay channels into dark sector states: initial and final state particles, maximum masses of the dark sector states that can be probed, and theoretical expectations for the branching fraction for $m_\chi = 1.0~\text{GeV}$ ($m_\chi = 0.95~\text{GeV}$ for $\Lambda^0\to\chi\pi^0$), $m_\phi = 0.95~\text{GeV}$ and $m_\xi = 0.04~\text{GeV}$. 
For decay channels to which several effective operators in Eq.~\eqref{eq:LEEFTchi} can contribute, we show the prediction for the operator that gives the largest branching ratio.
The predictions are normalized to the largest possible values of the Wilson coefficients given the LHC constraints on colored mediators.
We also display the current limit on $\Lambda \to \textrm{inv}$ from BESIII~\cite{BESIII:2021slv}. See also Section VIA in~\cite{Alonso-Alvarez:2021oaj} and Section~\ref{sec:hyppheno} in this review for qualitative discussion of expected sensitivities for these decay modes. 
\label{tab:hyperon:dark:sector:decays}
}
\end{table}

In Table~\ref{tab:hyperon:dark:sector:decays}, we list the most relevant decay modes of hyperons into dark sector states as triggered by the effective operators in Eq.~\eqref{eq:LEEFTchi}. The rate for each of these decays has been calculated in~\cite{Alonso-Alvarez:2021oaj} by matching these to the $SU(3)$ chiral Lagrangian~\cite{Claudson:1981gh}. All the necessary hyperon-decay transition matrix elements can be predicted in terms of known strong and electromagnetic properties of the baryons and of the proton decay constants that have been calculated on the lattice~\cite{Aoki:2017puj}. The detailed expressions  and predictions can be found in~\cite{Alonso-Alvarez:2021oaj}.

The Wilson coefficients involved in these decays, $C_{ud_a,d_b}^{R(L)}$ in Eq.~\eqref{eq:LEEFTchi}, are constrained by direct searches for the colored mediators at the LHC.  ATLAS~\cite{Aad:2020aze} and CMS~\cite{Sirunyan:2019ctn} SUSY searches for pair-produced squarks decaying into a neutralino and a quark rule out the existence of this kind of strongly interacting bosons in the mass region below $1.2~\textrm{TeV}$ with the current 139~fb$^{-1}$ of data.
For masses of the mediator in the range $1.2~\text{TeV} < M < 7~\text{TeV}$, searches for resonantly singly produced colored states decaying into quark pairs (dijet) or a quark and a dark baryon (jet+MET) can be used to place constraints on the Wilson coefficients for such masses. Although the limits depend on the exact flavor structure of the operator, an illustrative average value of the constraints on the Wilson coefficients is $C_{ud_a,d_b}^{R(L)} < 0.5~\mathrm{TeV}^{-2}$~\cite{Alonso-Alvarez:2021oaj}. In what follows, we normalize the branching ratios for the exotic hyperon decays to $C_{ud_a,d_b}^{R(L)} = 0.5~\mathrm{TeV}^{-2}$ in order to highlight the relevant window of parameter space in which hyperon searches can test uncharted regions of parameter space given LHC constraints. In addition, we note that some of the decay modes shown in Table~\ref{tab:hyperon:dark:sector:decays} are subject to astrophysical constraints arising from the observed neutrino signal from SN1987A, see~\cite{Alonso-Alvarez:2021oaj} for the constraints.

\begin{figure*}[t]
\centering
\begin{tabular}{cc}
		\label{fig:Decays1}
		\includegraphics[width=0.48\textwidth]{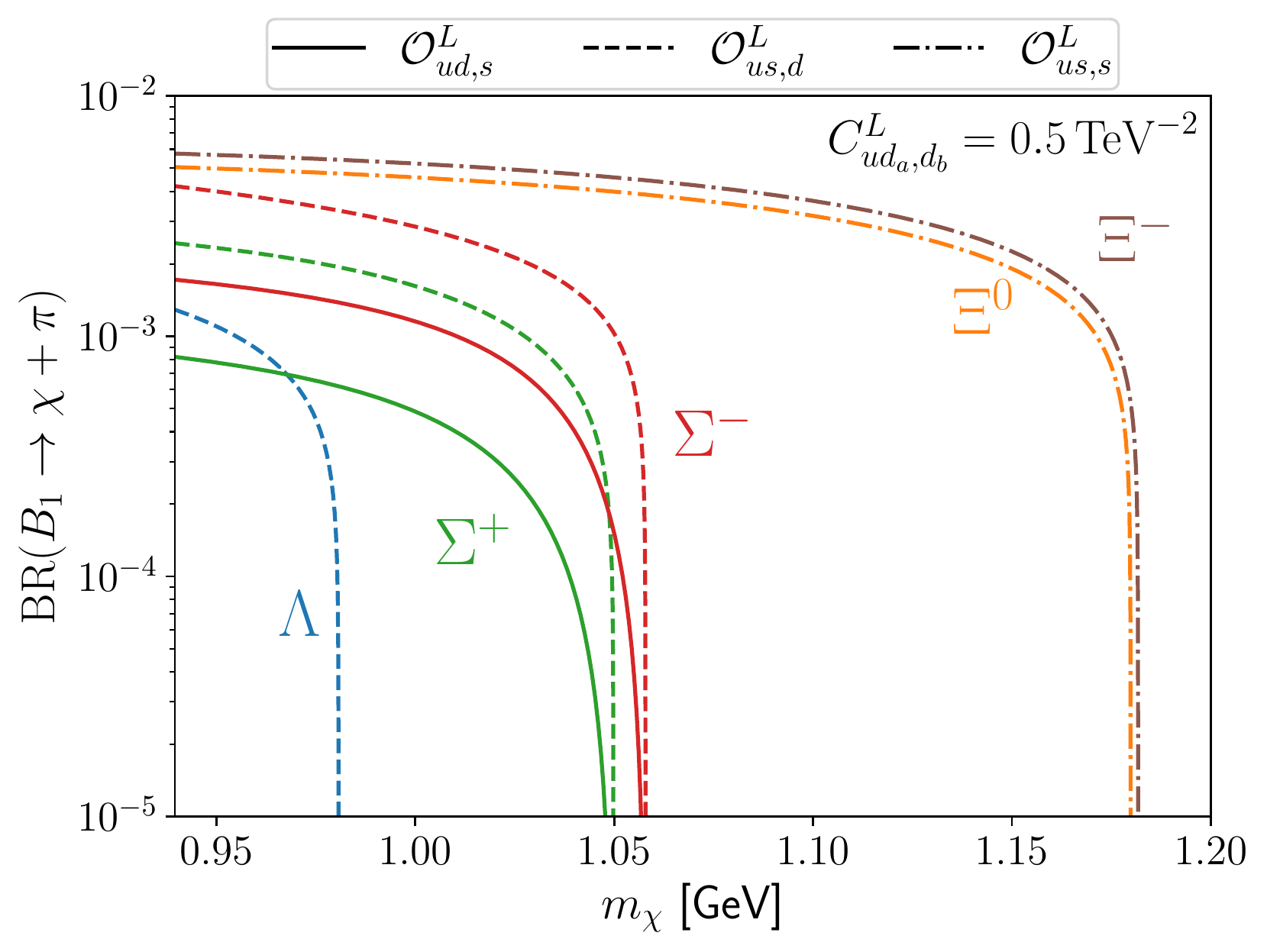}
&
		\label{fig:Decays2}
		\includegraphics[width=0.48\textwidth]{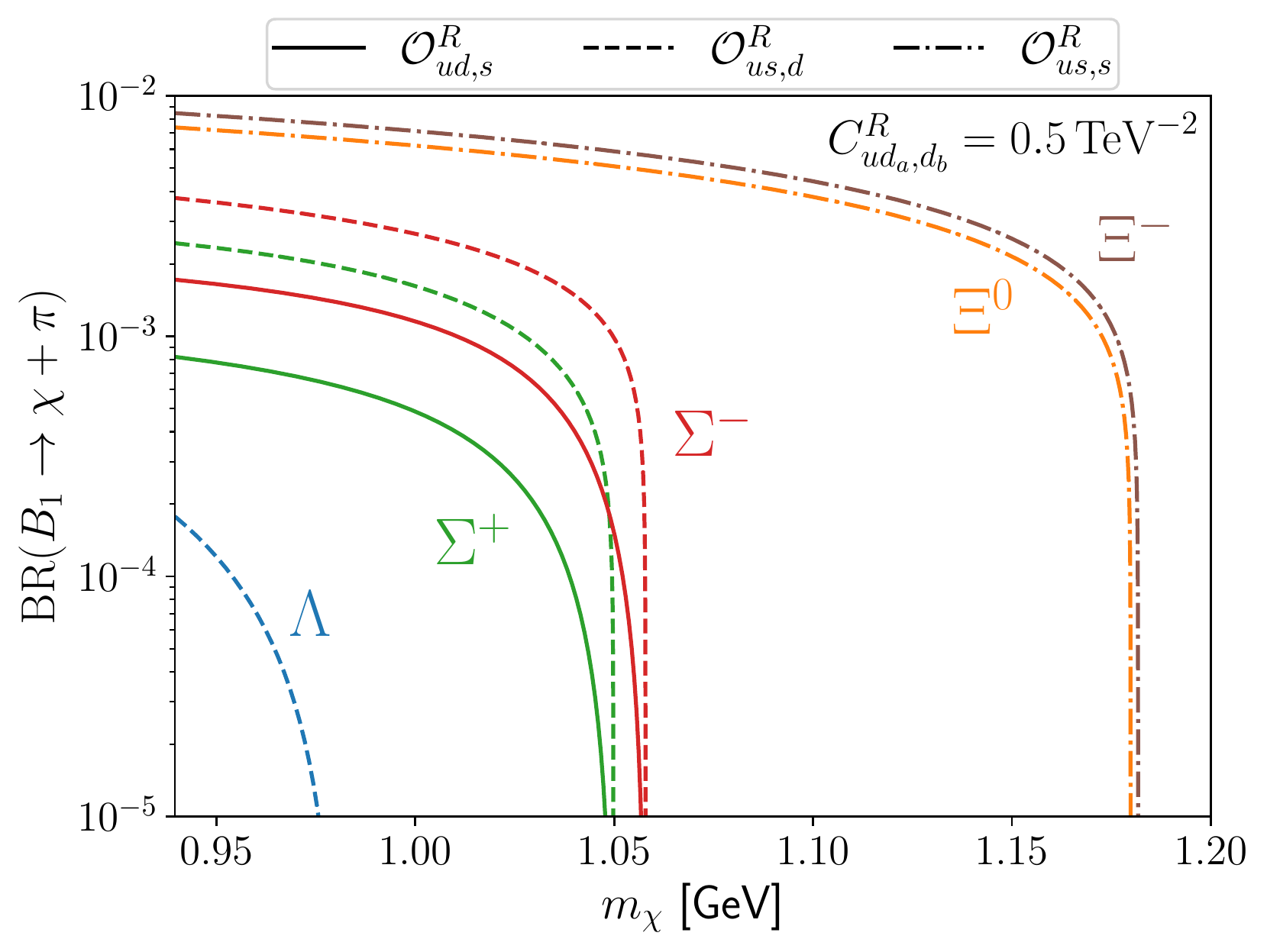}
\end{tabular}
\caption{\small Branching ratios for hyperon decays to a dark baryon $\chi$, and a pion as a function of the dark baryon mass, $m_\chi$.
The Wilson coefficients in \eqref{eq:LEEFTchi} are set to $C_{ud_a,d_b}^{R(L)}=0.5\,\text{TeV}^{-2}$, which roughly saturates the LHC constraints on the mediators, see the main text. The color coding corresponds to different initial hyperons as labeled, and the different line styles denote different flavor combinations in the Wilson coefficients as indicated.
The branching fractions for the other values of the Wilson coefficients follow from rescaling ${\cal B} \propto |C_{ud_a,d_b}^{R(L)}|^2$. 
\label{fig:DecaysBNV1}}
\end{figure*}

\begin{figure*}[t]
\centering
\begin{tabular}{cc}
\label{fig:Decays1:gamma}
\includegraphics[width=0.48\textwidth]{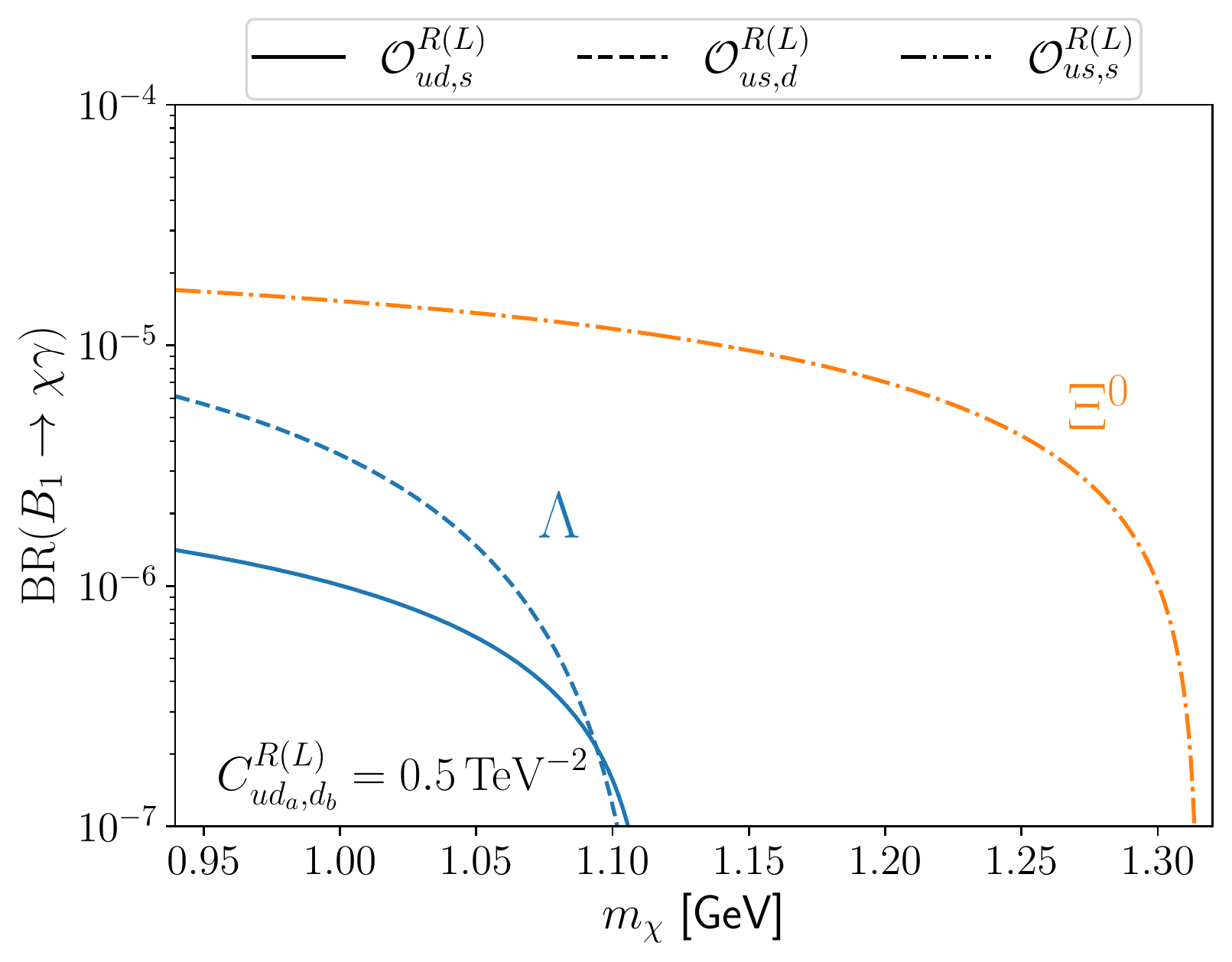}
&
\label{fig:Decays2:inv}
\includegraphics[width=0.48\textwidth]{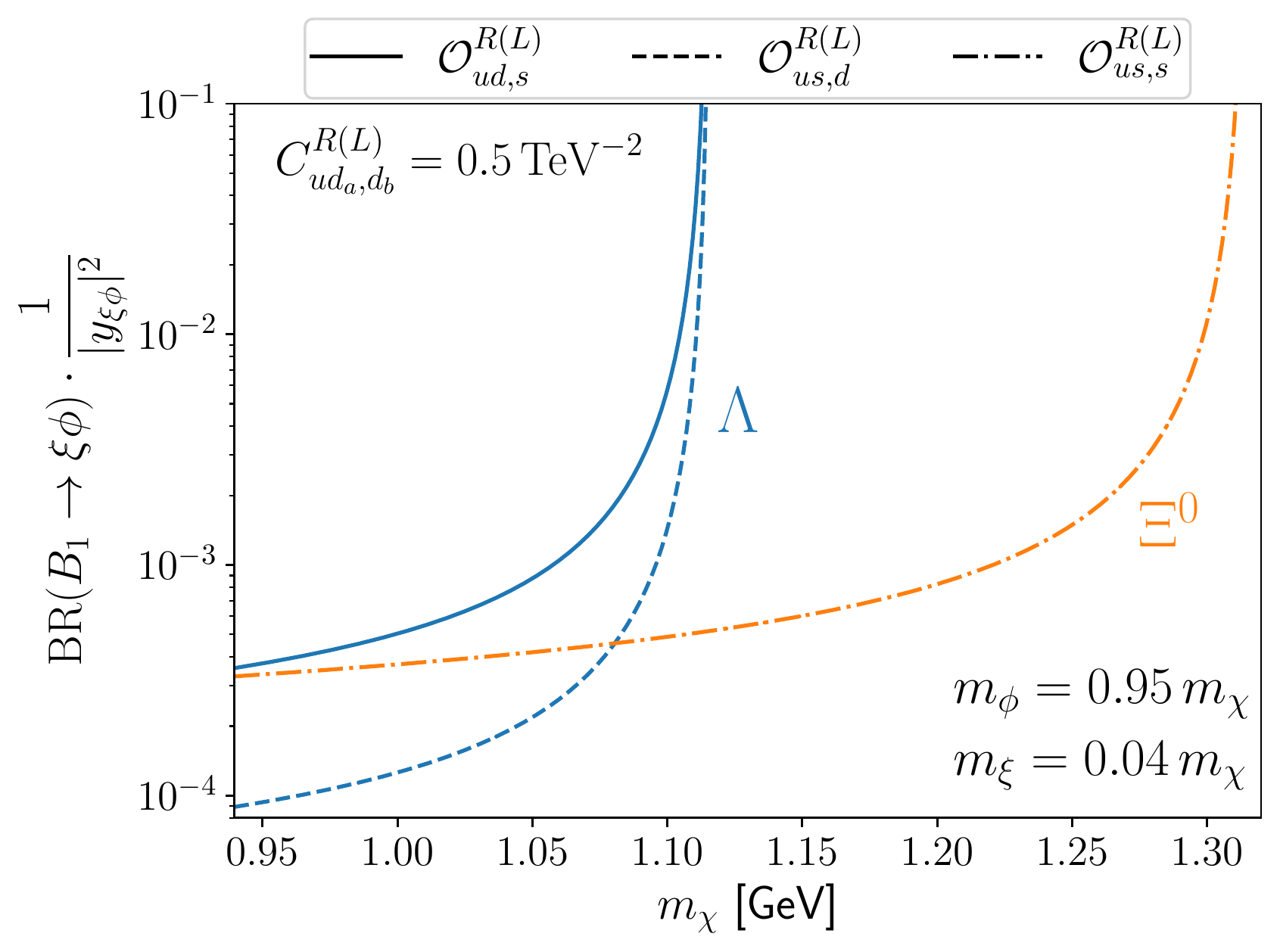}
\end{tabular}
\caption{\small Branching ratios for neutral hyperon decays to a dark baryon $\chi$ and a photon (left) and a purely invisible final state (right) as functions of dark baryon mass $m_\chi$. The dark sector particle masses and couplings are set in the right panel to $m_\phi=0.95m_\chi$, $m_\xi=0.04m_\chi$ and $y_{\xi\phi}=1$. Colors and line styles are the same as in Fig.~\ref{fig:DecaysBNV1}.
\label{fig:DecaysBNV2}}
\end{figure*}

In Table~\ref{tab:hyperon:dark:sector:decays} and in Figs.~\ref{fig:DecaysBNV1}, \ref{fig:DecaysBNV2} we show the predictions for the branching fractions for the $B_1\to\chi\pi$, $B_1 \to \chi  \gamma$, and $B_1 \to \xi  \phi$ decays for different hyperons $B_1$. 
To calculate the decay rates we set $C_{ud_a,d_b}^{R(L)} = 0.5\,\mathrm{TeV}^{-2}$ for a particular flavor choice $ud_a,d_b$ as denoted in the figures, while setting all the other Wilson coefficients to zero.
In the table we only show the largest possible branching ratio after marginalizing over the possible operators that can contribute to each decay mode.
For the purely invisible hyperon decays $B_1\to\xi \phi$, the predictions are given for some benchmark masses of the final state particles, $m_\phi=0.95m_\chi$ and $m_\xi=0.04m_\chi$.
The branching ratios quoted in Table~\ref{tab:hyperon:dark:sector:decays} are obtained for a representative value of the dark baryon mass $m_\chi=1$ GeV, while
$m_\chi$ is varied in Figs.~\ref{fig:DecaysBNV1}, \ref{fig:DecaysBNV2}.

Within the same class of decays, i.e., the decays induced by a nonzero value of a single Wilson coefficient $C_{ud_a,d_b}^{R(L)} $, the various decay channels have different sensitivities to this Wilson coefficient. For instance, the $\Sigma^-\to \chi\pi^-$ and  $\Sigma^+\to \chi\pi^+$ differ by a factor of a few when induced by the same $C_{ud,s}^{R(L)} $ operator (Fig.~\ref{fig:DecaysBNV1}). 
There are also differences between predictions for $C_{ud_a,d_b}^{R(L)}$ and the Wilson coefficient with swapped flavor indices, $C_{ud_b,d_a}^{R(L)} $, to the extreme that some decay channels are induced by one operator but not the other (which can be qualitatively understood using isospin symmetry). 
The resonant enhancement of $B_1\to \xi\phi$ at the end of the phase space (Fig.~\ref{fig:DecaysBNV2}, right) is due to the intermediate $\chi$ becoming on shell.
Note that the radiative decays $B_1\to \gamma\chi$ are suppressed by $\alpha_{\rm em}$. This explains why one typically needs to probe two orders of magnitude lower branching fractions than in $B_1\to \pi \chi$ decays in order to have the same sensitivity to the $C_{ud_a,d_b}^{L/R} $ operators.
However, radiative decays allow to test larger $m_\chi$, as near the endpoint region the $B_1\to \pi \chi$ decay is kinematically suppressed or even forbidden.
Finally, dark-baryon $\Sigma^0$ decays are not considered since their branching ratios are very suppressed due to the large width of the $\Sigma^0$ in the SM~\cite{Alonso-Alvarez:2021oaj}.

Decay signatures with multi-body final states such as $B_1\to\ell^+\ell^-\chi$, $B_1\to\pi\pi\chi$ or $B_1\to\ell^+\ell^-\pi \chi$ can be predicted in the same way as the two-body decays illustrated above. For light enough $\chi$ the $B_1\to\pi\pi\chi$  is dominated by the quasi-two-body decay, $B_1\to\rho \chi$. The corresponding branching ratio is thus expected to be roughly of the same size as for $B_1\to\pi \chi$, albeit somewhat smaller due to the smaller available phase space. For heavier $m_\chi$ the $B_1\to\pi\pi\chi$ becomes a genuine three body decay and thus carries a larger phase space suppression. The same comment applies to the 
case of the leptonic modes, but with an extra suppression by the $\alpha_{\rm em}$ factor. In spite of this, the leptonic decays could be interesting if the experiments are more efficient at detecting these modes than the others above. For instance, a decay such as $B_1\to\ell^+\ell^-\pi\chi$ produces a vertex that could enable a search at LHCb (see Section~\ref{sec:LHCb:hyperons:prospect} for projections).

\subsubsection{Discussion: relevance for the neutron decay anomaly and $B$-mesogenesis} 
Recently, it has been shown that the existence of a neutron dark decay channel, e.g., $n\to\chi\,\gamma$ or  $n\to \xi\,\phi$, with a branching ratio $\sim 1\%$ can resolve the  discrepancy in neutron lifetime measurements between bottle and beam experiments \cite{Fornal:2018eol}.
Phenomenologically consistent theories of this type  are based on Models 1--3 in Eqs.~(\ref{eq221})--(\ref{eq222}), with a minimally  extended dark sector \cite{Cline:2018ami,Karananas:2018goc,Grinstein:2018ptl}. Various experiments have searched for signatures of neutron dark decay, probing parts of the available parameter space \cite{Tang:2018eln,Sun:2018yaw,Klopf:2019afh,Pfutzner:2018ieu,Ayyad:2019kna} (for a review, see \cite{Fornal:2020gto}).
Within the framework of Model 2, Eq.\,(\ref{eq22}), the dark decay branching ratio ${\cal B}(n\to \chi\gamma) =1\%$ corresponds to
\begin{align}
C_{ud,d}^{R} \approx {2.7 \times 10^{-5}} \ {\rm TeV^{-2}} \ .
\end{align}
The natural mass scale for $\Phi$   is $M_\Phi \sim \mathcal{O}(100 \ {\rm TeV})$.
However, if  $|y_{ud} \, y_{\chi d}|\ll1$,  the  mass of $\Phi$ required to explain the neutron decay anomaly is much smaller, and  the LHC constraint $M_\Phi > 1.2 \ {\rm TeV}$ becomes relevant. For the benchmark $M_\Phi = 7 \ {\rm TeV}$, the required combination of couplings is $|y_{ud} \, y_{\chi d}| \sim 10^{-3}$.  
For the pure dark decay channel $n\to \xi\,\phi$, the $1\%$ branching ratio is achieved when 
\begin{align}
C_{ud,d}^{R}\cdot|y_{\xi\phi}| \approx {2 \times 10^{-8}} \ {\rm TeV^{-2}} \ .
\end{align}

If $\xi$ is the dark matter particle, the coupling $y_{\xi\phi} \approx 0.04$ allows for the annihilation $\xi \xi \to \phi \phi^*$ at a rate consistent with the observed dark matter relic density. Again, the natural scale  for $\Phi$ is high, $M_\Phi \sim \mathcal{O}(1000 \ {\rm TeV})$. In order to have $M_\Phi = 7 \ {\rm TeV}$, the couplings need to be $|y_{ud} \, y_{\chi d}| \sim 10^{-6}$. 
Since in Model 2 the interaction of $\Phi$ with second generation quarks is independent of $y_{ud}$ and $y_{\chi d}$, no concrete predictions for hyperon decays can be made, but a sizable hyperon decay rate  is possible. 
The situation is different for Model 1, for which the allowed branching ratio for $n\to\chi\gamma$ is smaller than $\sim 10^{-6}$~\cite{Fajfer:2020tqf}.  This is due to the antisymmetric nature of the Yukawa coupling which forbids  interactions  of $\Psi$ with only first generation quarks, and the stringent flavor  constraints from kaon mixing:  $|y_{\chi s}\, y_{\chi d}| \lesssim 10^{-3} \frac{M_\Psi}{1.5\text{TeV}} $ and $|y_{u s}\, y_{u d}| \lesssim 10^{-3} \frac{M_\Psi}{1.5\text{TeV}} $~\cite{Alonso-Alvarez:2021qfd}.
This implies that in the parameter region for which the $\Psi$ particle could be detected at the LHC, $M_\Psi \lesssim 10\,\text{TeV}$, the expected signal from  hyperon decays is too weak to be measured. On the other hand, if  $M_\Psi$ is sufficiently large, then a measurable ${\cal B}(\Lambda\to \chi\gamma) $ is expected.
Those constraints on Model 1 do not apply to the pure decay channel $n\to \xi\,\phi$, for which a measurable signal both at the LHC and in hyperon factories (such as BESIII) is possible. 

$B$-mesogenesis can explain the asymmetry between visible matter and antimatter and also the origin and nature of dark matter provided that $B$ mesons decay into dark sector antibaryons with a branching fraction $>0.01\,\%$~\cite{Elor:2018twp,Alonso-Alvarez:2021qfd}. In any of the Models 1, 2, 3 the decay of the $B$ meson is controlled by couplings to the $b$ quark but we expect couplings to other quarks to be present and could lead to non-standard hyperon decays. In particular, a combination of couplings that leads to an understanding of the origin of matter and dark matter of the Universe is~\cite{Alonso-Alvarez:2021qfd} $y_{db} = 0.3$, $y_{\chi u} = 2$, and $M_\Psi = 3\,\text{TeV}$ within Model 1. 
This corresponds to $B$ mesons decaying with a branching fraction of ${\cal B}({B\to \chi + {\rm hadrons}}) \simeq 0.03\%$. Since we know that consistency with $B$-Meson oscillations requires $|y_{db}\, y_{sd}^\star| \lesssim 0.2\, \frac{M_\Psi}{1.5\text{TeV}} $, we expect that the Wilson coefficients controlling hyperon decays could be as large as
\begin{align}
C_{ud,s}^R = -C_{us,d}^R \lesssim 0.02 \,\text{TeV}^{-2}.
\end{align}

By comparing this interaction strength with the decay $\Sigma^- \to \chi + \pi^- $ in Table~\ref{tab:hyperon:dark:sector:decays}, we can see that one could expect a branching fraction as large as $\sim  10^{-5}$, which could well be tested at the upcoming experiments. 
In addition, within $B$-Mesogenesis 
hyperons decaying fully invisibly are also expected. In this setting the $y_{\xi \phi}$ coupling should be sizeable in order to maintain the $\xi$ and $\phi$ populations coupled in the early Universe. For $y_{\xi \phi} = 1$, which would maintain the $\xi$ and $\phi$ populations in thermal equilibrium until the temperature of the early Universe is $T = 2\,\text{MeV}$, this would induce a decay rate for $\Lambda \to \xi + \phi$ at the level of $\lesssim 10^{-6}$, which should be compared with the recent constraint from BESIII~\cite{BESIII:2021slv}, ${\cal B}(\Lambda\to \textrm{inv}) < 7\times 10^{-5}$ at 90\% CL. These examples thus demonstrate that searches at BESIII could represent a relevant test of $B$-Mesogenesis in a way that is complementary to the LHC searches for resonant jets and missing energy at TeV energies, and direct searches to dark decays of b-flavored hadrons~\cite{Belle:2021gmc,Rodriguez:2021urv}.

\subsection{Other (more exotic) scenarios}
\label{sec:LPV}
{\em Authors: Bansal, Gori, Grossman,   Shuve, Sumensari, Tammaro, \underline{Zupan}}

In this section we collect several further, somewhat exotic possibilities for signals of new physics in the rare kaon or hyperon decays. The first possibility is a light NP particle that has flavor violating couplings to both quarks and leptons which induces $K\to\pi\mu e$ decays. We give two such examples, a flavor violating ALP (Section~\ref{sec:FV:ALP:mue}), and a light $Z'$ (Section~\ref{sec:light:FV:Z'}). A possible more complex dark sector which can give rise to dark jets is discussed in Section~\ref{sec:dark:jets}.


\subsubsection{Flavor violating ALP decays}
\label{sec:FV:ALP:mue}

In principle, flavor violating  couplings of ALP to the SM fermions, Eq.~(\ref{eq:ALP-f}), can be relevant not only in the ALP production (Section~\ref{subsec:QCDaxion:ALP}) but also in ALP decays. As we will see below, this is possible in a particular range of ALP parameters, giving rise to the $K\to\pi\mu e$ decay, a somewhat more exotic signature of a flavor violating ALP compared to the  signatures discussed in Section~\ref{subsec:QCDaxion:ALP}. Nevertheless it is interesting that despite the strong constraints from $K-\bar K$ mixing and rare muon decays, there is still the possibility of a NP discovery through such decays.
ALPs can also contribute at tree-level to the $K_L\to\mu^\pm e^\mp$ decays. Present experimental bounds on the $K^+\to\pi^+\mu^\pm e^\mp$ and $K_L\to\mu^\pm e^\mp$ decay branching fractions are ${\cal O}(10^{-11})$, as detailed in Section~\ref{sec:exp:lfv}.

The $K^+\to\pi^+\mu^\pm e^\mp$ differential decay rate is given by~\cite{Cornella:2019uxs}
\begin{equation}
\label{eq:ALP:emu:diffGamma}
\dfrac{\mathrm{d}\Gamma}{\mathrm{d}m_{e\mu}^2}= \dfrac{m_K m_\mu^2}{32 (4\pi)^3 f_a^4} \dfrac{m_{e\mu}^2 \lambda_K^{1/2} f_0 (m_{e\mu}^2)^2}{(m_{e\mu}^2-m_a^2)^2+m_a^2 \Gamma_a^2}\bigg(1-\dfrac{m_\mu^2}{m_{e\mu}^2}\bigg)^2\bigg(1-\dfrac{m_\pi^2}{m_K^2}\bigg)^2\Big[|C_{e\mu}^V|^2+|C_{e\mu}^A|^2\Big]|C^V_{sd}|^2,
\end{equation}
where $m_a$ and $\Gamma_a$ are the ALP mass and total decay width, while the coefficients $C_{ij}^{V/A}$ and the axion decay constant $f_a$ are defined in Eq.~\eqref{eq:ALP-f}. The $K\to\pi$ scalar form factor $f_0(m_{e\mu}^2)$~\cite{Aoki:2019cca} is a function of the di-lepton invariant mass squared, $m_{e\mu}^2$, while  $\lambda_K=\big[m_{e\mu}^2-(m_K-m_\pi)^2\big]\big[m_{e\mu}^2-(m_K+m_\pi)^2\big]$. Similar expressions hold for the decays $K_L\to \pi^0 e^-\mu^+$ and $K_S\to \pi^0 e^-\mu^+$, after the replacements $|C_{sd}^V|^2\to \mathrm{Im}[C_{sd}^V]^2$ and $|C_{sd}^V|^2\to \mathrm{Re}[C_{sd}^V]^2$, respectively. For the $K_L\to \mu^- e^+$ decays, the most general expression reads~\cite{Cornella:2019uxs}
\begin{equation}
\label{eq:ALP:emu:leptonic}
\Gamma (K_L \to \mu^- e^+) = \dfrac{m_K^5 }{128 \pi f_a^4}\dfrac{m_\mu^2 f_K^2}{(m_{K}^2-m_a^2)^2+m_a^2 \Gamma_a^2}\bigg(1-\dfrac{m_\mu^2}{m_K^2}\bigg)^2 \Big[|C_{e\mu}^V|^2+|C_{e\mu}^A|^2\Big]|\mathrm{Re}(C_{sd}^A)|^2. 
\end{equation}
An analogous expression can be obtained for $K_S\to\mu^-e^+$ by replacing $\mathrm{Re}[C_{sd}^A]^2\to \mathrm{Im}[C_{sd}^A]^2$.

\begin{figure}[t]
\centering
\includegraphics[width=0.33\linewidth]{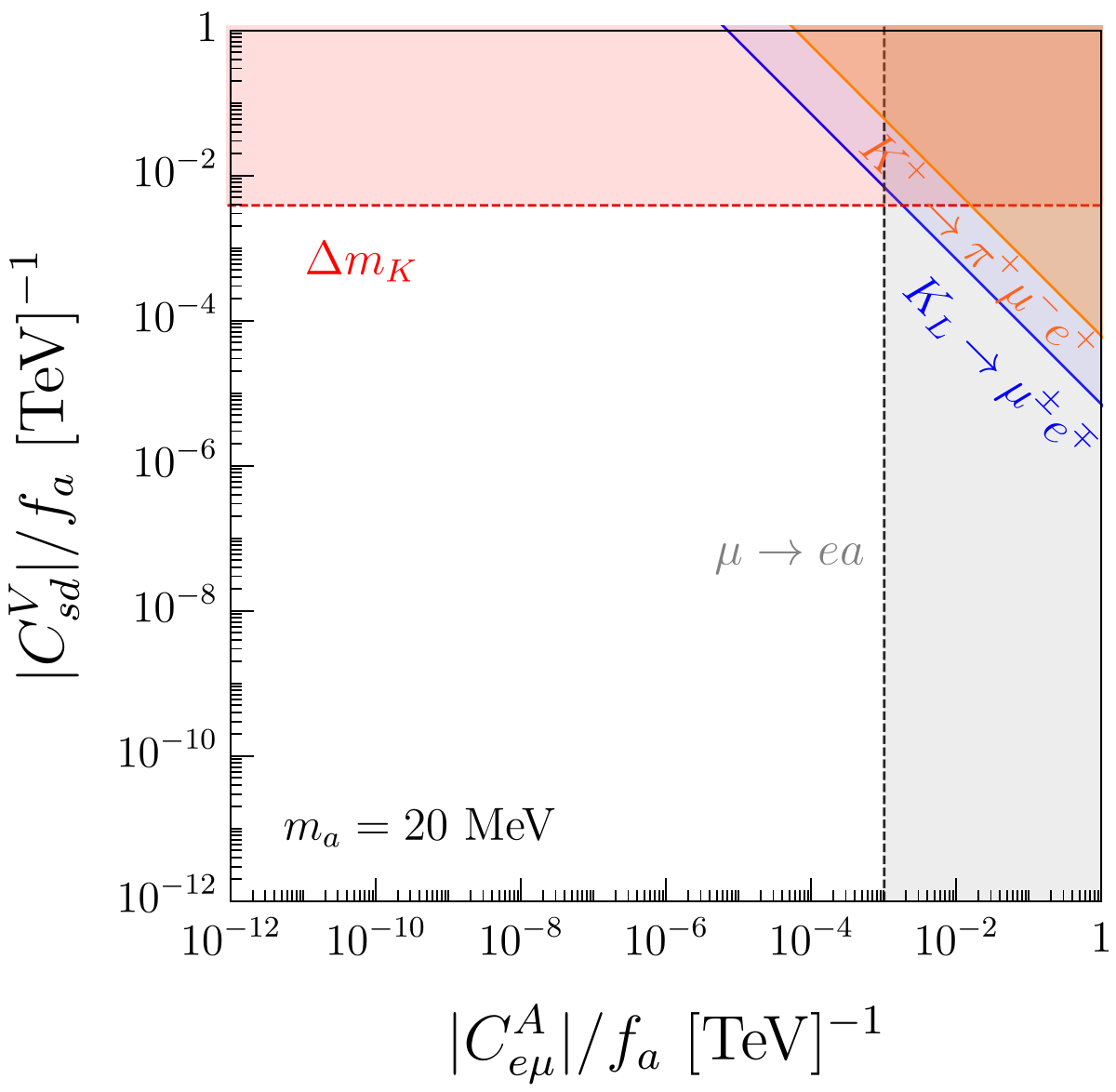}~~\includegraphics[width=0.33\linewidth]{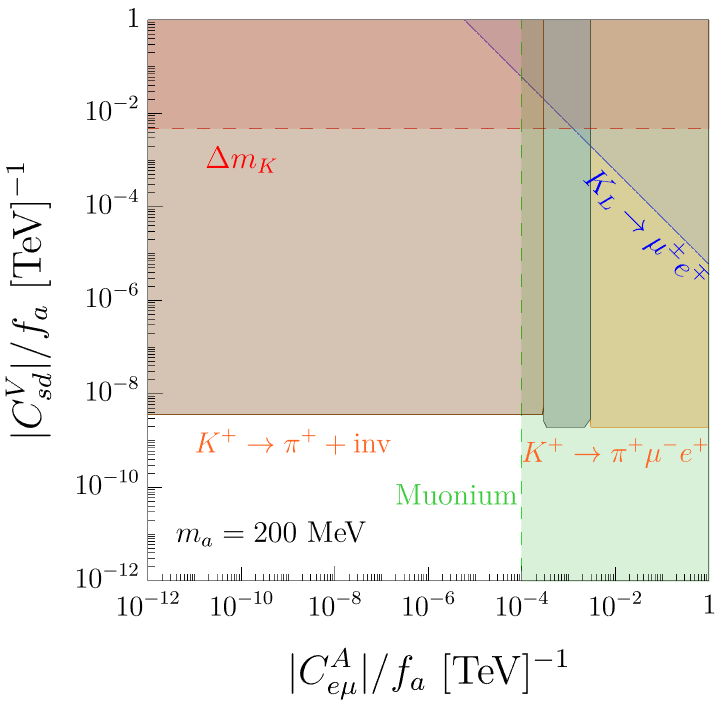}~~
\includegraphics[width=0.33\linewidth]{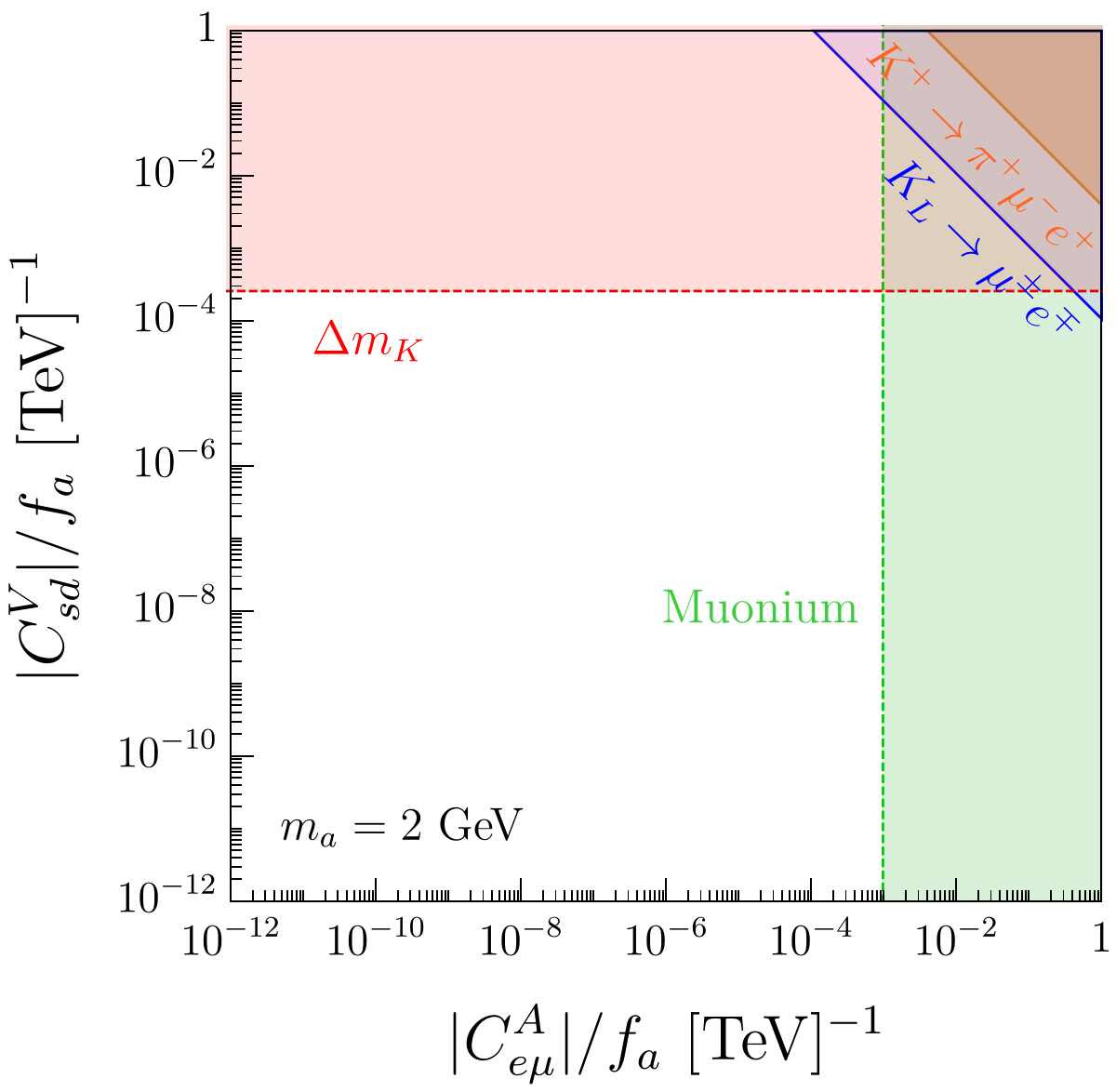}
\caption{\small Constraints on the plane $C_{e\mu}^A/f_a$ vs.~$C_{sd}^{V(A)}/f_a$ derived from $\mathcal{B}(K^+\to\pi^+ \mu^+ e^-)<1.3\times 10^{-11}$~\cite{Ambrose:1998us} (orange) and $\mathcal{B}(K_L\to \mu^\mp e^\pm)<4.7\times 10^{-12}$~\cite{Sher:2005sp} (blue) for three values of ALP mass: (i) $m_a=20$~MeV (left), (ii) $m_a=200$~MeV (center) and $m_a=2$~GeV. For simplicity, we assume that $C_{sd}^{V}=C_{sd}^{A}$, as predicted by a left-handed operator. We superimpose in the same plot constraints derived from $\mu\to e+\mathrm{inv}$~\cite{Calibbi:2020jvd} (gray), $\Delta m_K$~\cite{MartinCamalich:2020dfe} (red) and muonium-antimuonium oscillation~\cite{Masiero:2020vxk,Endo:2020mev} (green). 
}
\label{fig:lfv:alps}
\end{figure}

There are three regimes for the ALP mass, leading to distinct phenomenological features: 
\begin{itemize}
\item{{\bf Light ALP} ($m_a<m_\mu+m_e$)}. For such light ALPs the $K\to\pi\mu e$ transition is mediated by an off-shell ALP since the $a\to\mu e$ decays are kinematically forbidden. 
Numerically, 
\begin{align}
\mathcal{B}(K_L\to \pi^0\mu^+e^-) &\simeq 9.4 \cdot 10^{-22} \frac{|C_{e\mu}^V|^2+|C_{e\mu}^A|^2}{\big(10^{-3}\big)^2}\Big|\Im C_{sd}^V\Big|^2\biggr(\frac{10^6~{\rm GeV}}{f_a}\biggr)^4\,,
\\
\mathcal{B}(K^+\to\pi^+\mu^+e^-) &\simeq 2.1 \cdot 10^{-22} \frac{|C_{e\mu}^V|^2+|C_{e\mu}^A|^2}{\big(10^{-3}\big)^2}\Big|C_{sd}^V\Big|^2\biggr(\frac{10^6~{\rm GeV}}{f_a}\biggr)^4,
\end{align} 
obtained from the differential decay rates in \eqref{eq:ALP:emu:diffGamma} for $m_{e\mu}^2\gg m_a^2$, and choosing $C_{sd}^V/f_a$ ($C_{e\mu}^{V/A}/f_a$) at roughly the maximal value allowed by $K-\bar K$ mixing~\cite{MartinCamalich:2020dfe} (by using ${\cal B}(\mu\to e a)$ bounds, in the conservative case where $a$ escapes the detector \cite{Calibbi:2020jvd}). Under the same assumptions, the $K_L\to\mu^- e^+$ branching fraction reads
\begin{equation}
    {\cal B}(K_L\to\mu^-e^+) \simeq 5\times 10^{-20}\frac{|C_{e\mu}^V|^2+|C_{e\mu}^A|^2}{\big(10^{-3}\big)^2}\Big|\Re C_{sd}^A\Big|^2\biggr(\frac{10^6~{\rm GeV}}{f_a}\biggr)^4.
\end{equation}

Therefore for light ALPs, the LFV kaon branching fractions are constrained to be well below the reach of NA62 and KOTO and planned future experiments for both $K^+$ and $K_L$ decays, and BESIII for hyperon decays. The same conclusions apply to rates of decays with $\mu^-e^+ $ in the final state, which can be obtained from the above expressions by replacing $\mu\leftrightarrow e$.
 
\item{{\bf Intermediate ALP} ($m_\mu+m_e< m_a<m_K-m_\pi$)}. In this case the decays $K^+\to\pi^+ a$ followed by $a\to\mu^\pm e^\mp$ are considered. The decay width is given by
\begin{align}
\Gamma(K^+\to \pi^+ a)&= \frac{ \kappa_K m_K^3}{64 \pi f_a^2}  \Big|C_{sd}^V\Big|^2  &\Rightarrow&\quad \mathcal{B}(K^+\to \pi^+ a)\simeq 6.8 \times 10^{-12} \Big(\tfrac{10^{12}~{\rm GeV}}{f_a/ C_{sd}^V}\Big)^2,
\\
\Gamma(K_L\to \pi^0 a)&= \frac{ \kappa_K m_K^3}{64 \pi f_a^2}  \Big|\Im C_{sd}^V\Big|^2  &\Rightarrow&\quad \mathcal{B}(K_L\to \pi^0 a)\simeq 3.0 \times 10^{-11} \Big(\tfrac{10^{12}~{\rm GeV}}{f_a/ C_{sd}^V}\Big)^2,
\end{align}
where we collected the kinematical and form factor dependences in a single factor
\begin{equation}
\kappa_K= f_0(m_a^2)^2 (1-r_\pi^2)^2 \big[\big(1-(r_a+r_\pi)^2\big)\big(1-(r_a-r_\pi)^2\big)\big]^{1/2},    
\end{equation}
with $r_P=m_P/m_K$, and show the numerical value for $m_a=200$ MeV. Note that the $K-\bar K$ mixing bound, $f_a/|C_{sd}^V|\gtrsim 10^6{\rm~GeV}$ is much weaker than the sample values chosen above. Branching ratios of the above magnitude could be probed by NA62, KOTO and the future CERN $K_L$ experiment (Section~\ref{sec:exp:KLEVER}), as long as ALP decays are dominated by the $a\to\mu^\pm e^\mp$ channel, while the $a\to e^+e^-, \gamma\gamma$ and invisible decays are subleading. The ALP decay width is given by 
\begin{equation}
\Gamma_a=\dfrac{m_a m_\mu^2}{16\pi f_a^2} \bigg(1- \dfrac{m_\mu^2}{m_a^2}\bigg)^2 \big( |C_{e\mu}^V|^2+|C_{e\mu}^A|^2\big)\,,    
\end{equation}
which gives, for $m_a=200$ MeV,
\beq
c \tau_a\simeq 8.5 {\rm cm} \biggr(\frac{f_a}{10^{5}{\rm GeV}}\biggr)^2 \frac{1}{\big|C_{e\mu}^V\big|^2+\big|C_{e\mu}^A\big|^2}\,.
\eeq
The $K^+\to\pi^+ a(\to \mu e)$ signature therefore requires $C_{e\mu}^{V/A}\gtrsim 10^7 C_{sd}^V$, i.e., much stronger couplings to leptons than quarks.  
For further discussion of the present experimental constraints and future projected sensitivities, see Section~\ref{sec:exp:lfv}.

\item{{\bf Heavy ALP} ($m_a>m_K-m_\pi$)}. In this case the ALP is off-shell, and thus $K\to\pi\mu e$ is again a genuine three-body decay. The decay widths are the same as in the light ALP case, but taking $m_a\gg m_K-m_\pi$. The main phenomenological difference to the light ALP case is the absence of the $\mu\to ea$ constraints. There are still important muonium-antimuonium constraints constraints \cite{Endo:2020mev,Masiero:2020vxk}, shown in Fig.~\ref{fig:lfv:alps}. Parametrically, the branching ratios are given by
\begin{align}
\mathcal{B}(K_L\to\pi^0\mu^+e^-) &\simeq 3.9 \cdot 10^{-12}\Big( |C_{e\mu}^V|^2+|C_{e\mu}^A|^2\Big)\frac{\big|\Im C_{sd}^V\big|^2}{\big(10^{-3}\big)^2}\biggr(\frac{10^2 ~{\rm GeV}}{f_a}\cdot \frac{10~{\rm GeV}}{ m_a}\biggr)^4,
\\
\mathcal{B}(K^+\to \pi^+\mu^+e^-) &\simeq  8.1 \cdot 10^{-13}\Big( |C_{e\mu}^V|^2+|C_{e\mu}^A|^2\Big)\frac{\big| C_{sd}^V\big|^2}{\big(10^{-3}\big)^2}\biggr(\frac{10^2 ~{\rm GeV}}{f_a}\cdot \frac{10~{\rm GeV}}{ m_a}\biggr)^4,
\end{align} 
where we have chosen the values of $C_{sd}^V/f_a$ that are not yet excluded by $K-\bar K$ mixing constraints, and $C_{e\mu}^{V/A}/f_a$ that are still allowed by the muonium-antimuonium oscillations. For the leptonic kaon decay, the branching fraction is given by
\begin{align}
\mathcal{B}(K_L\to\mu^+e^-) &\simeq  2.9 \cdot 10^{-9}\Big( |C_{e\mu}^V|^2+|C_{e\mu}^A|^2\Big)\frac{\big| C_{sd}^V\big|^2}{\big(10^{-3}\big)^2}\biggr(\frac{10^2 ~{\rm GeV}}{f_a}\cdot \frac{10~{\rm GeV}}{ m_a}\biggr)^4\,,
\end{align} 
which is larger than for the three-body decays in this $m_a$ range.
 \end{itemize}
 
\subsubsection{Light flavor violating $Z'$}
\label{sec:light:FV:Z'}
The observed hierarchy of masses in the SM charged fermion sector could have a dynamical explanation. For instance, the hierarchical SM fermion masses could be a consequence of a spontaneously broken horizontal $U(1)_{\rm FN}$ symmetry \cite{Froggatt:1978nt}. 
If the horizontal $U(1)_{\rm FN}$ is gauged the associated $Z'$ has flavor violating couplings and can be searched for in FCNC processes. Which FCNC transition is the most sensitive depends on the value of $Z'$ mass, $m_{Z'}=g' v_f$, where $g'$ is the horizontal $U(1)$ gauge coupling, and $v_f$ the vev that breaks the $U(1)_{\rm FN}$.
A heavy $Z'$ can be discovered through its tree level off-shell contributions to the meson mixings, with the $K -\bar K$ mixing typically leading to the most stringent constraints. 
For small values of the gauge coupling, on the other hand, the $Z'$ can be light enough to be produced on-shell in rare $K$, $B$ and $D$ decays, which then give the most sensitive constraints \cite{Smolkovic:2019jow}. A particularly striking signature in this mass regime is the $K\to\pi Z'\to \pi e\mu$ decay, which is kinematically allowed when the $Z'$ mass is in the window $m_{Z'}\in[m_\mu+m_e,m_K-m_\pi]$. 

\begin{figure}[t]
\centering
\includegraphics[width=0.325\linewidth]{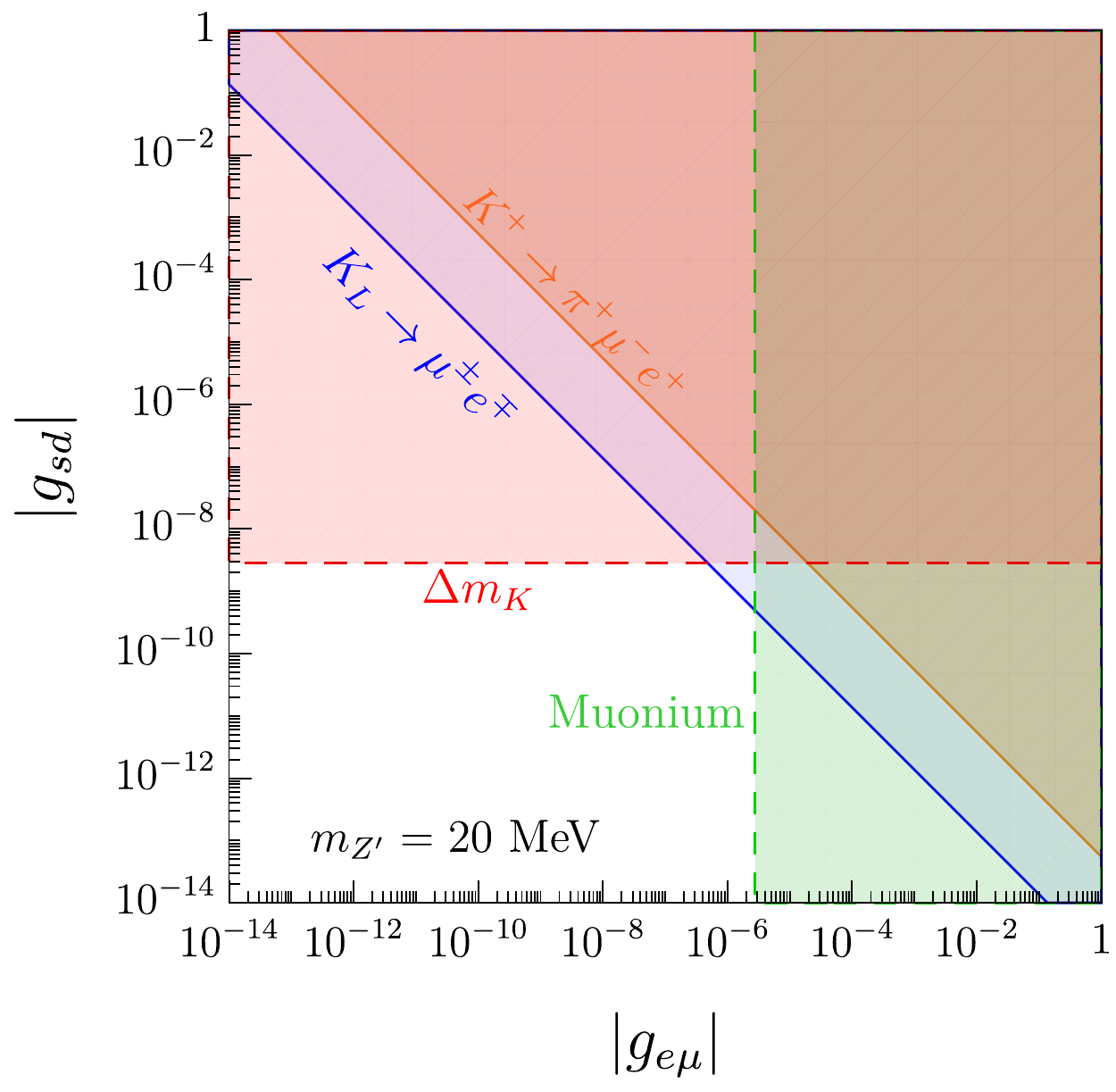}~~\includegraphics[width=0.325\linewidth]{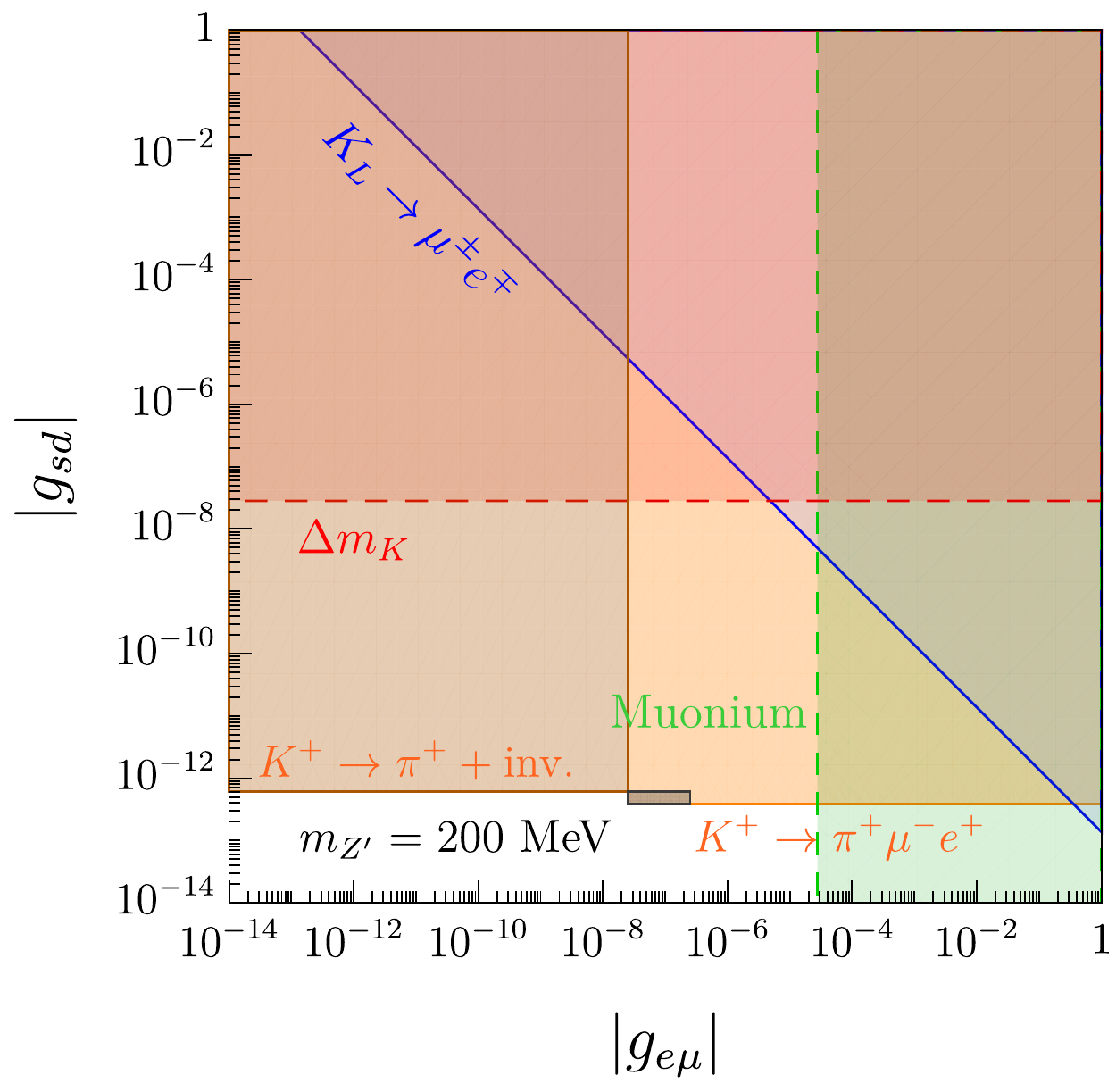}
\includegraphics[width=0.325\linewidth]{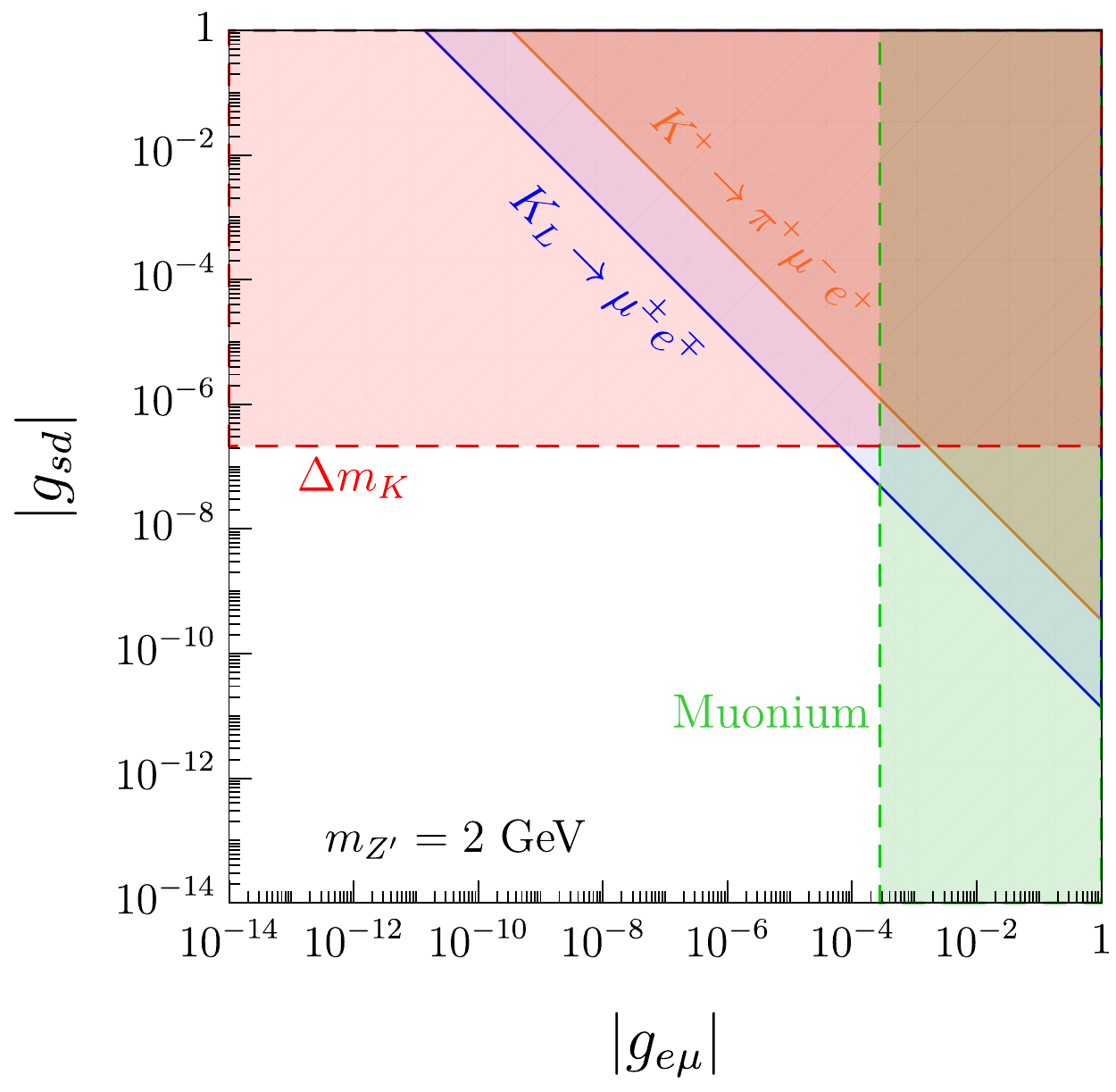}
\caption{\small Constraints on $Z'$ flavor-violating couplings to quarks, $g_{sd}$, and leptons, $g_{e\mu}$. Orange and blue lines represent the bounds from current limits on ${\cal B}(K^+\to\pi^+\mu^-e^+)$ and ${\cal B}(K_L\to\mu^\pm e^\mp)$, respectively (Section~\ref{sec:exp:lfv}). Red and green regions show the exclusions from $K-\bar K$ and $M-\bar M$ mixing, respectively. The dark grey region in the central plot shows the range of $g_{e\mu}$ values that could include displaced $Z'$ vertex decays.}  
\label{fig:LFV:Zprime}
\end{figure}

The main difference between the dark photon, discussed in Section \ref{sec:darkphoton}, and the light flavor violating $Z'$ is that the latter has renormalizable flavor violating couplings. For $Z'$ the low energy Lagrangian is therefore given by
\beq
\label{eq:lightZ:Lagrangian}
 \mathcal{L}_{\rm int} =  Z'_\mu \sum_{f,i,j} \left[ g^V_{f_i f_j} \big( \bar f_i \gamma^\mu f_j \big) +  g^A_{f_i f_j} \big( \bar f_i \gamma^\mu\gamma_5  f_j \big) \right],
\eeq
where  $f$ runs over the SM fermions, $i,j$ are generational labels, and in general the complex parameters $g_{f_if_j}^{V/A} = g_{f_jf_i}^{V/A}{}^*$ are nonzero also for $i\ne j$. In gauged horizontal $U(1)_{\rm FN}$ the $g^{V/A}_{f_i f_j}$ receive contributions both from the gauge charges of the SM fermions as well as from the flavor diagonal kinetic mixing of $Z'$ with the SM hypercharge. From low energy perspective the $g_{f_if_j}^{V/A}$ can be treated as free parameters.
We focus on the case where $g_{sd}^{V/A}$ and $g_{e\mu}^{V/A}$ dominate, while for other benchmarks motivated by the horizontal symmetry see Ref. \cite{Smolkovic:2019jow}.

Fig.~\ref{fig:LFV:Zprime} shows the bounds on flavor-violating $Z'$ couplings, $g_{e\mu} \equiv \big(|g_{e\mu}^V|^2 + |g_{e\mu}^A|^2\big)^{1/2}$ and a sample choice $g_{sd} \equiv  g_{sd}^V = g_{sd}^A $ taking it to be real,  for three illustrative $Z'$ masses, and setting all the other couplings to zero.  The muonium -- antimuonium ($M-\bar M$) oscillations bound larger values of $g_{e\mu}$ (green region), while large values of $g_{sd}$ are excluded by the $K-\bar K$ oscillations (red shaded region). The remainder of the parameter space is then first probed by $K_L\to\mu e$ (blue solid line) or $K^+\to\pi^+\mu^+e^-$ decays (orange solid line). For light $Z'$ (Fig.~\ref{fig:LFV:Zprime} left with $m_{Z'}=20$ MeV, but more generally for $m_{Z'} <m_\mu+m_e$) and for heavy $Z'$ (Fig.~\ref{fig:LFV:Zprime} right with $m_{Z'}=2$ GeV, but more generally for $m_{Z'} > m_K - m_\pi$) the $K^+\to \pi^+\mu e$ decay proceeds through an off-shell $Z'$ and thus the mixing constraints are in general more important then the constraint from the rare decay. In the intermediate mass regime, on the other hand, the $K^+\to\pi^+\mu^+e^-$ decay is due to a two-body decay $K^+\to\pi^+ Z'$, followed by $Z'\to\mu^+e^-$. This leads to very stringent constraints on $g_{sd}$ (in Fig.~\ref{fig:LFV:Zprime} center we assume 100\% branching ratio for $Z'\to\mu^+e^-$). For intermediate $Z'$ masses the $Z'$ becomes long lived for very small $g_{e\mu}$ couplings, $c\tau_{Z'} \simeq 6.3{\rm m}\, \big(10^{-7}/g_{e\mu}\big)^{2}$, and thus the $K^+\to \pi^+ Z'$ decay is seen as $K^+\to \pi^+ +$inv in the detector. The transition region would require special attention to recast the experimental constraints, and would include also decays with displaced vertices, which was not attempted in drawing Fig.~\ref{fig:LFV:Zprime} (right). Similar constraints are obtained if instead $g_{\mu e}$ is taken to be nonzero, but from $K^+\to\pi^+\mu^- e^+$ decays. In the numerical analysis we used the expressions for decay widths and the numerical inputs from Ref.~\cite{Smolkovic:2019jow}.

Experimental bounds on rare kaon decays branching fractions are reported in Section~\ref{sec:exp:lfv}. Two more processes are relevant for this study: $K - \bar K$ mixing and muonium - antimuonium ($M - \bar M$) oscillations. These are more relevant for light $Z'$, as the oscillation probability goes as $\sim g^4/m_{Z'}^4$, where $g = g_{sd}~(g_{e\mu})$ for $K - \bar K$ ($M - \bar M$) mixings. In the case of $K-\pi$ transitions only the vector part of the $Z'$ current, $V_\mu$, enters in the hadronic part of the matrix element $\langle \pi (p') | V_\mu | K(p) \rangle$, and is parametrized by two form factors, $f_{+,0}(q^2) = f_{+,0}(0) \big(1 + \lambda_{+,0} {q^2}/{m_\pi^2} \big)$. In the plots we use the central values for the parameters from~\cite{Aoki:2021kgd}.


\subsubsection{Dark jets}
\label{sec:dark:jets}

If a dark sector contains several relatively light states, the production of dark sector particles can result in dark jets with some of the dark sector particles decaying back to SM particles. Such decays are possible even if the interactions with the SM are very weak, i.e. the dark sector is in a ``hidden valley''~\cite{Strassler:2006im}. A well motivated example is a dark sector that contains ``dark quarks'' $q_{\rm dark}$ charged under a confining force $SU(N_{\rm dark})$. In this case the production of dark sector particles mimics the production of QCD jets in the SM, but potentially with a much lower confining scale than $\Lambda_{\rm QCD}$. 

Light dark quarks can be produced in rare kaon decays, either at tree level through flavor violating couplings between the dark sector mediator and the SM down quarks, or at loop-level through flavor conserving couplings to the top quark. The $s\to d q_{\rm dark} \bar q_{\rm dark}$ transitions such as $K^+\to \pi^+ q_{\rm dark}\bar q_{\rm dark}$, $K^0\to \pi^0 q_{\rm dark} \bar q_{\rm dark}$, $\Lambda/\Sigma^0 \to n q_{\rm dark} \bar q_{\rm dark}$, $\Sigma^+ \to p q_{\rm dark} \bar q_{\rm dark}$ would be followed by the emission of ``dark gluons'' splitting into other dark quarks. This decay chain thus results in two jets of dark particles. The dark gluons and dark quarks confine into dark mesons such as $\pi_{\rm dark}, \rho_{\rm dark}, \ldots$, and dark baryons, such as $n_{\rm dark}, n_{\rm dark}^*,$ and others. The dark mesons decay to the SM states, because the dark quarks have feeble interactions with the SM. The lightest dark baryon is often assumed to be stable and escapes the detector. 

The phenomenology of dark jets depends on the details of the dark sector model, both on the confining group, as well as on the dark sector field content and their masses (see Ref.~\cite{Knapen:2021eip} for some benchmarks in the case of heavier dark quarks). In general, the pattern of decays into SM particles depends on the flavor structure of the portal interactions, leading  to such signatures as {\em lepton jets} \cite{ArkaniHamed:2008qp,Baumgart:2009tn,Falkowski:2010cm}, {\em semi-visible  jets} \cite{Strassler:2008fv,Cohen:2015toa,Cohen:2017pzm}, {\em jet substructure from dark sector showers} \cite{Cohen:2020afv},  and {\em emerging jets} \cite{Schwaller:2015gea}. This complicated phenomenology simplifies significantly in the case of rare kaon decays, since kinematically only $\pi_{\rm dark}\to e^+e^-, \gamma\gamma$ (possibly with extra invisible final states) and $n_{\rm dark}\to e^+e^- +{\rm inv}, \ldots$, are possible for dark states that are light enough to form dark jets. 

If the decays of dark sector particles are prompt, this results in a shower of visible particles in the detector. Experimentally even more challenging is the possibility that the decays lead to displaced vertices. In this case the dark jets originate from many dark particles with varying decay times, resulting in a collection of displaced decays that appear in the detector as emerging jets forming far away from the interaction point~\cite{Schwaller:2015gea}.

\subsubsection{Lorentz violation}
\label{sec:Lorentz:viol}
Rare kaon decays can be used to test fundamental symmetries. Lorentz symmetry, for instance, forbids $K\to \pi \gamma$ and $K^0\to 3\gamma$ decays (for a massless photon), while these decays would be allowed in the presence of Lorentz symmetry breaking. A particular realization of Lorentz symmetry breaking is the noncommutative extension of the SM \cite{Calmet:2001na,Behr:2002wx,Melic:2005am,Melic:2005fm}, which implements the SM gauge group on a noncommutative space-time using the Seiberg-Witten map \cite{Seiberg:1999vs,Jurco:2001rq}. The size of Lorentz breaking is encoded in the constant antisymmetric tensor $\theta^{\mu\nu}=c^{\mu\nu}/\Lambda_{\rm NC}^2$, where $c^{\mu\nu}$ are dimensionless coefficients assumed to be ${\mathcal O}(1)$, while $\Lambda_{\rm NC}$ is the scale of non-commutativity. In this framework one expects ${\mathcal B}(K^+\to \pi^+\gamma)\simeq 0.8 \cdot 10^{-16} \big(1~{\rm TeV}/\Lambda_{\rm NC}\big)^4$ \cite{Melic:2005su}, and a similar result for the $K_{S,L}\to \pi^0\gamma$ decays. The present experimental bounds (Section~\ref{sec:K:pi:gamma}) then lead to a rather weak bound $\Lambda_{\rm NC}>20$~GeV, to be compared with $\Lambda_{\rm NC}>141$~GeV for noncommutative QED from $e^+e^-$ collisions~\cite{OPAL:2003eoc}, and $\Lambda_{\rm NC}>$ several TeV from LHC \cite{Hewett:2000zp,Hewett:2001im,Hinchliffe:2001im}. Significantly more stringent constraints on non-commutativity at the level of $10^{11}$ GeV or higher arise from several different low energy observables \cite{Mocioiu:2000ip,Anisimov:2001zc} (assuming that $\theta$ is constant at large distance and equal to the value at scale $\Lambda_{\rm NC}$).
 
In general, due to experimental challenges the searches for Lorentz violating neutral kaon decays, such as $K_{S,L}\to\pi\gamma$, will yield less stringent constraints on $\Lambda_{\rm NC}$ than their charged counterparts. The situation can be reversed, if there are new charge neutral light degrees of freedom, such as a new light (pseudo-)scalar, $X$. Lorentz violating decay, $K_L\to \gamma X$, followed by e.g. $X\to \gamma\gamma$, would then not have a charged kaon equivalent, making $K_L\to 3\gamma$ the leading experimental signature. See Section~\ref{sec:K:pi:gamma} for the experimental prospects. 


\subsection{Heavy new physics} \label{subsec:heavyNP}
{\em Authors: \underline{Brod}, Dery, Gorbahn, Grossman, Schacht, Stamou}

In this section we briefly summarize the SM predictions for some of the key rare kaon decays, and discuss possible contributions from new particles that are heavier than the electroweak scale.

\subsubsection{The SM prediction}
In the SM, the rare decays $K^+ \to \pi^+ \nu\bar\nu$ and $K_L \to
\pi^0 \nu\bar\nu$ are induced by $Z$-penguin and box diagrams. Since there is no photon-penguin contribution this results in a
hard (powerlike) GIM suppression of low-energy contributions to the
amplitude. The effective weak Hamiltonian for these decay modes can be
written as~\cite{Buchalla:1995vs}
\begin{equation}\label{eq:H:kpnn:sm}
  {\mathcal H}_\text{eff}
= \frac{4G_F}{\sqrt{2}} \frac{\alpha}{2\pi\sin^2\theta_w}
  \sum_{\ell=e,\mu,\tau} \big( \lambda_c X_\ell + \lambda_t X_t \big)
  (\bar s_L \gamma_\mu d_L) (\bar\nu_{\ell L} \gamma^\mu \nu_{\ell L})
  + \text{h.c.},
\end{equation}
where $\lambda_i = V_{is}^* V_{id}^{\phantom{*}}$ comprises the CKM
factors, and $X_\ell$ and $X_t$ are functions of the mass
ratios $x_c \equiv m_c^2/M_W^2$, $x_\tau \equiv m_\tau^2/M_W^2$
and $x_t \equiv m_t^2/M_W^2$~\cite{Buchalla:1995vs}. The former is strongly CKM suppressed for the CP-violating mode $K_L\to\pi^0
\nu\bar\nu$ and can be neglected in this case. These functions have been calculated up to NNLO in QCD and NLO in electroweak
corrections~\cite{Buras:2005gr, Buras:2006gb, Brod:2008ss, Brod:2010hi}. The SM predictions for the branching ratios are exceptionally clean since the requisite hadronic matrix elements can
be extracted from the well measured $K_{\ell 3}$ modes, including
higher-order chiral corrections~\cite{Mescia:2007kn}. Using the above SM predictions and input parameters from the PDG~\cite{ParticleDataGroup:2020ssz}, we find~\cite{Brod:2021hsj}:
\begin{align}
  {\cal B}(K^+\to\pi^+\nu\bar\nu) & = 7.73(16)(25)(54) \times 10^{-11},\\
  {\cal B}(K_L\to\pi^0 \nu \bar \nu) & = 2.59(6)(2)(28) \times 10^{-11}.
\end{align}
The errors in parentheses correspond to the remaining short-distance,
long-distance, and parametric uncertainties.

The weak Hamiltonian for the SD contribution to the leptonic mode $K_L \to \mu^+\mu^-$ has a similar form
\begin{equation}\label{eq:H:kmumu:sm}
  {\mathcal H}_\text{eff}
= \frac{4G_F}{\sqrt{2}} \frac{\alpha}{2\pi\sin^2\theta_w}
  \big( \lambda_c Y_\text{NL} + \lambda_t Y_t \big)
  (\bar s_L \gamma_\mu d_L) (\bar\mu_L \gamma^\mu \mu_L)
  + \text{h.c.},
\end{equation}
with different loop functions $Y_\text{NL} = Y_\text{NL}(x_c)$ and $Y
= Y(x_t)$. Also here, NNLO QCD and NLO electroweak corrections are
known~\cite{Gorbahn:2006bm, Bobeth:2013tba, Hermann:2013kca}. A
precise SM prediction of the branching ratio is complicated due to the
presence of poorly known long-distance contributions. However, the $K \to \mu^+\mu^-$ decay can still be used to probe the SM in a clean way~\cite{Dery:2021mct,DAmbrosio:2017klp}, if the time evolution of a beam made out of a different number of kaons and anti-kaons is measured. We discuss this possibility next.

\subsubsection{Clean SM test with $K \to \mu^+\mu^-$ decays}
\label{sec:KS:mumu}
Under the well-motivated assumption that the long-distance contribution to $K \to \mu^+\mu^-$ is CP-conserving, the CP-violating amplitude, $|A(K_S\to \mu^+\mu^-)_{\ell=0}|$, is a pure short-distance quantity (here, $\ell$ denotes the orbital angular momentum of the dimuon state). A measurement of $|A(K_S\to \mu^+\mu^-)_{\ell=0}|$ would provide a theoretically clean independent determination of $|\Im \lambda_t|$, and thus represents a sensitive probe of physics beyond the SM.
In the measurement of $\mathcal{B}(K\rightarrow \mu^+\mu^-)$ the two possible final state configurations, $l=0,1$, add incoherently. Because of that, a measurement of $\mathcal{B}(K_S\rightarrow \mu^+\mu^-)$ does not allow the extraction of the short-distance $l=0$ amplitude alone.
However, under certain assumptions, the same information can be extracted from the time dependent decay rate of a $K^0$ or $\bar K^0$ beam, as we discuss below.

The short-distance amplitude within the SM, resulting from the effective Hamiltonian of Eq.~(\ref{eq:H:kmumu:sm}), does not contribute to the $\ell=1$ final state. The same is true for any SM extension in which the leptonic operator is vectorial rather than a scalar.
This implies that within these models, 
\begin{equation}
    A(K_L\to \mu^+\mu^-)_{\ell=1}=0,
\end{equation}
leaving four non-vanishing theoretical parameters:
\begin{subequations}
\begin{align}
    &\, |A(K_L)_{\ell=0}|, \\
    &\, |A(K_S)_{\ell=0}|, \\
    &\, |A(K_S)_{\ell=1}|,
\end{align}
\end{subequations}
and 
\begin{equation}
    \varphi_0 \equiv \arg\big(A(K_L )_{\ell=0}^*A(K_S )_{\ell=0}\big).
\end{equation}
The complete system can be determined from the time-dependent decay rate, which in general is a sum of four functions of time:
\begin{equation}
    \left(\frac{\mathrm{d} \Gamma}{\mathrm{d} t} \right) = {\cal N}_f f(t),
\end{equation}
where ${\cal N}_f$ is a time-independent normalization factor, and 
\begin{equation}
    f(t) \equiv C_L e^{-\Gamma_L t}+
    C_S \,e^{-\Gamma_S t} + 2 \left[C_{\rm sin} \sin (\Delta m t) + C_{\rm cos} \cos
    (\Delta m t)\right]e^{-\Gamma t}. 
\end{equation}

For a $K^0$ ($\bar K^0$) beam, the experimental coefficients, $\{C_L, C_S,C_{\rm sin},C_{\rm cos}\}$ are given in terms of the theoretical parameters as
\begin{subequations}
\begin{align}
    C_L &=  |A(K_L)_{\ell=0}|^2, \\ 
    C_S &= |A(K_S)_{\ell=0}|^2 + \beta_{\mu}^2 |A(K_S)_{\ell=1}|^2 , \qquad \text{with} \quad \beta_\mu \equiv \left(1-4m_\mu^2/m_K^2\right)^\frac{1}{2},\\ 
    C_{\rm cos} &= \pm|A(K_S)_{\ell=0}^* A(K_L)_{\ell=0}|\,\cos\varphi_0,  \\
    C_{\rm sin} &= \pm |A(K_S)_{\ell=0}^* A(K_L)_{\ell=0}|\sin\varphi_0\,.
\end{align}
\end{subequations}
A measurement of the four coefficients is equivalent to the determination of all four theory parameters. In particular, the pure short-distance amplitude, $|A(K_S)_{\ell=0}|$, can be extracted using
\begin{equation}
    |A(K_S)_{\ell=0}|^2 = \frac{C_{\rm cos}^2+C_{\rm sin}^2}{C_L} \equiv \frac{C_{\rm int.}^2}{C_L}.
\end{equation}
It follows that the branching ratio can be determined from experimental observables via
\begin{eqnarray}\label{eq:BRexp}
    {\cal B}(K_S \to \mu^+\mu^-)_{\ell=0} = 
{\cal B}(K_L \to \mu^+\mu^-) \times {\frac{\tau_S}{\tau_L}} \times
\left(\frac{C_{\rm int.}}{ C_L}\right)^2\,.
\end{eqnarray}
The key ingredient is the sensitivity to the interference terms, represented by $C_{\rm int.}^2 = C_{\rm cos}^2 + C_{\rm sin}^2$.

The SM prediction arising from the Hamiltonian of Eq.~(\ref{eq:H:kmumu:sm}), is given by
\beq
\begin{split}
\label{eq:BRtheory}
    {\cal B}(K_S \to \mu^+\mu^-)_{\ell = 0} &= {\frac{\beta_\mu \,\tau_S}{16\pi m_K}}\left|
\frac{G_F}{\sqrt{2}} \frac{2\alpha}{ \pi\sin^2\theta_w} m_K
m_{\mu} Y_t \times   f_K \times \Im \lambda_t
 \right|^2 \\ 
&\approx 1.64 \cdot 10^{-13} \,\times \, \left|\frac{\Im \lambda_t}{(A^2\lambda^5\bar\eta)_\text{best fit}}\right|^2 \, ,
\end{split}
\eeq
where we used in the numerical evaluation
$(A^2\lambda^5\bar\eta)_\text{best fit}= 1.33\times 10^{-4}$.
The only hadronic uncertainty in this theory prediction comes from the decay constant, $f_K$, which can be related to the well-measured decay constant of charged kaons via isospin, introducing an uncertainty of ${\cal O}(1\%)$.
Eqs.~(\ref{eq:BRexp}), (\ref{eq:BRtheory}) demonstrate how a measurement of the interference terms in the time dependent rate of a beam with an unequal number of $K^0$ and $\bar K^0$ particles amounts to a very clean independent probe of $|\Im\lambda_t|$ from kaon physics.  


\subsubsection{Constraining heavy new physics}

The power-like GIM mechanism together with the structure of the CKM matrix makes the rare kaon decays extremely sensitive to contributions
of heavy new particles.
To make this explicit let us set the scale of new physics to
$\Lambda_{\mathrm{NP}} = 100$~TeV and parametrise the
contribution of heavy new physics through the effective interaction
\begin{equation}
  \label{eq:heff-heavy-np}
  \mathcal{H}_{\mathrm{eff}} \supset \frac{C_{lq}^{(1),sd}}{(100 \mathrm{TeV})^2}
  \sum_{\ell=e,\mu,\tau}
  (\bar s_L \gamma_\mu d_L) (\bar\nu_{\ell L} \gamma^\mu \nu_{\ell L})
  + \mathrm{h.c.}
\end{equation} 
Adding the current experimental and theoretical uncertainties in
quadrature, we can constrain physics up to a scale of $1/\sqrt{C_{lq}^{(1),sd}} \simeq 2$ in units of 100~TeV.
A measurement of ${\cal B}(K^+\to \pi^+\nu\bar\nu)$ with a 10\% uncertainty would test scales up to $1/\sqrt{C_{lq}^{(1),sd}} \simeq 4$ in units of 100~TeV.

This sensitivity can result in observable deviations from the SM prediction in a wide class of models of New Physics. See Ref.~\cite{Bordone:2017lsy} for a combined EFT and flavour symmetry
approach and Ref.\cite{Bobeth:2016llm} for effects that are generated from vector-quarks at tree-level.
Even at loop-level sizeable effects are possible in e.g. the Littlest Higgs model with T-parity, the MSSM or in models with an extra $Z'$ and
vector-quarks~\cite{Blanke:2015wba, Isidori:2006qy, Kamenik:2017tnu}.
While the analysis of tree-level contributions is relatively simple, loop effects are understood to be highly model dependent. In the SM, the (short-distance contributions to) the branching ratios are dominated by the gauge-invariant combination of penguin and box diagrams, $X_t$ and $Y_t$, involving the top quark. These loop functions can be generalized to include the contributions of any renormalizable model, while maintaining the properties of being finite and gauge independent. These last two properties can be shown with the help of coupling constant sum rules that are derived from Slavnov-Taylor identities which, in turn, encode the renormalizability of the considered model. This procedure can, in fact, be carried out for generic models involving heavy scalars, fermions, and vectors, without specification of the symmetry breaking mechanism that gives rise to the vector masses~\cite{Brod:2019bro, Bishara:2021buy}. This results in analytic expressions that specialise to the above model dependent result for the particle content of the respective model.

\newpage

\section{Present and future rare kaon and hyperon decay facilities}
\label{sec:facilities}
{\em Authors: Dettori, D\"obrich, Goudzovski, Kupsc, Lanfranchi, Martinez Santos, Massri, Moulson, Nanjo, Spadaro, Tung, Wang}

\subsection{The NA62 experiment at CERN}
\label{sec:exp:NA62}

The main goal of the NA62 experiment at CERN is the measurement of the $K^+\to\pi^+\nu\bar\nu$ decay rate to a 10\% precision using the decay in flight technique. The experiment operates a high-intensity $K^+$ beam with a momentum of 75~GeV/$c$ derived from primary 400~GeV/$c$ protons extracted from the SPS accelerator, and kaon decays in a long vacuum tank are detected~\cite{NA62:2017rwk}. The NA62 Run~1 dataset collected in 2016--2018, corresponding to $2.2\times 10^{18}$ protons on target, has led to the first observation of the $K^+\to\pi^+\nu\bar\nu$ decay based on 20~candidate events~\cite{NA62:2021zjw}. Several auxiliary trigger lines were in operation during Run~1 in addition to the main $K^+\to\pi^+\nu\bar\nu$ trigger line, with downscaling factors applied due to the overall trigger bandwidth limitation. The principal NA62 Run~1 kaon decay datasets correspond approximately to the following numbers of $K^+$ decays in the fiducial decay region~\cite{Collaboration:2713499}:
\begin{itemize}
\item the main $K^+\to\pi^+\nu\bar\nu$ trigger line: $N_K \approx 6\times 10^{12}$;
\item the di-muon trigger line: $N_K \approx 3\times 10^{12}$;
\item di-electron and electron-muon pair trigger lines: $N_K \approx 10^{12}$ each;
\item the minimum bias trigger providing a suitable dataset for all searches:
$N_K \approx {\rm few}\times 10^{10}$.
\end{itemize}

NA62 Run~2 started in July 2021, and is approved until Long Shutdown 3 (LS3). The goal of Run~2 is to increase the total dataset to an equivalent of 100 SM candidate $K^+\to\pi^+\nu\bar\nu$ events~\cite{Collaboration:2691873}. One can expect the auxiliary trigger lines listed above to be in operation throughout Run 2, with similar downscaling factors to those employed during Run~1. The proposed future $K^+$ decay program at CERN is discussed in Section~\ref{sec:exp:KLEVER}.

In 2003--2004, the NA48/2 experiment at CERN collected a dataset corresponding to $2\times 10^{11}$ charged kaon ($K^\pm$) decays in flight with the main trigger designed for collection of three-track vertices~\cite{NA482:2007ucr} originating from both negatively and positively charged kaons. In 2007, an early phase of the NA62 experiment collected a dedicated dataset with the primary goal of lepton universality test in $K^+\to\ell^+\nu$ decays~\cite{NA62:2012lny}. Due to the smaller longitudinal extent of the apparatus with respect to NA62 and the absence of three-track trigger downscaling, NA48/2 offers comparable or better sensitivity with respect to the NA62 Run~1 dataset for certain $K^+$ decays to multiple charged tracks.

The predecessors of the above experiments include NA48/1 which collected a sample of about $3\times 10^{10}$ $K_S$ decays using a high-intensity beam in 2002, and NA48 which published a number of $K_{S,L}$ rare decay measurements based on the 1997--2001 datasets~\cite{NA48:2007mvn}. Many of the rare kaon and hyperon measurements performed by these experiments are still world-leading.


\subsection{The future kaon program at CERN}
\label{sec:exp:KLEVER}

A plan for a comprehensive program to further study the rare decay modes of both $K^+$ and $K_L$ mesons is currently taking shape~\cite{NA62:2020upd}, to be carried out after LS3 with high-intensity kaon beams from the CERN SPS in multiple phases, including the following:
\begin{itemize}
\item an experiment to measure ${\cal B}(K^+\to\pi^+\nu\bar\nu)$ at the 5\% level ($\sim 400$ SM signal events) with a primary beam of intensity four times that of NA62;
\item an experiment to measure ${\cal B}(K_L\to\pi^0\nu\bar\nu)$ at the 20\% level ($\sim 60$ SM signal events) with a primary beam of intensity six times that of NA62;
\item a hybrid phase, with the neutral beam for the $K_L$ phase and the downstream tracking and particle identification detectors for the $K^+$ phase, to allow measurements of $K_L$ decays to charged particles, most importantly the very rare $K_L\to \pi^0\ell^+\ell^-$ decays.
\end{itemize}

This plan relies on the availability of a high-intensity, slow-extracted 400-GeV proton beam from the SPS to the current NA62 experimental hall, which is ideally suited for the next-generation kaon experiments. Up to $10^{19}$ protons on target per year will be required. Many of the existing detectors would be used in all of the program stages (especially the main calorimeter and photon vetoes), and new detectors will be developed to meet the specifications of the entire program.

The $K_L\to\pi^0\nu\bar\nu$ phase, known as KLEVER, has been extensively discussed within the context of the Physics Beyond Colliders initiative at CERN~\cite{Ambrosino:2019qvz}. The boost from the high-energy neutral beam facilitates the rejection of background channels such as $K_L\to\pi^0\pi^0$ by detection of the additional photons in the final state. Background from $\Lambda\to n\pi^0$ decays must be kept under control, and recent work has focused on the possibility of a beamline extension and other adaptations of the experiment to ensure sufficient rejection of this channel. The layout also poses challenges for the design of the small-angle vetoes, which must reject photons from $K_L$ decays escaping through the beam exit. Given the additional complexity implied by the beamline extension, together with the desirability of gaining experience with the neutral beam in earlier phases of the experiment, either the $K^+$ phase or the hybrid phase will most likely be the first to run after LS3.


\subsection{The KOTO experiment and the future program at J-PARC}
\label{sec:exp:KOTO}

The main goal of the KOTO experiment at J-PARC is a search for the $K_L\to\pi^0\nu\bar\nu$ decay. An intense $K_L$ beam is generated with 30-GeV protons hitting a production target in the J-PARC Hadron Experimental Facility.
Two collimators in a 20-m long beam line shape the beam in a solid angle of $7.8~\mu\mathrm{sr}$ at an extraction angle of $16^\circ$,
which provides $K_L$ beam with a peak momentum of 1.4~GeV/$c$ in a $8\times 8$~cm$^2$ cross-section at the beam line exit.
The contamination of short-lived and charged particles in the beam is suppressed with the 20-m long beam line and a sweeping magnet,
while the photon flux is reduced using a 7-cm long lead absorber. 
The KOTO detector and the decay volume are located downstream of the beam line.
To identify the $K_L\to\pi^0\nu\bar\nu$ signal,
the two observable particles in the final state are the photons from the $\pi^0\to\gamma\gamma$ decay. These are detected with a calorimeter, and the $\pi^0$ decay vertex is reconstructed assuming that the two are emitted from a $\pi^0$ decay, and the the decay vertex is located on the beam axis. The reconstructed vertex position and transverse momentum of the $\pi^0$ are used to define the signal region. 
The $K_L\to2\gamma$ decay is suppressed by requiring a high transverse momentum of the reconstructed $\pi^0$. Other backgrounds from $K_L$ decays involving charged particles or additional photons are reduced using the charged particle counter upstream and the hermetic veto system surrounding the decay region.

The KOTO experiment started data collection from 2013, and collected a data sample equivalent to $3.8\times 10^{11}$ $K_L$ decays. corresponding to $5.3\times 10^{19}$ protons on target, in 2015--2018. KOTO has established an upper limit of $3\times 10^{-9}$ on the branching ratio of the $K_L\to\pi^0\nu\bar\nu$ decay at 90\% CL with the data collected in 2015~\cite{KOTO:2018dsc}, and has achieved a single event sensitivity of $7.8\times 10^{-10}$ with the data collected in 2016--2018~\cite{Ahn:2020opg}.
The target single event sensitivity of the KOTO experiment is ${\mathcal O}(10^{-11})$, to be achieved by 2025. The next generation KOTO step-2 experiment with a target single event sensitivity of $O(10^{-13})$ has been discussed in Ref.~\cite{KOTOstep2, KOTOstep2InWhitePaperOfHEFex}.
Higher $K_L$ flux and harder momentum spectrum for step-2 will be provided by a new beam line with a $5^\circ$ extraction angle. A longer decay region and a larger diameter calorimeter will be used to increase the data sample.


\subsection{The LHCb experiment at CERN}
\label{sec:exp:LHCb}

The LHCb experiment has collected a dataset corresponding to an integrated luminosity of about 9~fb$^{-1}$ during Run 1 in 2010--2012 and Run 2 in 2015--2018. This dataset provides the leading sensitivity for decays of short-lived strange particles, i.e. $K_S$ and hyperon decays. Therefore LHCb is complementary to the dedicated $K^+$ and $K_L$ decay experiments described in Sections~\ref{sec:exp:NA62}, \ref{sec:exp:KLEVER}, \ref{sec:exp:KOTO}. Generically, final states with missing energy are difficult to tag in the LHCb environment, making the exploration of light weakly coupled new physics quite challenging. The biggest potential of LHCb in kaon and hyperon physics lies in the measurement of rare decays into opposite-sign muon pairs ($K_S\to\mu^+\mu^-$, $\Sigma\to p\mu^+\mu^-$). This owes to the dedicated trigger, which had an efficiency of about 15\% during Run 2, and is expected to reach an efficiency close to 100\% from Run 3 onward. The experiment is set to collect large datasets in future. The timeline for operation and upgrades of the LHCb experiment in the coming years, including the proposed Phase-II upgrade, is presented in Table~\ref{tab:LHCbTime}. 
\begin{table}[]
    \centering
    \begin{tabular}{cccccccc}
    \hline\hline
        Year &22--24&25--27&28--30&31&32--34&
        35&36--39\\
        \hline
        LHC status &  Run 3 & LS3 & Run 4 & LS4 & Run 5 & LS5 & Run 6\\
        LHCb peak lum. (cm$^{-2}$s$^{-1}$)&$2\times10^{33}$ & &$2\times10^{33}$ & &$2\times10^{34}$& & $2\times10^{34}$   \\
        LHCb Integrated lum. (fb$^{-1}$) & & & 50 & & & & 300\\
        \hline\hline
    \end{tabular}
    \caption{Data collection plans of the LHCb upgrade (2022--2030) and the proposed Phase-II upgrade (2032--2039). For a wider experimental landscape see~\cite{EuropeanStrategyforParticlePhysicsPreparatoryGroup:2019qin}.}
    \label{tab:LHCbTime}
\end{table}


\subsection{Kaon decays in beam-dump experiments}
\label{sec:ThickTarget}

Kaon factories typically obtain their beams as secondaries from primary protons striking a thin target. On the other hand, sizable kaon fluxes can be obtained from proton beams impinging on thick targets, i.e. with kaon factories operating in the beam dump mode. This offers an opportunity of searching for exotic kaon decays to hidden sector particles, which in turn decay to visible final states.
Beam-dump data have been collected by past and current experiments: CHARM collected $2.4\times 10^{18}$ protons on target (pot); NA62 collected $3\times 10^{16}$~pot in Run 1 and $1.5\times 10^{17}$~pot in optimised conditions in 2021, aiming for collection of $10^{18}$~pot by the end of Run~2~\cite{Collaboration:2691873}; KOTO has collected $2.2\times 10^{17}$~pot so far.

Production of hidden-sector particles in kaon decays in thick-target experiments results from two competing effects. Firstly, due to no kaon momentum selection in the beam dump mode, a large initial number of kaons is produced, e.g. about 0.9 per proton for the 400~GeV beam used by CHARM and NA62~\cite{Atherton:1980vj}.
However only a small fraction of these kaons are relevant for hidden-sector searches. Namely, only a small fraction of kaons decay before they are re-absorbed in the target material, since the hadronic absorption length $l_{\rm H}\simeq 15$~cm is much smaller than the typical beam dump targets thickness. For example, the probability for 75-GeV kaon decaying within the first absorption length of the target, before re-absorption, is $\Gamma l_{\rm H}/\gamma\sim 10^{-5}$. This large effect has been identified recently~\cite{Winkler:2018qyg} but omitted in much of the previous literature. The determination of the reach of a beam dump experiment for exclusive rare decay searches thus requires a detailed analysis in order to convert experimental observations to the new physics parameter space. Benefits of beam dump operation include the production of new states from $B$ decays (with about $10^{-7}$~$B$ mesons produced per proton on target) allowing to probe higher masses, or direct production of hidden-sector particles in proton collisions. Examples of constraints on exotic kaon decays obtained in beam-dump experiments include those on the Higgs portal scalar from the CHARM experiment (Fig.~\ref{fig:higgs_portal_summary}) and HNLs from the T2K experiment (Fig.~\ref{HNLLims}).
 

\subsection{The BESIII experiment}
\label{sec:exp:BESIII}

The BEPCII $e^+e^-$ collider is a charm-tau factory with a peak luminosity of $1.0 \times 10^{33}$ cm$^{-2}$s$^{-1}$ running at the center of mass energy in the 2--5~GeV range, the unique features of which allow for a rich physics program~\cite{Asner2008}. BESIII is a hermetic detector with a geometrical acceptance of $93\%$ (out of the complete $4\pi$) in a 1~T solenoid~\cite{bes3detector}. It has accumulated world's largest datasets of 
$1.0\times 10^{9}$
$J/\psi$ events (corresponding to 3.1 fb$^{-1}$ integrated luminosity)~\cite{BESIII:2021cxx}, $3\times 10^{9}$ $\psi(3686)$ events, and 2.9~fb$^{-1}$ integrated luminosity at $\psi(3773)$, all of which are produced directly from the $e^+e^-$ collisions. A number of searches for rare decays and physics beyond the SM have been performed by BESIII, utilizing the clean experimental environment~\cite{Chen:2021fcb,Wang:2020gzc}.

Hyperons are mainly produced at BESIII via $\psi\to B_1 \bar{B}_1$, where $B_1\bar{B}_1$ denotes a polarized, quantum-entangled hyperon pair. The $\psi$ mesons are produced in the annihilation of unpolarized electron-positron beams and therefore the spin-density matrix of the $\psi$ depends only on the scattering angle $\theta_B$ between the electron beam direction and the hyperon momentum. The coherent production of $B_1\bar{B}_1$ pairs from the decay, with the subsequent weak decays of the $B_1$ and $\bar{B}_1$, therefore constitute a spin-entangled quantum system where the final state is specified by two global real parameters. These features add an exciting new dimension to the study of CP violations and could enhance the related new physics search sensitivity with relatively modest statistics compared with the other hyperon sources. The projections for BESIII reach for rare and forbidden hyperon decays are reported in Ref.~\cite{Li:2016tlt}.

BESIII is optimizing its data-taking plans over the coming years to accumulate even larger data sets with an ever-improving understanding of the detector performance. It can continue operation for 8 years or even longer, with a high quality physics program~\cite{Ablikim:2019hff}. There is also a plan of increasing the luminosity and the center of mass energy to expand the physics reach further~\cite{Zhang2021}.

\subsection{Super charm-tau factories}
\label{sec:exp:super:tau:charm}

Super charm-tau factories (SCTFs) are proposed next-generation electron-positron colliders in the charm energy region with a luminosity two orders of magnitude larger than at BEPCII (Section~\ref{sec:exp:BESIII}), aiming to collect data samples of more than $1\times 10^{12}$ $J/\psi$ events per year of data taking. The two current SCTF proposals are those in  Russia~\cite{Epifanov:2020elk,Vorobiev:2021sct} and China~\cite{Luo:2018njj}. If one of such proposals is accepted, the collected number of hyperon decays from $J/\psi$ decays will increase by at least a factor of $100$ compared to the BESIII data. This implies an order of magnitude reduction in statistical uncertainties for all processes. Two improvements are being considered for the SCTF projects~\cite{Salone:2022lpt}: the center-of-mass energy spread compensation, and the longitudinal polarization of the electron beam.

The first option will allow the center-of-mass energy spread to be less than $\sim 0.9$ MeV, which will better match  the $J/\psi$ natural width of $\Gamma=90$~keV. This will effectively increase the number of the $J/\psi$ events produced for a given integrated luminosity and reduce the out-of-resonance backgrounds, improving the sensitivity to rare processes. This can be achieved by a collision scheme where electrons (positrons) with higher momenta collide with positrons (electrons) with lower momenta~\cite{Renieri:1975wt,Avdienko:1983mee,Telnov:2020rxp}. 

As far as the second option is concerned, it is estimated that at $J/\psi$ energies an electron beam polarization of about $80\%$  can be obtained without reduction in the beam current~\cite{Koop:2019sct}. Such a polarization is much larger than the observed natural transverse polarization of the baryon pairs from the $J/\psi$ decays (which is roughly $\sim 11\%$) and will increase the sensitivity of some hyperon studies.

\newpage

\section{Experimental signatures in rare kaon decays}
\label{sec:exp:kaons}
{\em Authors: Dettori, \underline{Goudzovski}, Martinez Santos, Moulson, Nanjo, Tung}

Experimental status and prospects for the rare kaon decay modes of interest for hidden-sector searches are discussed below. Each class of decay modes is presented in a dedicated subsection, linked to the relevant models in Section~\ref{sec:models}.


\subsection{The $K\to\pi+{\rm inv}$ decays}
\label{sec:sig:Kpi}
{\it These arise from $s\to d(\varphi/a/A')$ quark level transitions with light new physics particles escaping the detector. The relevant models are Higgs mixed scalar, $\varphi$, in Section~\ref{sec:higgs}, ALP, $a$, in Section~\ref{sec:ALP}, dark photon, $A'$, in Section~\ref{sec:darkphoton}, and modified kinematics from emission of light scalars or vectors coupled to neutrinos in Section~\ref{sec:neutrinos}. While not the main topic of Section~\ref{sec:2X}, invisible three-body decays can be a natural signature of the models with two dark sector particles. Models leading to enhanced $K_L\to \pi^0 X_{\rm inv}$ with little or no effect on $K^+\to \pi^+ X_{\rm inv}$ are discussed in Section~\ref{sec:GNVgamma}. The SM decay $K\to\pi\nu\bar\nu$ is theoretically clean, and represents a well-known probe for heavy new physics (Section~\ref{subsec:heavyNP}).}

Using the complete Run~1 dataset, the NA62 experiment established upper limits on ${\mathcal B}(K^+\to\pi^+X_{\rm inv})$ at the level of ${\cal O}(10^{-11})$ in the mass ranges 0--110~MeV and 154--260~MeV~\cite{NA62:2021zjw,NA62:2020xlg}, improving on the previous limits from the BNL E949 experiment~\cite{Artamonov:2009sz}. The search represents a direct extension of the $K^+\to\pi^+\nu\bar\nu$ measurement, which determines the $m_X$ range scanned. The event selection requires a $K^+$ from an incoming beam and a $\pi^+$ in the final state, in the absence of any other in-time activity in the detector. The largest (and irreducible) background is due to the $K^+\to\pi^+\nu\bar\nu$ decay. The two-body $K^+\to\pi^+X_{\rm inv}$ decay is characterised by a peak in the reconstructed missing mass ($m^2_{\textrm{miss}}$) distribution, with a width of ${\cal O}(10^{-3})~{\rm GeV}^2$ determined by the experimental resolution, on top of the continuous $K^+\to\pi^+\nu\bar\nu$ spectrum. The upper limits on the decay rate, assuming $X$ to be invisible or having a lifetime $\tau_X\gg 10$~ns~\cite{NA62:2021zjw}, are displayed in Fig.~\ref{fig:BRKpiinv}. Weaker limits, obtained by assuming visible decays and shorter lifetimes, are presented in Ref.~\cite{NA62:2021zjw}. The upper limit on the decay branching fraction weakens by up to a factor of two towards $m_X=0$, due to the $m^2_{\textrm{miss}}>0$ selection condition used to define the signal region for the $K^+\to\pi^+\nu\bar\nu$ analysis. The resulting constraints on the Higgs portal and long-lived ALP phase space are discussed in Sections~\ref{sec:higgs} and \ref{eq:longlived}, and shown in Figs.~\ref{fig:higgs_portal_summary} and \ref{fig:plotma}, respectively. Above the di-muon threshold, large values of the Higgs-scalar mixing angle $\theta$ are not excluded since the dark scalar decay length becomes comparable to the length of the experimental setup.

The above search for $K^+\to\pi^+X_{\rm inv}$ with the NA62 Run~1 dataset is almost background-free: the background is ${\cal O}(1)$ events in each $X$ mass hypothesis (determined by the experimental resolution), and decreases as a function of $m_X$. However the background is expected to become significant for the large future datasets (Sections~\ref{sec:exp:NA62}, \ref{sec:exp:KLEVER}). The sensitivity to the decay branching fraction can be expected to improve by up to a factor of four with the NA62 Run~2 data, and by at least another factor of two at the future CERN high-intensity kaon beam facility. The sensitivity to the Higgs portal and ALP parameters is determined from the scaling ${\cal B}(K^+\to\pi^+X)\propto \theta^2$ and ${\cal B}(K^+\to\pi^+X)\propto 1/f_a^2$.

The NA62 experiment has also established upper limits on ${\cal B}(K^+\to\pi^+X_{\rm inv})$, at the ${\cal O}(10^{-9})$ level in the 110--155~MeV mass range, in the vicinity of the $\pi^0$ mass, from a dedicated analysis based on $10\%$ of the Run~1 minimum bias dataset~\cite{NA62:2020pwi} (Figs.~\ref{fig:higgs_portal_summary}, \ref{fig:BRKpiinv}). An upper limit on the invisible neutral pion decay rate is also reported: ${\cal B}(\pi^0\to{\rm inv})<4.4\times 10^{-9}$ at $90\%$~CL. The sensitivity of this search is limited by the $K^+\to\pi^+\pi^0$ background with two undetected photons from the $\pi^0\to\gamma\gamma$ decay. The search resulted in an observation of 12 candidate events, compared to the expected background of $10^{+22}_{-8}$ events, and was thus consistent with the background-only hypothesis.
To improve the sensitivity, the event selection has been optimised to reduce the probability of missing both photons produced in the $\pi^0$ decay, where a probability of $(2.8^{+5.9}_{-2.1})\times 10^{-9}$ has been achieved. The obtained upper limits weaken as a function of $|m_X-m_{\pi^0}|$ because the same definition of the signal region  in terms of the reconstructed missing mass squared $m_{\rm miss}^2$ (optimized for the $\pi^0\to{\rm inv}$ search and centered on the $\pi^0$ mass) is used for each $m_X$ value. The analysis procedure can be modified to use a signal region centered on $m_X$ for each mass hypothesis, which would lead to sensitivity that would be improving as a function of $|m_X-m_{\pi^0}|$. The sensitivity to be obtained with larger datasets is potentially limited by the systematic uncertainties on the expected $K^+\to\pi^+\pi^0$ background.

A search for the $K^+\to\pi^+X$ decay in the 260--350~MeV mass range (above the $m_X$ values probed by the beam-dump experiments) is in principle possible with the NA62 data. This mass range is used as a control region for the $K^+\to\pi^+\nu\bar\nu$ analysis, therefore a background estimation technique is not readily available. The sensitivity above the di-pion threshold of 280~MeV is limited by the $K^+\to\pi^+\pi^+\pi^-$ background with two undetected pions (see Fig.~8 of Ref.~\cite{NA62:2021zjw}). Nevertheless, an ${\cal O}(10^{-9})$ sensitivity to ${\cal B}(K^+\to\pi^+X_{\rm inv})$ can be reasonably expected. In the Higgs portal scenario, the range of $\theta$ values probed would shrink towards larger $m_X$ values (a trend already seen in Fig.~\ref{fig:higgs_portal_summary}), due to the dark scalar decay length and therefore the acceptance of the event selection 
falling as a function of $m_X$ as seen from Eq.~(\ref{eq:higgs_portal_widths}). As a result, it is unlikely to obtain limitations on the Higgs portal above the di-pion threshold. In contradistinction, extension of the search to $m_X>260$~MeV is highly relevant for the ALP scenario (Fig.~\ref{fig:plotma}). More generally, constraining the transitions to an invisible $X$ final state over the entire mass range is a crucial probe for dark sectors as the invisible decay might be the dominant one. A possible search for visible dark scalar decay $\varphi\to\mu^+\mu^-$ at NA62, which is possibly competitive with the search for this decay at LHCb, is discussed in Section~\ref{sec:sig:Kpill}.

Measurements of the $K_L\to\pi^0\nu\bar\nu$ decay naturally provide limits on $\mathcal{B}(K_L\to\pi^0X_{\rm inv})$. 
The $K_L\to\pi^0\nu\bar\nu$ decay itself is the dominant background to the $K_L\to\pi^0X_{\rm inv}$ search, with any backgrounds to $K_L\to\pi^0\nu\bar\nu$ also contributing to the $K_L\to\pi^0X_{\rm inv}$ search. 
Since no missing mass reconstruction is performed, there is no significant reduction in the acceptance for $X$ in the $\pi^0$ mass region. The limits obtained from $K_L$ experiments are therefore complementary to those from the $K^+$ experiments.  

The KOTO experiment has reported upper limits of ${\cal B}(K_L\to\pi^0 X_{\rm inv})$ down to $2.4\times 10^{-9}$ at 90\% CL in the $m_X$ range 0--260~MeV with the 2015 dataset, without degradation in the $\pi^0$ mass region~\cite{KOTO:2018dsc}. More recently, KOTO has published a search for the $K_L\to\pi^0\nu\bar\nu$ decay with the 2016--2018 dataset, reporting three signal candidate events for an expected background of $1.22\pm 0.26$ events~\cite{Ahn:2020opg}. KOTO has not updated the $K_L\to\pi^0 X_{\rm inv}$ decay limits since no improvement on the 2015 dataset was expected due to the background contribution. However, while still preliminary, the KOTO 2016--2018 analysis was re-interpreted to derive the $\mathcal{B}(K_L\to\pi^0X_{\rm inv})$ limits~\cite{Kitahara:2019lws}. Applying the same method to the KOTO 2016--2018 result published in Ref.~\cite{Ahn:2020opg}, we obtain the upper limits of ${\cal O}(10^{-3})$ on the mixing angle $\theta$ presented in Fig.~\ref{fig:higgs_portal_summary}. These limits are very similar to those originally reported by KOTO using the 2015 dataset~\cite{KOTO:2018dsc}. KOTO sensitivity is expected to improve in the short term, using datasets with reduced background to be collected with an upgraded setup.

Future $K_L\to\pi^0\nu\bar\nu$ experiments (KOTO step-2, KLEVER) promise further improvements, however the sensitivity will be limited by backgrounds. In particular, the KLEVER sensitivity to $\mathcal{B}(K_L\to\pi^0X_{\rm inv})$ for $m_X<200$~MeV is estimated to be $0.8\times10^{-11}$, corresponding to a sensitivity of $4\times 10^{-5}$ on the Higgs portal mixing parameter $\theta$ (Fig.~44 of Ref.~\cite{Agrawal:2021dbo}). The KOTO step-2 sensitivity can be expected to be similar. 

Fig.~\ref{fig:higgs_portal_summary} shows that a sensitivity of $\mathcal{O}(10^{-13})$ to ${\cal B}(K\to\pi X_{\rm inv})$ would cover the entire parameter space of the Higgs portal scalar below 200~MeV, down to the BBN bound. This sensitivity, however, is out of reach of the present and planned experiments.

\subsection{The $K \to 2\pi+{\rm inv}$ decays}
\label{sec:sig:K2pi}
{\it The quark level transition is $s\to d (\varphi /a/A')$. Compared to $K\to \pi X$ decays, these probe a different set of couplings: the CP-violating coupling for the Higgs mixed scalar $\varphi$ (Section~\ref{sec:higgs}) and the axial coupling for ALP (Section~\ref{sec:ALP}). For the dark photon, $K\to 2\pi A'$ is one of the major discovery channels (Section~\ref{sec:darkphoton}).
}

The strongest constraints to date on the branching ratio of the $K^+\to\pi^+\pi^0 X_{\rm inv}$ decay  come from a sample of about $10^8$ $K^-$ decays analysed by the ISTRA+ collaboration~\cite{Tchikilev:2003ai}, improving on an earlier search at the BNL E787 experiment~\cite{E787:2000iwe}. Upper limits of the decay branching fraction of approximately $0.9\times 10^{-5}$ are reported in the $m_X$ range of 0--200~MeV, assuming $X$ to be a pseudoscalar sgoldstino~\cite{Gorbunov:2000cz}. These upper limits have only a weak dependence on $m_X$, with the exception of the $\pi^0$ mass region  where the limits degrade by up to a factor of four due to the $K^-\to\pi^-\pi^0\pi^0$ background. These results are used to set a limit on the axial flavor violating ALP coupling (Section~\ref{sec:fv_alp}). The NA62 Run~1 minimum bias dataset is expected to improve on the ISTRA+ sensitivity by 1--2 orders of magnitude in the above $m_X$ range. The sensitivity will depend on the background contamination, in particular from decays with undetected photons: $K^+\to\pi^+\pi^0\gamma$ (relevant for low $m_X$) and $K^+\to\pi^+\pi^0\pi^0$ (affecting mostly $m_X$ hypotheses in the $\pi^0$ mass region). 

The strongest limit on the $K_L\to\pi^0\pi^0\nu\bar\nu$ branching fraction ($8.1\times 10^{-7}$ at 90\% CL) has been reported at the KEK E391a experiment~\cite{E391a:2011aa}. Upper limits on the $K_L\to\pi^0\pi^0X_{\rm inv}$ branching fraction, also obtained from this analysis, vary from $7\times 10^{-7}$ for $m_X=50$~MeV to $4\times10^{-5}$ for $m_X=200$~MeV. The degradation of sensitivity towards larger $m_X$ values is caused by the reduction in acceptance. The KOTO experiment is conducting a search for $K_L\to\pi^0\pi^0\nu\bar\nu$ and $K_L\to\pi^0\pi^0X_{\rm inv}$ decays with a single event sensitivity of ${\mathcal O}(10^{-9})$, and is expected to extend the search to $m_X=0$ thereby probing the massless ALP and dark photon. Future searches at KOTO step-2 and KLEVER might reasonably be expected to improve this sensitivity by up to two orders of magnitude.


\subsection{The $K\to\pi\gamma +{\rm inv}$ decays}
\label{sec:Kpiginv}
{\it The quark level transition that leads to this signature is $s\to d \gamma (\varphi/ a/ A')$, where the photon is radiated either from the initial kaon or final state pion. Other possibilities are: the photon arising from a decay of a dark sector particles, e.g.,  $s\to d(\varphi/a\to\gamma A') $, or from the decay of a neutral pion, $K\to \pi(\pi^0\to \gamma A')$. Phenomenologically the most important example is the dark photon, see Section~\ref{sec:darkphoton}, since the extra $\gamma$ in the final state allows for the $K\to\pi\gamma A'$ transition even in the massless case. }

A search for $K^+\to\pi^+\pi^0$, $\pi^0\to\gamma A'$ decay chain involving an invisible dark photon in the mass range 30--130~MeV has been performed by NA62 using a fraction of Run 1 dataset corresponding to fewer than $10^{11}$ kaon decays~\cite{NA62:2019meo}; the trigger chain used for the analysis is not available in the main NA62 Run~1 dataset. The background from the $\pi^0\to\gamma\gamma$ decay with an undetected photon limits the sensitivity for $A'$ mass below about 40~MeV, while the search is almost background free for higher $m_{A'}$ values. The upper limits on the dark photon coupling parameter $\varepsilon^2$ reach the $10^{-7}$ level, and improve on the previous searches at NA64 and BaBar in the 60--110~MeV mass range. An upper limit of ${\cal B}(\pi^0\to\gamma\nu\bar\nu)<1.9\times10^{-7}$ is also reported from this study. The NA62 sensitivity to the $K^+\to\pi^+\gamma A'$ decay is yet to be established. The above processes could in principle be studied at future high statistics $K^+$ experiments, most likely using minimum bias datasets.

The $K_L\to\pi^0\gamma A'$ decay is being studied by KOTO with a dataset
of $2.8\times 10^{10}$ pot collected using a trigger requiring three clusters in the calorimeter, already used for the $K_L\to\pi^0\gamma$ search~\cite{KOTO:2020bhx}. Two photons from the $\pi^0\to\gamma\gamma$ decay can be efficiently selected from the three calorimeter clusters assuming the kinematics of the decay involving a dark photon~\cite{Su:2020xwt}.
The target single event sensitivity for this search is ${\cal O}(10^{-7})$. KOTO step-2 and KLEVER are expected to improve in sensitivity but may ultimately be limited the $K_L\to\pi^0\pi^0$ background.


\subsection{The $K\to 2\pi\gamma+{\rm inv}$ decays}
\label{sec:Kpipiginv}

{\it The simplest example is $K\to 2 \pi (\pi^0\to \gamma A')$, with $\pi^0$ decay to the dark photon due to the kinetic mixing with the SM photon (Section~\ref{sec:darkphoton}). Other, more exotic transitions, are also possible. These include a quark level transition  $s\to  d (\varphi /a \to \gamma A')$ where Higgs-mixed scalar $\varphi$ has CP-violating couplings, or ALP $a$ has axial couplings, and in each case the mediator has large couplings to the dark photons.}

A search for production of the invisible dark photon in the $K^+\to\pi^+\pi^0\pi^0$, $\pi^0\to\gamma A'$ decay chain is not competitive with the search for the $K^+\to\pi^+\pi^0$, $\pi^0\to\gamma A'$ process discussed in Section~\ref{sec:Kpiginv}. The NA62 sensitivity to the $K^+\to\pi^+\pi^0\gamma A'$ decay is yet to be established, and is expected to be limited by the $K^+\to\pi^+\pi^0\pi^0$ background, at least for low $A'$ masses. The minimum bias NA62 Run~1 dataset are available for this search.

The $K_L\to\pi^0\pi^0\gamma A'$ decay is accessible by the KOTO experiment. However the expected  sensitivity is  limited to ${\cal O}(10^{-3})$ due to the $K_L\to3\pi^0$ background, which will also ultimately limit the sensitivity of the future KOTO step-2 and KLEVER experiments in this channel.


\subsection{The $K \to \pi \gamma \gamma$ decays}
\label{sec:sig:Kpigg}
{\it The quark level transition is $s\to d (\varphi/a\to \gamma\gamma)$. For Higgs mixed scalar, the $\varphi\to \gamma\gamma$ decay rate is negligibly small (Section~\ref{sec:higgs}) but could be important for more general scalars, for instance a dilaton. For ALP, the $a\to\gamma\gamma$ decay can be one of the discovery channels (Section~\ref{sec:ALPdiphotondecay}). The $K\to\pi\gamma\gamma$ decay channel is important for the examples of GN violating models for which the violation is only apparent due to experimental loopholes (Section~\ref{sec:GNVgamma}).}

Owing to an experimental signature similar to that of the dominant $K^+\to\pi^+\pi^0$ decay, the $K^+\to\pi^+\gamma\gamma$ process is typically studied using donwscaled minimum bias datasets. A sample of $K^\pm\to\pi^\pm\gamma\gamma$ decays obtained from minimum bias datasets collected in 2003--2007, corresponding to $2.5\times 10^9$ kaon decays in total, has been analysed by the NA48/2 and NA62 experiments~\cite{NA482:2013faq, NA62:2014ybm}. In total, 380 candidates with a 10\% background contamination from $K^\pm\to\pi^\pm\pi^0\gamma$ and $K^\pm\to\pi^\pm\pi^0\pi^0$ decays are observed in the kinematic range $m_{\gamma\gamma}>220$~MeV. The combined model-dependent~\cite{DAmbrosio:1996cak} measurement of the decay rate in the full kinematic range based on the two datasets is reported to be ${\cal B}(K^\pm\to\pi^\pm\gamma\gamma)=(1.003\pm0.056)\times 10^{-6}$~\cite{NA62:2014ybm}.
Below the $\pi^0$ peak, the BNL E949 experiment has set an upper limit of ${\cal B}(K^+\to\pi^+\gamma\gamma) < 8.3 \times 10^{-9}$ at 90\% CL for $m_{\gamma\gamma}<108.5$~MeV based on the observation of no data events in this mass range, compared to the SM expectation of ${\cal B}=(6.10^{+0.16}_{-0.12})\times 10^{-9}$~\cite{E949:2005qiy}. This result is based on ${\cal O}(10^{12})$ stopped kaons collected with a dedicated trigger line. None of the experiments has performed dedicated peak searches on the di-photon mass. Interpretation of these measurements in terms of searches for ALP decays is discussed in Section~\ref{sec:ALPdiphotondecay}.

The NA62 Run~1 minimum bias dataset is estimated to contain about $3\times 10^3$ $K^+\to\pi^+\gamma\gamma$ candidates in the range $m_{\gamma\gamma}>220$~MeV, which is the world's largest sample by far. Moreover, the $m_{\gamma\gamma}$ resolution is improved with respect to the earlier datasets thanks to the beam tracker. A dedicated peak search at NA62 would improve the  sensitivity to ALP decays substantially, and allow for an extension of the lower limit of the accessible $m_{\gamma\gamma}$ range below 220~MeV. Due to the absence of a dedicated trigger line, NA62 is currently unable to improve on the E949 limit below the $\pi^0$ mass. The larger datasets to be collected at CERN (Sections~\ref{sec:exp:NA62} and \ref{sec:exp:KLEVER}) would improve the sensitivity further, although the search is already background-dominated.

The $K_L\to \pi^0\gamma\gamma$ decay rate has been measured to a 3\% precision by NA48~\cite{NA48:2002xke} and KTeV~\cite{KTeV:2008nqz} using samples of ${\cal O}(10^3)$ signal candidates. The measured branching ratio of $1.3\times 10^{-6}$ is in agreement with the ${\cal O}(p^6)$ chiral perturbation theory prediction. The main background to the SM process comes from the $K_L\to 3\pi^0$ decay with two undetected photons. Neither NA48 nor KTeV searched for peaks in the di-photon mass spectra. Interpretation of these measurements in terms of searches for ALP decays is discussed in Section~\ref{sec:ALPdiphotondecay}. The KOTO sensitivity to the $K_L\to \pi^0\gamma\gamma$ decay is limited by the $K_L\to3\pi^0$ background with the six photons from $\pi^0\to\gamma\gamma$ decays leading to four calorimeter clusters due to cluster fusion. In the longer term, both KLEVER and KOTO step-2 should be able to improve on the precision of the existing measurements. However, a peak search in the diphoton mass spectrum will be limited by the  $K_L\to3\pi^0$ and the $K_L\to\pi^0\gamma\gamma$ backgrounds.


\subsection{The $K\to\pi \ell_\alpha^+\ell_\alpha^-$ decays}
\label{sec:sig:Kpill}

{\it The quark level transitions are  $s\to d (\varphi/a/A'\to \ell_\alpha^+\ell_\alpha^-)$, where the flavor of the final state leptons is the same, while their charges are opposite. 
The relevant models are the Higgs mixed scalar (Section~\ref{sec:higgs}), promptly decaying ALP (Section~\ref{sec:ALP}) and dark photon (Section~\ref{sec:darkphoton}).}

Low-background $K^\pm\to\pi^\pm e^+e^-$ samples of $1.03\times 10^4$ and $0.73\times 10^4$ candidates in the mass range $m_{ee}>140$~MeV have been analyzed by the BNL E865 and CERN NA48/2 collaborations,
respectively~\cite{E865:1999ker, NA482:2009pfe}, leading to the PDG average model-dependent branching fraction in the full kinematic range of ${\cal B}=(3.00\pm0.09)\times 10^{-7}$~\cite{ParticleDataGroup:2020ssz}. No dedicated searches for resonances in the $m_{ee}$ spectrum have been performed by these experiments, while the sensitivity to ${\cal B}(K^+\to\pi^+X)$ assuming prompt $X\to e^+e^-$ decays is estimated to be ${\cal O}(10^{-9})$ . On the other hand, a search for resonances has been performed at BNL in the mass range $m_{ee}<100$~MeV, leading to ${\cal B}(K^+\to\pi^+X)<4.5\times 10^{-7}$ at 90\% CL assuming prompt $X\to e^+e^-$ decays~\cite{PhysRevLett.59.2832}. The above results lead to stringent constraints on ALP production and prompt decay, presented in Sections~\ref{sec:ALPdilepton} and \ref{sec:lldecaysAlves} (Figs.~\ref{fig:cggcwweeplots} and \ref{fig:K_to_ee_prompt}).

The $K^+\to\pi^+e^+e^-$ decay sample collected by the NA62 experiment during Run~1 using the (downscaled) di-electron trigger is larger than the previous ones~\cite{na62-lfv-2022}, and is essentially background free in both mass ranges, $m_{ee}<100$~MeV and $m_{ee}>140$~MeV~\cite{NA62:2019eax}. In the vicinity of the $\pi^0$ mass in the $m_{ee}$ spectrum, a large contribution from the $K^+\to\pi^+\pi^0$ decay (followed by $\pi^0\to e^+e^-$ and possibly $\pi^0_D\to\gamma e^+e^-$, $\pi^0_{DD}\to e^+e^-e^+e^-$ decays) is expected. A comprehensive analysis of the $m_{ee}$ spectrum in the full kinematic range is yet to be performed with the NA62 Run~1 dataset, including a $\pi^0\to e^+e^-$ measurement competitive with the KTeV one (see Section~\ref{sec:sig:Kpipiee}). While the blind spot in the vicinity of the $\pi^0$ mass can be avoided in the $m_{ee}$ resonance search, the sensitivity near the $\pi^0$ mass would be limited by the enhanced background, and by the uncertainty in the SM prediction of the $\pi^0\to e^+e^-$ decay rate~\cite{Dorokhov:2007bd}.
The NA62 dataset also allows for a systematic search for the $K^+\to\pi^+X$, $X\to e^+e^-$ process with displaced $X$ decay vertices, providing sensitivity to $X$ lifetimes of up to 10~ns.

Searches for dark photon production and prompt decay $K^\pm\to\pi^\pm A'$, $A'\to e^+e^-$ (Section~\ref{sec:darkphoton}) are not competitive with the world data due to the suppression of the $K^\pm\to\pi^\pm A'$ process~\cite{Davoudiasl:2014kua}. This has been demonstrated by the NA48/2 analysis in the 140--210~MeV mass range~\cite{NA482:2015wmo}, leading to ${\cal O}(10^{-5})$ sensitivity to the mixing parameter $\varepsilon^2$, to be compared with the limits shown in Fig.~\ref{fig:darkphoton_bounds} (right).

The NA48/2 dataset has been used to obtain a sample of $1.69\times 10^7$ fully reconstructed Dalitz decays of the neutral pion, $\pi^0_D\to\gamma e^+e^-$ decays, produced mainly in the $K^\pm\to\pi^\pm\pi^0$ process. This sample has been used to search for a promptly decaying dark photon via the $\pi^0\to\gamma A'$, $A'\to e^+e^-$ decay chain~\cite{NA482:2015wmo}. Upper limits of ${\cal O}(10^{-6})$ have been established on the mixing parameter $\varepsilon^2$ in the dark photon mass range of 9--120~MeV (Fig.~\ref{fig:darkphoton_bounds}, right). Using a similar analysis technique, a measurement of the slope of the $\pi^0$ electromagnetic transition form factor, $F(x)= 1+ax$, where $x=(m_{ee}/m_{\pi0})^2$, has been performed in the time-like momentum transfer region with the $\pi^0_D\to\gamma e^+e^-$ decay sample collected in 2007 using the NA48/2 setup, yielding $a = (3.68 \pm 0.57)\times 10^{-2}$~\cite{NA62:2016zfg}. The NA62 Run~1 sample of fully reconstructed $\pi^0_D\to\gamma e^+e^-$ decays is of a similar size to the NA48/2 sample, however the sensitivity to dark photon production is expected to degrade towards smaller $m_{A'}$ values, due to the highly forward detector geometry.

A low-background sample of 3120 $K^\pm\to\pi^\pm\mu^+\mu^-$ candidates has been analyzed by the NA48/2 experiment, including a branching fraction measurement, ${\cal B}=(0.96\pm0.03)\times 10^{-7}$~\cite{NA482:2010zrc}, and a dedicated search for promptly decaying resonances in the $m_{\mu\mu}$ distribution~\cite{NA482:2016sfh}. The trigger logic allowed for a prompt resonance decay search only, leading to optimal acceptances for lifetimes below 20~ps. For lifetimes in excess of 100~ps, the sensitivity weakens proportionally to the assumed lifetime. The upper limits obtained on the product ${\cal B}(K^+\to\pi^+X)\cdot{\cal B}(X\to\mu^+\mu^-)$ assuming prompt decay are ${\cal O}(10^{-9})$, limited by the SM background. Similarly to the electron case, these results lead to stringent constraints on ALP production and prompt decay presented in Sections~\ref{sec:ALPdilepton} and \ref{sec:lldecaysAlves} (Figs.~\ref{fig:cggcwweeplots} and \ref{fig:K_to_mumu_prompt}), while the search for dark photon emission and prompt decay (Section~\ref{sec:darkphoton}) in this process is not competitive. Interpretation of these results in terms of Higgs mixed dark scalar production and decay (Section~\ref{sec:higgs}) is not competitive with the LHCb limits (Fig.~\ref{fig:higgs_portal_summary}).

A preliminary study of $2.8\times 10^4$ $K^+\to\pi^+\mu^+\mu^-$ candidates obtained from the NA62 Run~1 di-muon dataset has been reported recently~\cite{Bician:2020ukv}. An extension of this analysis to search for $K^+\to\pi^+X$ followed by prompt or displaced $X\to\mu^+\mu^-$ decays is yet to be performed. For the prompt decay, 
we expect the sensitivity to the product ${\cal B}(K^+\to\pi^+X)\cdot{\cal B}(X\to\mu^+\mu^-)$ to improve on average by about a factor of four with respect to NA48/2. In contrast with NA48/2, the NA62 trigger logic does not inhibit the displaced vertex search, which is almost background-free and would provide ${\cal O}(10^{-10})$ sensitivity for $X$ lifetimes in the 0.1--10~ns range. This makes NA62 potentially competitive with LHCb in terms of the dark scalar search via the $K^+\to\pi^+\varphi$, $\varphi\to\mu^+\mu^-$ decay chain for $m_{\varphi}>2m_\mu$ (Section~\ref{sec:higgs}, Fig.~\ref{fig:higgs_portal_summary}). The signal yield is proportional to $\sin^4\theta$, with two powers due to ${\cal B}(K^+\to\pi^+\varphi)$ and two powers due to the finite decay region length, the $\varphi\to e^+e^-$ decay rate is negligible, while the sensitivity to $\varphi\to\pi^+\pi^-$ (for $m_\varphi>2m_\pi$) is limited by the absence of a dedicated trigger line.

The most sensitive searches to date for the ultra-rare $K_L\to\pi^0\ell^+\ell^-$ decays with ${\cal O}(10^{-11})$ expected SM branching fractions~\cite{Mescia:2006jd} have been performed by the KTeV experiment. In both cases, the numbers of candidate events are consistent with the expected background, which is ${\cal O}(1)$ events for both searches. The upper limits at 90\% CL of the decay branching fraction are reported to be~\cite{KTEV:2000ngj,KTeV:2003sls}
\begin{equation}
{\cal B}(K_L\to\pi^0e^+e^-) <  2.8 \times10^{-10}, \quad
{\cal B}(K_L\to\pi^0\mu^+\mu^-) <  3.8 \times10^{-10}.
\end{equation}
The resulting constraints on ALP production and prompt decay, shown in Fig.~\ref{fig:cggcwweeplots}, are similar to or weaker than those obtained from the $K^+\to\pi^+\ell^+\ell^-$ decays. First observation of these decay modes can be expected within the hybrid phase of the future CERN kaon program (Section~\ref{sec:exp:KLEVER}). Unlike dedicated $K_L\to\pi^0\nu\bar\nu$ experiments, the hybrid phase detector would be equipped with a tracker for measurements of charged particles produced in $K_L$ decays, enabling these and other measurements.

\subsection{The $K\to\pi\pi\ell_\alpha^+\ell_\alpha^-$ decays}
\label{sec:sig:Kpipiee}
{\it The relevant models are a Higgs mixed scalar with CP violating couplings (Section~\ref{sec:higgs}), ALP with axial couplings (Section~\ref{sec:ALP}) and the dark photon (Section~\ref{sec:darkphoton}).}

First observation and study of the $K^\pm\to\pi^\pm\pi^0 e^+e^-$ decay has been reported by the NA48/2 experiment~\cite{NA482:2018gxf}. Based on 4919 candidates with a 4.9\% background contamination, the decay branching fraction is measured to be ${\cal B}(K^\pm\to\pi^\pm\pi^0 e^+e^-) = (4.24\pm0.14)\times 10^{-6}$. This result has been used to evaluate the sensitivity to prompt ALP decays (Section~\ref{sec:lldecaysAlves}, Fig.~\ref{fig:K_to_ee_prompt}).
The NA62 dataset is not expected to be competitive with NA48/2 due to the highly forward geometry leading to a reduced acceptance, and the downscaling applied to the di-electron trigger chain.

The $K_{S,L}\to\pi^+\pi^-e^+e^-$ decays have been measured by the NA48 and NA48/1 experiments from low-background samples of ${\cal O}(10^4)$ candidates~\cite{NA48:2003pwz,NA481:2011gxv}, leading to PDG averaged branching fractions~\cite{ParticleDataGroup:2020ssz}
\begin{eqnarray}
{\cal B}(K_{S}\to\pi^+\pi^-e^+e^-) &=& (4.79 \pm 0.15)\times 10^{-5},\\
{\cal B}(K_{L}\to\pi^+\pi^-e^+e^-) &=& (3.11 \pm 0.19)\times 10^{-7}.
\end{eqnarray}
It is expected that the future LHCb Run~3 dataset (Section~\ref{sec:exp:LHCb}) would allow observation of the $K_S\to\pi^+\pi^-e^+e^-$ decay, possibly surpassing the NA48/1 sensitivity~\cite{MarinBenito:2193358}. Improvements on the $K_L\to\pi^+\pi^-e^+e^-$ decay are expected within the hybrid phase of the future CERN kaon program (Section~\ref{sec:exp:KLEVER}).  

Upper limits on the $K_L\to\pi^0\pi^0\ell^+\ell^-$ decay rates have been set by the KTeV experiment from background-free analyses, including a search for resonances in the $\mu^+\mu^-$ system at the ${\cal O}(10^{-10})$ sensitivity level~\cite{KTeV:2002tpo, KTeV:2011xrc}. Sensitivity might be improved within the hybrid phase of the future CERN kaon program. KTeV has also measured the $\pi^0\to e^+e^-$ decay branching ratio to be $(6.44\pm0.33)\times 10^{-8}$ from a sample of 794 $K_L\to\pi^0\pi^0\pi^0_{ee}$ candidates~\cite{KTeV:2006pwx}.


\subsection{The $K\to \pi \ell_\alpha^+\ell_\alpha^- \ell_\beta^+\ell_\beta^-$ decays}
\label{sec:sig:5track}
{\it The quark level transition is $s\to d X X$, with both $X$ decaying visibly, $X\to 2e, 2\mu$. In most models the quark transition is due to a decay chain $s\to d X_{\rm heavy}^{(*)}, X_{\rm heavy}^{(*)}\to 2X$. For on-shell $X_{\rm heavy}$ one can search for resonances in $2\ell$ and $4\ell$ mass spectra. The kinematically allowed decays are $K\to \pi 4 e$, $K\to \pi 2 e 2\mu$, where $X\to e^+e^-$ and/or $X\to \mu^+\mu^-$. One may also consider the flavor violating options $K\to \pi 3 e \mu$, $K\to \pi e 3\mu$, which would arise from a combination of $X\to e^+e^-$, $X\to \mu^+\mu^-$ and $X\to \mu^\pm e^\mp$ decays. A related signature is also $K\to \pi 4\ell+$inv. The models are discussed in Section~\ref{sec:2X}.}

The NA62 Run~1 di-electron dataset offers an opportunity for the first search for the $K^+\to\pi^+A'A'$ decay followed by prompt $A'\to e^+e^-$ decays, as well as the prompt
decay chain including an intermediate resonance ($K^+\to\pi^+\varphi$, $\varphi\to A'A'$, $A'\to e^+e^-$), and the SM $K^+\to\pi^+e^+e^-e^+e^-$ decay (Section~\ref{sec:HiggsedU1-0}). One can expect single event sensitivities to the decay branching fractions at the ${\cal O}(10^{-9})$ level. The background is likely to be dominated by the $K^+\to\pi^+\pi^0_D\pi^0_D$ decays (where $\pi_D^0$ denotes the Dalitz decay, $\pi^0_D\to\gamma e^+e^-$) with two undetected soft photons, as well as by coincidences in time of multiple $K^+\to\pi^+\pi^+\pi^-$ decays and possibly other kaon decays. The backgrounds can be expected to be low but are yet to be evaluated.

Further possibilities of multi-track processes with muons (Section~\ref{sec:HiggsedU1-0})
or missing energy (Section~\ref{sec:miss:eng:manybody}) have not been considered by experiments yet, though they could be searched for in principle with the NA62 Run~1 di-lepton datasets.


\subsection{The $K_L\to\gamma+{\rm inv}$ decays}
\label{sec:KL:gamma:inv}
{\it The relevant model is dark photon in Section~\ref{sec:darkphoton}.}

The $K_L\to\gamma A'$ decay to a massless dark photon $A'$ (Section~\ref{subsec:darkphoton}) is being searched for by KOTO. A dataset corresponding to $7\times 10^{16}$ pot (i.e. about 0.1\% of the total KOTO exposure to date) has been collected requiring a single cluster in the calorimeter. The expected single event sensitivity is ${\mathcal O}(10^{-6})$, and the main background is caused by the beam neutrons. The $K_L\to\gamma A'$ decay also represents a natural target for KOTO step-2 and KLEVER experiments.


\subsection{The $K\to\pi\gamma$ and $K_L\to 3\gamma$ decays}
\label{sec:K:pi:gamma}
{\it The decays $K\to\pi \gamma$ and $K_L\to 3\gamma$ are forbidden by Lorentz and gauge invariance. Since the observational bounds on the photon mass  are very severe, the relevant models induce such decays through Lorentz violation in Section~\ref{sec:Lorentz:viol}. }

The BNL E949 experiment has set an upper limit ${\mathcal B}(K^+\to \pi^+\gamma)<2.3\times 10^{-9}$ at $90\%$ CL, obtained as a by-product of a search for $K^+\to\pi^+\gamma\gamma$ near the pion momentum endpoint~\cite{E949:2005qiy}. This search is based on ${\cal O}(10^{12})$ stopped kaons collected with a dedicated trigger line, and is background-free. In contrast, NA62 has only donwscaled minimum bias datasets available for this search, with an estimated Run~1 single-event sensitivity at the ${\cal O}(10^{-9})$ level.

A search with the KOTO 2016--2018 dataset has led to an upper limit of
${\cal B}(K_L\to \pi^0\gamma) <  1.7\times 10^{-7}$ at 90\% CL~\cite{KOTO:2020bhx}. This analysis can be extended to searches for Lorentz violating $K_L\to 3\gamma$ decays, for instance due to $K_L\to\gamma X$ process followed by a prompt $X\to\gamma\gamma$ decay. KOTO step-2 and KLEVER are expected to improve the sensitivity for both processes.

\subsection{The $K_L\to\gamma\gamma +{\rm inv}$ decays}
\label{sec:KL:2gamma:inv}
{\it This is an important channel for the GN violating models discussed in Section~\ref{sec:GNVgamma}. The models can be grouped into {\it i)} the $\pi^0$ impostor decays, $K_L\to X_2 X_1, X_2\to \gamma\gamma$ with $X_1$ invisible, {\it ii)} dipole portal transitions $K_L\to X_2 X_2, X_2\to X_1\gamma $, and {\it iii)} the non-standard $\pi^0$ production, $K_L\to X_2 X_1 ,  X_2\to X_1\pi^0$, as well as $K_L\to X_2^* \to  X_1 \pi^0$ or $K_L\to X_2^* \to  X_1X_1 \pi^0$.  
}

A search for the $K_L\to\gamma\gamma +{\rm inv}$ decay can be performed as a by-product of $K_L\to\pi^0\nu\bar\nu$ search in the KOTO experiment~\cite{Ahn:2020opg}. This consideration also applies to the future KOTO step-2 and KLEVER experiments. The sensitivity to $K_L\to\gamma\gamma +{\rm inv}$ can be improved by optimising the signal region definitions with respect to the $K_L\to\pi^0\nu\bar\nu$ analysis, according to the decay kinematics.


\subsection{The $K_{S,L}\to \ell^+\ell^-\!\! (+{\rm inv})$ and $K_{S,L}\to \ell^+\ell^-\gamma\gamma$ decays}
\label{sec:KL2l:inv}
{\it The GN violating models discussed in Section~\ref{sec:fundamendal:viol}, $K_L\to X_2 X_1$, lead also to the $K_L\to \ell^+\ell^-\!\! +{\rm inv}$ signature if $X_2\to \ell^+\ell^-$ decay is sizable, with $X_1$ invisible, or to $K_L\to \ell^+\ell^-\gamma\gamma$, if $X_1\to\gamma\gamma$ is appreciable. The two dark sector models in Section~\ref{sec:2X} can also lead to $K_L\to \ell^+\ell^-\!\! (+{\rm inv})$ if dark photon has an invisible decay mode open. The $K_S\to aa$ decay for MeV QCD axion discussed in Section~\ref{sec:2X} can lead to $K_S\to\ell^+\ell^-+{\rm inv}$,  $K_S\to\ell^+\ell^-\gamma\gamma$, $K_S\to 4\ell$, if $a$ has also the invisible decay channel. Measurement of CP violating amplitude $A(K_S\to \mu^+\mu^-)_{\ell =0}$ represents a clean probe for heavy new physics (Section~\ref{sec:KS:mumu}).}
\label{sec:exp:kleplep}

The $K_L\to\mu^+\mu^-$ decay rate has been measured to be ($7.18\pm0.17)\times 10^{-9}$~\cite{E871:2000wvm}, and the ultra-rare $K_L\to e^+e^-$ decay with an ${\cal O}(10^{-11})$ branching fraction has been observed with four signal candidates~\cite{BNLE871:1998bii}, by the BNL E871 experiment. The 
$K_L\to \ell^+\ell^-\gamma\gamma$ decays have been observed by the KTeV experiment~\cite{KTeV:2000kyj,KTEV:2000mky}. Substantial experimental progress on the $K_L\to\ell^+\ell^-(X_{\rm inv})$ and $K_L\to\ell^+\ell^-\gamma\gamma$ processes can be expected within the hybrid phase of the future CERN kaon program (Section~\ref{sec:exp:KLEVER}). The $K_{S,L}\to\pi^0\ell^+\ell^-$ decays are discussed Sections~\ref{sec:sig:Kpill} and \ref{sec:exp:ks}.


\subsection{The $K_{S,L}\to \ell_\alpha^+\ell_\alpha^- \ell_\beta^+\ell_\beta^-$ decays}
\label{sec:K0:4l} 
{\it The relevant models are the GN violating models (Section~\ref{sec:GNVgamma}), and the production of multiple hidden-sector particles in models (Section~\ref{sec:2X}).}

Searches for the $K_{S,L}\to \ell_\alpha^+\ell_\alpha^- \ell_\beta^+\ell_\beta^-$ signatures (here $\alpha$ and $\beta$ are generational indices) are sensitive to neutral kaon decays completely into dark-sector mediator particles, if these mediators decay back through visible channels (Sections \ref{sec:fundamendal:viol} and~\ref{sec:HiggsedU1}). The expected SM branching ratios for these decays range from ${\cal O}(10^{-14})$ for $K_S\to 2(\mu^+\mu^-)$ to ${\cal O}(10^{-8})$ for $K_L\to 2(e^+e^-)$~\cite{DAmbrosio:2013qmd}, allowing for low-background searches for new physics.

The $K_L\to 2(e^+e^-)$ decay has been measured from clean samples of ${\cal O}(100)$ candidates by the KTeV and NA48 experiments~\cite{AlaviHarati:2001ab,Lai:2005kw}, giving the PDG averaged branching ratio of $(3.56\pm 0.21)\times 10^{-8}$~\cite{ParticleDataGroup:2020ssz}. The $K_L\to \mu^+\mu^-e^+e^-$ decay has been measured by the KTeV experiment from a clean sample of 132~candidates, and its branching fraction is reported to be $(2.69\pm 0.27)\times 10^{-9}$~\cite{KTeV:2002kut}. Both measurements are in agreement with the SM predictions~\cite{DAmbrosio:2013qmd}. There are no experimental data on the ultra-rare $K_L\to 2(\mu^+\mu^-)$ decay. Further improvements on the $K_L$ measurements are expected to come from the hybrid phase of the future CERN kaon program (Section~\ref{sec:exp:KLEVER}).

The ultra-rare $K_S\to 2(e^+e^-)$ and $K_S\to\mu^+\mu^-\ell^+\ell^-$ decays have not been measured yet, and the latter final state represents a possible future target for the LHCb  experiment~\cite{Alves:2018npj}. The $K_{S,L}\to\pi^02(e^+e^-)$ decays (Section~\ref{sec:HiggsedU1-0}) have not been searched for.


\subsection{The $K^+\to \ell_\alpha^++{\rm inv}$ decays}
\label{sec:sig:Kl}
{\it The quark level transitions are $\bar s\to \bar u \ell_\alpha^+N$ where $N$ is the heavy neutral lepton (Section~\ref{sec:HNL}), or $\bar s\to \bar u \ell_\alpha^+\nu$ $ \ell_\alpha^+\nu$ with an extra emission of $a/A'$ from the either the charged lepton or SM neutrino leg, see Section~\ref{sec:muonicf} for leptonic force models and Section~\ref{sec:neutrinos} for strongly interacting neutrino models.}

A search for heavy neutral lepton production in $K^+\to e^+N$ decays, where $N$ is invisible (Section~\ref{sec:HNL}), has been performed with the NA62 Run~1 sample using the main $K^+\to\pi^+\nu\bar\nu$ trigger line. Upper limits of ${\cal O}(10^{-9})$ are established on the mixing parameter $|U_{eN}|^2$ in the HNL mass range of 144--462~MeV~\cite{NA62:2020mcv},  complementing the search in the 62--135~MeV range in $\pi^+\to e^+N$ decay~\cite{PIENU:2017wbj}.

Searches for the $K^+\to\mu^+N$ decay, with $N$ invisible, have been performed with prescaled minimum bias datasets since a single muon is the most common $K^+$ decay final state. The BNL E949 and NA62 (Run~1) experiments have established ${\cal O}(10^{-8})$ upper limits on $|U_{\mu N}|^2$ in complementary HNL mass ranges of 178--300~MeV and 200--384~MeV, respectively~\cite{E949:2014gsn,NA62:2021bji}. The constraints on HNL mass and mixing parameters $|U_{\ell N}|^2$ established by these searches are shown in Fig.~\ref{HNLLims}.

Searches for HNL production in $K^+\to\ell^+N$ decays are limited by backgrounds. In particular, the abundant $K^+\to\mu^+\nu$ decay followed by the $\mu^+\to e^+\nu\bar\nu$ decay in flight represents an irreducible background to the $K^+\to e^+N$ process. Another background source to the $K^+\to e^+N$ decay is the $\pi^+\to e^+\nu$ decay of the beam pions (which are the majority particle in the NA62  beam). The peaking nature of the $K^+\to\ell^+N$ signals in terms of the reconstructed missing mass allows for data driven background evaluation, reducing the systematic uncertainties in the background estimates. It is expected that the future high-intensity kaon beam facility at CERN (Section~\ref{sec:exp:KLEVER}) would improve the sensitivity to the mixing parameters $|U_{\ell N}|^2$ by up to a factor of five with respect to NA62 Run~1. The sensitivity of the search for $\pi^+\to e^+N$ decays of beam pions at NA62 and its successors is expected to improve on the PIENU limits~\cite{PIENU:2017wbj} in the mass range $m_{N}<m_\pi$. The gap in $m_N$ can be covered by the search for $K^+\to\pi^0e^+N$ decay, although with a limited sensitivity~\cite{Tastet:2020tzh}.

The first search for the $K^+\to\mu^+\nu X_{\rm inv}$ decay  (Sections~\ref{sec:muonicf} and \ref{sec:neutrinos}) has been performed by using the NA62 Run~1 minimum bias data~\cite{NA62:2021bji}. In contrast to HNL production in two-body decays, the $K^+\to\mu^+\nu X_{\rm inv}$ signal has a continuous missing mass distribution. The background evaluation is therefore based on the full simulation (rather than the sidebands of missing mass spectrum for each $m_N$ hypothesis), leading to larger uncertainties in the estimated background. The irreducible background due to the $K^+\to\mu^+\nu\nu\bar\nu$ decay is negligible, with expected ${\cal B}(K^+\to\mu^+\nu\nu\bar\nu)=1.62\times 10^{-16}$~\cite{Gorbunov:2016tbk}. The sensitivity is limited by the $K^+\to\mu^+\nu\gamma$ and $K^+\to\pi^0\mu^+\nu$ backgrounds with undetected soft photons. The background estimate is affected by large systematic uncertainties due to the accuracy of modelling of non-gaussian tails of missing mass reconstruction, photon veto inefficiency and photon conversions. To reduce the background, a limited reconstructed missing mass region, $m_{\rm miss}^2>0.1~{\rm GeV}^2$, is analysed. A mass range of 10--370~MeV is considered for a scalar and a vector $X$ particle, and upper limits on the decay branching fraction ranging from ${\cal O}(10^{-5})$ for low $m_X$ to ${\cal O}(10^{-7})$ for high $m_X$ are established assuming the model of Ref.~\cite{Krnjaic:2019rsv}. An upper limit on the SM decay rate is also established: ${\cal B}(K^+\to\mu^+\nu\nu\bar\nu)<1.0\times 10^{-6}$ at 90\% CL. These limits are much weaker than the values accounting for the $(g-2)_\mu$ discrepancy (Section~\ref{sec:muonicf}), despite the single event sensitivity of ${\cal O}(10^{-9})$ being in the region of interest. Reduction of systematic uncertainties in the background estimate in order to  make progress towards the region in parameter space of interest for $(g-2)_\mu$ is very challenging.

NA62 is expected to able to perform a search for the $K^+\to e^+\nu X_{\rm inv}$ decays at the ${\cal O}(10^{-6})$ sensitivity level with the Run~1 minimum bias dataset. The current upper limit ${\cal B}(K^+\to e^+\nu\nu\bar\nu)<6\times 10^{-5}$~\cite{Heintze:1977kk} is also likely to be improved by NA62 by 1--2 orders of magnitude. Similarly to the $K^+\to e^+N$ case, this search would be strongly limited by background.


\subsection{The $K^+\! \to \ell_\alpha^+ \ell_\beta^- \ell_\beta^+\!+{\rm inv}$ decays}
\label{sec:3l:inv}

{\it This signature can arise from $K^+\to \ell_\alpha^+ (N \to \ell_\beta^+\ell_\beta^-\nu)$ decays, where the heavy neutral lepton decays through a non-standard neutral current interaction such as an extra $U(1)_d$ (Sections~\ref{sec:HNL} and~\ref{sec:2X}), or from $K^+\!\to \ell_\alpha^+\nu (a/A' \to \ell_\beta^+\ell_\beta^-)$, such as in leptonic force models (Section~\ref{sec:muonicf}).}

Out of the four SM four-lepton rare $K^+$ decay modes~\cite{Bijnens:1992en}, all with the exception of $K^+\to\mu^+\mu^-\mu^+\nu$ have been measured by the BNL E865 experiment~\cite{Poblaguev:2002ug, Ma:2005iv}. The NA62 Run~1 dilepton dataset offers an opportunity to measure all these processes. Due to the excellent resolution on the vertex position and kinematic variables, NA62 has observed for the first time $K^+\to\mu^+\mu^-\mu^+\nu$ decay with the SM branching fraction of $1.35\times 10^{-8}$~\cite{Bijnens:1992en} affected by the background from the abundant $K^+\to\pi^+\pi^+\pi^-$ process followed by pion decays in flight~\cite{Boretto:2716696}.

Search for the leptonic force mediator (Section~\ref{sec:muonicf}) at NA62 in the prompt $K^+\to\mu^+\nu X$, $X\to\mu^+\mu^-$ decay chain is promising due to the low background. Assuming the total background to be of the same order of magnitude as the contribution from the rare $K^+\to\mu^+\mu^-\mu^+\nu$ decay (which is supported by the preliminary study of the data~\cite{Boretto:2716696}), the analysis of Ref.~\cite{Krnjaic:2019rsv} showed that a di-muon search with the NA62 Run~1 data would completely cover the $(g-2)_\mu$ and DM benchmarks for most masses between the $\mu^+\mu^-$ threshold and the kinematic endpoint for $X$ production. The sensitivity of kaon factories complements the search at BaBar~\cite{TheBABAR:2016rlg}, which rules out the $(g-2)_\mu$ preferred parameter space for di-muon decays of vectors but not of scalars \cite{Batell:2016ove}.

Search for the leptonic force mediator in the prompt $K^+\to\mu^+\nu X$, $X\to e^+e^-$ decay chain is expected to have low background in the mass range $m_{X}>140$~MeV where the main background is the $K^+\to\mu^+\nu e^+e^-$ decay with the SM branching fraction of $8.5\times 10^{-8}$~\cite{Bijnens:1992en}. The search might be background-dominated below the $\pi^0$ mass, where a large contribution from the $K^+\to\pi^0\mu^+\nu$, $\pi^0_D\to\gamma e^+e^-$ decay chain is expected. A search involving a displaced $X\to e^+e^-$ vertices would be affected by this background to a much smaller degree. These considerations also apply to the searches for short-lived HNL (Section~\ref{sec:HNL}) and dark photon (Section~\ref{sec:darkphoton}) decays.


\subsection{The $K^+\! \to \ell_\alpha^+ \gamma\gamma+{\rm inv}$ decays}
\label{sec:Klgg+inv}

{\it This signature can be due to $K^+\! \to \ell_\alpha^+ \nu (a\to \gamma\gamma)$ decays, if the leptonic or neutrino force mediator, $a$, decays to photons (Sections~\ref{sec:muonicf} and \ref{sec:neutrinos}). The other possibility is $K^+\! \to \ell_\alpha^+ N$, where the HNL decays via the $N\to \gamma\gamma \nu$ channel. An example is the $N\to a \nu$, $a\to\gamma\gamma$ decay chain, see Section~\ref{sec:HNL} and also models in Section \ref{sec:2X}. The two photons may also be a signal of the $\pi^0$ decay in the $K^+\! \to \pi^0 \ell_\alpha^+ N$ chain. 
}

The $K^+\to\mu^+\nu X$, $X\to\gamma\gamma$ decay chain, involving both prompt and displaced $X$ decay vertices, allows for a search for the muonic force mediator (Section~\ref{sec:muonicf}). Reconstruction of a visible resonance improves the sensitivity with respect to the case of an invisible $X$. The expected single-event sensitivity (i.e. the sensitivity in the assumption of no background) provided by the NA62 Run~1 minimum bias dataset is ${\cal O}(10^{-9})$. However this search is blind in the region of the $\pi^0$ mass populated by the $K^+\to\pi^0\mu^+\nu$ background, and is likely to be limited by backgrounds also outside the $\pi^0$ mass. The expected background sources are non-gaussian mass tails of the semileptonic $K^+\to\pi^0\mu^+\nu$ decay, and $K^+\to\pi^+\pi^0(\pi^0)$ decays, possibly followed by pion decays in flight. 

Among the searches for HNL production (Section~\ref{sec:HNL}), the $K^+\to\pi^0\ell^+N$ transition is not competitive in general with the purely leptonic $K^+\to\ell^+N$ decays (Section~\ref{sec:sig:Kl}). The three-body decay is suppressed with respect to the two-body decay (with the exception of $m_N\lesssim 20$~MeV in the positron case), and offers a limited kinematically allowed HNL mass range. The NA62 sensitivity for the positron channel is further limited by the downscaling of the minimum bias trigger (in contrast to the $K^+\to e^+N$ analysis based on the main $K^+\to\pi^+\nu\bar\nu$ trigger chain) and the radiative $K^+\to \pi^0e^+\nu\gamma$ background. Nevertheless, the $K^+\to\pi^0 e^+N$ decay uniquely provides the sensitivity to $|U_{eN}|^2$ in the mass gap 135--144~MeV not currently covered by either the $K^+\to e^+N$ search~\cite{NA62:2020mcv} (due to the background conditions) or the $\pi^+\to e^+N$ search~\cite{PIENU:2017wbj} (due to the kinematic endpoint), as seen in Fig.~\ref{HNLLims}~(left). The estimated sensitivity to $|U_{eN}|^2$ achievable with the NA62 Run~1 dataset in this mass range is $2\times 10^{-6}$~\cite{Tastet:2020tzh}.


\subsection{Lepton flavor violating kaon decays}
\label{sec:exp:lfv}
{\it The decays are $K_L\to \mu^\pm e^\pm$ and $K^+\to \pi^+\mu^\pm e^\mp$, for instance from $s\to d a/Z'$ quark level transition, followed by the flavor violating decay of ALP (light $Z'$). The ALP can also be produced off-shell. The restrictions on such models are discussed in Section~\ref{sec:LPV}.}

NA62 has obtained an upper limit of ${\cal B}(K^+\to\pi^+\mu^-e^+)<6.6\times 10^{-11}$ at 90\% CL with the Run~1 di-lepton dataset, under the assumption of uniform phase space distribution~\cite{NA62:2021zxl}. This analysis has also led to the upper limit of ${\cal B}(\pi^0\to\mu^-e^+)<3.2\times 10^{-10}$ at 90\% CL, using the $K^+\to\pi^+\pi^0$ process as the source of tagged neutral pions. For the $K^+\to\pi^+\mu^+e^-$ decay, the strongest limit to date of $1.3\times 10^{-11}$ at 90\% CL comes from the BNL E777/865 experiments~\cite{Sher:2005sp}. The NA62 Run~1 sensitivity for this mode is comparable, and a new search with the Run~1 data is not justified. The sensitivity of NA62 and follow-up experiments (Section~\ref{sec:exp:KLEVER}) is expected to improve almost linearly with the size of the future datasets thereby reaching ${\cal O}(10^{-12})$, and searches for displaced $X\to\mu^\pm e^\mp$ vertices are possible at NA62 in principle. The LHCb experiment is projected to reach a similar ${\cal O}(10^{-12})$ sensitivity to the $K^+\to\pi^+\mu^\pm e^\mp$ decays in the long term~\cite{Borsato:2018tcz}.

The BNL E871 experiment has established an upper limit of ${\cal B}(K_L\to \mu^\pm e^\mp)<4.7\times 10^{-12}$ at 90\% CL~\cite{BNL:1998apv}. The KTeV experiment has established the following upper limits at 90\% CL: ${\cal B}(K_L\to\pi^0\mu^\pm e^\mp)<7.56\times 10^{-11}$, ${\cal B}(K_L\to\pi^0\pi^0\mu^\pm e^\mp)<1.64\times 10^{-10}$~\cite{KTeV:2007cvy}, and ${\cal B}(K_L\to e^\pm e^\pm \mu^\mp\mu^\mp)<4.1\times 10^{-11}$~\cite{KTeV:2002kut}. Further improvements on LFV decays of the $K_L$ are foreseen at the hybrid phase of the future CERN kaon program (Section~\ref{sec:exp:KLEVER}).


\subsection{Lepton number violating kaon decays}
\label{sec:exp:lnv}

{\it If the heavy neutral lepton is a Majorana fermion, then the quark level transition $u \bar s\to \ell_\alpha^+ (N^* \to d\bar u \ell_\beta^+) $ is possible, resulting in the decay $K^+\to \pi^-\ell_\alpha^+\ell_\beta^+$ (see Section~\ref{sec:HNL}).}

NA62 has reported the following upper limits at 90\% CL of the decay branching fractions using Run~1 di-lepton datasets: ${\cal B}(K^+\to\pi^-e^+e^+)<5.3\times 10^{-11}$ and ${\cal B}(K^+\to\pi^-\pi^0e^+e^+)<8.5\times 10^{-10}$~\cite{na62-lfv-2022}; ${\cal B}(K^+\to\pi^-\mu^+\mu^+)<4.2\times 10^{-11}$~\cite{NA62:2019eax}; ${\cal B}(K^+\to\pi^-\mu^+e^+)<4.2\times 10^{-11}$~\cite{NA62:2021zxl}. And independent indirect upper limit of $O(10^{-11})$ on the last process has been derived from the bound on muon conversion $\mu^- + (Z,A) \to e^+ + (Z-2,A)$~\cite{Littenberg:2000fg}. The above results are based on the full Run~1 dataset, with the exception of the $K^+\to\pi^-\mu^+\mu^+$ bound based on about 30\% of the dataset. The results are obtained under the assumption of uniform phase space distributions, which is customary for LNV decay searches. The searches are not currently limited by backgrounds, and the sensitivity is expected to improve in the future almost linearly with the size of the dataset, possibly with the exception of the $K^+\to\pi^-\mu^+\mu^+$ mode. One can expect NA62 and follow-up experiments at CERN (Section~\ref{sec:exp:KLEVER}) to reach the sensitivity of ${\cal O}(10^{-12})$ for the branching fractions of the above lepton number violating $K^+$ decays, as well as the $K^+\to\ell_\alpha^-\nu\ell_\beta^+\ell_\beta^+$ modes. The LHCb sensitivity to the $K^+\to\pi^-\ell^+\ell^+$ decays is expected to be similar in the long term~\cite{Borsato:2018tcz}.

The NA48/2 experiment at CERN has reported an analysis of the $K^\pm\to\pi^\pm\mu^\mp\mu^\mp$ final state including a scan in the $\pi^\pm\mu^\mp$ mass, corresponding to the emission and decay of a short-lived Majorana neutrino (Section~\ref{sec:HNL})~\cite{NA482:2016sfh}. The trigger logic allowed for prompt decay search only, leading to optimal acceptances for lifetimes below 20~ps. The search is background free, and the upper limits of ${\cal O}(10^{-10})$ are obtained on the product ${\cal B}(K^+\to\mu^+N)\cdot{\cal B}(N\to\pi^-\mu^+)$ assuming prompt decays. Considering the low background, NA62 Run 1 can improve on these results by up to an order of magnitude, and the search can be extended to lifetimes up to 10~ns without significant loss of acceptance using the displaced vertex approach.


\subsection{Rare $K_S$ decays}
\label{sec:exp:ks}
{\it Most of the rare $K_L$ decays have their $K_S$ counterparts, but with $\tau_{K_S}/\tau_{K_L}={\mathcal O}(10^{-3})$ smaller branching ratios, which means that significantly more $K_S$ decays are required for a similar reach. The signatures with potentially interesting phenomenological reach are $K_S\to \pi 2\ell$ that can arise for Higgs mixed scalar (Section \ref{sec:higgs}), promptly decaying ALP (Section~\ref{sec:ALP}) and dark photon (Section~\ref{sec:darkphoton}), as well as $K_S\to 2 \ell 2\gamma$ (see also Section~\ref{sec:KL2l:inv}) and $K_S\to 2\ell$ (see Section~\ref{sec:K0:4l}). The observable specific to  $K_S$ is the CP violating amplitude $A(K_S\to \mu^+\mu^-)_{\ell =0}$ which represents a clean probe for heavy new physics (Section~\ref{sec:KS:mumu}).}

The NA48/1 experiment at CERN has reported the first evidence for the rare decays $K_S\to\pi^0\ell^+\ell^-$ using the full dataset collected in 2002, measuring their branching fractions to be~\cite{NA481:2003cfm,NA481:2004nbc}
\begin{displaymath}
{\cal B}(K_S\to\pi^0e^+e^-) = 
\left(3.0^{+1.5}_{-1.2}\right)\times 10^{-9}, ~~~~
{\cal B}(K_S\to\pi^0\mu^+\mu^-) = \left(2.9^{+1.5}_{-1.2}\right)\times 10^{-9},
\end{displaymath}
in agreement with the SM expectations. In the former case, the measurement is performed in the accessible  kinematic range $m_{ee}>165~{\rm MeV}/c^2$. The LHCb experiment is expected to overtake the NA48/1 sensitivity for $K_S\to\pi^0\mu^+\mu^-$ channel with the future Run 3 dataset~\cite{Chobanova:2195218}.

The first observation of the rare $K_S\to\pi^0\gamma\gamma$ decay reported by the NA48 experiment based on the dataset collected in 2000 is also in agreement with the SM expectation~\cite{NA48:2003ydp}:
\begin{displaymath}
{\cal B}(K_S\to\pi^0\gamma\gamma) = (4.9\pm1.8)\times 10^{-8}.
\end{displaymath}

The LHCb experiment has recently reported an upper limit at 90\% CL on the $K_S\to\mu^+\mu^-$ branching fraction using the combined Run 1+2 dataset~\cite{LHCb:2020ycd}:
\begin{displaymath}
{\cal B}(K_S\to\mu^+\mu^-) < 2.1\times 10^{-10},
\end{displaymath}
to be compared to the SM prediction of ${\cal B}_{\rm SM}(K_S\to\mu^+\mu^-) = (5.18\pm1.50)\times 10^{-12}$~\cite{Ecker:218371, DAmbrosio:2017klp}. The ultimate expected LHCb sensitivity for this decay is close to the SM branching ratio, and is expected to be limited by the statistical uncertainty on the background subtraction and the signal yield~\cite{Alves:2018npj}.

Further discussion of rare $K_S$ decays can be found in Sections~\ref{sec:sig:Kpipiee} and \ref{sec:K0:4l}.


\subsection{Dark showers}
\label{sec:exp:darkjet}
{\it Exotic scenarios are discussed in Section~\ref{sec:LPV}.}

The most likely dark shower (Section~\ref{sec:dark:jets}) signatures in kaon decays are $K\to\pi j_{\ell}$ and $K\to\pi j_{\ell} j_{\ell }$, where $j_{\ell} $ denotes a lepton jet consisting of $e^+e^-$ pairs, photons and potentially missing energy. Depending on the mediator and final dark sector particle masses and lifetimes, the experimental signature can deform continuously from the prompt lepton jet signal in the case of short lifetimes all the way to just missing energy in case the lifetimes are sufficient for the dark sector particles to traverse the detector. A search for prompt lepton jets can be performed using the NA62 di-electron dataset (Section~\ref{sec:exp:NA62}). In case of pure missing energy signal, the results of the $K\to\pi X_{\rm inv}$ searches  (Section~\ref{sec:sig:Kpi}) can be reinterpreted in terms of constraints on the parameters of the more complex dark sector. The intermediate region, with lifetimes in the dark sector leading to displaced decays, leads to emergent lepton jet signatures, and would require dedicated analysis techniques to fully exploit the potential reach of the experiments. Explicit phenomenological models are essential for facilitation and interpretation of these searches.


\subsection{Inclusive signatures}
\label{sec:exp:inclusive}

For dark-sector scenarios leading to multiple signatures in kaon decays with comparable branching fractions, the sensitivity can be improved by combining searches in several channels. An example is a possible search for ALP production and prompt decay (Section~\ref{sec:ALPdilepton}) in the $K^+\to\pi^+e^+e^-$ and $K^+\to\pi^+\mu^+\mu^-$ modes at NA62, both collected via dedicated trigger lines (Section~\ref{sec:sig:Kpill}).
One could obtain the bounds on ${\mathcal B}(K^+\to\pi^+ a)$ as functions of ALP mass $m_a$, ALP lifetime $\tau_a$, and branching ratios ${\mathcal B}(a\to e^+e^-)$, ${\mathcal B}(a\to \mu^+\mu^-)$ from a combined analysis. When both $a\to e^+e^-$ and $a\to \mu^+\mu^-$ channels are have non-zero rates, the limits on the ALP production branching ratio would be stronger from a combined analysis than from the individual analyses.

\newpage

\section{
Complementary probes in rare hyperon decays}
\label{sec:hyppheno}
{\em Authors: Dettori, Fornal, Geng,  Kupsc, \underline{Martin Camalich}, Martinez Santos, Shi, Wang}

Hyperons are the strange baryonic siblings of the nucleons and can lead to strangeness-changing decay signatures complementary to those of the kaons. Their lifetimes are $\mathcal O(10^{-10})$~s, significantly shorter than the $K^+$ and $K_L$ lifetimes $\mathcal O(10^{-8})$~s~\cite{ParticleDataGroup:2020ssz}. If kaonic and hyperonic amplitudes for the $s\to d X$ transition are of similar size ($X$ are ``dark sector'' particles),  we can expect the rare hyperon-decay to be about two orders of magnitude less sensitive for a given dataset, due to the shorter lifetimes. On the other hand, hyperons do carry nonzero spin and baryon number, and are as such  sensitive to the new physics interactions that are difficult or impossible to search for using kaons~\cite{Chang:2014iba}. Furthermore, non-negligible abundances of hyperons are expected in some stellar systems, such as neutron stars. The existence of $s\to d X$ transitions could then lead to modified stellar cooling dynamics, the absence of which places stringent astrophysical bounds on these interactions~\cite{MartinCamalich:2020dfe,Camalich:2020wac}. Experimentally, hyperons produced in $e^+e^-$ colliders, such as BESIII, are polarized~\cite{Ablikim:2018zay}, providing an additional handle to suppress the SM backgrounds for the dark-sector searches in hyperon decays. 

In this section we investigate the most prominent hyperon-decay signatures: the neutrino modes $B_1\to B_2 \nu\bar\nu$, where $B_1$ ($B_2$) is the initial (final) baryon, induced by the SM (or heavy BSM) $s\to d\nu\bar\nu$ transition; and the two-body decays of the type $B_1\to B_2 X$, where $X$ is an axion/ALP or a massless dark photon. The corresponding models have been discussed in Sections~\ref{subsec:heavyNP}, \ref{subsec:QCDaxion:ALP} and \ref{subsec:darkphoton}. The baryon-number-violating signatures like $B_1\to \pi+\textrm{inv}$, $B_1\to \gamma+\textrm{inv}$ and $B_1\to\textrm{inv}$ are discussed in detail in Section~\ref{subsec:dark-baryon} and in this section we discuss other such signatures that are accessible at the LHCb. In this report we focus on the transitions between baryons in the lowest-lying $1/2^+$ baryon octet. There may be other decays of phenomenological interest, for instance the decays of the spin-3/2 decuplet baryon $\Omega^-(sss)$, which however, we do not review here.


\subsection{Baryonic matrix elements}

Accurate predictions for most of the hyperon decay rates of interest can be made through the use of flavor $SU(3)$ symmetry, supplemented with data and lattice QCD computations of baryonic (mostly nucleonic) form factors, along with the calculations of the $SU(3)$-breaking corrections using chiral perturbation theory (ChPT).  
The $SU(3)$-breaking expansion is particularly useful because it allows for a systematic expansion of the rates and reduces the number of nonperturbative inputs needed for given desired accuracy~\cite{Chang:2014iba}. For $B_1\to B_2$ decays the phase space factor is given by the mass difference of the two hyperons, $\delta=(M_1-M_2)/M_1$, which can be taken as the relevant $SU(3)$-breaking parameter.

The baryonic matrix elements required for the $ B_1\to  B_2$ decays involving dark sectors are 
\begin{align}
\label{eq:baryons:MDM}
\langle B_2(p^\prime) | &\bar{d} \sigma^{\mu\nu} s | B_1 (p)\rangle 
=g_T\bar u_2(p^\prime)\sigma^{\mu\nu}u_1(p),
\\
\label{eq:baryons:vector}
\langle   B_2(p^\prime) | &\bar{d} \gamma_\mu s |  B_1 (p)\rangle
=
\bar{u}_2 (p^\prime)  \Big[
f_1(q^2)  \,  \gamma_\mu  
+i \frac{f_2(q^2)}{m_{  B_1}}   \, \sigma_{\mu \nu}   q^\nu  
+ \frac{f_3(q^2)}{m_{  B_1}}   \,  q_\mu  
\Big]  
 u_1 (p), 
 \\ 
 \label{eq:baryons:axial:vector}
\langle   B_2(p^\prime) | &\bar{d} \gamma_\mu \gamma_5 s |  B_1 (p)\rangle
=
\bar{u}_2(p^\prime)  \Big[
g_1(q^2)    \gamma_\mu
+i \frac{g_{2} (q^2)}{m_{  B_1}}   \sigma_{\mu \nu}   q^\nu  
+ 
\frac{g_{3} (q^2)}{m_{  B_1}}   q_\mu  
\Big]  \,\gamma_5  u_1(p),
\end{align}
where $q=p-p'$. The matrix element in the first line \eqref{eq:baryons:MDM} is needed for the  magnetic dipole moment (MDM)   induced production of the massless dark photon.\footnote{The electric dipole moment (EDM) matrix element is obtained via the relation $\sigma^{\mu\nu}\gamma_5=i/2\varepsilon^{\mu\nu\rho\sigma}\sigma_{\rho\sigma}$.} The relevant form factor is therefore evaluated at $q^2=0$, and equals the parameter  $g_T$, i.e., the tensor coupling of the transition. The  vector and axial-vector form factors in \eqref{eq:baryons:vector} and \eqref{eq:baryons:axial:vector} can be related to the form factors entering the semileptonic hyperon decay amplitudes, i.e., for the $B_1\to B_2\ell^-\bar\nu$ transition~\cite{Cabibbo:2003cu}. As we show explicitly below, only the vector and axial vector charges, $f_1(0)$ and $g_1(0)$, are needed for our predictions.\footnote{The form factors can be expanded in powers of $q^2/m_X^2\sim\delta^2$, where $m_X\sim$1 GeV is the hadronic scale related to the mass of the resonance coupling to a given current~\cite{Ecker:1989yg}.} Table \ref{tab:baryonFFs} collects the values for the hadronic parameters used in our numerical analyses. The values of $f_1=f_1(0)$ and $g_1=g_1(0)$ are accurate up to $\mathcal O(\delta^2)$ corrections~\cite{Cabibbo:2003cu,Ledwig:2014rfa,Geng:2021fog}, while the values for $g_T$ are obtained using $SU(3)$ symmetry from the tensor charges of the nucleons that were obtained in Ref.~\cite{Gupta:2018lvp} using lattice QCD. As the momenta carried by the final particles is parametrically also ${\mathcal O}(\delta)$, we perform the $SU(3)$-breaking expansion directly at the level of the decay rates and observables.

\begin{table}[t]
\renewcommand{\arraystretch}{1.5}
  \setlength{\arrayrulewidth}{.25mm}
\centering
\small
\setlength{\tabcolsep}{0.5 em}
\begin{tabular}{  c  c  c  c  c  c  }
    \hline
    \hline
&$\Lambda\to n$ & $\Sigma^+ \to p$ & $\Xi^-\to \Sigma^-$ & $\Xi^0\to \Sigma^0$ & $\Xi^0\to\Lambda$\\
\hline
$f_1$ &$-1.22(6)$ & $-1.00(5)$ & $1.00(5)$ &$-0.71(4)$&$1.22(6)$\\
$g_1$ &$-0.88(2)$&$0.33(2)$&$1.22(4)$&$-0.86(3)$&$0.21(4)$ \\
$g_T$ &$-0.73(12)$&$0.20(4)$&$0.99(10)$&$-0.70(7)$&$0.24(4)$ \\
    \hline\hline
\end{tabular}
\caption{Hyperon decay form factor values considered in this work. The predictions for the vector charges, $f_1$, are obtained using flavor $SU(3)$ symmetry, and are protected from $\mathcal O(\delta)$ corrections by the Ademollo-Gatto theorem~\cite{Cabibbo:2003cu}. The predictions for the axial charges include the $\mathcal O(\delta)$ corrections calculated in chiral perturbation theory~\cite{Ledwig:2014rfa,Geng:2021fog}. The tensor charges are the $SU(3)$-symmetric predictions that follow from the LQCD determination of  tensor charges for nucleons~\cite{Gupta:2018lvp}. 
\label{tab:baryonFFs}}
\end{table}

\subsection{$B_1\to B_2+\textrm{inv}$ signatures}
\label{sec:B1toB2inv}

Searches for light new physics in the hyperon $B_1\to B_2+\textrm{inv}$ decays are complementary to the $K\to\pi(\pi)+\textrm{inv}$ decays, discussed in Sections~\ref{sec:sig:Kpi} and~\ref{sec:sig:K2pi}. The SM background from $B_1\to B_2\pi^0$ decays, with two missing photons in $\pi^0\to\gamma\gamma$ decay, reduces the sensitivity to new physics in the kinematic region $p_{\rm miss}^2\sim m_{\pi^0}^2$, relevant, e.g., for the massive ALP searches. This is analogous to the $\pi^0$ blind spot of the $K\to\pi+\textrm{inv}$ searches. The $p_{\rm miss}^2\simeq0$ region relevant for the QCD axion and massless dark photon searches is affected by the SM background from weak radiative $B_1\to B_2\gamma$ decays, with an undetected photon.

The SM backgrounds can be reduced, if hyperons are produced with a certain degree of polarization, $\vec P$. The differential decay rate for the SM decay $B_1\to B_2\pi^0$,  as a function of the angle $\theta$ between the polarization vector $\vec P$ and the recoil direction of the daughter baryon $B_2$, is given by 
\begin{align}
\frac{d\Gamma}{d\cos\theta}=\frac{\Gamma}{2}\left(1+P\alpha_\pi\cos\theta\right), 
\end{align}
and similarly for the $B_1\to B_2\gamma$ decay, with $\alpha_\pi\to \alpha_\gamma$. 
The parameters $\alpha_\pi$ and $\alpha_\gamma$  describe the interference between the parity conserving and parity violating amplitudes. They are decay channel dependent but well known experimentally~\cite{ParticleDataGroup:2020ssz}. In general, the angular dependence of the SM background decays will differ from the decays involving dark-sector particles.  The differential distributions can thus be used to discriminate signal from background. Finally, we note in passing that the $\Xi^{0,-}\to\Sigma^{0,-} \pi^0$ decays are kinematically not allowed. This means that there is no corresponding SM background in the ALP searches using the $\Xi^{0,-}\to\Sigma^{0,-} +$invisible decays. 


\subsubsection{Heavy new physics and the neutrino decay modes}

In the presence of heavy NP the dimension-6 weak Hamiltonian for the $s\to d\bar\nu\nu$ transition is given by (in the notation of Eq.~(\ref{eq:H:kpnn:sm}))
\begin{align}
\label{eq:H:hypenunu}
\mathcal H_{\rm eff}=\frac{4\alpha G_F}{\sqrt{2}}\lambda_t\sum_{\ell=e,\mu,\tau}\left(C_{\nu_\ell}^L(\bar d_L\gamma^\mu s_L)(\bar\nu_\ell\gamma_\mu\nu_\ell)+C_{\nu_\ell}^R(\bar d_R\gamma^\mu s_R)(\bar\nu_\ell\gamma_\mu\nu_\ell)\right)+{\rm h.c.}
\end{align}
This generalizes the weak Hamiltonian in Eq.~\eqref{eq:H:kpnn:sm} by allowing for the possibility of NP induced right-handed currents, $C_{\nu_\ell}^R\ne 0$. In the SM the Wilson coefficients are given by
\begin{eqnarray}
\label{eq:stodnunuWC}
C_{\nu_\ell}^L=\frac{1}{2\pi\sin^2\theta_w}\left(\lambda_c X_\ell/\lambda_t+X_t\right), \qquad C_{\nu_\ell}^R=0.
\end{eqnarray}

The decay width for the $B_1\to B_2\bar\nu\nu$ decays,  expanded up to NLO in the $SU(3)$-breaking parameter $\delta$, is~\cite{Geng:2021fog}
\beq
\begin{split}
\Gamma(B_1\to B_2\bar \nu\nu)=\sum_{\ell=e,\mu,\tau}\frac{\alpha^2G_F^2|\lambda_t|^2f_1^2\Delta^5}{60\pi^3}
&\Big[\left(1-\tfrac{3}{2}\delta\right)\left|C_{\nu_{\ell}}^L+C_{\nu_{\ell}}^R\right|^2
\\
&+3\left(1-\tfrac{3}{2}\delta\right)\frac{g_1^2}{f_1^2}
\left|C_{\nu_{\ell}}^L-C_{\nu_{\ell}}^R\right|^2+{\cal O}\big(\delta^2\big)\Big],
\label{eq:B1toB2nunuGamma}
\end{split}  
\eeq
where $\Delta=M_1-M_2$. Note that the hyperon decays are sensitive to both the vector, $C_{\nu_{\ell}}^R-C_{\nu_{\ell}}^L$, and the axial-vector, $C_{\nu_{\ell}}^R+C_{\nu_{\ell}}^L$, quark currents.

\begin{table}[t]
\renewcommand{\arraystretch}{1.5}
  \setlength{\arrayrulewidth}{.25mm}
\centering
\small
\setlength{\tabcolsep}{0.5 em}
\begin{tabular}{ l   c  c  c  c  c  }
    \hline
    \hline
Predictions/Projections
&$\Lambda \to n \bar \nu \nu$ 
& $\Sigma^+\to  p \bar \nu \nu$
& $\Xi^-\to \Sigma^- \bar \nu \nu$ 
& $\Xi^0\to \Sigma^0 \bar \nu \nu$
& $\Xi^0\to \Lambda \bar \nu \nu$
\\
\hline 
$10^{13}\times{\cal B}(B_1\to B_2\nu\bar\nu)^{\rm SM}$~\cite{Geng:2021fog}& $6.26(16)$ &    $3.49(16)$ &
 $1.10(1)$ &
 $0.89(1)$&
 $5.52(13)$\\
\hline
$10^{6}\times{\cal B}(B_1\to B_2\nu\bar\nu)^{\rm BESIII}$~\cite{Li:2016tlt} & $<0.3$ & $<0.4$ &
-- &
$<0.9$ & $<0.8$ \\
$10^3\times|C_{\nu_\ell}^L+C_{\nu_\ell}^R|^{\rm BESIII}$&
$<1.6$
& $<1.7$ 
& --
& $<10$ 
& $<1.8$\\ 
\hline
\hline
    
\end{tabular}
\caption{SM predictions for $B_1\to B_2\nu\bar\nu$ branching ratios up to NLO in the $SU(3)$ breaking  parameter $\delta$ (uncertainty estimates are only due to the missing $\mathcal{O}(\delta^2)$ terms), projected bounds on the branching ratios achievable by BESIII~\cite{Li:2016tlt}, and the corresponding expected limits on the combination of the Wilson coefficients $|C_{\nu_\ell}^L+C_{\nu_\ell}^R|$.   
\label{tab:hypnunu}}
\end{table}

In Table~\ref{tab:hypnunu} we show the SM predictions for the branching ratios ${\cal B}(B_1\to B_2\nu\bar\nu)$ for several hyperon decays (with $B_1 B_2=\{\Lambda n, \Sigma^+ p,\Xi^-\Sigma^-, \Xi^0\Sigma^0, \Xi^0\Lambda\}$), the current sensitivity projections from BESIII~\cite{Li:2016tlt}, and the corresponding expected limits on the quark-axial combination $|C_{\nu_\ell}^L+C_{\nu_\ell}^R|$ of the Wilson coefficients, assuming that only this combination is nonzero. A detailed discussion of the implications of these projections can be found in Ref.~\cite{Geng:2021fog} (see also \cite{Tandean:2019tkm,Li:2019cbk,Su:2019tjn}).  

\subsubsection{Two-body decays $B_1\to B_2X$}
\label{sec:baryon_twobody}

We consider two types of new physics models that lead to the two body hyperon decays, $B_1\to B_2 X$, where $X$ is a light ALP or a massless dark photon.
The ALP with flavor violating couplings to SM quarks, Eq.~\eqref{eq:ALP-f}, results in a quark level transition $s\to d a$. This induces the hyperon decay, $B_1\to B_2 a$, with the partial decay width
\begin{align}
\label{eq:width_ALP}
\Gamma(B_1\to B_2 a)=\frac{\bar p\,\bar\omega^2}{2\pi}\left(\frac{f_1^2}{|F^V_{ds}|^2}+\left(1-\frac{x_a^2}{\delta^2}\right)\frac{g_1^2}{|F^A_{ds}|^2}\right)+\mathcal O (\delta^2), 
\end{align}
where  $x_a=m_a/M_1$, while $\bar\omega$ ($\bar p$) is the energy (magnitude of the 3-momentum) of the ALP in the parent-hyperon rest frame. The ALP coupling constants were absorbed into the effective axion decay constant, $F_{ds}^{V,A}=2f_a/C_{ds}^{V,A}$, see also Eq.~\eqref{eq:ALP-f}. 

The decay into massless dark photon are induced by the MDM- and EDM-type flavor violating interactions given by dimension-5 operators in Eq.~\eqref{dipole}. The corresponding decay rate is  
\begin{align}
\Gamma(B_1\to B_2 A')=\frac{4g_T^2\bar\omega^3}{\pi\Lambda^2}\left(|\mathbb{D}_M^{sd}|^2+|\mathbb{D}_E^{sd}|^2\right).
\end{align}

In Table \ref{tab:hyp2D} we show numerical expressions for the branching fractions ${\cal B}(B_1\to B_2X)$ as functions of the relevant NP scales, assuming the couplings to be equal to unity.

\begin{table}[t]
\renewcommand{\arraystretch}{3.}
  \setlength{\arrayrulewidth}{.25mm}
\centering
\small
\setlength{\tabcolsep}{0.5 em}
\begin{tabular}{ c   c  c  c  }
    \hline
    \hline
    & $B_1\to B_2 a$ $(m_a=0)$ & $B_1\to B_2 a$ $(m_a\simeq m_{\pi^0})$ &  $B_1\to B_2 A'$\\
    \hline
$\Lambda n$ & $0.40\left(\frac{10^6~\text{GeV}}{|F_{ds}^V|}\right)^2+0.21\left(\frac{10^6~\text{GeV}}{|F_{ds}^A|}\right)^2$ &$0.28\left(\frac{10^6~\text{GeV}}{|F_{ds}^V|}\right)^2+0.06\left(\frac{10^6~\text{GeV}}{|F_{ds}^A|}\right)^2$  &$1.15\left(\frac{10^6~\text{GeV}}{|\Lambda|}\right)^2$\\  
$\Sigma^+ p$ & $0.22\left(\frac{10^6~\text{GeV}}{|F_{ds}^V|}\right)^2+0.02\left(\frac{10^6~\text{GeV}}{|F_{ds}^A|}\right)^2$ &$0.19\left(\frac{10^6~\text{GeV}}{|F_{ds}^V|}\right)^2+0.02\left(\frac{10^6~\text{GeV}}{|F_{ds}^A|}\right)^2$  &$0.07\left(\frac{10^6~\text{GeV}}{|\Lambda|}\right)^2$ \\
$\Xi^-\Sigma^-$ &$0.07\left(\frac{10^6~\text{GeV}}{|F_{ds}^V|}\right)^2+0.10\left(\frac{10^6~\text{GeV}}{|F_{ds}^A|}\right)^2$&$0.02\left(\frac{10^6~\text{GeV}}{|F_{ds}^V|}\right)^2+0.002\left(\frac{10^6~\text{GeV}}{|F_{ds}^A|}\right)^2$&$0.52\left(\frac{10^6~\text{GeV}}{|\Lambda|}\right)^2$\\
$\Xi^0\Sigma^0$ &$0.06\left(\frac{10^6~\text{GeV}}{|F_{ds}^V|}\right)^2+0.08\left(\frac{10^6~\text{GeV}}{|F_{ds}^A|}\right)^2$&$0.01\left(\frac{10^6~\text{GeV}}{|F_{ds}^V|}\right)^2+0.0006\left(\frac{10^6~\text{GeV}}{|F_{ds}^A|}\right)^2$&$0.44\left(\frac{10^6~\text{GeV}}{|\Lambda|}\right)^2$\\
$\Xi^0\Lambda$ &$0.66\left(\frac{10^6~\text{GeV}}{|F_{ds}^V|}\right)^2+0.03\left(\frac{10^6~\text{GeV}}{|F_{ds}^A|}\right)^2$&$0.53\left(\frac{10^6~\text{GeV}}{|F_{ds}^V|}\right)^2+0.01\left(\frac{10^6~\text{GeV}}{|F_{ds}^A|}\right)^2$ &$0.20\left(\frac{10^6~\text{GeV}}{|\Lambda|}\right)^2$\\
\hline
\hline
\end{tabular}
\caption{Numerical expressions for the branching fractions ${\cal B}(B_1\to B_2X)$, where $X$ is an axion/ALP or a massless dark photon, as functions of the relevant NP scales, assuming the couplings to be equal to unity. For the ALP case we take $m_a=135$~MeV for $\Lambda n$, $\Sigma^+p$ and $\Xi^0\Lambda$ transitions and $m_a=120$~MeV for $\Xi^-\Sigma^-$ and $\Xi^0\Sigma^0$ transitions.
\label{tab:hyp2D}}
\end{table}

\subsection{Experimental prospects}
\subsubsection{Status and prospects at BESIII and SCTFs}
At the electron-positron colliders a hyperon-antihyperon pair is produced from $J/\psi$ decays with $\mathcal O(10^{-3})$ branching fractions, meaning that with $1.0 \times 10^{6}$ $J/\psi$ events collected 
by BESIII~\cite{BESIII:2021cxx}, considering the typical detection efficiency, $\mathcal O(10^6)$ hyperon-antihyperon pairs could be well reconstructed.
This allows to tag hyperon (antihyperon) production in the process $e^+e^-\to J/\psi\to B_1\bar B_1$ by using common decay mode of $B_1(\bar B_)$ and to investigate absolute branching ratio of rare decays modes of the $\bar B_1(B_1)$. First measurements and searches using this method have been performed by BESIII~\cite{BESIII:2020iwk, BESIII:2021ynj, BESIII:2021slv, BESIII:2021emv}. 

These results were obtained using collisions of unpolarized electron and positron beams at the center-of-mass energy corresponding to the $J/\psi$ resonance. The baryon-antibaryon pair production mechanism at electron-positron colliders is symmetric and allows for simultaneous determination of the decays properties of (anti)hyperons. Two analysis methods are possible: the exclusive (double tag) method with the decay chains of both baryon and antibaryon fully reconstructed, and the inclusive (single tag) method with only the decay chain of either the baryon or the antibaryon reconstructed. 

The inclusive method can be readily used for studies of the hyperon decay modes with invisible signatures, such as the channels listed in Section~\ref{sec:B1toB2inv}. The BESIII projections for $B_1\to B_2\bar\nu\nu$ decays are listed in Table \ref{tab:hypnunu}. These simplified projections from Ref.~\cite{Li:2016tlt} may be improved upon using several  published results and some preliminary simulations. For example, in the search for the lepton-number-violating $\Sigma^-$ decays with $1.3 \times 10^{9}$ $J/\psi$ events~\cite{BESIII:2020iwk}, the actual obtained experimental upper limits are  $\mathcal{B}(\Sigma^{-} \to p e^{-} e^{-}) < 6.7 \times 10^{-5}$
and $\mathcal{B}(\Sigma^{-} \to \Sigma^{+} X_{\rm inv}) < 1.2 \times 10^{-4}$, respectively. Given that this analysis is almost background-free, one would expect from scaling naively with the statistics the bounds  $\mathcal{B}(\Sigma^{-} \to p e^{-} e^{-}) < 8.7 \times 10^{-6}$ and $\mathcal{B}(\Sigma^{-} \to \Sigma^{+} X_{\rm inv}) < 1.6 \times 10^{-5}$  from the $10^{10}$ $J/\psi$ sample. These bounds are 14.5 and 16 times less stringent than the simplified estimates of Ref.~\cite{Li:2016tlt}. The preliminary simulations of the decay channels  $\Sigma^+ \to p + {\rm inv}$ and $\Xi^0 \to \Lambda + {\rm inv}$ also shows similar scale factors with respect to the estimations obtained in Ref.~\cite{Li:2016tlt}. Full invisible decays of hyperons are also within the reach with BESIII. For example, $\Lambda$ invisible decay is searched for the first time, and its branching ratio is found to be less than $7.4 \times 10^{-5}$ at $90\%$ CL with the $J/\psi$ sample~\cite{BESIII:2021slv}. 

Realistic projections for SCTF can be obtained by using BESIII results and scaling to the expected at least two orders of magnitude data sets.

\subsubsection{Status and prospects at LHCb}
\label{sec:LHCb:hyperons:prospect}

The LHCb experiment has reported the first evidence for the $\Sigma^+\to p\mu^+\mu^-$ decay with Run 1 dataset,  measuring the branching fraction to be~\cite{LHCb:2017rdd}
\begin{equation}
{\cal B}(\Sigma^+\to p\mu^+\mu^-) = \left(2.2_{-1.3}^{+1.8}\right)\times 10^{-8}.
\end{equation}
This result is inconsistent with the HyperCP anomaly~\cite{HyperCP:2005mvo} and excludes the central HyperCP branching fraction value, while being consistent with the SM expectation. A yield of ${\cal O}(10^3)$ candidates per year is envisaged in the LHCb upgrade, thanks to a more efficient trigger chain~\cite{Alves:2018npj}. This would lead to a sensitivity to the decay rate at a few percent level, and allow for a differential and forward-backward asymmetry measurements. In addition, CP violation can be studied by measuring the  difference between the $\Sigma^+\to p\mu^+\mu^-$ and $\Sigma^-\to\bar p\mu^+\mu^-$ rates, 
and a full angular analysis probing the Lorentz structure in the $s\to d\mu\mu $ currents is foreseen.

The $\Sigma^+\to p e^+e^-$ decay is expected to be accessible by LHCb, leading to a lepton flavour universality test with respect to the muonic channel. Decays violating lepton and baryon number or lepton flavour conservation, such as $\Sigma^+\to\bar p \mu^+ \mu^+$ and $\Sigma^+\to p \mu^\pm e^\mp$ will be possibly probed down to branching fractions of $10^{-9}$ or below, depending on the background level. 

The LHCb experiment also has the potential for producing leading measurements of semimuonic hyperon
decays~\cite{Alves:2018npj}.
The collaboration is currently
analysing existing data for ${\cal B}(\Lambda^0\to p\mu^-\bar\nu)$ measurement~\cite{Brea_Rodr_guez_2020}. The experimental sensitivity of hyperon decays to dark baryons predicted by $B$-mesogenesis (Section \ref{sub:dark:baryon:models}) has been evaluated for this paper, using the same fast simulation and methodology as done for $B$ decays
in Ref.~\cite{Rodriguez:2021urv}. We conclude that the expected sensitivities are at the $10^{-6}$ level for ${\cal B}(\Xi^0\to\pi^+\pi^- X)$, and ${\mathcal O}(10^{-10})$ for ${\cal B}(\Xi^-\to\mu^+\mu^-\pi^-X)$, where $X$ is a dark sector particle escaping detection. The latter process is expected to have a very small background, and it is possible that the search will be statistically limited. The former decay on the other hand could well be systematically limited, in case a more detailed analysis reveals that the background is difficult to model accurately.

\newpage

\section{Flagship measurements}
\label{sec:flagship}

The following searches for light dark sectors in kaon and hyperon decays would bring major improvements.
Many of these searches can be performed with existing datasets.
\begin{itemize}
\item Searches for the $K^+\to\pi^+X_{\rm inv}$ decay (performed with NA62 Run~1 dataset, Section~\ref{sec:sig:Kpi}) should be extended into the mass range $m_X > 260$~MeV, and pursued with future datasets. Also desirable would be the improvement of signal acceptance for $m_X=0$, which is currently limited at NA62 by the definition of the $K^+\to\pi^+\nu\bar\nu$ signal and control regions. These measurements represent a unique probe into the Higgs mixed dark scalar (Section~\ref{sec:higgs}) and ALP (Section~\ref{sec:ALP}) phase space, and other models in which $X$ particle does not decay preferentially to visible states. In particular, for the Higgs mixed scalar an improvement on ${\mathcal B}(K^+\to \pi^+ X_{\rm inv})$ by two orders of magnitude would close the gap to the BBN excluded region up to $m_{\varphi}=200$ MeV (Fig.~\ref{fig:higgs_portal_summary}). A similar improvement would also close the low mass range of the ALP parameter space, with the caveat that a comprehensive study of the BBN bounds for different ALP couplings is still missing. Searches for the $K_L\to\pi^0 X_{\rm inv}$ decay (performed by KOTO, Section~\ref{sec:sig:Kpi}) should be pursued with future datasets: they are complementary to the $K^+\to\pi^+X_{\rm inv}$ for $m_X$ in the vicinity of the $\pi^0$ mass.
\item Searches for resonances in the $m_{\ell\ell}$ spectra of $K^+\to\pi^+\ell^+\ell^-$ decays (Section~\ref{sec:sig:Kpill}), and the $m_{\gamma\gamma}$ spectrum of the $K^+\to\pi^+\gamma\gamma$ decays (Section~\ref{sec:sig:Kpigg}), including scans in the lifetime of the intermediate particle, should be pursued with the world's largest samples of these decays collected by NA62 to advance the study of the ALP phase space. Searches at future experiments would benefit from the development of dedicated trigger lines. Studies of the $K_L\to\pi^0\gamma\gamma$ decays (Section~\ref{sec:sig:Kpigg}) should be pursued with the KOTO dataset and future $K_L$ datasets. Improvement in sensitivity to resonance production by two orders of magnitude would close the gap to the constraints from beam dump searches for a significant ALP mass range (Figs.~\ref{fig:ALPvis}, \ref{fig:ALPtoeetarget}), thus making $K\to\pi X_{\rm inv}$ the only phenomenologically viable ALP signature.
\item
Improvement in sensitivity to the two-body $K^+\to\ell^+X_{\rm inv}$ branching ratios by two orders of magnitude would start probing the seesaw neutrino mass models with ${\cal O}(100~{\rm MeV})$ sterile neutrinos (Fig.~\ref{HNLLims}).
Improvement in sensitivity to the three-body $K^+\to\mu^+\nu X_{\rm inv}$ branching ratio by an order of magnitude would probe fully the preferred region for self-interacting neutrinos that may alleviate the Hubble tension (Fig.~\ref{fig:Hubble:tension:neutrinos}). Progress can be achieved with the future NA62 datasets (Section~\ref{sec:sig:Kl}).
\item First direct searches for the leptonic force mediator (Section~\ref{sec:muonicf}) in the $K^+\to\mu^+\nu X_{\rm inv}$ decays, with $X\to\mu^+\mu^-$ and $X\to\gamma\gamma$ should be performed with the available NA62 dataset (Sections~\ref{sec:3l:inv}, \ref{sec:Klgg+inv}). These channels can probe a region of parameter space where these mediators can explain the muon $g-2$ anomaly. Searches at future experiments would benefit from the development of dedicated trigger lines. 
\item Searches for emission of multiple dark sector mediators in kaon decays such as the $K^+\to\pi^+2(e^+e^-)$ should be performed with the NA62 dataset (Section~\ref{sec:sig:5track}) as they would be the first probes of a simple Higgsed $U(1)_d$ model for a dark photon in a particular region of parameter space (Section~\ref{sec:2X}).
\item 
The $K\to\pi\pi X_{\rm inv}$ (Section~\ref{sec:sig:K2pi}) and $B_1\to B_2 X_{\rm inv}$ (Section~\ref{sec:baryon_twobody}) decays of kaons and hyperons probe different couplings to those involved in the $K\to\pi X_{\rm inv}$ transitions, such as the CP violating couplings in the Higgs-mixed scalar model (Section~\ref{sec:higgs}) and the axial couplings of the ALP (Sections~\ref{sec:ALP} and \ref{sec:hyppheno}). The missing energy signature due to the ALP escaping the detector can also be replaced by the signatures due to a promptly decaying ALP, $a\to \gamma\gamma$ or $a\to \ell^+\ell^-$. Kaon measurements should be pursued with the available NA62 and KOTO datasets, and hyperon measurements with the BESIII and LHCb datasets.
\end{itemize}


\section{Conclusions}
\label{sec:conclusions}

\begin{sidewaystable}
  \centering
  \resizebox{\textwidth}{!}{
\begin{tabular}{lccccccccccc}
\hline\hline
 \multirow{2}{*}{Decay \textbackslash \ Model}                 & \ref{sec:higgs} Higgs    &  \multirow{2}{*}{\ref{subsec:QCDaxion:ALP}  ALP}             & \ref{sec:HNL} Heavy                        & \ref{sec:darkphoton} Dark                & \ref{sec:muonicf} Leptonic  	& \ref{sec:neutrinos} Strongly  & \ref{sec:GNVgamma} GN  & \ref{sec:2X} Two dark       & \ref{sub:dark:baryon:models} Dark & \ref{sec:LPV} More  & \ref{subsec:heavyNP} Heavy   
                   \\
                    					     & portal 	  &                					&  Neutral Lepton                        & Photon                & Force $(X)$  	&  Int. Neutrino & Violation  & sector particles      & Baryons & exotic &  New Physics  
                   \\
                   \hline       
\ref{sec:sig:Kpi} $K\to\pi$+inv       &  \checkmark                & \checkmark       &                  $-$                               &               \checkmark                 &          $-$              &         \checkmark                          &  \checkmark                 &    \checkmark                              &        $-$         &     $-$             &  \checkmark                        \\
\ref{sec:sig:K2pi} $K\to\pi\pi$+inv       &     CP viol.             & axial coupl.        &                      $-$                           & \checkmark even massless                     &           $-$             &                 $-$                  &        $-$           &          $-$                        &         $-$        &       $-$           &            $-$              \\
 \multirow{2}{*}{\ref{sec:Kpiginv} $K\to\pi\gamma$+inv}       &      possible in           &       possible  in             &          \multirow{2}{*}{$-$ }                                        & \multirow{2}{*}{\checkmark even     massless}              &       \multirow{2}{*}{$-$ }                  &          \multirow{2}{*}{$-$ }                        &       \multirow{2}{*}{$-$ }            &           \multirow{2}{*}{$-$ }                        &     \multirow{2}{*}{$-$ }             &     \multirow{2}{*}{$-$ }             &       \multirow{2}{*}{$-$ }                   \\
										         &      extensions           &       extensions                &        				                                          &                &                       &                                 &                  &                                  &                 &                  &                          \\
\ref{sec:Kpipiginv} $K\to2 \pi\gamma$+inv       &     $-$             &        $-$               &            $-$                                     & $\pi^0\to \gamma A'$                &        $-$                &           $-$                        &         $-$          &            $-$                      &      $-$           &    possible              &      $-$                    \\
 \multirow{2}{*}{\ref{sec:sig:Kpigg} $K\to \pi \gamma\gamma$}            & negligible   & \multirow{2}{*}{\checkmark prompt }               &             \multirow{2}{*}{$-$ }                                     &         \multirow{2}{*}{$-$ }                         &           \multirow{2}{*}{$-$ }               &            \multirow{2}{*}{$-$ }                         & lifetime  &     \multirow{2}{*}{$-$ }                               &          \multirow{2}{*}{$-$ }         &         \multirow{2}{*}{$-$ }           &              \multirow{2}{*}{$-$ }              \\
										       & 			 (\checkmark dilaton) 	               &                                              	 &                                		   &                       		 &                    &               &  loophole &                                  &                 &                  &                          \\
 \multirow{2}{*}{\ref{sec:sig:Kpill} $K\to\pi\ell_\alpha\ell_\alpha$}         &   \multirow{2}{*}{\checkmark  prompt}            & \multirow{2}{*}{\checkmark prompt}               & \multirow{2}{*}{$-$}                   &            \multirow{2}{*}{\checkmark}                    &        \multirow{2}{*}{$-$}                &           \multirow{2}{*}{$-$}                        & lifetime &                                  &      \multirow{2}{*}{$-$}           &      \multirow{2}{*}{$-$}            &        \multirow{2}{*}{$-$}                  \\
										            &                  &					               & 								                   &                                &                        &                                   & loophole &                                  &                 &                  &                          \\
 \multirow{2}{*}{\ref{sec:sig:Kpipiee} $K\to\pi \pi \ell_\alpha\ell_\alpha$}        &      \multirow{2}{*}{CP viol.}            & axial coupl.  &               \multirow{2}{*}{$-$}                                   &            \multirow{2}{*}{\checkmark}                     &          \multirow{2}{*}{$-$}               &         \multirow{2}{*}{$-$}                           &     \multirow{2}{*}{$-$}               &                 \multirow{2}{*}{$-$}                  &       \multirow{2}{*}{$-$}           &        \multirow{2}{*}{$-$}           &                \multirow{2}{*}{$-$}           \\
										         &                  &  \& prompt &                                                 &                                &                        &                                   &                   &                                  &                 &                  &                          \\
\multirow{3}{*}{\ref{sec:sig:5track} $K\to\pi\ell_\alpha\ell_\alpha \ell_\beta \ell_\beta$}       &      \multirow{3}{*}{$-$}            &         \multirow{3}{*}{$-$}               &       \multirow{3}{*}{$-$}               &               \multirow{3}{*}{$-$}                  &           \multirow{3}{*}{$-$}              &             \multirow{3}{*}{$-$}                       &                   & $A'$,        &         \multirow{3}{*}{$-$}         &          \multirow{3}{*}{$-$}         &          \multirow{3}{*}{$-$}                 \\
 													      &                  &                       &                                                 &                                &                        &                                   &                   & MeV axion, also         &                 &                  &                          \\
													  &                  &                       &                                                 &                                &                        &                                   &                   & $K\to\pi2\ell_\alpha 2\ell_\beta $inv        &                 &                  &                          \\
\ref{sec:KL:gamma:inv} $K_L\to\gamma$+inv      &     $-$             &          $-$              &                   $-$                               & \checkmark                         &               $-$          &                    $-$                &         $-$           &                 $-$                  &          $-$        &       $-$            &            $-$               \\
\ref{sec:K:pi:gamma} $K\to \pi \gamma, 3\gamma$      &     $-$             &          $-$              &                   $-$                               & $-$                        &               $-$          &                    $-$                &         $-$           &                 $-$                  &          $-$        &       Lorentz viol.            &            $-$               \\
\ref{sec:KL:2gamma:inv} $K_L\to\gamma\gamma$+inv      &        $-$           &          $-$              &                          $-$                        &               $-$                  &        $-$                 &                 $-$                   & \checkmark (Table \ref{tab:GNV})           &           $-$                        &      $-$            &  $-$                  &     $-$                      \\
\ref{sec:KL2l:inv} $K_{S,L}\to\ell^+\ell^-$+inv  &        $-$          &            $-$           &                  $-$                               &           $-$                     &             $-$           &             $-$                      & possible             & possible                            &       $-$           &          $-$         & $K_S\to \mu \mu$  \\
\ref{sec:KL2l:inv} $K_{S,L}\to2 \ell 2 \gamma$      &     $-$              & $-$             &     $-$                                             &              $-$                   &       $-$                  &                 $-$                   & possible             & possible                            &      $-$            &         $-$          &            $-$               \\
\ref{sec:K0:4l} $K^0 \to4 \ell $     			 &        $-$           &     $-$         &         $-$                                         &                 $-$                &           $-$              &                   $-$                 &     possible        & possible                            &         $-$         &             $-$      &                 $-$          \\
\ref{sec:sig:Kl} $K^+\to\ell^+$+inv    &     $-$              &            $-$            & \checkmark                                         &          $-$                       & \checkmark ($X\to$inv)    &  \checkmark                                 &          $-$         &                   $-$               &        $-$         &        $-$         &            $-$              \\
\ref{sec:3l:inv} $K^+\to 3\ell$+inv &        $-$          &       $-$                &                               possible                 &           $-$                       & \checkmark ($X\to \ell\ell$)     &                 $-$                    &       $-$              & $U(1)$+HNL                         &        $-$         &        $-$          &        $-$                  \\
 \multirow{2}{*}{\ref{sec:Klgg+inv} $K^+\to \ell\gamma\gamma$+inv }    &      \multirow{2}{*}{$-$}             &         \multirow{2}{*}{$-$}              & $K^+\to \pi^0 \ell^+ N$  &               \multirow{2}{*}{$-$}                 & possible      &            \multirow{2}{*}{possible}                       &     \multirow{2}{*}{$-$}              &         \multirow{2}{*}{possible}                           &      \multirow{2}{*}{$-$}            &    \multirow{2}{*}{$-$}               &      \multirow{2}{*}{$-$}                     \\
 										         &                  &                       & $(m_N \lesssim 20\text{\,MeV})$ &                             &  	($X\to 2\gamma$)					    &                                   &                   &                                  &                 &                  &                          \\
\ref{sec:exp:lfv} LFV          &      $-$             &        $-$                &               $-$                                   &                $-$                 &           $-$              &                 $-$                   &         $-$           &                  $-$                 &       $-$           & FV ALP, $Z'$               & FV ALP                        \\
\multirow{2}{*}{\ref{sec:exp:lnv} LNV}          &       \multirow{2}{*}{$-$}           &      \multirow{2}{*}{$-$}                 & \checkmark ($K^+\to \ell^+ N$,    &            \multirow{2}{*}{$-$}                    &             \multirow{2}{*}{$-$}           &             \multirow{2}{*}{$-$}                      &         \multirow{2}{*}{$-$}          &     \multirow{2}{*}{$-$}                             &       \multirow{2}{*}{$-$}          &  \multirow{2}{*}{$-$}                &            \multirow{2}{*}{\checkmark (Maj. HNL)}                \\
					         &                  &                       & $N\to \pi^- \ell^+$ )  &                                &                        &                                   &                   &                                  &                 & 				                  &                          \\
 \multirow{2}{*}{\ref{sec:exp:ks} Rare $K_S$ decays}      	&      \multirow{2}{*}{  $K_S\to \pi (\pi) 2\ell$  }         &             $K_S\to \pi (\pi) 2\ell,$             &      \multirow{2}{*}{  $-$  }                                        & $K_S\to A'\gamma$                       	 &         \multirow{2}{*}{  $-$  }               &            \multirow{2}{*}{  $-$  }                       &           \multirow{2}{*}{  $-$  }            &  \multirow{2}{*}{$K_S\to 4\ell$}  &      \multirow{2}{*}{  $-$  }               &         \multirow{2}{*}{  $K_S\to 2\gamma+$inv  }         &  \multirow{2}{*}{ $K_S\to \mu \mu$} \\
                  					  &                  &       $\to \pi (\pi) 2\gamma  $              &                                                 &    $\to A'\gamma\pi$      &                        &                                   &                   &                                   &                 &                  &                          \\				         
\ref{sec:exp:darkjet} Dark Shower  &         $-$         &         $-$               &                 $-$                                 &              $-$                   &              $-$           &               $-$                     &           $-$         &                    $-$               &       $-$           &  \checkmark                &             $-$              \\
 \multirow{2}{*}{\ref{sec:hyppheno} Hyperon}         &       \multirow{2}{*}{  $B_1\to B_2 \varphi$}          &    Table \ref{tab:hyp2D}                   &            \multirow{2}{*}{  $-$}                                      & Table \ref{tab:Kaon decay} &                        &            \multirow{2}{*}{  $-$}                         &        \multirow{2}{*}{  $-$}             &                         \multirow{2}{*}{  $-$}         		  & Table \ref{tab:hyperon:dark:sector:decays}        &        \multirow{2}{*}{  $-$}            &       \multirow{2}{*}{  $-$}                     \\
					        &                  &   $B_1\to B_2 a$                    &                                                 & $B_1\to B_2 A'$ &                        &                                   &                   &                                 &      $B\to \gamma/M$+inv    &                  &                          \\
\hline\hline
\end{tabular}
}

\caption{Summary of different models that can lead to new physics signatures in rare kaon and hyperon decays.}
\label{tab:summary}
\end{sidewaystable}

Rare kaon decays are among the most sensitive probes of both heavy and light new physics, as a combined consequence of both experimental and theoretical considerations. On one hand, large datasets are available in the current and future kaon experiments, two to three orders of magnitude larger than the usable $B$ and $D$ meson datasets at Belle, BaBar and LHCb. At the same time the kaon decay width is power suppressed, $\Gamma_K\propto m_K^5/m_W^4$, compared to the $B$ and $D$ mesons, $\Gamma_{B,D}\propto m_{B,D}^5/m_W^4$. For example, the total $K^+$ decay width is almost four orders of magnitude smaller than for the $B^+$ meson, $\Gamma_{B^+}/\Gamma_{K^+}\simeq 7.5 \times 10^3$. 

The small total decay width enhances the sensitivity to NP decay channels, ${\mathcal B}(K\to X_{\rm NP})={\Gamma}(K\to X_{\rm NP})/\Gamma_K$, so that the sensitivity to the same branching ratio value in reality corresponds to a sensitivity to a smaller NP partial decay width for rare kaon decays. This is particularly important for searches for kaon decays to light new physics particles. In models where the NP particles couple to the SM via renormalizable interactions, and thus dimensionless couplings such as the Higgs-scalar mixing angle, $\theta$, the NP decay width is ${\Gamma}(K\to\pi \varphi)\propto \theta^2 m_K$ and  thus ${\mathcal B}(K\to \pi \varphi) \propto \theta^2 (m_W/m_K)^4$. This is to be compared with heavy meson decays: ${\mathcal B}(B\to K \varphi) \propto \theta^2 (m_W/m_B)^4$. For light new physics that couples to the SM through dimension-5 operators, such as ALPs, the scaling changes to ${\mathcal B}(K\to \pi a) \propto  (m_W^2 /f_a m_K)^2$, to be compared to ${\mathcal B}(B\to K a) \propto (m_W^2 /f_a m_B)^2$. Rare kaon decays therefore represent the most sensitive probes for light new physics (assuming no particular flavor structure for the light new physics couplings), as long as the decays to light new physics are kinematically allowed. 

In this manuscript we provided a comprehensive overview of the new physics models that can be probed by the rare kaon decays, as well as a model-independent discussion of possible new physics signatures. Since kaons are relatively light there is only a limited set of final state particles to be considered: only decays involving photons, electrons/positrons, muons, pions, and missing energy are possible. These final state building blocks can still result in relatively complex signatures, such as lepton jets, with many electron positron pairs and missing energy (Section~\ref{sec:LPV}). Barring such rather exotic possibilities, the list of possible signatures is rather manageable and is collected in Table~\ref{tab:summary}. 

Furthermore, for a number of channels and/or models the improved sensitivities in the next generation of kaon experiments could lead to a qualitative leap in the phenomenological implications. For instance, with two to three orders of magnitude larger datasets one could close the gap for Higgs-mixed scalar all the way to the BBN floor (Fig.~\ref{fig:higgs_portal_summary}). The searches for promptly decaying ALPs in $K\to\pi a$ decays, in either the $a\to\gamma\gamma$ or the $a\to e^+e^-$ channel, could be improved such that the gap to constraints from beam dump searches would be closed (Figs.~\ref{fig:ALPvis}, \ref{fig:ALPtoeetarget}). This would either lead to a discovery of the ALP, or leave $K\to\pi a_{\rm inv}$, where $a_{\rm inv}$ is and ALP that escapes the detector, as the only remaining phenomenologically viable possibility. An improvement in sensitivity to ${\cal B}(K^+\to\ell^+N)$ by two orders of magnitude would start probing the minimal (vanilla) seesaw neutrino mass models for sterile neutrino masses in the ${\mathcal O}(100~{\rm MeV})$ regime (Fig.~\ref{HNLLims}). An order of magnitude improvement on ${\cal B}(K^+\to\mu^+\nu X_{\rm inv})$ would probe fully the preferred region for self-interacting neutrinos that may alleviate the Hubble tension (Fig.~\ref{fig:Hubble:tension:neutrinos}). The $K_L$ decays are in general less sensitive to the above models, since they are experimentally more challenging. In contrast, $K_L\to\pi^0\nu\bar\nu$ is theoretically the cleanest and would thus provide the highest sensitivity to heavy NP. Furthermore, $K_L$ decays can also probe a number of Grossman-Nir violating models without being excluded by charged kaon decays (Section \ref{sec:GNVgamma}).

The experimental limits on rare hyperon decays are less stringent. Even so, the measurements of rare hyperon decays can well lead to a new physics discovery. This is possible because such decays probe different couplings than the $K\to \pi$ transitions. For instance, $B_1\to B_2 X_{\rm inv}$ would probe CP violating couplings in Higgs mixed scalar model, and axial couplings for ALPs. The other interesting possibility is offered by searches for decays that are otherwise not kinematically allowed, such as the $B\to \gamma/M X_{\rm inv}$ decays in the dark baryon models (Section~\ref{sub:dark:baryon:models}). 

In short, measurements of rare kaon and hyperon decay branching ratios, including dedicated studies of the differential distributions, represent crucial probes of light dark sectors. In the present manuscript we provided the status of current searches in the different channels, and identified new search strategies for under-explored signatures. The impact of current experiments on the parameter space of various light dark sector scenarios was summarized, and future projections presented.

\subsubsection*{Acknowledgements}

We are grateful to Brigitte Bloch-Devaux for sharing tabulated NA48/2 results on rare kaon decay spectra, and for the discussion of the analysis.
BD acknowledges support through ERC-2018-StG-802836 (AxScale project).
The work of BS is supported by the US NSF under Grant PHY-1820770 and by the Research Corporation for Science Advancement through a Cottrell Scholar Award.
DSMA acknowledges support from Los Alamos National Laboratory's Science Program Office, and from the DOE Office of Science High Energy Physics under contract number DE-AC52-06NA25396.
DW is supported by Joint Large-Scale Scientific Facility Funds of the NSFC and CAS under Contract No.~U1832207, and National Key Research and Development Program of China under Contract No.~2020YFA0406400. 
EG acknowledges the IPPP Associateship which has enabled fruitful collaboration leading to this work, and support from the STFC (United Kingdom).
GE is supported by the Cluster of Excellence Precision Physics, Fundamental Interactions and Structure of Matter (PRISMA${}^+$ -- EXC~2118/1) within the German Excellence Strategy (project ID 39083149).
JB is supported in part by DOE grant de-sc0011784.
JMC acknowledges support from the Spanish MINECO through the ``Flavor in the era of the LHC'' grant PGC2018-102016-A-I00 and also from the ``Ram\'on y Cajal'' program RYC-2016-20672.
The work of JTC is supported by the Ministerio de Ciencia e Innovaci\'on under FPI contract PRE2019-089992 of the SEV-2015-0548 grant.
JZ acknowledges support in part by the DOE grant de-sc0011784 and NSF OAC-2103889.
KT is supported in part by the US Department of Energy grant DE-SC0010102 and by JSPS KAKENHI 21H01086.
The work of ME is supported by a Fellowship of the Alexander von Humboldt Foundation.
MT acknowledges the financial support from the Slovenian Research Agency (research core funding No. P1-0035).
RXS acknowledges support from the National Natural Science Foundation of China under Grants No.12147145 and Project funded by China Postdoctoral Science Foundation No.~2021M700343.
RZ acknowledges support from the European Union's Horizon 2020 research and innovation programme under the Marie Sk\l odowska-Curie grant agreement No 860881-HIDDeN
SH is supported in part by the DOE Grant DE-SC0013607, and in part by the Alfred P.~Sloan Foundation Grant No.~G-2019-12504.
SS is supported by a Stephen Hawking Fellowship from UKRI under reference EP/T01623X/1 and the Lancaster-Manchester-Sheffield Consortium for Fundamental Physics, under STFC research grant ST/T001038/1.
The work of TK is supported by the Japan Society for the Promotion of Science (JSPS) Grant-in-Aid for Early-Career Scientists (Grant No.~19K14706) and the JSPS Core-to-Core Program (Grant No.~JPJSCCA20200002). 
VVG acknowledges support from the European Research Council through grant 724777 RECEPT.



\bibliographystyle{report}
\bibliography{\bibfiles}
\end{document}